# Z(3) Interfaces in Lattice Gauge Theory

STEPHEN T. WEST
St. John's College

*Theoretical Physics*
*Department of Physics*
*Oxford University*

Oxford University

# Z(3) Interfaces in Lattice Gauge Theory

Stephen T. West
St. John's College

*Theoretical Physics*
*Department of Physics*
*Oxford University*

Oxford University

ABSTRACT

We study various properties of the $Z(3)$ interface which forms between the different ordered phases of pure $SU(3)$ gauge theory above a critical temperature. We simulate the theory on a computational lattice of two space and one time dimensions, at various temperatures above this critical point. At high temperatures, we perform an accurate measurement of the interface tension, which shows good agreement with the prediction of perturbation theory. Near the critical temperature, we show that it is possible to observe the behaviour of the interface amidst the phase turbulence near its collapse, and we characterise its fluctuations in terms of various displacement moments. We show that the interface behaves like a fluctuating string, and model its behaviour in terms of an interacting scalar field theory. We also examine other properties of the interface, including its intrinsic and Debye electric screening widths, its formation and movement on the lattice, and its wetting.

*Finally Cutangle spoke, very slowly and carefully.*

*'I look at it like this,' he said. 'Before I heard him talk, I was like everyone else. You know what I mean? I was confused and uncertain about all the little details of life. But now,' he brightened up, 'while I'm still confused and uncertain it's on a much higher plane, d'you see, and at least I know I'm bewildered about the really fundamental and important facts of the universe.'*

*Treatle nodded. 'I hadn't looked at it like that,' he said, 'but you're absolutely right. He's really pushed back the boundaries of ignorance. There's so much about the universe we don't know.'*

*They both savoured the strange warm glow of being much more ignorant than ordinary people, who were ignorant of only ordinary things.*

Terry Pratchett — "Equal Rites" (1987)

# To my parents

# Acknowledgements

I should like to thank my supervisor, John Wheater, for his guidance and support during my time as a graduate student. I am also grateful to Mike Teper for the chance to examine his SU(2) simulation code, and to both him and Amanda Michels for profitable discussions; also, to James Carpenter for introducing me to modern statistical methods.

I should also like express my thanks to the P.P.A.R.C. for supporting my research, to the denizens of room 1.12 and many others in Theoretical Physics for making it a pleasant experience, and to my parents for keeping me sane.

A short summary of this work is about to appear in the paper "Properties of the $Z(3)$ Interface in (2+1)-D $SU(3)$ Gauge Theory", by S.T. West and J.F. Wheater, to be published in the "Proceedings of Lattice '95", *Nucl. Phys.* **Proc. Suppl. B** (1996) .

# Contents

# Figure Captions

NESCIRE QUAEDAM MAGNA PARS SAPIENTAE EST

Grotius

*For in much wisdom is much grief:*
*and he that increaseth knowledge increaseth sorrow.*

Ecclesiastes 1.18

IMPOSSIBILIUM NULLA OBLIGATIO EST

Celsus

# Chapter 1.

# Phases in SU(3) Gauge Theory: Theoretical Preliminaries

## 1.1. Prologue: Phases in Physics

MANY PHYSICAL systems possess assorted phases, undergoing transitions between the different phases under certain conditions, for instance at a particular temperature. Obvious examples include $H_2O$, with phases as ice, water and steam and transitions at $0^oC$ and $100^oC$, and ferromagnets, with ferro- and paramagnetic phases. Phase transitions in systems such as these have been studied extensively in the past. Where different phases coexist, interfaces can form between them, and the interfaces have also been objects of previous study[1].

Quantum chromodynamics (QCD) is now generally accepted as the theory responsible for the "strong force" of particle physics. The gluon sector of this theory is described by the $SU(3)$ gauge theory. This gauge theory itself exhibits several phases at finite temperatures, as will be detailed in the following sections. Interfaces can form between the different phases, and this thesis is concerned with the study of several properties of one particular type of interface, which may be of relevance to cosmology and to the heavy ion collision experiments of nuclear physicists at the Relativistic Heavy Ion Collider.

This chapter will begin with some fundamentals of field theories at finite temperature, in particular $SU(3)$ gauge theory. It will then continue with a discussion of the types of interface present in the theory, the mathematical framework for their description, and some of the predicted properties of one of the types. Subsequent chapters will proceed to consider this one particular type of interface, first at high temperatures, where it is most stable, and then at lower temperatures, where it is close to collapse.

## 1.2. Introduction: Phases in SU(3) Theory

IT has been known for some time that pure gauge theories have a non-confining high-temperature phase[2]. This was first shown non-perturbatively, in the strong-coupling limit of Hamiltonian lattice gauge theory[3]. In this situation, for instance, Abelian gauge theory can be shown to reduce to the "Villain model". This is a well studied statistical-mechanical system which is known to be disordered at high "Villain temperature", with a pair of opposite charges experiencing a confining linear potential; and ordered at low temperature, with a normal Coulomb potential. The Villain temperature is actually inversely proportional to the temperature, $T$, defined in gauge theory, which is used from now on. Therefore, any theory in a confining phase at $T = 0$ must have a phase transition at some finite temperature, called the critical temperature ($T_c$), separating the confining and non-confining phases. These ideas are supported by numerical simulations[4].

A consideration of QCD at finite temperatures reveals that the pure gauge $SU(3)$ sector has the two expected phases. At low temperatures, there is the "disordered" phase where the colour charge of QCD is confined, and the vacuum is symmetric under the group $Z(3)$, the "centre" of the $SU(3)$ group (as discussed further in section 1.7). This is why no free quarks are seen in the relatively cool universe of today. At high temperatures, though, one finds a non-confining, $Z(3)$-breaking "ordered" phase, corresponding to the free quark-gluon "plasma" believed to exist in the hot early universe. In fact, owing to the $Z(3)$ breaking, there exist three different high-temperature (high-$T$) phases, namely three degenerate vacua, corresponding to the three members of $Z(3)$, the cube roots of unity. Fermionic matter breaks the vacuum degeneracy, albeit on a small scale, so we consider only the pure gauge theory in what follows.

It is apparent that two different types of interface are possible in the theory. First, there is one between the ordered and disordered phases; this type is only stable at the critical temperature, since only then do the temperature and pressure match on both of its sides. Second, an interface can form between two of the ordered phases with different $Z(3)$ vacua, and this is usually called an "order-order interface", or "$Z(3)$ interface". All thermodynamic quantities are the same on both sides of a $Z(3)$ interface, but they dip slightly across the object itself. Clearly, this second type of interface can only exist above the critical temperature, as only there does one find the ordered phases themselves.

The aim of this thesis is to study some properties of the $Z(3)$ interfaces, by means of Monte-Carlo simulations on a Euclidean lattice, in two very different temperature regimes: first, in the high-$T$ limit, where analytic predictions have been made concerning properties such as the surface tension and Debye screening mass of the interface; and second, just above the critical temperature, where the interfaces are in the process of collapse.

## 1.3. Cosmological Implications

POSSIBLE cosmological significance has been claimed for these interfaces in the past, in the rôle of domain walls in the early universe, which could be responsible for matter formation[5][6]. The suggested process is that of bubble nucleation, forming hadronic matter during the phase transition from the high-$T$ quark-gluon plasma to the low-$T$ hadronic phase. It has been believed for some time[7], and has been shown in simulations[8][9], that the interfaces of $SU(3)$ show something called "complete wetting". The meaning of this is that a $Z(3)$ interface between two ordered phases really consists of two order-disorder interfaces stuck back to back, leaving a very thin slice of disorder between two ordered domains. One possible scenario for the formation of hadronic matter in the early universe envisages many different domains of ordered phase at high temperature, separated by $Z(3)$ interfaces. As the temperature drops towards the critical value, the interfaces start to widen; and when the temperature hits the critical value, the constituent order-disorder interfaces split apart, spreading the slice of disordered hadronic phase outwards.

However, it has more recently been emphasised[10][11][12] that the order parameter distinguishing the $SU(3)$ phases is fundamentally a Euclidean object, with no counterpart in Minkowski space. Thus, whilst the $Z(3)$ symmetry exists, and hence $Z(3)$ domains and interfaces exist, in the Euclidean path integral, it is less certain that the interfaces can exist as physical objects in the universe. The claim has even been made[13] that only one true physical phase exists, even in Euclidean space, at high temperatures, and we shall address this when discussing our results later. In Minkowski space, the interfaces have been shown[11], on current understanding, to have unphysical thermodynamic quantities when a certain number of fermion families is considered in addition to the gauge fields, resulting from the unusual fact that the ordered phases occur at *high* temperatures. The free energy, $F$, is proportional to the fourth power of the temperature with a *positive* coefficient for many $SU(N)$ systems with varying numbers of fermion families present: $F = |\gamma|T^4$. Thus, the pressure, $P = -|\gamma|T^4$; the internal energy, $E = -3|\gamma|T^4$; the specific heat, $C = -12|\gamma|T^3$; and the entropy, $S = -4|\gamma|T^3$.

As pointed out in [12], no physical systems of this type can exist with positive temperature, but metastable states can exist with inverse population, *i.e.* *negative* temperature, flipping the sign of the specific heat and entropy as desired, and forming a type of quark-gluon "laser". The problem may be assuaged by embedding the interface within a system whose positive entropy exceeds the negative entropy of the interface itself, and such a system has recently been postulated[14], with the interface contained within a larger volume of metastable vacuum in the early universe. However, it is noted in [12] that any statistical description of a system containing even a subsystem of negative entropy must fail, since no states are available to the subsystem. For instance, the standard electroweak model well above the QCD phase transition has negative free energy density[5][6], thanks to the contributions of the leptons, Higgs and weak gauge bosons; but if we consider the case where the quark-gluon subsystem has $F_{qg} = |\gamma| T^4$, then the weak coupling of the other particles cannot rectify the situation. Even setting this problem aside, the problem of interpretation of the interface remains, as the order parameter is non-local in Euclidean time, corresponding to an imaginary time-like gauge field in Minkowski space, and thus an imaginary chemical potential for the colour charge[15]. *Adhuc sub judice lis est...* In any event, by virtue of their contribution to the partition function and to expectation values calculated using the Euclidean path integral, the interfaces must be included in a non-perturbative analysis of the thermodynamics of QCD, if only (like the instanton) as a calculational device without physical form in our universe.

In order to make analytical predictions about the interfaces, a mathematical description of them is necessary. For this reason, the next sections introduce some elements of quantum field theory at finite temperatures which will be of use in future chapters.

## 1.4. The Partition Function and Thermodynamics

ONE now introduces a Schrödinger field operator at time $t$, $\hat{\phi}(\mathbf{x}, t)$, and its conjugate momentum operator, $\hat{\pi}(\mathbf{x}, t)$[16]. For convenience, one labels the eigenstates of the field operator at $t = 0$ as $|\phi >$, satisfying

$$\hat{\phi}(\mathbf{x}, 0)|\phi >= \phi(\mathbf{x})|\phi >,$$

with eigenvalue $\phi(\mathbf{x})$. The amplitude for a transition from initial state $|\phi_i >$ at $t = t_i$ to final state $|\phi_f >$ at $t = t_f$ is then given, as usual, by the following functional integral in $(d-1)$ space and one time ($d$ or "$(d-1)+1$") dimensions:

$$< \phi_f | e^{-iH(t_f - t_i)} | \phi_i >= \int \mathcal{D}\pi \mathcal{D}\phi \; e^{i \int_{t_i}^{t_f} dt \int d^{d-1}x (\pi \dot{\phi} - \mathcal{H}(\pi, \phi))}. \tag{1.1}$$

The integral over $\pi$ is without restriction, but the integral over $\phi$ is constrained by the requirements that $\phi(t = t_i) = \phi_i$ and $\phi(t = t_f) = \phi_f$. $H$ is the Hamiltonian, and $\mathcal{H}$ the Hamiltonian density.

Now, the partition function of a statistical mechanical system in thermodynamic equilibrium is given by

$$Z = \text{Tr } e^{-\beta_T H} = \int d\phi < \phi | e^{-\beta_T H} | \phi >, \tag{1.2}$$

where $\beta_T$ is the inverse temperature, $T^{-1}$, and the integral sums over all states. From $Z$, all thermodynamic properties can be calculated:

$$F = -\frac{1}{\beta_T}\ln Z \qquad P = -\frac{\partial F}{\partial V}\bigg|_T \qquad S = -\frac{\partial F}{\partial T}\bigg|_V \qquad etc... \tag{1.3}$$

Also, the expectation value of any operator is simply given by:

$$< \hat{\mathcal{O}} >= \frac{1}{Z}\mathrm{Tr}\big(\hat{\mathcal{O}}e^{-\beta_T H}\big). \tag{1.4}$$

We now generalise (1.1) and (1.2) to the logical conclusion of a path integral over gauge fields, before turning to consider how to distinguish between the different phases of a gauge theory at finite temperatures.

## 1.5. Finite-Temperature Field Theory

One can make (1.1) look more suggestive for $Z$ by introducing an imaginary time coordinate $\tau = it$ and putting $t_i = 0$ and $it_f = \beta_T$, giving

$$< \phi_f|e^{-\beta_T H}|\phi_i >= \int \mathcal{D}\pi\mathcal{D}\phi\; e^{\int_0^{\beta_T} d\tau \int d^{d-1}x(i\pi\dot{\phi}-\mathcal{H}(\pi,\phi))},$$

where previous functions of $t$ are now functions of $\tau$, and $\dot{\phi} \equiv \frac{\partial\phi}{\partial\tau}$. This enables one to write

$$Z = \int \mathcal{D}\pi \int_{\text{periodic}} \mathcal{D}\phi\; e^{\int_0^{\beta_T} d\tau \int d^{d-1}x(i\pi\dot{\phi}-\mathcal{H}(\pi,\phi))},$$

where the periodic boundary conditions on $\phi$ are that $\phi(\mathbf{x}, 0) = \phi(\mathbf{x}, \beta_T)$, arising from the trace in (1.2). This periodicity restricts the energy values associated with the fields to a set of discrete values.

For a general renormalisable scalar field Hamiltonian,

$$\mathcal{H} = \pi^2 - \mathcal{L}(\phi, \partial_\mu\phi),$$

with $\mathcal{L}$ being the Lagrangian. One can perform the integration over $\pi$ to give

$$Z = \int_{\text{periodic}} \mathcal{D}\phi\; e^{\int_0^{\beta_T} d\tau \int d^{d-1}x\mathcal{L}(\phi,\partial_\mu\phi)},$$

where the (irrelevant) normalisation constant has been discarded, and the Lagrangian is Euclidean.

This technique can be extended straightforwardly to gauge fields. When applied to quantum electrodynamics (QED), for instance, one obtains the following functional integral for the electromagnetic field alone, given for an arbitrary gauge specified by $\mathcal{F} = 0$, where $\mathcal{F}$ is a general function of the vector potential, $A^\mu$, and its derivatives:

$$Z = \int_{\text{periodic}} \mathcal{D}A^\mu \delta(\mathcal{F}) \det\left(\frac{\partial \mathcal{F}}{\partial \alpha}\right) e^{\int_0^{\beta_T} d\tau \int d^{d-1}x(-\frac{1}{4}F_{\mu\nu}F^{\mu\nu})}.$$

The electromagnetic field tensor $F^{\mu\nu} = \partial^\mu A^\nu - \partial^\nu A^\mu$, no colour indices being contained within $A^\mu$ for QED, and $\alpha(\mathbf{x}, t)$ specifies possible gauge transformations.

For a general gauge group $\mathcal{G}$, with generators $T^a$ and structure constants $f_{abc}$, the usual vector potential acquires a colour index, $A^\mu_a$, and thence the field strength becomes

$$F^{\mu\nu}_a = \partial^\mu A^\nu_a - \partial^\nu A^\mu_a - g f_{abc} A^\mu_b A^\nu_c,$$

introducing dimensionless coupling $g$. For a general gauge group $SU(N)$, $a$ runs from 1 to $N^2 - 1$. In a convenient shorthand, we shall later write the product over colour indices as $A^\mu = \mathbf{A}^\mu \cdot \mathbf{T}$. The infinitesimal gauge transformations are now given by $\alpha_a(\mathbf{x}, t)$, and there are $N^2 - 1$ gauge-fixing functions specified by $\mathcal{F}^a$, so

$$Z = \int_{\text{periodic}} \mathcal{D}A^\mu_a \delta(\mathcal{F}^b) \det\left(\frac{\partial \mathcal{F}^c}{\partial \alpha^d}\right) e^{\int_0^{\beta_T} d\tau \int d^{d-1}x(-\frac{1}{4}F^a_{\mu\nu}F^{\mu\nu}_a)}. \tag{1.5}$$

To work in a physical gauge, one can simply choose $\delta(\mathcal{F}^b)$ to pick such a gauge within the generating functional above. However, it is often more convenient to work in an unphysical gauge such as the covariant gauge, where ghost fields $\eta_a, \overline{\eta}_a$ are introduced to give

$$Z = \int_{\text{periodic}} \mathcal{D}A^\mu_a \mathcal{D}\overline{\eta}_a \mathcal{D}\eta_a \; e^{\int_0^{\beta_T} d\tau \int d^{d-1}x(-\frac{1}{4}F^a_{\mu\nu}F^{\mu\nu}_a - \frac{1}{2\xi}(\partial_\mu A^\mu_a)^2 + g f^{abc}\overline{\eta}_a \partial_\mu A^\mu_b \eta_c + \partial_\mu \overline{\eta}_a \partial^\mu \eta_a)}.$$

## 1.6. The Polyakov Line and Confinement

The phases of a thermal gauge theory are characterised by the free energies of static configurations of quarks and antiquarks[17]. To calculate such free energies, one needs to consider creation and annihilation operators, $\psi^\dagger_a(\mathbf{x}_i, t)$ and $\psi_a(\mathbf{x}_i, t)$, for static quarks with colour $a$ at position $\mathbf{x}_i$ and time $t$, together with charge conjugates, $\psi^{\dagger c}_a$ and $\psi^c_a$, for antiquarks. All equal-time anticommutators of these fields vanish with the exception of

$$\{\psi_a(\mathbf{x}_i, t), \psi^\dagger_b(\mathbf{x}_j, t)\} = \delta_{ij}\delta_{ab} \tag{1.6}$$

and its charge conjugate.

For the general gauge group $\mathcal{G}$ of the previous section, using the shorthand defined there for $A^\mu$, the quark fields obey the static time-evolution equation

$$\left(-i\frac{\partial}{\partial \tau} - gA^0(\mathbf{x}_i, \tau)\right)\psi(\mathbf{x}_i, \tau) = 0,$$

whence integration yields

$$\psi(\mathbf{x}_i,\tau) = \mathcal{T}\big(e^{ig\int_0^\tau d\tau' A^0(\mathbf{x}_i,\tau')}\big)\psi(\mathbf{x}_i,0). \tag{1.7}$$

Consider states $|\Psi>$ containing heavy quarks at $\mathbf{x}_1,\ldots,\mathbf{x}_{N_q}$ and antiquarks at $\mathbf{x}'_1,\ldots,\mathbf{x}'_{N_{\overline{q}}}$. The free energy of a collection of $N_q$ quarks and $N_{\overline{q}}$ antiquarks is given by a sum over all such states, from (1.2) and (1.3):

$$e^{-\beta_T F(\mathbf{x}_1,\ldots,\mathbf{x}_{N_q},\mathbf{x}'_1,\ldots,\mathbf{x}'_{N_{\overline{q}}})} \sim \sum_{|\Psi>} <\Psi|e^{-\beta_T H}|\Psi> . \tag{1.8}$$

Using the quark field operators, one can rewrite this expression in terms of new states $|\widetilde{\Psi}>$ without heavy quarks:

$$\begin{aligned} e^{-\beta_T F_{N_q N_{\overline{q}}}} \sim \sum_{|\widetilde{\Psi}>} \Big\langle \widetilde{\Psi} \,\Big|\, &\sum_{\{a,b\}} \Psi_{a_1}(\mathbf{x}_1,0)\ldots\psi_{a_{N_q}}(\mathbf{x}_{N_q},0)\psi^c_{b_1}(\mathbf{x}'_1,0)\ldots\psi^c_{b_{N_{\overline{q}}}}(\mathbf{x}'_{N_{\overline{q}}},0) \\ &\times e^{-\beta_T H}\psi^\dagger_{a_1}(\mathbf{x}_1,0)\ldots\psi^\dagger_{a_{N_q}}(\mathbf{x}_{N_q},0)\psi^{\dagger c}_{b_1}(\mathbf{x}'_1,0)\ldots\psi^{\dagger c}_{b_{N_{\overline{q}}}}(\mathbf{x}'_{N_{\overline{q}}},0) \,\Big|\, \widetilde{\Psi}\Big\rangle. \end{aligned} \tag{1.9}$$

Remembering that $e^{-\beta_T H}$ will evolve an operator $\hat{O}(\tau)$ in Euclidean time ($\hat{O}(\tau+\beta_T) = e^{\beta_T H}\hat{O}(\tau)e^{-\beta_T H}$), one now introduces pairs of exponentials $e^{-\beta_T H}e^{\beta_T H}$ to the left of each destruction operator in (1.9). Evolving these operators to time $\beta_T$ and using the anticommutation properties to group operators into pairs at the same location, one obtains

$$\begin{aligned} e^{-\beta_T F_{N_q N_{\overline{q}}}} \sim \sum_{|\widetilde{\Psi}>} \Big\langle \widetilde{\Psi} \,\Big|\, &\sum_{\{a,b\}} e^{-\beta_T H}\psi_{a_1}(\mathbf{x}_1,\beta_T)\psi^\dagger_{a_1}(\mathbf{x}_1,0)\ldots\psi_{a_{N_q}}(\mathbf{x}_{N_q},\beta_T)\psi^\dagger_{a_{N_q}}(\mathbf{x}_{N_q},0) \\ &\times\psi^c_{b_1}(\mathbf{x}'_1,\beta_T)\psi^{\dagger c}_{b_1}(\mathbf{x}'_1,0)\ldots\psi^c_{b_{N_{\overline{q}}}}(\mathbf{x}'_{N_{\overline{q}}},\beta_T)\psi^{\dagger c}_{b_{N_{\overline{q}}}}(\mathbf{x}'_{N_{\overline{q}}},0) \,\Big|\, \widetilde{\Psi}\Big\rangle. \end{aligned} \tag{1.10}$$

Using (1.6) and (1.7), one can finally write

$$e^{-\beta_T F_{N_q N_{\overline{q}}}} = \mathrm{Tr}\big(e^{-\beta_T H}L(\mathbf{x}^q_1)\ldots L(\mathbf{x}^q_{N_q})L^\dagger(\mathbf{x}^{\overline{q}}_1)\ldots L^\dagger(\mathbf{x}^{\overline{q}}_{N_{\overline{q}}})\big), \tag{1.11}$$

where the trace is taken over states of the pure gluon theory, and $L(\mathbf{x})$ is the "Polyakov line", sometimes called the "Wilson line" ($\mathcal{T}$ denotes time ordering):

$$L(\mathbf{x}) = \mathrm{Tr}\Omega(\mathbf{x}), \qquad \text{with} \qquad \Omega(\mathbf{x}) \sim \mathcal{T}e^{ig\int_0^{\beta_T} d\tau A^0(\mathbf{x}_i,\tau)} \tag{1.12}$$

In particular, the normalisation in (1.8) for gauge group $SU(N)$ is $\frac{1}{N^{N_q+N_{\overline{q}}}}$, and

$$\Omega(\mathbf{x}) = \frac{1}{N}\mathcal{T}e^{ig\int_0^{\beta_T} d\tau A^0(\mathbf{x}_i,\tau)}.$$

Recalling (1.4), one divides (1.11) by the vacuum, *viz.* the equivalent with $N_q = N_{\overline{q}} = 0$, to give

$$e^{-\beta_T \Delta F_{N_q N_{\overline{q}}}} =< L(\mathbf{x}_1^q) \dots L(\mathbf{x}_{N_q}^q) L^\dagger(\mathbf{x}_1^{\overline{q}}) \dots L^\dagger(\mathbf{x}_{N_{\overline{q}}}^{\overline{q}}) >,$$

where $\Delta F = F_{N_q N_{\overline{q}}} - F_{00}$.

In particular, the expectation value of a single Polyakov line gives the self-energy of a single quark in the gluon medium ($F_q \equiv \Delta F_{10}$):

$$e^{-\beta_T F_q} =< L(\mathbf{x}) > .$$

Since these quarks are static and infinitely heavy, $F_q$ is meaningless in itself. However, the correlation function of two Polyakov lines, $\Gamma$, gives the static quark-antiquark free energy ($F_{q\overline{q}} \equiv \Delta F_{11}$):

$$e^{-\beta_T F_{q\overline{q}}(\mathbf{x}-\mathbf{y})} =< L(\mathbf{x}) L^\dagger(\mathbf{y}) >= \Gamma(\mathbf{y}-\mathbf{x}).$$

Static quarks will be confined if this correlation function vanishes as $|\mathbf{x}-\mathbf{y}| \to \infty$, but cluster decomposition demands that

$$\Gamma(\mathbf{x}) =< L(0) L^\dagger(\mathbf{x}) > \underset{|\mathbf{x}|\to\infty}{\longrightarrow} < L(0) >< L^\dagger(\mathbf{x}) >= |< L(0) >|^2$$

so that $F_{q\overline{q}}(\mathbf{x}) \underset{|\mathbf{x}|\to\infty}{\longrightarrow} 2F_q$, and so this condition is equivalent to $< L(\mathbf{x}) >= 0$, showing that the insertion of a single quark requires infinite energy, *i.e.* the theory is in a confining phase. In contrast, if $< L(\mathbf{x}) >\neq 0$ then $F_{q\overline{q}}(\mathbf{x})$ will be finite and the colour charge deconfined. Thus, the Polyakov line gives an effective test for confinement.

## 1.7. The Centre Symmetry and its Breaking

Turning again to the expression for the generating functional in (1.5), one is now led to consider the symmetries of the Euclidean action,

$$\mathcal{S}_E = \int d\tau d^{d-1}x F^a_{\mu\nu} F^{\mu\nu}_a,$$

and the behaviour of the Polyakov line under the symmetry-related transformations.

Returning to the shorthand notation of section 1.5, one sees that under a general non-Abelian gauge transformation

$$A^\mu \to U A^\mu U^{-1} + iU\partial^\mu U^{-1}, \qquad U(\mathbf{x},t) \in SU(N) \tag{1.13}$$

the Polyakov line will transform in the following manner:

$$L(\mathbf{x}) \to \mathrm{Tr}\big(U(\mathbf{x},0)\Omega(\mathbf{x})U^\dagger(\mathbf{x},\beta_T)\big). \tag{1.14}$$

This means that $L$ is invariant if $U$ is periodic in time. However, $\mathcal{S}_E$ is actually invariant under a larger group than merely the periodic gauge transformations; the only physical restraint is that the gauge potentials remain periodic in time. Let $U(\mathbf{x}, t+\beta_T) = zU(\mathbf{x}, t)$ in (1.13); since $A^i$ must be single-valued, this tells one that $z$ must be independent of $\mathbf{x}$, satisfying $z^{-1}A^i z = A^i$. This almost always implies that $z$ is an element of $\mathcal{C}$, the "centre" of the gauge group (the set of elements which commute with all members of the group): $\mathcal{C} = \{z \in G | zgz^{-1} = g \text{ for all } g \in G\}$. This global symmetry is an invariance under gauge transformations which are only periodic up to an element of the centre; it leaves local observables unchanged, and it remains after all gauge fixing. However, (1.14) reveals that the topologically non-trivial Polyakov line, which wraps around the periodic boundary condition in the time direction, is not invariant under these transformations, but rather is rotated by an element of the centre:

$$L(\mathbf{x}) \to zL(\mathbf{x}).$$

For the $SU(N)$ groups, the centre is $Z(N)$, whose elements are the $N$th roots of unity, $e^{2\pi iq/N}$ $(q = 0, 1, \ldots, N-1)$. This $Z(N)$ symmetry may not be dynamically realised, however: if it is impossible to interpolate from one $Z(N)$ vacuum to another through field configurations of finite energy and measure, the symmetry will be spontaneously broken. In the absence of such breaking, $< L(\mathbf{x}) >$ must vanish, corresponding to $F_q = \infty$. This will lead to an exponential decay of the correlation function at large distances:

$$\Gamma(\mathbf{x}) \sim e^{-\sigma|\mathbf{x}|/T}.$$

From the expression for the free energy of a quark-antiquark pair given earlier, we obtain a linear potential with string tension $\sigma(T)$: a confining phase. However, the symmetry could be broken at some finite temperature, with $N$ possible broken-symmetry ground states (vacua) for the pure gauge system.

How can the symmetry be broken at high temperatures? Consider the continuum generating functional again, for a moment, absorbing the dimensionless coupling through $gA^\mu \to A^\mu$:

$$Z = \int_{\text{periodic}} \mathcal{D}A^\mu e^{-\frac{1}{g^2}\int_0^{\beta_T} d\tau \int d^{d-1}x \frac{1}{4}\mathrm{Tr}F^2_{\mu\nu}}. \tag{1.15}$$

If one rescales $\tau \to \beta_T \tau$, $A^0 \to A^0/\beta_T$, this can be rewritten as

$$\int \mathcal{D}A^\mu e^{-\frac{1}{g^2 T}\int_0^1 d\tau \int d^{d-1}x \frac{1}{2}\mathrm{Tr}(T^2\mathbf{E}^2+\mathbf{B}^2)}, \qquad \text{with} \qquad A^0(\tau=1) = A^0(\tau=0), \tag{1.16}$$

where the electromagnetic tensor has been written explicitly in terms of its electric and magnetic components. As $T \to \infty$, the only non-zero contributions will come from field configurations with $\mathbf{E} = 0$, implying that $A^i$, the spatial components of the gauge potential, must be static up to a gauge transformation. This, together with the boundary conditions on $A^0$, implies that $\Omega(\mathbf{x})$ must be an element of the centre group[18]. Thus, as $L(\mathbf{x}) = \mathrm{Tr}\Omega(\mathbf{x})$, one sees that

$$<L>= e^{2\pi iq/N} \neq 0, \qquad (q = 0, 1, \ldots, N-1), \tag{1.17}$$

drawing the conclusion that the global centre symmetry is broken at high temperatures, corresponding to a deconfined phase, with the $N$ degenerate vacua being labelled by the $N$ possible distinct expectation values of $L$, $< L >= e^{2\pi iq/N} L_0$. This spontaneous breaking of the discrete global centre symmetry ($<L>\neq 0 \Rightarrow F_q$ finite) will lead[19] to a correlator

$$\Gamma(\mathbf{x}) \sim |<L>|^2 (1 + \beta_T e^2 e^{-\mu|\mathbf{x}|}/|\mathbf{x}|^p), \tag{1.18}$$

giving a short-range static potential $F_{q\overline{q}} \sim -e^2 e^{-\mu|\mathbf{x}|}/|\mathbf{x}|^p$: the Debye screening of an electric plasma.

The point to emphasise is that $< L(\mathbf{x}) >$ is acting as an order parameter for the centre symmetry, distinguishing between broken and unbroken phases, and between the different broken phases. This has been seen perturbatively[20][21] and by simulation[17].

It should also be noted, from (1.16), that this high-$T$ behaviour of a theory in $d$ dimensions, with bare coupling $g^2$, is equivalent to that of a $T = 0$ theory (where the electric field component of the action will also vanish) in $(d-1)$ dimensions (because of the static configurations) with effective coupling $g^2 T$. This effective coupling will become infinitely strong as $T \to \infty$, leading to strong-coupling behaviour. Thus, the Wilson loops mentioned in the next chapter show confining behaviour at all temperatures, and do not provide a valid test for confinement.

## 1.8. The Background Field Method

When we move beyond the classical level and seek to include quantum corrections, the manifest gauge invariance of a gauge field theory is usually broken by the gauge-fixing procedure. The "background field"[22][23] quantises the field theory in the presence of a background field whose transformation properties are chosen to restore certain symmetries of the generating functional, and we shall rely on this method in the next chapter.

The procedure is to write the gauge field from the classical action as $A^\mu = A^\mu_B + a^\mu$, where $A^\mu_B$ is the background field and $a^\mu$ the quantum field over which the functional integral is taken. The "background field gauge" is then used, breaking the gauge invariance of $a^\mu$, but retaining that of $A^\mu_B$. External sources are coupled only to $a^\mu$, thus allowing quantum calculations without losing explicit gauge invariance.

First, conventional quantisation: adding a source $J$ in the conventional fashion to the expression for the generating functional in (1.5), and writing $x = (\tau, \mathbf{x})$ so that $d^d x \equiv d\tau d\mathbf{x} \equiv d\tau d^{d-1}x$, we have

$$Z[J] = \int_{\text{periodic}} \mathcal{D}A^\mu_a \det\left(\frac{\partial \mathcal{F}^c(A)}{\partial \alpha^d}\right) e^{\int d^dx\ (\mathcal{L}(A) - \frac{1}{2\xi}(\mathcal{F}_a(A))^2 - J^a_\mu A^\mu_a)}. \tag{1.19}$$

Here,

$$\mathcal{L}(A) = -\frac{1}{4}F^a_{\mu\nu}F^{\mu\nu}_a \qquad \text{and} \qquad \mathcal{F}_a(A) = \partial_\mu A^\mu_a \qquad \text{(covariant gauge)},$$

and $(\frac{\partial \mathcal{F}^c(A)}{\partial \alpha^d})$ is the derivative under the infinitesimal transformation

$$\delta A^\mu(x) = -D^\mu \alpha(x), \tag{1.20}$$

where we define covariant derivatives for the standard and background fields as

$$D^\mu = \partial^\mu - ig[A^\mu,\ \ ] \qquad \text{and} \qquad D^\mu_B = \partial^\mu - ig[A^\mu_B,\ \ ]. \tag{1.21}$$

The functional derivatives of $Z[J]$ with respect to $J$ are the disconnected Green's functions of the theory, with the connected functions generated by $W[J] = \ln Z[J]$. We define the effective action by the Legendre transformation

$$\Gamma[\overline{A}] = W[J] - \int d^dx\ J^\mu_a \overline{A}^a_\mu, \qquad \text{where} \qquad \overline{A}^a_\mu = \frac{\delta W}{\delta J^\mu_a},$$

and generate the one-particle-irreducible Green's functions by taking derivatives of the effective action with respect to $\overline{A}$.

Now, the background field: we define $\widetilde{Z}, \widetilde{W}$ and $\widetilde{\Gamma}$ by analogy with $Z, W$ and $\Gamma$:

$$Z[J, A_B] = \int_{\text{periodic}} \mathcal{D}a^\mu_a \det\left(\frac{\partial \mathcal{F}^c(a, A_B)}{\partial \alpha^d}\right) e^{\int d^dx\ (\mathcal{L}(A_B + a) - \frac{1}{2\xi}(\mathcal{F}_a(a, A_B))^2 - J^a_\mu a^\mu_a)}, \tag{1.22}$$

$$\widetilde{W}[J, A_B] = \ln \widetilde{Z}[J, A_B], \qquad \widetilde{\Gamma}[\widetilde{a}, A_B] = \widetilde{W}[J, A_B] - \int d^dx\ J^\mu_a \widetilde{a}^a_\mu, \qquad \widetilde{a}^a_\mu = \frac{\delta \widetilde{W}}{\delta J^\mu_a}.$$

The single transformation of $A$ in (1.20) now corresponds to connected transformations of $A_B$ and $a$. Invariance of the action under (1.20) implies invariance under "quantum" transformations, where $\delta A^\mu_B = 0$, and "background" transformations, where $\delta A^\mu_B = -D^\mu_B \alpha(x)$, with $\delta a^\mu$ being chosen to give (1.20) in each case. To maintain explicit gauge invariance of the action with respect to the background transformations, we choose $\mathcal{F}_a(a, A_B)$ to transform covariantly under them, for example

$$\mathcal{F}_a(a, A_B) = (D^\mu_B a_\mu)_a. \tag{1.23}$$

Now, $(\frac{\partial \mathcal{F}^c(a, A_B)}{\partial \alpha^d})$ is the derivative under the quantum transformations, for which $\delta a^\mu = -D^\mu \alpha(x)$, so a consideration of $\mathcal{F}_a(a + \delta a, A_B)$ reveals that

$$\frac{\partial \mathcal{F}^c(a, A_B)}{\partial \alpha^d} = -(D^\mu_B D_\mu)^c_d. \tag{1.24}$$

By changing variables $a \to a - A_B$ in (1.22), we find the connection between $\widetilde{\Gamma}$ and $\Gamma$:

$$\widetilde{\Gamma}[\widetilde{a}, A_B] = \Gamma[\widetilde{a} + A_B] \qquad \Rightarrow \qquad \widetilde{\Gamma}[0, A_B] = \Gamma[A_B]$$

The latter is invariant under the background transformation $\delta A_B^\mu = -D_B^\mu \alpha(x)$, so is a gauge-invariant functional of $A_B^\mu$.

We can now use (1.23) and (1.24) to rewrite (1.22), introducing complex Grassmann ghost fields $\eta$ in a relation of the form

$$\det\left(\frac{\partial \mathcal{F}^c(a, A_B)}{\partial \alpha^d}\right) \sim \int \mathcal{D}\eta \mathcal{D}\overline{\eta}\; e^{\int d^d x\ d^d y\ \overline{\eta}_c(x)(\frac{\partial \mathcal{F}^c(x)(a,A_B)}{\partial \alpha^d(y)})\eta^d(y)},$$

so that

$$Z[J, A_B] = \int_{\text{periodic}} \mathcal{D}a_a^\mu \mathcal{D}\eta \mathcal{D}\overline{\eta}\; e^{\int d^d x\ (\mathcal{L}(A_B + a) - \frac{1}{2\xi}(D_B^\mu a_\mu)^2 + \overline{\eta}(-D_B^\mu D_\mu)\eta - J_\mu^a a_a^\mu)}. \tag{1.25}$$

# Chapter 2.

# Predictions for Z(3) Interfaces in SU(3) Gauge Theory

## 2.1. The Z(N) Interface

In the deconfined phase, where the $Z(N)$ symmetry has been spontaneously broken, we have seen that the system can exist in any one of $N$ degenerate vacua. It is possible to arrange boundary conditions of the system so that different parts of it exist in different vacua, *i.e.* different $Z(N)$ phases. The existence of these distinct domains forces the appearance of "domain walls", or "$Z(N)$ interfaces", where they meet[24][21]. Within these interfaces, the gauge fields interpolate between the different vacua, as does the expectation value of the Polyakov line.

The calculation of the interface tension is clearly an instanton problem: the field is interpolating from one $Z(3)$ vacuum to another. An effective one-dimensional theory will describe the interface profile. The instanton is the solution of the classical equations of motion, with action $\sim 1/g^2$ as in (1.15), and so we might naïvely expect the interface tension to have the same behaviour. To test this guess, we can construct an effective action from the classical piece, acting as a kinetic term, and a quantum piece, a potential term formed by integrating out fluctuations to one-loop order. The $Z(N)$ instanton is the stationary point of this total action.

In a case which will be relevant for future chapters, consider a system of finite extent at a temperature $T$, as illustrated in fig. 2.1. The Euclidean time dimension runs from 0 to $\beta_T = 1/T$, and let us suppose there are two space dimensions, one ($z$) much longer than the other ($x$), and both much longer than the time dimension: $L_z \gg L_x \gg \beta_T$. We assume that the system is in one $Z(N)$ phase at the $z = 0$ end of the longitudinal dimension and another at the $z = L_z$ end, forcing the appearance of a $Z(N)$ interface somewhere along the $z$ direction. Where the Polyakov line was previously parametrised by the discrete values $q = 0, 1, \ldots, N-1$ in (1.17), we shall now make $q$ a continuous function of longitudinal position: $q \to q(z)$.

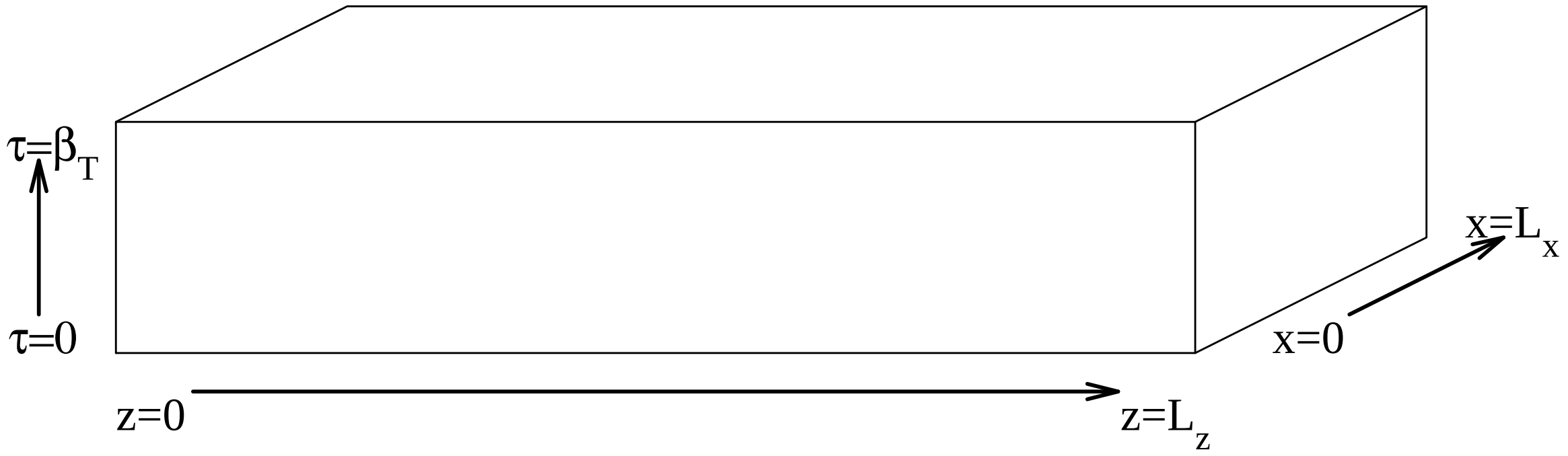

**Fig. 2.1:** A system of finite extent, with two space and one Euclidean time dimensions.

As we have seen, the order parameter for the $Z(N)$ symmetry is the Polyakov line, given in the fundamental representation by (1.12), where the gauge coupling is now written explicitly:

$$L(z) = \frac{1}{N} \mathrm{Tr} \mathcal{T} e^{ig \int_0^{\beta_T} A^0(z) d\tau}. \tag{2.1}$$

We assume that the gauge fields obey strictly periodic boundary conditions. Since the time direction is finite, states with constant $A^0$ are inequivalent. Whilst we could work in a gauge with $A^0$ set to zero at zero temperature, the periodic boundary conditions in the time direction make this impossible for $T \neq 0$: were we to set $A^0 = 0$ by a gauge transformation of the functional integral, the $A^i$ would violate the boundary conditions. Therefore, following the method outlined in [24][21], we choose to make a global colour rotation so that $A^\mu = (A^0, \mathbf{0})$, with $A^0$ a time-independent, diagonal $N \times N$ matrix:

$$A^0(z) = \frac{2\pi T q(z)}{gN} \begin{pmatrix} \mathbb{I}_{N-1} & \mathbf{0} \\ \mathbf{0} & -(N-1) \end{pmatrix} \equiv \frac{2\pi T q(z)}{gN} \begin{pmatrix} 1 & & & \\ & \ddots & & \\ & & 1 & \\ & & & -(N-1) \end{pmatrix}. \tag{2.2}$$

In the presence of this $A^0$ field, the Polyakov line is given by

$$L(z) = e^{2\pi i q(z)/N} \left( 1 - \frac{1}{N} [1 - e^{-2\pi i q(z)}] \right).$$

One can now understand the $Z(N)$ structure of the vacuum: the $N$ degenerate vacua occur for $q(z) = 0, 1, \ldots, N-1$, corresponding to $L(z) = e^{2\pi i q(z)/N}$. The ordinary perturbative vacuum has $A^0 = q(z) = 0$, with $L(z) = 1$. It has been shown[21] that the path taken by a gauge field when interpolating between different vacua has minimal action for this form of $A^0(z)$, as desired.

Returning to the generating functional in (1.25), without source term, and extracting the dimensionless coupling explicitly again, the Euclidean action is given by

$$\mathcal{S}_E = \int d\tau d^d x \left( \frac{1}{2g^2} \mathrm{Tr} F_{\mu\nu}^2 - \frac{1}{2\xi} (D_B^\mu a_\mu)^2 + \overline{\eta}(-D_B^\mu D_\mu)\eta \right).$$

The first term is the (classical) kinetic part, and the second and third terms form a potential, $V$.

For the $A^\mu \equiv A^0(z)$ field of (2.2), with transverse space dimensions of volume $V_{trans}$, the kinetic part reduces to

$$\frac{\beta_T V_{trans}}{g^2} \int \mathrm{Tr}(\partial_z A^0(z))^2 dz = \frac{4\pi^2 T V_{trans}(N-1)}{g^2 N} \int \left(\frac{dq}{dz}\right)^2 dz.$$

Classically, there is no potential for $A^0$, and all values of $A^0 \neq 0$ are degenerate. However, by computing one-loop quantum effects about the background field

$$A_B^\mu = (A^0, \mathbf{0}), \qquad \text{where } A^0 \text{ is as given in (2.2),} \tag{2.3}$$

we shall see that the degeneracy is broken. We treat the background $A^0$ field as constant in $z$, justifying this later. Expanding $V$ to quadratic order in the quantum fluctuations $a^\mu$, we can then integrate out both these gauge fields and the ghost fields, to give (to one-loop order):

$$V = \Big\{ \underbrace{\tfrac{1}{2}\mathrm{Tr}\ln\Big[ -(D_B)^2 \delta_{\mu\nu} + \Big(1 - \frac{1}{\xi}\Big) D_B^\mu D_B^\nu \Big]}_{\text{from gauge fields}} - \underbrace{\mathrm{Tr}\ln\Big( -(D_B)^2 \Big)}_{\text{from ghost fields}} \Big\}.$$

The potential is independent of $\xi$, so we use the Feynman gauge, $\xi = 1$, in our $d$ dimensions, to reduce the previous expression to

$$V = \left(\frac{d}{2} - 1\right) \mathrm{Tr}\ln(-(D_B)^2).$$

Now, each matrix function $a(x)$ on which $D_B^\mu$ acts can be written as a product of a scalar function of space-time with a vector in an $(N^2-1)$-dimensional matrix space, where the basis matrices are the $(N^2-1)$ generators of $SU(N)$. To find $\mathrm{Tr}\ln(-(D_B)^2)$, we should diagonalise $D_B$ in function space and then sum over its eigenvalues. A suitable basis is given by the plane waves $e_p(x) = e^{ip_\mu x^\mu}$, with 4-momentum $p_\mu = (p_0, \mathbf{p})$; the coordinates of $a(x)$ are its Fourier components $\widetilde{a}(p)$, and the action of $D_B$ on these comes from (1.21):

$$D_B^\mu a(x) = \partial^\mu a(x) - ig[A_B^\mu, a(x)] \qquad \Rightarrow \qquad D_\mu^B \widetilde{a}(p) = ip_\mu \widetilde{a}(p) - ig[A_B^\mu, \widetilde{a}(p)].$$

Our choice of $A_B^\mu$ earlier in (2.3) eases the eigenvalue calculation. For the space components, the eigenvalues of $-(D_B^i)^2$ are simply $(p^i)^2$. For the time component, the periodic boundary conditions force $p_0 = 2\pi n_0/\beta_T = 2\pi n_0 T$ for integer $n_0$. A suitable set of $SU(N)$ generators enables the eigenvalues of $-(D_B^0)^2$ to be calculated rapidly; for instance, we can split the $(N^2-1)$ generators into $(N-1)$ diagonal generators similar to (2.2),

$$T_i \sim \begin{pmatrix} \mathbb{I}_{i-1} & & \\ & -(N-2) & \\ & & \mathbb{I}_{N-i} \end{pmatrix}, \qquad (i = 1, \ldots, N-1),$$

and $(N-1)!$ pairs of off-diagonal ladder generators[25], $T^{\pm}_{i,j}$, with all elements zero except for one, whose value is common to the pair, at the $j$th row and $i$th column of $T^{+}_{i,j}$ and the $i$th row and $j$th column of $T^{-}_{i,j}$ ($i = 2, \ldots, N$ and $j = 1, \ldots, i-1$). When these generators are commuted with $A_B^0$, as given by (2.3), it should be apparent that only the $(N-1)$ pairs of generators $T^{\pm}_{N,j}$ will give a non-zero commutation bracket (proportional to $\pm T^{\pm}_{N,j}$ respectively), with the other $(N-2)! = \frac{1}{2}(N-2)(N-1)$ off-diagonal pairs and $(N-1)$ diagonal generators commuting with $A_B^0$. Thus, the eigenvalues of $-(D_B^0)^2$ are seen to be:

$$[2\pi T(n_0 \pm q)]^2 \quad : \quad (N-1) \text{ copies of each (charged generators);}$$
$$[2\pi T n_0]^2 \quad : \quad (N-1)^2 \text{ copies (neutral generators).}$$

The part of the trace over space-time now simply gives the total space-time volume, $V_{tot} = \beta_T L_z V_{trans}$; however, this is conventionally omitted from the "potential", so we shall ignore it until we insert $V(q)$ into the action. Hence, putting $V(q) = \mathcal{V}(q) - \mathcal{V}(0)$ so that $V(0) = 0$,

$$\mathcal{V}(q) = (N-1)(\tfrac{d}{2}-1)\ T\sum_{n_0}\int \frac{d^{d-1}\mathbf{p}}{(2\pi)^{d-1}}\Big\{(N-1)\ln((2\pi T n_0)^2 + \mathbf{p}^2)$$
$$+ \ln((2\pi T(n_0 - q))^2 + \mathbf{p}^2) + \ln((2\pi T(n_0+q))^2 + \mathbf{p}^2)\Big\}.$$

Note that for finite temperatures, we have replaced the integral over $p_0$ with a discrete sum:

$$\int \frac{d^d p}{(2\pi)^d} \to T\sum_{n_0}\int \frac{d^{d-1}\mathbf{p}}{(2\pi)^{d-1}}.$$

Since $n_0$ and $\mathbf{p}$ run over all negative and positive values, the $(n_0 \pm q)$ terms give the same contribution, and the subtraction of $\mathcal{V}(0)$ yields

$$V(q) = 2(N-1)(\tfrac{d}{2}-1)\ T\sum_{n_0}\int \frac{d^{d-1}\mathbf{p}}{(2\pi)^{d-1}}[\ln((2\pi T(n_0+q))^2 + \mathbf{p}^2) - \ln((2\pi T n_0)^2 + \mathbf{p}^2)].$$

Now, we can use the following two integral formulae to make progress:

$$\int_0^\infty \frac{e^{-ax} - e^{-bx}}{x}dx = \ln b - \ln a \qquad \text{and} \qquad \int_{-\infty}^{\infty} e^{-ax^2} = \sqrt{\frac{\pi}{a}}.$$

The first enables us to rewrite $V(q)$ as

$$V(q) = 2(N-1)(\tfrac{d}{2}-1)\ T\sum_{n_0}\int \frac{d^{d-1}\mathbf{p}}{(2\pi)^{d-1}}\int_0^\infty \frac{dt}{t}e^{-t\mathbf{p}^2}[e^{(2\pi T n_0)^2 t} - e^{(2\pi T(n_0+q))^2 t}],$$

before the second gives the result of Gaussian integration over **p**:

$$V(q) = 2(N-1)(\tfrac{d}{2}-1)\ T\sum_{n_0}\int_0^\infty \frac{dt}{t}\frac{1}{(4\pi t)^{d/2}}[e^{(2\pi T n_0)^2 t} - e^{(2\pi T(n_0+q))^2 t}].$$

A combination of the Poisson summation formula and a standard exponential integral leads to a useful relation:

$$\int_{-\infty}^{\infty} e^{-(ax^2+bx+c)}dx = \sqrt{\tfrac{\pi}{a}}e^{(b^2-4ac)/4a} \quad \text{and} \quad \sum_{n=-\infty}^{\infty} F(n) = \sum_{r=-\infty}^{\infty}\left\{\int_{-\infty}^{\infty} e^{2\pi i r x}F(x)dx\right\}$$

$$\Rightarrow \qquad \sum_n e^{-a(n+b)^2} = \sum_r \sqrt{\frac{\pi}{a}}\ e^{-\pi(r^2/a)-2\pi i b r}.$$

Along with a change of variable to $\hat{t} = 4\pi T^2 t$, this leads to

$$V(q) = 2(N-1)(\tfrac{d}{2}-1)\ T^d \sum_{r=-\infty}^{\infty}\int_0^\infty \frac{d\hat{t}}{\hat{t}}\hat{t}^{-d/2}e^{-\pi r^2/\hat{t}}(1-e^{2\pi i r q}).$$

A change of variables $\hat{t} \to u = 1/\hat{t}$, followed by exponential integration over $u$, gives the final result:

$$V(q) = 4(N-1)(\tfrac{d}{2}-1)T^d\frac{\Gamma(\frac{d}{2})}{\pi^{d/2}}B_d(q), \tag{2.4}$$

where, re-expressing the sum over all $r$ as a sum over positive $r$,

$$B_d(q) = \sum_{r=1}^{\infty}\frac{1}{r^d}[1-\cos(2\pi r q)].$$

For $SU(3)$ theory in $2+1$ dimensions, $V(q)$ is illustrated in fig. 2.2. The potential is clearly periodic: $V(q) \equiv V(q \bmod 1)$. The effective potential is minimised, as mentioned before, when $q$ takes an integer value, *i.e.* it vanishes when $< L > \in Z(N)$, and $< L >$ is indeed an order parameter for the $N$ degenerate vacua.

It should be noted in passing that this picture changes with the addition of fermions, as mentioned in section 1.3. This is because a gauge transformation of the form (1.13) causes the fermion wavefunction to transform simply as $\psi \to U\psi$. Thus, whereas the gauge fields are invariant under a $Z(N)$ transformation, the fermion wavefunctions are generally not: $\psi \to e^{2\pi i q/N}\psi$. This means that the boundary conditions, which should be periodic for the gauge fields and anti-periodic for the fermions, actually correspond to "twisted" conditions for the fermions: $\psi(\beta_T) = e^{-2\pi i q/N}\psi(0)$. Thus, the $Z(N)$ symmetry is lost, and the structure of fig. 2.2 changes[10], with the central maximum and minima rising as fermion families are added but the outer maxima and $q = 0$ minimum falling, the latter becoming the true vacuum. For $N = 3$ and four or more fermion families, the metastable minima at $q = 1$ and $q = 2$ actually correspond to *positive* free energy density of the form mentioned in section 1.3, and the peculiar thermodynamic properties discussed there set in.

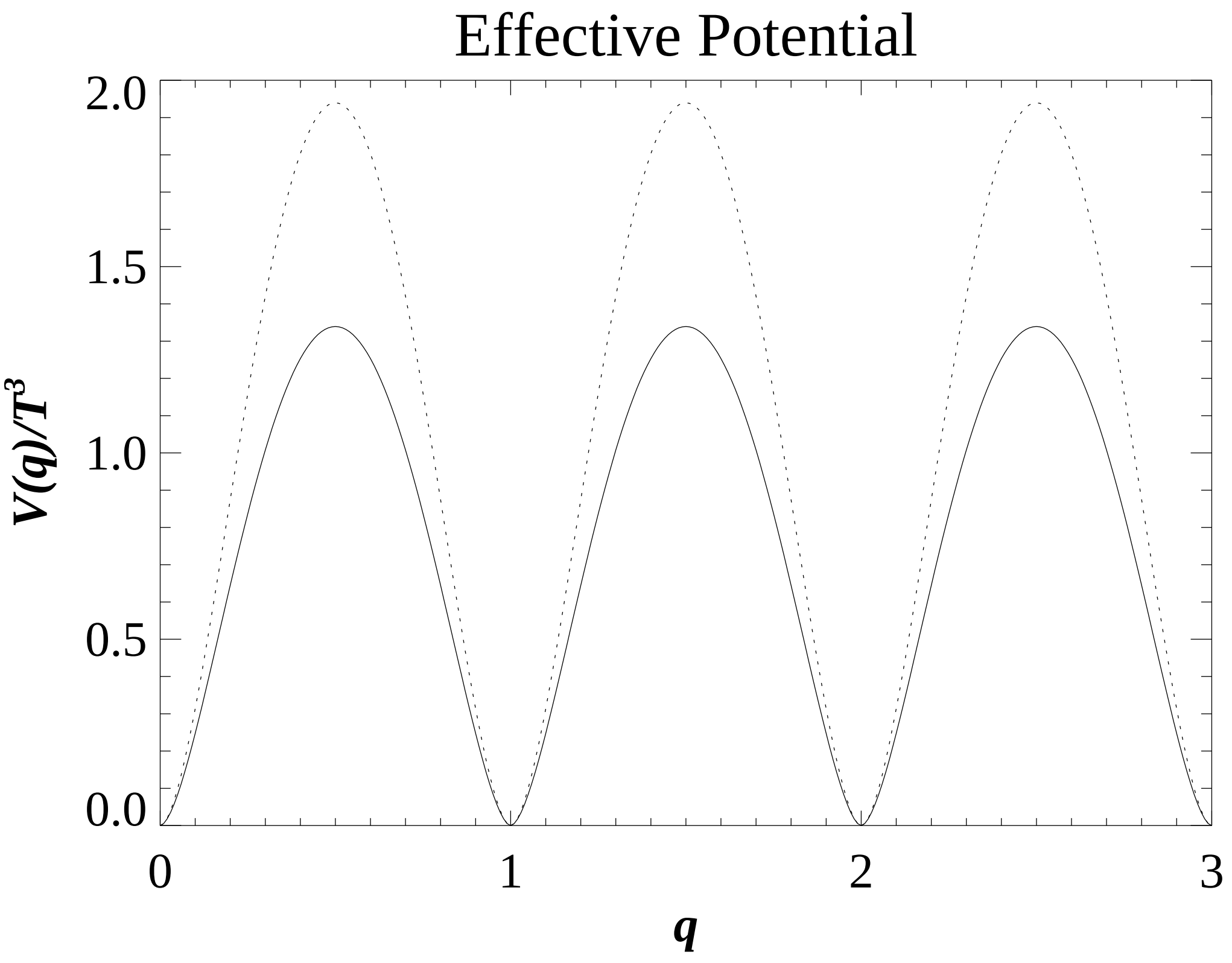

**Fig. 2.2:** The effective potential, with temperature dependence factored out, for $d = 3$ and $N = 3$. This quantity is equal to $2B_3(q)/\pi$. The solid line is the continuum version given here; the dotted one is the lattice version for $N_t = 2$, as calculated in section 2.3.

We now have an expression for the effective action in $q$, remembering to integrate $V(q(z))$ over the total space-time volume, $V_{tot} \to \beta_T V_{trans} \int dz$, and using $\beta_T = 1/T$:

$$\mathcal{S}_{eff}(q) = V_{trans} T \int dz \left\{ \frac{4\pi^2(N-1)}{g^2 N} \left( \frac{dq}{dz} \right)^2 + 4T^{d-2}(N-1)(\tfrac{d}{2}-1) \frac{\Gamma(\frac{d}{2})}{\pi^{d/2}} B_d(q) \right\}.$$

For $d = 3$, we convert to the dimensionless variable $z' \equiv \sqrt{NT} gz$ :

$$\mathcal{S}_{eff} = L_x \frac{T^{1.5}}{g} \frac{4\pi^2(N-1)}{\sqrt{N}} \int dz' \left\{ \left( \frac{dq}{dz'} \right)^2 + \frac{B_3(q)}{4\pi^3} \right\}.$$

Treating the integrand as a standard Lagrangian with "time" $z'$, we know that the equivalent Hamiltonian (the energy, $\mathcal{E} = (\frac{dq}{dz'})^2 - \frac{B_3(q)}{4\pi^3}$) is a constant of the motion. In the simplest case, we set $q = 0$ at one end of the system and $q = 1$ at the other, with $\frac{dq}{dz'} = 0$ at both. With these boundary conditions, the instanton has $\mathcal{E} = 0$, so

$$\begin{aligned}\mathcal{S}_{eff} &= L_x \frac{T^{1.5}}{g} \frac{4\pi^2(N-1)}{\sqrt{N}} \int dz' \left\{ \left( \frac{dq}{dz'} - \sqrt{\frac{B_3(q)}{4\pi^3}} \right)^2 + \left( \frac{dq}{dz'} \right) \frac{\sqrt{B_3(q)}}{\pi^{3/2}} \right\} \\ &= L_x \frac{T^{1.5}}{g} \frac{4\pi^{1/2}(N-1)}{\sqrt{N}} \int dq \sqrt{B_3(q)}.\end{aligned}$$

The interface tension, $\alpha$, is defined to be the effective action per unit area of the interface:

$$\mathcal{S}_{eff} = \beta_T L_x \alpha.$$

This gives a prediction for the surface tension of

$$\begin{aligned} \alpha &= \frac{T^{2.5}}{g} \frac{4\pi^{1/2}(N-1)}{\sqrt{N}} \int dq \sqrt{B_3(q)} \\ &= \alpha_0 \frac{T^{2.5}}{g}, \end{aligned} \tag{2.5}$$

as the integral is just a number, 1.018. Evaluating (2.4) for $d = 3$, we see that $V(q)/T^3 = (N-1)B_3(q)/\pi$, so, in a result that we shall use again later,

$$\alpha_0 = 4\pi \sqrt{\frac{N-1}{N}} \int dq \sqrt{V(q)/T^3}. \tag{2.6}$$

For the $SU(3)$ system we shall go on to consider, this means that

$$\alpha_0 = 8.33. \tag{2.7}$$

It only remains to justify our original assumption that $q$ could be treated as independent of $z$. In fact, the instanton must have width of order 1 in terms of the dimensionless coordinate $z'$, as there is no length scale, and this corresponds to $1/g\sqrt{T}$ in terms of $z$. For weak coupling, this is much larger than the natural length scale of the system, $1/T$. Thus, the instanton field varies slowly enough in $z$ that it can be treated as constant to leading order.

## 2.2. Lattice Formalism

Although some of the above properties can be examined in the continuum, many calculations are performed on the lattice[26][27]. The advantage of this approach is that the short-distance cut-off enters the definition of the field theory in a natural way, allowing non-perturbative calculations to be made. In particular, lattice theories lend themselves to strong-coupling expansions and Monte-Carlo simulations, allowing the statistical evaluation of path integrals to a specified accuracy. In particular, the confining and non-confining behaviour of quarks can be seen in the lattice formalism, together with the transition between these two regimes. Using numerical simulations on the lattice, one can also approach the continuum limit and compare results with analytic predictions.

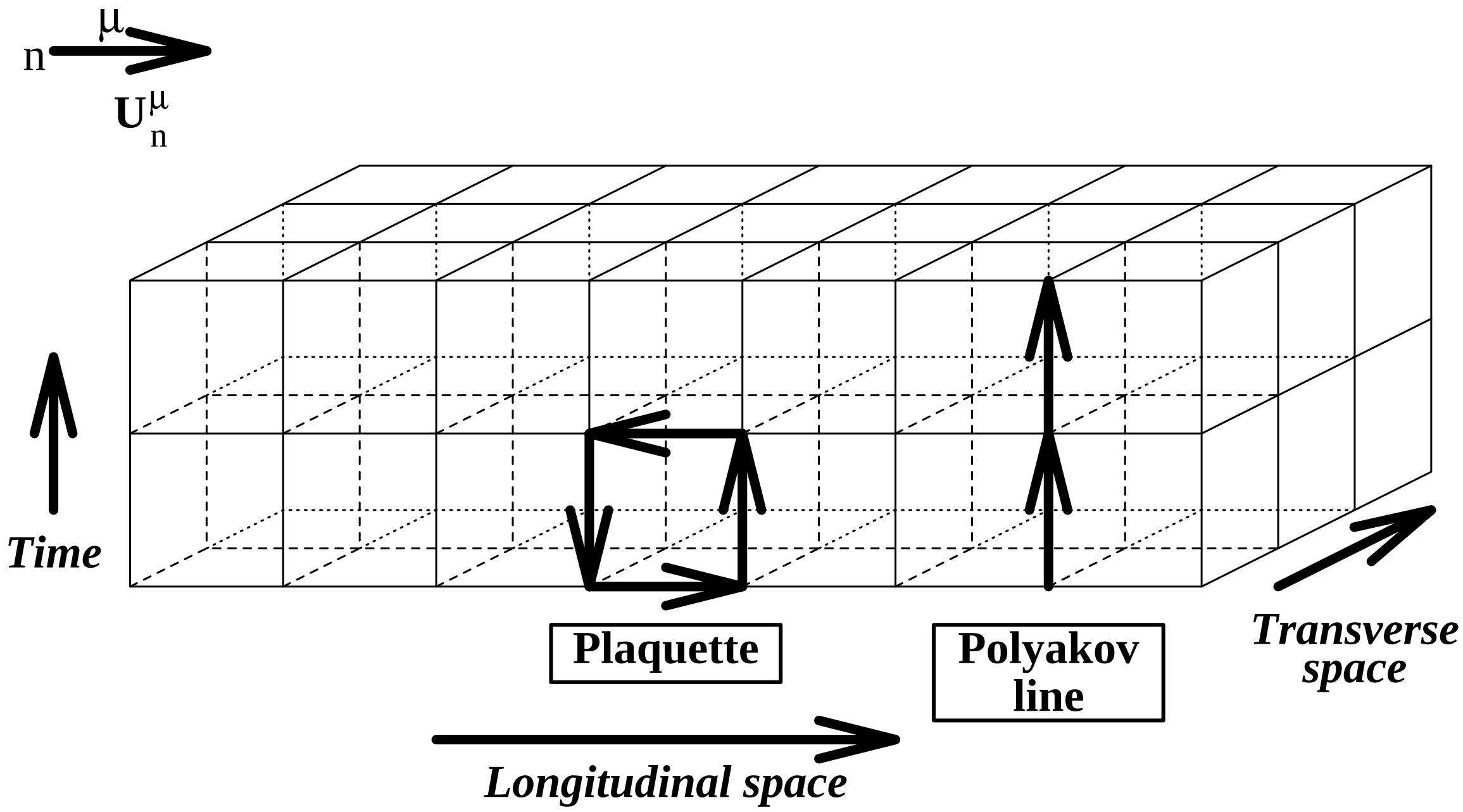

**Fig. 2.3:** The Euclidean lattice in two space and one time ("2+1") dimensions. Link variables, $U$, on the lattice are illustrated by bold arrows, as indicated by the example to the top left. A plaquette and a Polyakov line are shown; the Polyakov line wraps around the boundary condition in the time direction, but this is not shown.

In the Euclidean formalism, one considers a hypercubic lattice whose sites are separated by lattice spacing $a$. This is the only quantity on the lattice which has dimension, and acts as an ultraviolet cut-off for the theory on the lattice. If we specify $N_t$ links in the time direction, this corresponds to a temperature

$$T = 1/\beta_T = 1/N_t a. \tag{2.8}$$

Similarly, we have $N_z$ links in the longitudinal direction and $N_x$ in the transverse, so

$$L_z = N_z a \qquad \text{and} \qquad L_x = N_x a.$$

For some gauge group $\mathcal{G}$, we can define a gauge field

$$U_n^\mu = e^{ia\mathbf{T}\cdot\mathbf{A}_{n\mu}} \in \mathcal{G}, \tag{2.9}$$

residing on the link leaving site $n$ in direction $\mu$. The exponential of the field strength, $e^{iF_{\mu\nu}}$, is now equivalent to the product of fields around a plaquette (four connected links forming the smallest possible square, as illustrated in fig. 2.3):

$$U_n^{\mu\nu} = U_n^\mu U_{n+\hat{\mu}}^\nu U_{n+\hat{\nu}}^{\mu\dagger} U_n^{\nu\dagger}. \tag{2.10}$$

The usual procedure is now to consider the Wilson action, which only depends on plaquette variables ($N$ is the order of symmetry, as in section 1.5):

$$S_W = \frac{2N}{g^2} \sum_{\substack{n \\ \mu>\nu}} (1 - \tfrac{1}{N} \Re e \mathrm{Tr} U_n^{\mu\nu}). \tag{2.11}$$

The conventional bare (dimensionless) coupling, $g$, is related to the lattice parameter $\beta$ by

$$\beta = 2N/a^{4-d} g^2 \tag{2.12}$$

in $d$ dimensions. For our future purposes, $d = 3$ and $N = 3$, so that $\beta = 6/ag^2$.

This action is, naturally, invariant under local gauge transformations, $U_n^\mu \to V_n U_n^\mu V_{n+\hat{\mu}}^\dagger$, for $V_n \in \mathcal{G}$, as well as the global centre symmetry discussed in section 1.7. Using the gauge fields on the links, we can define "Wilson loops", gauge-invariant observables given by the trace of an ordered product around a closed curve $C$:

$$W[C] = \mathrm{Tr} \prod_{n\mu \in C} U_n^\mu. \tag{2.13}$$

These topologically trivial quantities are not aware of the centre symmetry and $Z(3)$ phases.

Owing to the periodicity of the lattice in time, topologically non-trivial loops also exist, winding around the lattice one or more times; these cannot be shrunk to a point. The Polyakov line is one such, defined, as in (2.1) for the continuum, by considering a path parallel to the time axis at some spatial location $\mathbf{x}$:

$$L(\mathbf{x}) = \mathrm{Tr} \prod_{n=1}^{N_t} U^0_{\mathbf{x}+n\hat{t}}. \tag{2.14}$$

An example of a Polyakov line is shown in fig. 2.3. Many of these non-trivial quantities *are* aware of the $Z(3)$ phase structure, as we saw for the Polyakov line in section 1.7.

## 2.3. The Lattice Surface Tension

SINCE we shall be examining the properties of an interface in $SU(3)$ theory in $2+1$ dimensions, with $N_t = 2$, it behoves us to repeat the procedure giving the interface tension, this time on the lattice. This simply involves replacing continuum quantities in the calculation with their lattice equivalents, following [28].

If we consider the space dimensions to be essentially infinite, taking only the time dimension to be small, then the continuum momenta are restricted as follows on the lattice:

$$p_i \to \frac{2}{a} \sin\left(\frac{p_i a}{2}\right), \qquad -\frac{\pi}{a} < p_i < \frac{\pi}{a};$$

$$p_0 \to \frac{2}{a} \sin\left(\frac{p_0 a}{2}\right), \qquad p_0 = 2\pi n_0 T,\ 2\pi(n_0 \pm q)T, \qquad n_0 = 0, 1, \ldots, N_t - 1.$$

The techniques used are largely the same as for the continuum calculation, but we leave the integration over the momenta until last, as the restriction is now $-\pi < p_\mu < \pi$ (setting $a$=1 for convenience) rather than $-\infty < p_\mu < \infty$. These integrals over momenta give modified Bessel functions:

$$I_n(t) = \frac{1}{2\pi}\int_{-\pi}^{\pi} dk\ e^{t\cos k + ink} \tag{2.15}$$

The effective potential on the lattice is finally written (C.F. (2.4)) as:

$$V_{lat}(q) = 4(N-1)(\tfrac{d}{2}-1)\sum_{r=1}^{\infty}(1-\cos(2\pi rq))F(rN_t,d), \tag{2.16}$$

where the newly introduced function $F$ is defined in terms of the modified Bessel functions, $I_n$:

$$F(n,d) = \int_0^\infty \frac{dt}{t}e^{-dt}I_n(t)[I_0(t)]^{d-1}.$$

For $d=3$ and $N_t=2$, the values of $F(2r,3)$ are given in Appendix A. For these values, the lattice quantity, $V_{lat}(q)N_t^d$, is plotted in fig. 2.2 alongside the continuum version, $V(q)/T^d$. The lattice expression for the surface tension is given by (2.6):

$$\alpha_0 = 4\pi\sqrt{\frac{N-1}{N}}\int_0^1 dq\sqrt{V_{lat}(q)N_t^3}.$$

For $N_t=2$ and $N=3$, this gives the prediction (C.F. (2.7)) that

$$\alpha_0 = 9.82. \tag{2.17}$$

To check that the correct continuum limit is recovered as $N_t \to \infty$, we need to find the behaviour of $F(n,d)$ as $n\to\infty$. We use the method of steepest descent, which determines the asymptotic ($s\to\infty$) behaviour for an integral of the form

$$\mathcal{I}(s) = \int_{\mathcal{C}} dz g(z)e^{sf(z)},$$

as follows:

- Find the saddle point of $f(z)$, *i.e.* $z_0$ such that the first derivative, $f'(z_0)=0$;
- Expand around this point using $z = z_0 + \delta e^{i\alpha}$, and choose $\alpha$ so that $\Re e\{f(z)\}$ is a maximum, giving the path of steepest descent;
- Substitute into the following formula, the approximation sought for large $s$:

$$\mathcal{I}(s) \approx \frac{\sqrt{2\pi}g(z_0)e^{sf(z_0)}e^{i\alpha}}{|sf''(z_0)|^{1/2}}.$$

From (2.15), we can write

$$I_n(nt) = \frac{1}{2\pi}\int_{-\pi}^{\pi} dk\ e^{n(t\cos k + ik)}, \qquad I_0(nt) = \frac{1}{2\pi}\int_{-\pi}^{\pi} dk\ e^{n(t\cos k)}$$

The forms of $I_n(nt)$ and $I_0(nt)$ enable us to find saddle points at $k = \cos^{-1}(\frac{\sqrt{t^2+1}}{t})$ and $k = 0$ respectively, and the steepest descent in each case is for $\alpha = 0$, giving

$$I_n(nt) \approx \frac{e^{n[\sqrt{t^2+1}-\sinh^{-1}(1/t)]}}{\sqrt{2\pi nt}(1+1/t^2)^{1/4}}, \qquad I_0(nt) \approx \frac{e^{nt}}{\sqrt{2\pi nt}}.$$

Substituting into our formula for $F(n,d)$,

$$\begin{aligned} F(n,d) &= \int_0^\infty \frac{d(nt)}{nt} e^{-dnt} I_n(nt)[I_0(nt)]^{d-1} \\ &= \int_0^\infty \frac{dt}{t} \frac{e^{n[t(\sqrt{1+1/t^2}-1)-\sinh^{-1}(1/t)]}}{(2\pi nt)^{d/2}(1+1/t^2)^{1/4}}. \end{aligned}$$

This exponent has its stationary point at $t = +\infty$, and, since the dominant contribution to the integral will occur around this value of $t$, we can replace the integrand by its approximate form as $t \to +\infty$. A final transformation of variable, $t \to u = 1/t$, then reveals that

$$F(n,d) = \int_0^\infty du e^{-nu^2/2} \frac{u^{d/2-1}}{(2\pi n)^{d/2}},$$

so, by the definition of the gamma function,

$$F(n,d) \to \frac{\Gamma(d/2)}{\pi^{d/2} n^d} \qquad \text{as} \qquad n \to \infty,$$

as desired from a comparison of (2.4) with (2.16).

## 2.4. The Debye Mass

The electric ("Debye") mass $m$, or inverse Debye screening length, governs gauge-invariant correlation functions of the time-like component of the gluon field ($A_0$) at large distances and high temperatures. The free energy of a quark-antiquark pair, over and above the sum of their separate free energies, vanishes as the quarks become infinitely far apart:

$$F_{q\bar{q}}(\mathbf{x}) - 2F_q = V(|\mathbf{x}|, T) \underset{|\mathbf{x}|\to\infty}{\longrightarrow} 0 \qquad \text{for } T > T_c.$$

From perturbation theory, we know that quarks are screened at a distance of the order of the inverse electric mass, so rather than being logarithmic in $\mathbf{x}$, the interaction potential takes, in the manner of (1.18), the form

$$V(|\mathbf{x}|, T) \underset{|\mathbf{x}|\to\infty}{\sim} -Ce^{-2m|\mathbf{x}|} \qquad \text{for } T > T_c.$$

The factor of two arises because gauge invariance leads to an exchange of two gluons being the lowest-order contribution in perturbation theory. The electric mass is one of the fundamental parameters of a gauge theory at finite temperature. One can perform a self-consistent and gauge-invariant calculation of its value, through the subtracted correlation function

$$\ll L(0)L^\dagger(\mathbf{x}) \gg \quad \equiv \quad < L(0)L^\dagger(\mathbf{x}) > - < L(0) >< L^\dagger(\mathbf{x}) >,$$

and the Feynman diagrams contributing to this correlation function have been calculated[29] [30], with the dimensional regularisation mass scale set to $m$. We shall quote the expression found for $m^2$ later in this section, but it may be instructive beforehand to examine screening in a simple model: a plasma of bosons, each with mass $m_g$ and charge $\pm g$[31].

A positive test charge inserted into the plasma will attract (repel) negatively (positively) charged bosons, and the net induced charge density will result in an electrostatic potential $\phi(r)$. The energy of a given boson in the absence of the test charge, $E = \sqrt{p^2 + m_g^2}$, will change in its presence by $\pm g\phi$, so we can write a consistent expression for this induced charge density using the bosonic density of states:

$$\rho \sim -\int d^{d-1}p \Big( \frac{g}{e^{\beta_T(E+g\phi)} - 1} + \frac{-g}{e^{\beta_T(E-g\phi)} - 1} \Big). \tag{2.18}$$

Far away from the test charge, $\phi(r \to \infty) \to 0$, and we can expand $e^{\beta_T g\phi} \approx 1 + \beta g\phi$. Changing variables so as to pull factors of $\beta_T$ out of the resulting integral, we find that $\rho \sim -g^2T^2\phi$ for $d = 4$. The inclusion of the test charge in $\rho$ gives exponential screening with this expression for the Debye mass:

$$\phi \sim \frac{e^{-mr}}{r}, \qquad m^2 \sim g^2T^2. \tag{2.19}$$

For $d = 3$, we find that $\rho \sim -g^2T\phi \ln \frac{T}{g\phi}$, and $\phi$ shows Gaussian rather than exponential screening. In fact, the effective one-loop potential for small $q$ shows the following behaviour:

$$V_{eff} \sim q^2 \ln q,$$

leading to a divergence in the Debye mass. However, the full loop calculation mentioned earlier, treating the Debye mass self-consistently, reveals that the logarithmic divergence is halted at an infrared cut-off, resulting in the return of exponential screening. In our simple plasma model, this can be seen by an approximate evaluation of (2.18) for non-zero mass $m_g$, putting $e^{\beta_T E} \approx 1 + \beta_T E$ for $m_g < E < \beta_T^{-1}$:

$$\begin{aligned} \rho &\sim -g^2\beta_T\phi \Big( \int_{m_g}^{\beta_T^{-1}} EdE \frac{(1+\beta_T E)}{(\beta_T E)^2} + \int_{\beta_T^{-1}}^{\infty} EdE \frac{e^{\beta_T E}}{(e^{\beta_T E} - 1)^2} \Big) \\ &\sim -g^2T\phi \ln \left( \frac{T}{m_g} \right) \qquad \text{to leading order.} \end{aligned}$$

This corresponds to the return of exponential screening:

$$\phi \sim \frac{e^{-mr}}{\sqrt{r}}, \qquad m^2 \sim g^2 T \ln\left(\frac{T}{m_g}\right).$$

The prediction of the full loop calculation[29][30] mentioned previously is that

$$V_{eff} \sim q^2 m^2,$$

with a self-consistent solution for the Debye mass:

$$m^2 = \frac{g^2 NT}{4\pi} \ln(T/g^2) \quad + \quad O\Big[g^2 T \ln(\ln(T/g^2))\Big]. \tag{2.20}$$

Clearly, this mass is a non-analytic function of coupling constant and temperature. This formula is only strictly applicable for $T/g^2 \gg 1$.

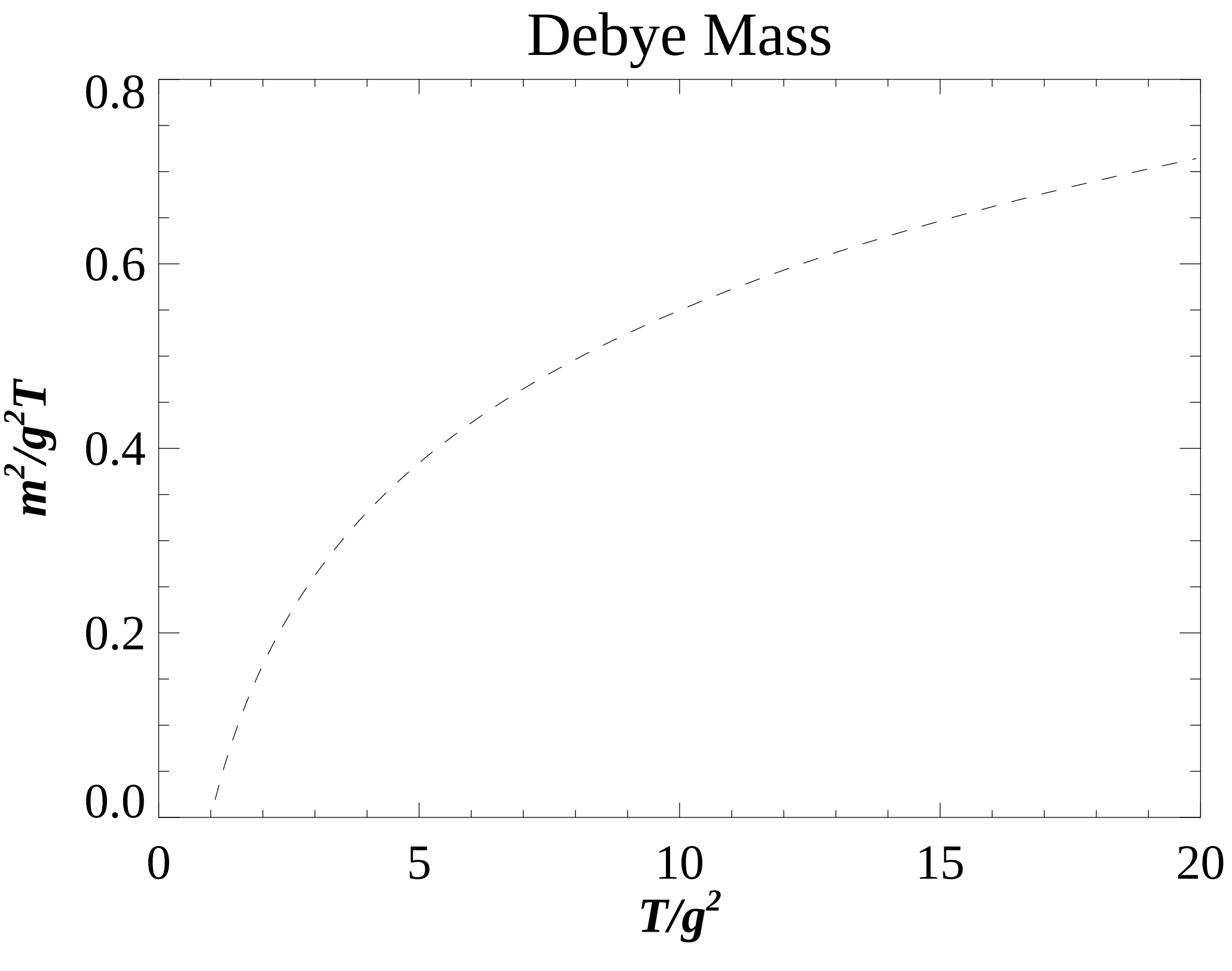

**Fig. 2.4:** The predicted lowest-order behaviour of the Debye mass, $m$.

The prediction for $m$ is shown in fig. 2.4. In terms of the lattice parameter, (2.20) becomes

$$\frac{m^2}{g^2 T} = \frac{3}{4\pi} \ln(\beta/12) + O\Big[\ln(\ln(\beta/12))\Big]. \tag{2.21}$$

Since the one-loop $V(q) \sim q^2 \ln(1/q)$ rather than $q^2$ for small $q$, we are not able to extract an expression for the Debye mass from $\frac{\partial^2 V_{lat}(q)}{\partial q^2}\big|_{q=0}$.

## 2.5. The Deconfinement Transition

We can re-express the $d$-dimensional functional integral over gauge fields, seen in section 1.6, in terms of the Polyakov line, $L$, by inserting a delta function conforming to (2.1), and so construct an effective action in terms of $L$ [19]:

$$e^{-S_{eff}[L]} = \int \mathcal{D}A_\mu e^{-S[A_\mu]} \prod_{\mathbf{x}} \delta\Big(L(x) - \mathrm{Tr}\mathcal{T}e^{i\int dtA_0(\mathbf{x},t)}\Big).$$

Integrating out the space-like gauge fields, $A^i$, their correlation length, $\xi$, sets the range of a set of coupling- and temperature-dependent kernels, $S_2, S_4, \ldots$, in

$$\begin{aligned} S_{eff}[L] = \int d\mathbf{x}\ V(L(\mathbf{x}))\ &+ \int d\mathbf{x}\ d\mathbf{y}\ L^\dagger(\mathbf{x}) S_2(\mathbf{x}-\mathbf{y}) L(\mathbf{y}) \\ &+ \int d\mathbf{x}\ d\mathbf{y}\ d\mathbf{z}\ d\mathbf{w}\ L^\dagger(\mathbf{x}) L^\dagger(\mathbf{y}) S_4(\mathbf{x},\mathbf{y},\mathbf{z},\mathbf{w}) L(z) L(w)\ + \ldots. \end{aligned}$$

It can be shown[19] that $\xi$ is finite at low *and* high temperatures, with space-like Wilson loops at both showing an area-law behaviour. Assuming that this will also hold at the phase transition itself, the kernels should be short-ranged.

The deconfinement-confinement transition now appears as a $(d-1)$-dimensional order-disorder transition in statistical mechanics, the time dimension having gone. Integrating out the short-range interactions up to length $\xi$ leaves a simple Landau-Ginzburg effective action:

$$S_{eff} = \int dx \Big( (\partial_i L)^2 + V(L) \Big).$$

$S_{eff}$ retains the invariance under the centre symmetry $L \to zL$, so that the potential $V(L)$ must do the same. However, we know little else about $V(L)$, so we need to study it in mean-field theory (MFT), or by Monte Carlo simulation, to find out details of the order of the phase transition, as seen in [32] for the strong-coupling limit.

In MFT we neglect fluctuations, so the equilibrium state of the system is represented by the absolute minimum of $V$. If the minimum is at $L = 0$, we have the symmetric phase; at the transition, the transition is shown to be second order by the minimum moving away from zero smoothly, and first by it moving discontinuously. If the transition is continuous, renormalisation group (RG) theory tells us that an RG fixed point will govern the critical behaviour. Whole "universality" classes of theories have their critical points mapped onto the same fixed point, with the same exponents. This can be very powerful: if, in the space of $(d-1)$-dimensional theories with short-range interactions invariant under the same centre symmetry, there is only a single fixed point, it reveals to us that a $d$-dimensional gauge theory will have the same critical behaviour as a $d$-dimensional spin system with the same symmetry.

MFT predicts that systems with a global $Z(3)$ symmetry should have first-order transitions. The symmetry allows a general effective potential depending on the cubic invariant $\Re e L^3$, which we can then expand around the origin, choosing $L$ to be real:

$$V(L) = f(|L|^2) + g(|L|^2, \Re e L^3) = aL^2 + bL^3 + cL^4 + \dots$$

A sufficiently large value of $a$ will give a unique minimum at $L = 0$, with a metastable minimum at $L \neq 0$. The true minimum will jump to this second value when $a$ is reduced towards zero, so the cubic term has produced a first-order transition. This argument has been confirmed for $d \geq 4$, but is invalidated for $d = 3$ by topological excitations. For our interests, $SU(3)$ in $d = 3$ dimensions should have the same critical behaviour as the Z(3) Potts spin system, *viz.* a second-order transition described by the following critical exponents:

Table 2.1: Critical Exponents

| Exponent | Definition | Regime | Value |
|---|---|---|---|
| *Magnetisation* | $< L(\mathbf{x}) > \sim \|T - T_c\|^{\beta_M}$ | $T \geq T_c$ | $\beta_M = 0.11$ |
| *Susceptibility* | $\chi = \sum \Gamma(\mathbf{x}) \sim \|T - T_c\|^{-\gamma}$ | $T \leq T_c$ | $\gamma = 1.44$ |
| *Specific heat* | $C = T\partial^2 \ln Z/\partial T^2 \sim \|T - T_c\|^{-\alpha}$ | $T \leq T_c$ | $\alpha = 0.33$ |
| *Correlation length* | $\xi = T/\sigma \sim \|T - T_c\|^{-\nu}$ | $T \leq T_c$ | $\nu = 0.83$ |
| *Critical correlation* | $\Gamma(\mathbf{x}) \sim \|\mathbf{x}\|^{-(d-2+\eta)}$ | $T = T_c$ | $\eta = 0.27$ |

The critical coupling, for $SU(3)$ theory in $2 + 1$ dimensions, has previously been estimated[33] as

$$\beta_c = 8.175 \pm 0.002 \tag{2.22}$$

in the infinite-volume limit of a lattice with $N_t = 2$, from measurements of $\beta_c \approx 8.04, 8.11, 8.12$ and $8.14$ for $L_z = L_x = 20, 30, 40$ and $60$ respectively. The coupling is related to the temperature, as previously, by $\beta = 12T/g^2$.

It has also been predicted[30] that

$$< L >= (m^2a^2)^{(N^2-1)g^2/16\pi T_c}, \qquad T \to T_c^+$$

and, leading from this, that

$$\beta_M = \frac{(N^2-1)g^2}{16\pi T_c} = \frac{6}{\pi\beta_c} \approx 0.23 \qquad \text{for } SU(3),$$

but this is based on an approximate analytic calculation predating universality arguments, and so cannot be treated as reliable.

# Chapter 3.

# Monte-Carlo Techniques & Results at High Temperatures

## 3.1. Computer Simulation (2+1 dimensions)

In order to test the predictions of the previous chapter for the properties of the $Z(3)$ interface, it is possible to simulate pure gauge $SU(3)$ theory on a computational lattice. In what follows, we have simulated the theory in $2+1$ dimensions, *i.e.* two space and one Euclidean time dimensions, as in fig. 2.3. One would prefer to work in the real world of $3+1$ dimensions, of course, but it should be remembered that the interface profile is a one-dimensional object, interpolating between one $Z(3)$ vacuum at $z = 0$ and another at $z = L_z$. All other spatial dimensions are therefore transverse, simply specifying the dimensions of the interface itself: in $2+1$ dimensions, the interface is one-dimensional (a string), and in $3+1$ dimensions, it is two-dimensional (a sheet). An interface in a $(2+1)$-dimensional system, as well as being worthy of study in its own right, may tell us much about properties in the higher dimension. The most obvious reason to work in $2+1$ dimensions, apart from the benefit to be had from studying the simpler, one-dimensional interface, is that by working in three Euclidean dimensions rather than four, we are able to considerably reduce the computational time needed for results of a desired accuracy.

The $SU(3)$ system with $2+1$ dimensions provides an extremely useful testing ground, and is interesting in its own right. However, it differs from a similar system in $3+1$ dimensions in at least one fundamental respect: the nature of the phase transition at the critical temperature. In $2+1$ dimensions, the transition is second order, meaning that the Polyakov line diverges smoothly from zero in the disordered phase, at and below the critical temperature, to its non-zero values in the different $Z(3)$ phases above $T_c$. Thus, an interface present in this theory above the critical temperature would be expected to shrink continuously to nothing as $T \to T_c$. However, the phase transition in the pure glue theory has been confirmed[34] to be *first* order in $3+1$ dimensions, so that there is a discontinuity in the Polyakov line at the critical temperature, *i.e.* a jump from zero to a non-zero value as the temperature increases through $T_c$, followed by a further smooth change away from zero as the $T \to \infty$. Hence, while many inferences may be drawn from results in 2+1 dimensions, it is not clear that properties such as those describing the collapse of the interface at $T_c$ will apply equally to $3+1$ dimensions. It should be borne in mind, though, that a second-order phase transition in $3+1$ dimensions *is* seen in an $SU(2)$ system, and so those properties which do not carry directly over to an $SU(3)$ system in $3+1$ dimensions may, instead, be relevant to an $SU(2)$ system.

In the computer simulation of pure gauge $SU(3)$ theory, the fundamental variables are simply the link variables of (2.9). The variable $U_n^\mu$ is attached to the link leaving site $n$ in the $\mu$ direction, where $\mu = 0, 1, 2$, corresponding to our one time and two space dimensions, and now $U \in SU(3)$. In an obvious extension of notation, $U_n^{\mu\dagger} = U_{n+\hat{\mu}}^{-\mu}$. We can construct physical observables from these link variables, as in (2.13), and then use Monte-Carlo procedures to compute a statistical average over all possible configurations of the system. Leaving the details of Monte-Carlo techniques used for the next section, we first deal with the creation of an interface on the lattice.

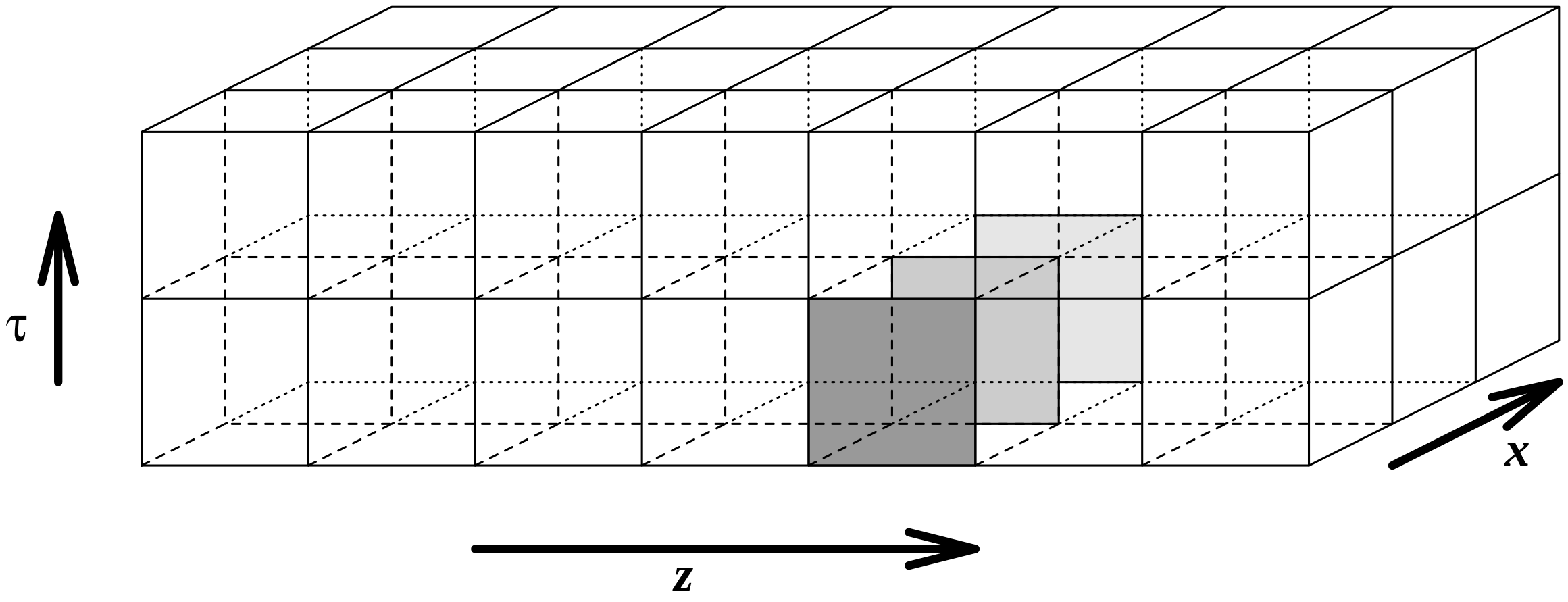

**Fig. 3.1:** The Euclidean lattice in $2+1$ dimensions: there are $N_t$ links in the time dimension, $\tau$; the longitudinal space dimension, $z$, is considerably longer than the transverse dimension, $x$. The shaded plaquettes indicate the position of the twist, when present.

The type of lattice used is shown in fig. 3.1, with two space and one (Euclidean) time dimensions. To study the phases of the system, we construct a Polyakov line at each point in the space dimensions (the $x - z$ plane), using (2.14). We impose periodic boundary conditions in all directions, and set the temperature, specified through the lattice parameter, to be $\beta \gg \beta_c$, where $\beta_c$ corresponds to the temperature of the deconfinement phase transition of (2.22). Left to its own devices, the entire system will settle into just one of the $Z(3)$ phases, this being the configuration of lowest energy. An example of such a configuration, as may be verified from (2.11), is for all $U_n^\mu = 1$. We need some way to force the system into producing a $Z(3)$ interface, and this is achieved by judicious use of a "twist"[8].

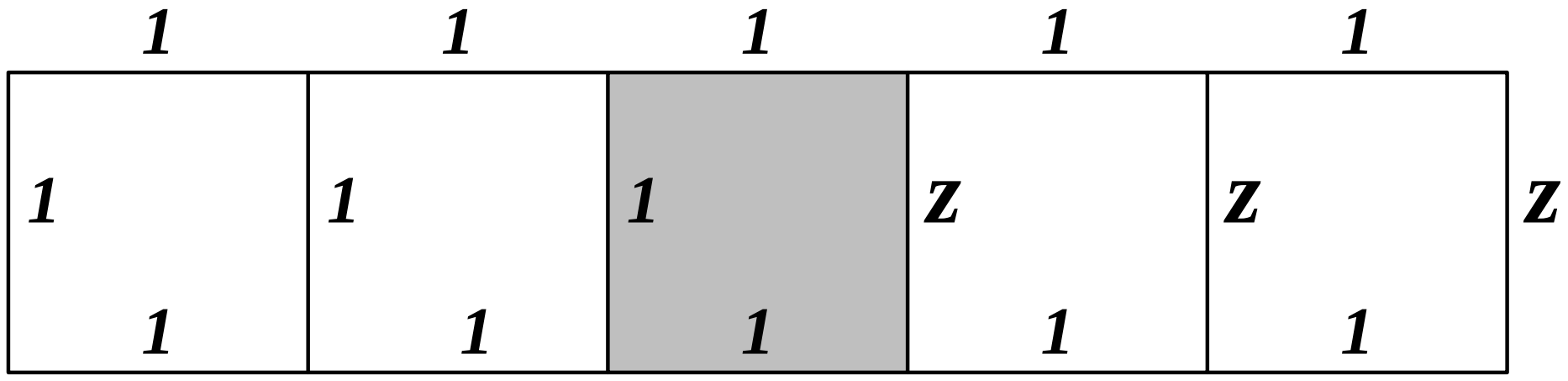

**Fig. 3.2:** The effect of the twist: part of the $\tau - z$ plane of fig. 3.1 for fixed $\tau, x$. Polakov lines to the left of the shaded (twisted) plaquette show a different $Z(3)$ phase from those to the right.

The twist is introduced by modifying the action (2.11) so that every plaquette in the $\tau - z$ plane at some particular value of $\tau$ and $z$ is to be premultiplied by an element, $z^{-1}$, of the centre group when it appears in the action. This is illustrated in fig. 3.1, with the twisted plaquettes shaded. The action becomes

$$S_W = \frac{6}{g^2}\left( \sum_{shaded\ \mathbb{P}} (1 - \tfrac{1}{3}\Re e(z^{-1}\mathrm{Tr}U_{\mathbb{P}})) + \sum_{other\ \mathbb{P}} (1 - \tfrac{1}{3}\Re e\mathrm{Tr}U_{\mathbb{P}})\right),$$

where $\mathbb{P}$ represents a plaquette ($U_{\mathbb{P}}$ is the product of fields around a plaquette, as in (2.10)), and $z$ is a member of the centre group: $z \in Z(3)$. Whereas the original action favours all $U = 1$ at high temperatures, so that $U_{\mathbb{P}} = 1$, the twisted part now favours $U_{\mathbb{P}} = z$ to compensate for the pre-factor. This means that if we order the system so that all $U = 1$ to the left of the twist, giving Polyakov lines in the $\mathbb{Z}(3)$ phase given by $< L >= 1$, then the twist will cause the link variables around the twisted plaquettes to behave as in fig. 3.2, with the right-hand time-like link given by $U = z$. All equivalent time-like links to the right of the shaded plaquette at the same $\tau$ will also take this value, so that so that $U_{\mathbb{P}} = 1$ again, making $< L >= z$ to the right of the twist. No extra energy is associated with the twist itself; it merely acts as a change of variables, transforming one $Z(3)$ vacuum into another. However, the periodic boundary conditions force the appearance of a real physical $Z(3)$ interface somewhere else in the $z$ direction, where the Polyakov lines interpolate from $z$ back to 1. Thus, we have forced the creation of two $Z(3)$ domains on the lattice, with one $Z(3)$ interface between them. We shall place the twist at the boundary in our simulations, so that the left-hand end of the system is in one phase and the right-hand end in another.

One might ask why the interface should form along the $z$ direction rather than the $\tau$. In our simulation, the fact that $N_t = 2$ precludes the formation of an interface in the $\tau$ direction; the dimension is simply not long enough for any interpolation to occur. Thus, thermodynamics favour an interface in the $z$ direction, as long as $\beta_T \ll L_z$.

We thus set the initial conditions of our simulation to be that all link variables are unity except those in $\{\tau = 0 \ , \ 0 \leq z \leq L_z/2 \ , \ x\}$, where they are $e^{2\pi i/3}$. The system is then allowed to equilibrate before measurements start, using the Monte-Carlo techniques described below.

## 3.2. Monte-Carlo Techniques

THE object of Monte-Carlo methods [26] is to evaluate the expectation value of a quantity not by means of a path integral of high dimension,

$$< \mathcal{O} >= \frac{\int \mathcal{D}U \mathcal{O} e^{-S}}{\int \mathcal{D}U e^{-S}},$$

but by a statistical average over gauge configurations $\{U\}_i$, generated according to the Boltmann distribution $e^{-S}$:

$$< \mathcal{O} >= \frac{1}{i_{max}} \sum_{i=1}^{i_{max}} \mathcal{O}(\{U\}_i). \tag{3.1}$$

Each $\{U\}_i$ represents a set of link variables, $U$, on all links of the lattice. This technique is one of "importance sampling", where the effort has shifted from the evaluation of the path integral, with many configurations contributing little, to the generation of $\{U\}_i$ with the Boltzmann weight. The statistical fluctuations in the mean value for $\mathcal{O}$ fall as $1/\sqrt{i_{max}}$ as $i_{max} \to \infty$. In practice, many tens of thousands of configurations are generally needed for a good estimate.

In the "heat-bath" algorithm, one starts with a configuration $\{U\}$, and then changes one of the link variables to generate a new configuration $\{U'\}$. This new configuration replaces the old if the rules of the algorithm are satisfied. One "sweep" of the lattice is accomplished by applying this procedure to each link in turn. The object of the exercise is to generate, first of all, a configuration at thermal equilibrium: $\{U\}_1$, a member of the Boltzmann distribution. One then sweeps the lattice again until a new, statistically independent member of the Boltzmann distribution is generated: $\{U\}_2$. Repeating this procedure many times, and calculating $\mathcal{O}(\{U\})$ for each, one obtains $< \mathcal{O} >$ through (3.1).

The heat-bath algorithm works by touching a heat bath to each link variable in turn. Consider a particular link $l$ with link variable $U$: all the links which interact with $l$ have fixed $U$ matrices, providing a background. On $l$, one chooses $U'$ from the gauge group $\mathcal{G}$ with a probability density proportional to the Boltzmann factor:

$$dP(U') = e^{-\beta S(U')} dU'. \tag{3.2}$$

This algorithm satisfies detailed balance, ensuring that any ensemble will be brought to equilibrium eventually. Moreover, since any other algorithm which varies one link at a time will simulate the heat bath after repeated application, this method must be the fastest to attain equilibrium in terms of the number of iterations. The challenge, therefore, is to ensure that the computational time necessary for a heat-bath iteration is on a par with that of alternative algorithms.

A suitably fast algorithm for $SU(2)$ was originally suggested by Creutz[35], and then an alternative by Kennedy and Pendleton[36]. One can write the Wilson action (2.11) in the form

$$\beta S_W(U) = (\text{constant}) - \beta' \Re e \text{Tr}(\Sigma h), \tag{3.3}$$

with $\beta' = \beta/3$, $\Sigma$ a complex $2 \times 2$ matrix equal to the sum of products of link variables neighbouring $l$, and $h$ the $SU(2)$ matrix (link variable) associated with the link $l$ to be updated. Thus, from (3.2), the problem reduces to the generation of an $SU(2)$ matrix obeying the distribution

$$Q(h)(dh) \propto e^{\beta' \Re e \text{Tr}(\Sigma h)}(dh),$$

where $(dh)$ is the group-invariant Haar measure. We can parametrise all $2 \times 2$ matrices in terms of the $2 \times 2$ unit matrix, $\mathbb{I}$, and the vector of Pauli spin matrices, $\sigma$:

$$a = a_0 \mathbb{I} + i\mathbf{a} \cdot \sigma, \qquad (a_\mu \in \mathbb{R}, \quad a^2 = a_0^2 + \mathbf{a} \cdot \mathbf{a} = 1).$$

In fact, we define $a = uh$ and $(u_0 = \xi^{-1} \Re e \Sigma_0,\ \mathbf{u} = \xi^{-1} \Re e \mathbf{\Sigma})$, where $a, u \in SU(2)$. In this notation, $\Re e \text{Tr}(\Sigma h) = \xi \text{Tr} a$. Furthermore, the Haar measure takes a simple form, which can be rewritten in Polar coordinates:

$$(da) = \delta(1 - a^2) d^4 a = \tfrac{1}{2} \sqrt{1 - a_0^2}\ \delta(r - \sqrt{1 - a_0^2})\ \sin\theta\, da_0\, dr\, d\theta\, d\phi.$$

Thus, we need to generate $a_0$ with the (normalised) probability distribution

$$P(a_0) da_0 = \frac{\alpha}{\pi I_1(\alpha)} \sqrt{1 - a_0^2}\ e^{\alpha a_0} da_0,$$

where $\alpha = 2\beta' \xi$, the factor of two coming from $\text{Tr}\mathbb{I}$, and $I_1(\alpha)$ is a modified Bessel function. We then generate $a_i$ uniformly on a two-sphere of radius $\sqrt{1 - a_0^2}$.

The algorithm of Creutz generates such $a_0$ values with an exponential distribution, and then imposes the square-root factor by means of an accept/reject procedure. However, for our purposes, $\alpha$ is generally large, so that the normalisation pre-factor is small; this leads to a high rate of rejection. The Kennedy-Pendleton algorithm improves on this by working in terms of a new variable $\delta = \sqrt{1 - a_0}$, so that

$$P'(\delta) d\delta \propto \sqrt{1 - \tfrac{1}{2}\delta^2}\ \delta^2 e^{-\alpha \delta^2} d\delta, \qquad (0 \leq \delta \leq \sqrt{2}).$$

This algorithm is faster than that of Creutz for $\alpha \geq 1.7$. It can be implemented as follows [36]:

- Generate four uniformly distributed pseudo-random numbers in the unit interval: $R, R', R'', R'''$;
- Set $X = -\frac{1}{\alpha}\ln R, \quad X' = -\frac{1}{\alpha}\ln R', \quad C = \cos^2(2\pi R'')$;
- Set $\overline{\delta} = X' + XC$;
- If $R'''^2 > 1 - \frac{1}{2}\overline{\delta}$, start again; otherwise, set $a_0 = 1 - \overline{\delta}$.

The extension of this algorithm to produce $N \times N$ unitary matrices with distribution (3.2) becomes more and more complicated as $N$ increases, reducing the speed of the procedure, and favouring alternative algorithms to the heat bath, such as the Metropolis method[37]. However, Cabibbo and Marinari[38] proposed a faster algorithm for $SU(N)$ based on updating $SU(2)$ subgroups. Their method consists of first selecting a set $\{F : SU(2)_k, \quad k = 1, \ldots, N-1\}$ of $SU(2)$ subgroups of $SU(N)$ such that there is no left ideal, *i.e.* only the whole $SU(N)$ group is invariant under left multiplication by $F$. The choice of $F$ that we use has elements of the form

$$a_k = \begin{pmatrix} 1 & & & & & & \\ & \ddots & & & & & \\ & & 1 & & & & \\ & & & \alpha_k & & & \\ & & & & 1 & & \\ & & & & & \ddots & \\ & & & & & & 1 \end{pmatrix},$$

where $\alpha_k$ is an $SU(2)$ matrix located at the $k$th and $(k+1)$th rows and columns. In each step of the iteration, the new link variable is obtained by premultiplying the previous value by one matrix from each of the $N-1$ subgroups:

$$U' = a_{N-1}a_{N-2}\ldots a_1 U, \qquad a_k \in SU(2)_k, \qquad (k = 1, \ldots, N-1).$$

We obtain the matrices $a_k$ by using the Kennedy-Pendleton algorithm to choose the appropriate $\alpha_k$. It was shown in [37] that $U'$ emerges with a Boltzmann distribution if $U$ has one; thus, the procedure is ergodic.

The Monte-Carlo study of lattice gauge theory often suffers from slow evolution of physical variables, owing to large correlations between configurations separated only by a few sweeps. Near a second-order phase transition, "critical slowing down" means that the autocorrelation time of the Monte-Carlo evolution diverges as the critical point is approached and physical correlation lengths increase. One way of circumventing this problem is to use "over-relaxation" to accelerate the heat-bath evolution.

This use of over-relaxation was first suggested by Adler[39] for fields with quadratic coupling in the action. Brown & Woch[40] and Creutz[41] extended the procedure to more general theories. The idea is to pick a trial change for a link variable so that the system goes to an area of phase space as far as possible from that of the original value without increasing the energy by a large amount. This is done by finding roughly the locus of minimum energy for the variable, and then picking a trial value on the "opposite" side of this minimum. Consider a trial change of link variable $h \to h'$. If a group element $h_0$, independent of $h$, minimises the system energy, then we wish $h'$ to be on the "opposite side" of $h_0$ to $h$, for instance

$$h' = h_0 h^{-1} h_0.$$

One can use a Metropolis algorithm to accept the new value if the configuration energy is unchanged or reduced by the change $h \to h'$, and accept it with probability $e^{-\Delta S}$ if it is increased.

For an action of the form (3.3), a natural choice for $h_0$ is the inverse of the group element obtained by projecting $\Sigma$ onto the group $SU(2)$. For $SU(2)$, $\Sigma$ is always proportional to a group element, $\widetilde{\Sigma}$. The energy of the new configuration will be the same as that of the old for $SU(2)$, so the Metropolis algorithm will always accept this change; the algorithm is deterministic and microcanonical. To apply this to our $SU(3)$ system, we again use the $SU(2)$ subgroup procedure of Cabibbo and Marinari, over-relaxing each of the subgroups of our $SU(3)$ matrices in turn. A previous study[42] suggested that for $SU(3)$ gauge theory in $3+1$ dimensions below the critical temperature, a mixture of seven Brown-Woch over-relaxation steps followed by one heat-bath step appeared optimal for reducing the autocorrelation time. Above the critical temperature, their results did not indicate any particular optimal ratio. For our simulations, therefore, there appears to be no particular optimal ratio to choose; we use seven Brown-Woch steps for every one heat bath, after many initial heat-bath sweeps to equilibrate the system. The link matrices are renormalised after each sweep, to ensure that rounding errors do not cause violation of their unitarity. Measurements of physical operators take place only every four sweeps, to reduce the correlations caused by adjacent configurations.

## 3.3. The Wandering Interface

We can see in fig. 3.3 an interface on a $2 \times 24 \times 96$ lattice ($N_t \times N_x \times N_z$). The longitudinal direction goes across the page, and the transverse direction into it. The "height" in each case is determined by the value of the Polyakov line, and has nothing to do with the time dimension, which has been integrated over to form the Polyakov line. The twist has been placed at the boundary of the $z$ direction, as mentioned before, so the left-hand end of the lattice is in one $Z(3)$ phase with $< L > \approx e^{2\pi i/3}$, and the right-hand end in another with $< L > \approx 1$. The temperature is fairly high: $\beta = 50 \gg \beta_c$. Thus, the interface is fairly well-defined and stable, since its free energy is correspondingly high, suppressing the formation of fluctuations in phase. In fig. 3.4, the same system is seen *without* a twist.

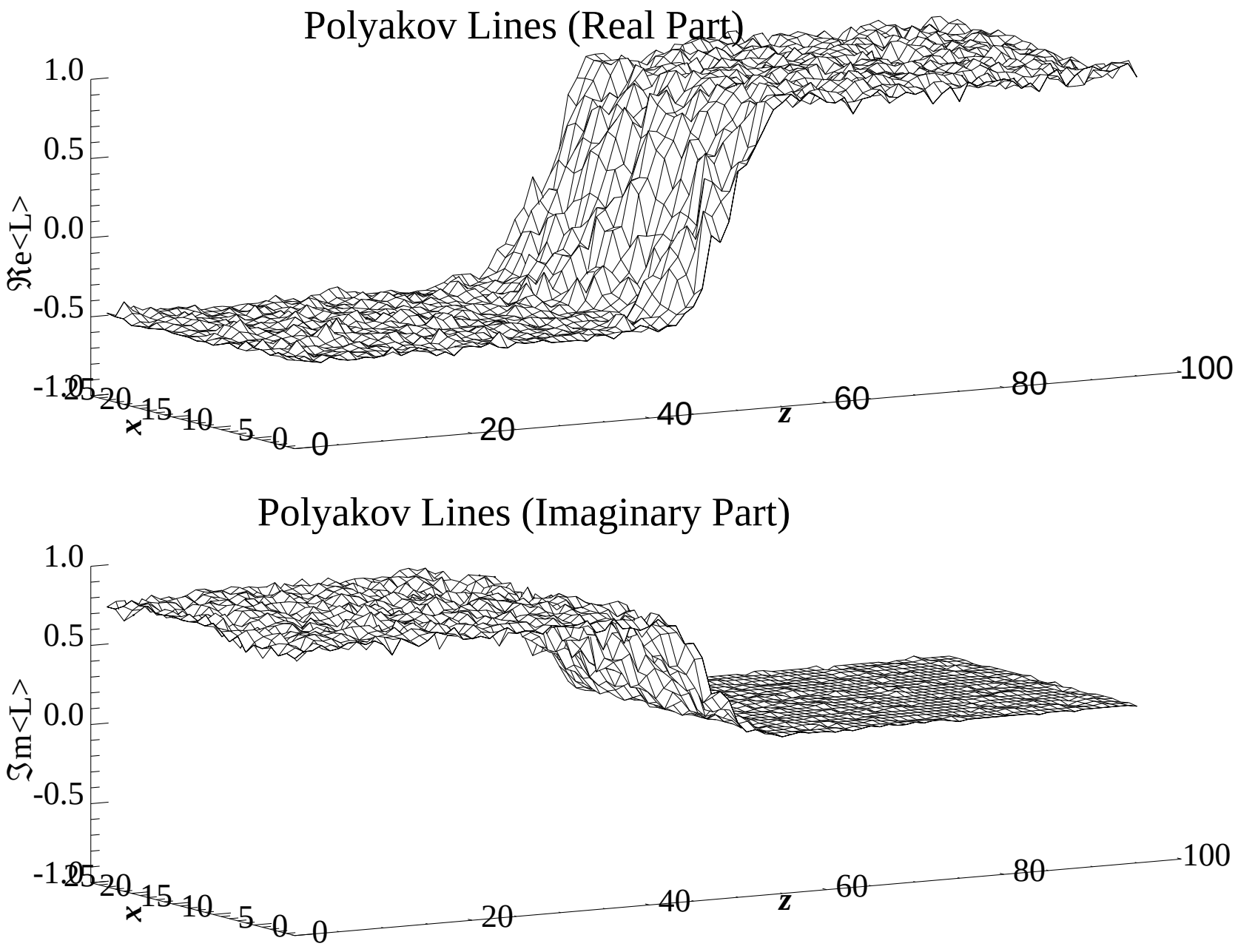

**Fig. 3.3:** The real (top) and imaginary (bottom) parts of Polyakov lines on a $2 \times 24 \times 96$ lattice with twist. The interface can clearly be seen between the $< L >\approx e^{2\pi i/3}$ domain on the left and the $< L >\approx 1$ domain on the right. The twist is at the boundary between $z = 0$ and $z = L_z$, and should not be confused with the position of the interface. The temperature is fairly high ($\beta = 50$), resulting in a high suppression of the fluctuations in phase.

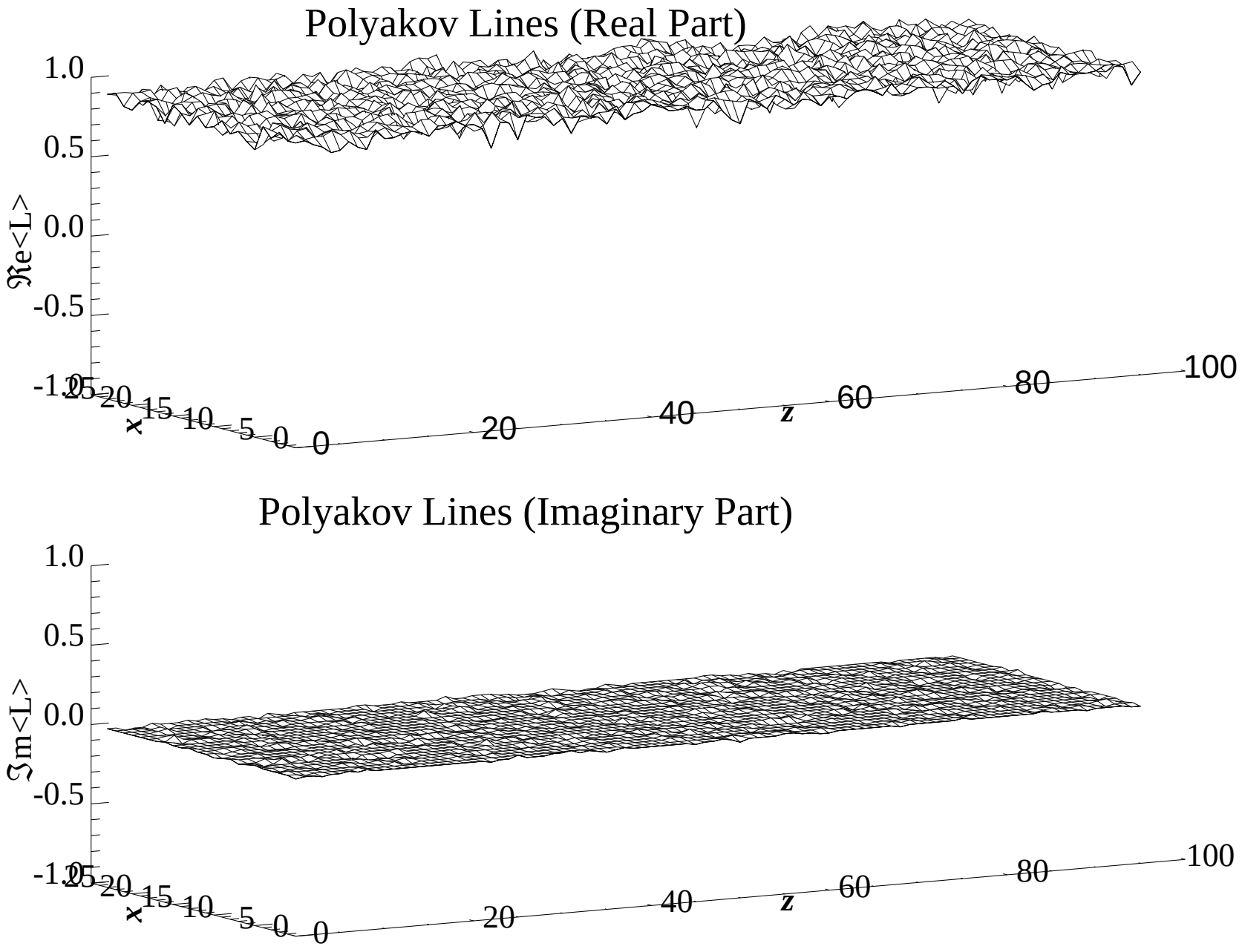

**Fig. 3.4:** The real (top) and imaginary (bottom) parts of Polyakov lines on a $2 \times 24 \times 96$ lattice without twist, also with $\beta = 50$.

In this picture, the interface is roughly in the centre of the lattice. However, the interface is actually translation invariant; there is nothing in the action to specify that it should be located at any particular value of $z$. We shall always start the interface in the centre of the lattice, as mentioned in section 2.1, but for this chapter, we shall allow it to wander freely along the lattice. We can keep track of the interface most easily by examining a transverse-averaged profile of the Polyakov lines.

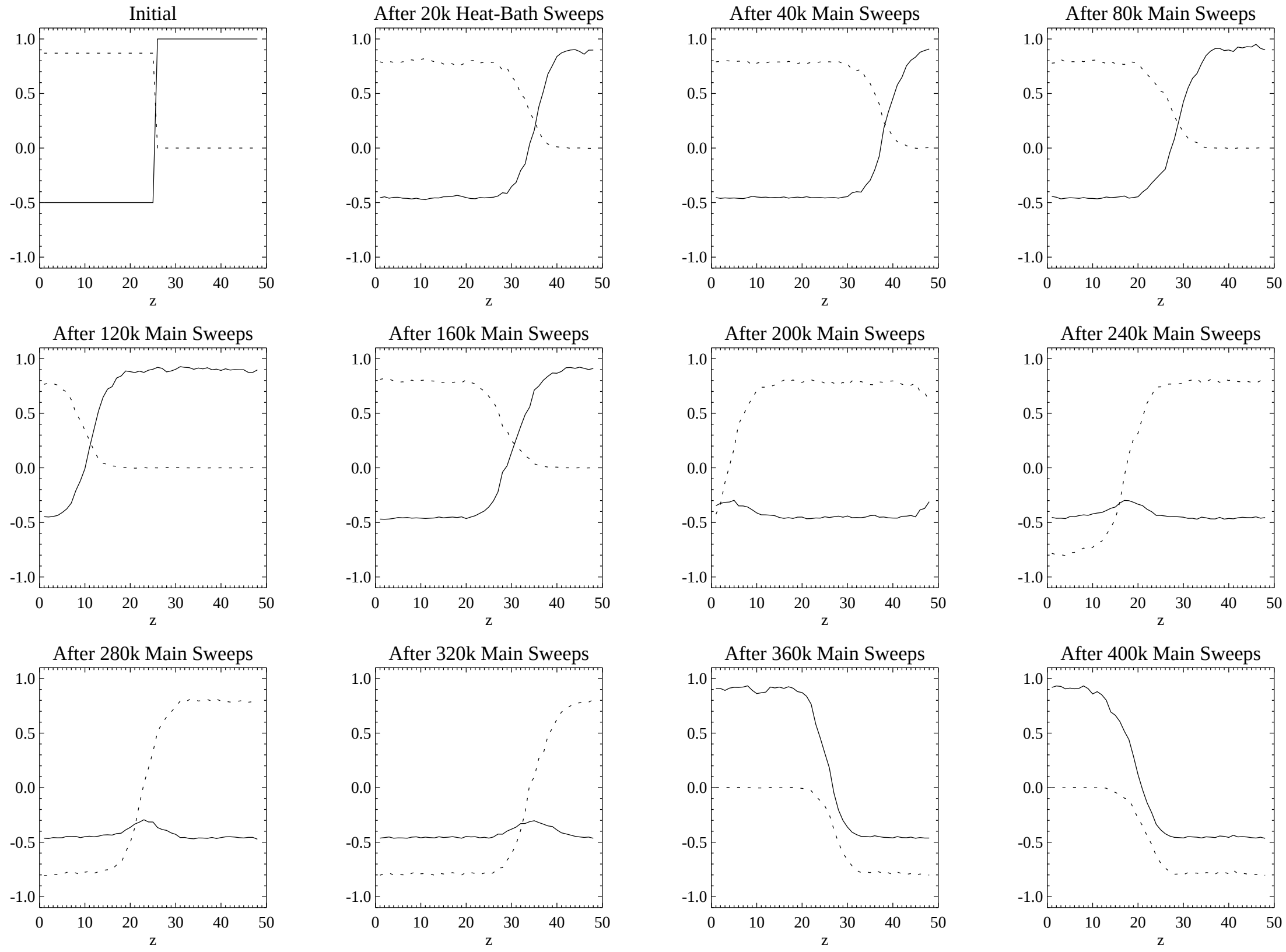

**Fig. 3.5:** Profiles of the Polyakov lines on a $2\times16\times48$ lattice for $\beta = 50$, averaged across the transverse direction. The real (solid) and imaginary (dotted) parts of the Polyakov lines are shown. The first picture shows the initial configuration before Monte-Carlo evolution begins; the second, the configuration after 20k heat-bath sweeps; subsequent ones, 40k sweeps apart.

We can see the interface walking along the lattice in fig. 3.5. Placed at the centre of the lattice before Monte-Carlo evolution begins (first picture), the interface has taken up its shape of least energy after twenty thousand ("20k") equilibration sweeps (second picture). It has also moved along the lattice, and continues to do so in subsequent pictures, which are 40k sweeps apart. In particular, the interface often approaches the twist. In the 200k picture, the interface is passing through the twist, and in the 240k picture, the system has tunnelled into a different $Z(3)$ vacuum: now, $< L >= e^{2\pi i/3}$ on the left and $< L >= e^{-2\pi i/3}$ on the right. After 260k sweeps, the system has tunnelled again.

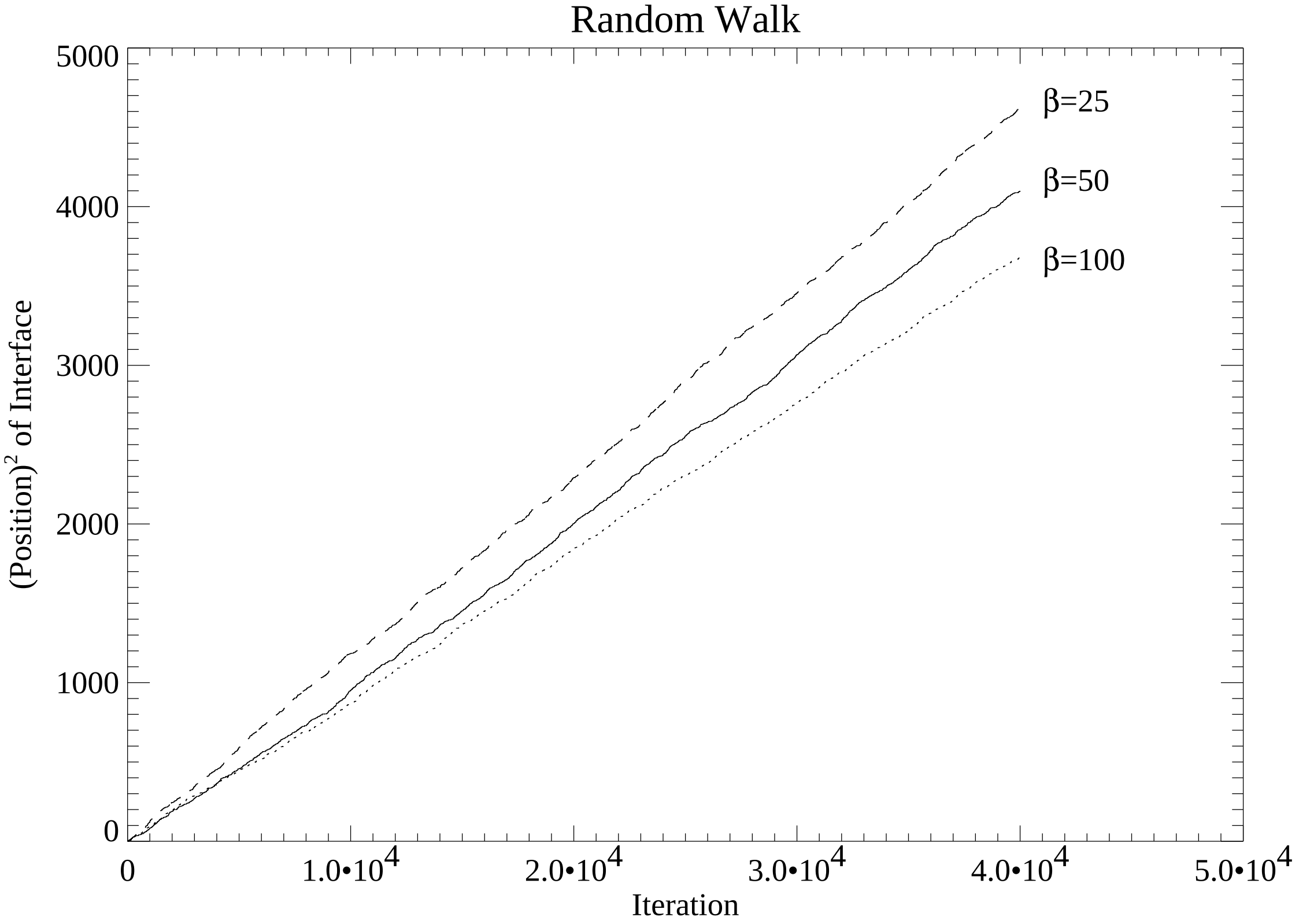

**Fig. 3.6:** The movement of the interface with computer 'time', for a $2 \times 24 \times 96$ lattice and various $\beta$.

The movement of the centre of the interface as proceeds is shown in fig. 3.6, a plot of the square of the position of the interface centre against the iteration number. To make this particular measurement, we allow the interface to move each sweep, but then slide the whole gauge configuration along the lattice so that the interface is put back in the middle of the $z$ axis; essentially, we have put the interface on a treadmill, so that it is allowed to move, without really going anywhere. This technique will be used throughout the next chapter, and will be described in more detail there. It is used here simply to cure problems that would arise in identifying the interface as it passed through the twist.

In a random walk with step size $a$, the lattice spacing in our case, the probability that at time $t$ the object will be at position $z$ is

$$P(z,t) = \frac{1}{2}\sqrt{\frac{2}{\pi cat}}e^{-z^2/2cat}2a,$$

implying that

$$< z^2(t) >= ca^2 t,$$

where the speed of movement is given by $c$. The interface, from fig. 3.6, clearly obeys this linear relation for all the values of $\beta$ measured, demonstrating that it is indeed executing a random walk, in accordance with translation invariance. As the temperature decreases, the speed of the walk is seen to increase slightly. This is also expected, as the energy penalty suppressing fluctuations in phase will decrease as the temperature drops, allowing the interface to move more swiftly by combining with, and budding off, fluctuations nearby. We shall look again at the random walk of the interface in the next chapter, at much lower temperatures.

## 3.4. The Surface Tension

The free energy of a system is $F = -T \ln Z$, as in (1.3), so

$$\frac{\partial}{\partial\beta}\left(\frac{F}{T}\right) = \frac{1}{\beta} < S(U) > .$$

This gives us the behaviour of the free energy in terms of the average plaquette action. We can perform simulations for the same lattice with and without twisted plaquettes, *i.e.* with and without a $Z(3)$ interface, and the difference in free energy between the two gives us the free energy of the interface itself, from which we can measure its surface tension:

$$F_{interface} = F_{twisted} - F_{untwisted} = \alpha A,$$

where $\alpha$ is the surface tension, and $A$ is the 'area' of the interface (in our case, the transverse size of the lattice). Thus,

$$\frac{\partial}{\partial\beta}\left(\frac{\alpha A}{T}\right) = \frac{1}{\beta} < \Delta S(U) > .$$

Therefore, recalling (2.5) and substituting $A = L_x a$,

$$\frac{\Delta S}{\beta} = \frac{\partial}{\partial\beta}\left(\frac{\alpha_0 T^{2.5}}{g}\frac{L_x a}{T}\right) = \frac{L_x}{2\sqrt{6\beta}\; N_t^{1.5}}\,\alpha_0, \tag{3.4}$$

where we have used (2.8) and (2.12) to substitute for $T$ and $g$.

The prediction (2.17) that $\alpha_0 = 9.82$ for a system with $N_t = 2$ was made in the infinite-temperature limit. Hence, to test this prediction it is necessary to estimate $\alpha_0$ in this limit. The best way to do this is to carry out a survey of $\alpha_0$ for various values of $\beta$, and then to extrapolate the results to the infinite limit.

To estimate $\alpha_0$ for a particular $\beta$, we need to measure the difference in total plaquette action between two simulations which are identical except for the presence or absence of a twist. Several tens of thousands of heat-bath sweeps were used to ensure complete equilibration of the lattice before measurements began. "Main sweep" refers to a sweep after equilibration is complete. As mentioned in section 3.2, the ratio of over-relaxed to heat-bath sweeps was seven to one, and measurements of physical quantities were taken every four sweeps. A DEC 2100 A500MP machine was used for most of the computation, and the simulation details are listed in the following table. The CPU time listed was needed for runs with or without twist, and therefore needs to be doubled to give the total time required for an estimate of $\alpha_0$.

Table 3.1: Simulation Details

| $\beta$ | Lattice Size | Heat-Bath + Main Sweeps (000's) | CPU Time (hrs) |
|---|---|---|---|
| 16 | $2 \times 12 \times 30$ | 15 + 400 | 40 |
| 28.125 | $2 \times 12 \times 36$ | 20 + 400 | 60 |
| 50 | $2 \times 16 \times 48$ | 20 + 400 | 130 |
| 112.5 | $2 \times 24 \times 72$ | 40 + 400 | 400 |
| 200 | $2 \times 32 \times 96$ | 40 + 400 | 800 |

For each of these runs, with and without twist, we now list the results for the average action for a single plaquette, $< S^{\mathbb{P}} >$. This has to be multiplied by the total number of plaquettes ($3 \times N_t \times N_z \times N_x$) to give the total action of the configuration. The difference between the two values is then used in (3.4) to obtain the measurements of $\alpha_0$ in the last column.

Table 3.2: Simulation Results

| $\beta$ | $< S^{\mathbb{P}} >_{twisted}$ | $< S^{\mathbb{P}} >_{untwisted}$ | $\alpha_0$ |
|---|---|---|---|
| 16 | 0.17510(1) | 0.17379(1) | 13.1(1) |
| 28.125 | 0.097575(6) | 0.096832(6) | 11.79(13) |
| 50 | 0.054310(3) | 0.053916(3) | 11.12(12) |
| 112.5 | 0.0239740(7) | 0.0238074(7) | 10.58(6) |
| 200 | 0.0134549(5) | 0.0133644(5) | 10.21(8) |

One would not expect the finite size of the lattices used to have a great effect on the untwisted results, as the whole lattice will be in one phase, but it is less clear how great the effect would be in the presence of the twist. The shape of the interface may not flatten out completely at opposite ends of the longitudinal dimension if the lattice is too short, skewing our measurement of the total interface free energy. The lattice sizes above were chosen to be large enough for the finite-size effects to be negligible, by studying results for $\beta = 50$ on various sizes of lattice and scaling the spatial dimensions with the interface width, which is proportional to $\sqrt{\beta}$, as we saw at the end of section 2.1, and which we choose as our measure of physical correlation lengths. The only exception to this was the measurement for $\beta = 16$, for which the lattice could not be safely made any smaller. The results are given in the following table, each from 2+20k sweeps. Similar runs without twist gave a result of $< S^{\mathbb{P}} >_{untwisted} \approx 0.05390(1)$, and this value was assumed for each lattice in the calculation of $\alpha_0$:

Table 3.3: Finite-Size Survey: $\beta = 50$

| Lattice Size | $< S^{\mathbb{P}} >_{twisted}$ | $\alpha_0$ |
|---|---|---|
| $2 \times 16 \times 32$ | 0.05449(1) | 11.1(3) |
| $2 \times 16 \times 40$ | 0.05438(1) | 11.3(3) |
| $2 \times 16 \times 48$ | 0.05432(1) | 11.9(3) |
| $2 \times 16 \times 64$ | 0.05422(1) | 12.0(3) |
| $2 \times 24 \times 48$ | 0.05432(1) | 11.9(3) |
| $2 \times 24 \times 60$ | 0.05423(1) | 11.6(3) |
| $2 \times 24 \times 72$ | 0.05418(1) | 11.9(3) |

This suggested, given the errors, that a lattice size of $2 \times 16 \times 48$ was adequate for $\beta = 50$.

Now, as mentioned above, the prediction of (2.17) is only valid in the limit $\beta \to \infty$, so the question arises of how to extrapolate our results to test this prediction. This is really a question of calculating infra-red divergences. In $3 + 1$ dimensions, one would expect the one-loop correction to (2.17) to go like $1/\beta$ in this limit, since it was seen in (2.19) that the Debye mass, $m \sim T$. However, in $2 + 1$ dimensions, it is conceivable that a dependence on $\ln \beta$ would need to be taken into account, as this appears in the infra-red correction for the Debye mass in this case (2.20). For this reason, it is difficult to predict the form of the $\beta$-dependence. In any case, it would be quite possible for the logarithmic dependence to masquerade as a power law over a restricted range of $\beta$.

In the absence of a definite prediction for the form of the finite-$\beta$ correction, we perform the extrapolation using a best fit to a power of $1/\beta$. A computer-generated fit to the data, using the *IDL* package and its "COMFIT" function and leaving the power as a free parameter, suggests a power of 0.85 with $\chi^2 = 2.39$. This gives a $\beta = \infty$ limit of $\alpha_0 = 9.91(1)$, as extrapolated in fig. 3.8. *IDL* fits for various imposed powers give a lowest $\chi^2$ of 2.34 for a power of 0.80, with $\alpha_0 = 9.83(1)$. However, $\chi^2 \leq 2.6$ for powers between 0.75 and 0.90. Therefore, in the absence of a definite prediction for the power of $1/\beta$, we allow for a reasonable error in the fit, and use 0.85(10), the range for which extrapolations are illustrated in fig. 3.7. Thus, our infinite-temperature extrapolation for the interface tension is

$$\alpha_0 = 9.91(1)(14),$$

in very good agreement with the instanton calculation result of (2.17). The first error is associated with the fit for $1/\beta^{0.85}$, and the second with the uncertainty in that value for the power.

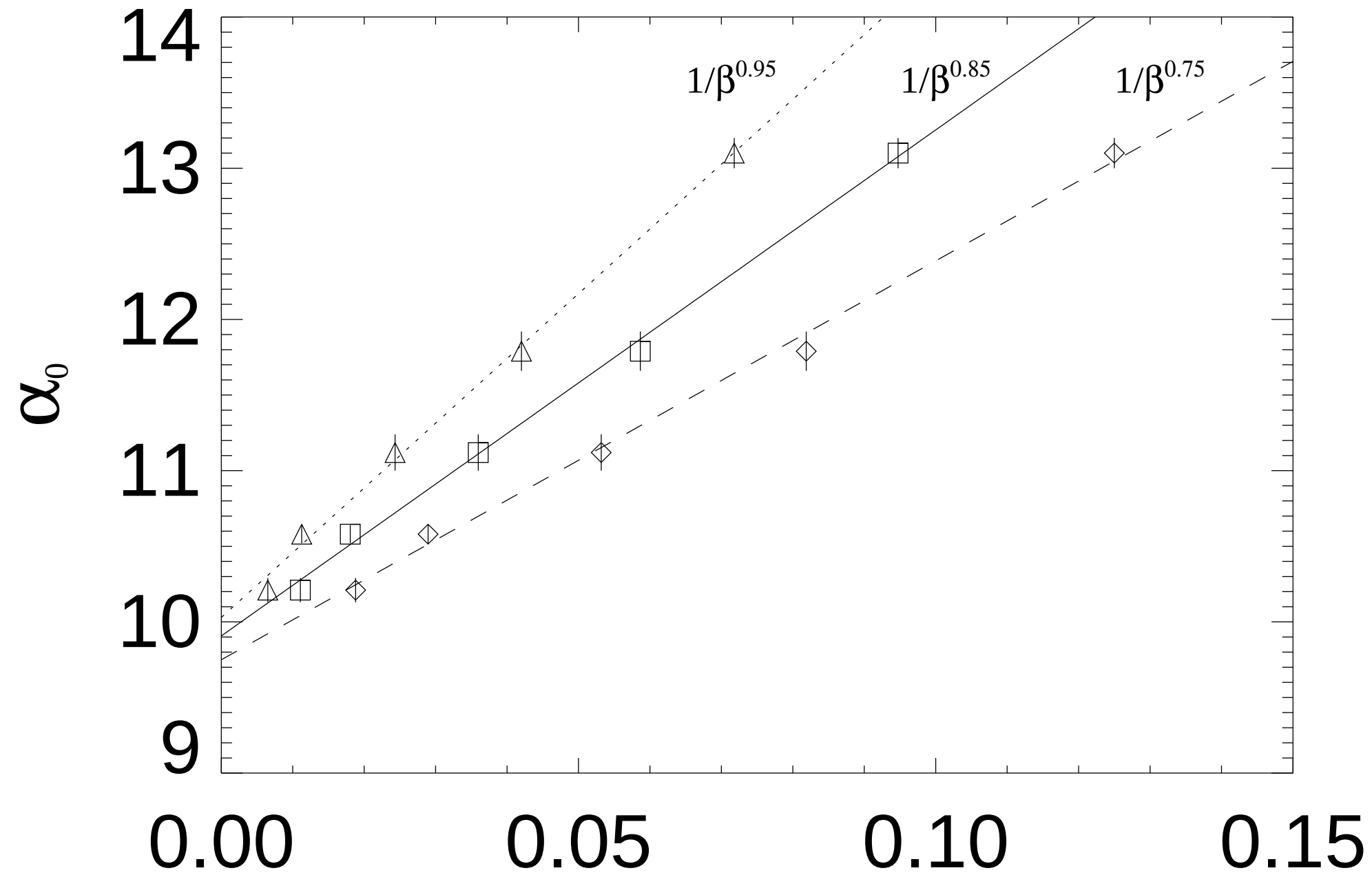

**Fig. 3.7:** Extrapolations of $\alpha_0$ are shown as $\beta \to \infty$ for various powers of $\beta$. For each power, the same five data points are plotted with a particular symbol, and the accompanying line of extrapolation is marked with a corresponding label for the x-axis.

It has been argued[13] that the different $Z(3)$ vacua distinguished by different values of the Polyakov line actually correspond to one and the same physical state. This claim comes from considering the rôle of infra-red divergences in the calculation of the surface tension, and, if true, would imply that the surface tension is actually zero (since no physical interface can exist). The agreement of our results for $\alpha_0$ with perturbation theory tends to disprove this argument. It is, of course, conceivable that our measurements for $\alpha_0$ would dramatically fall towards zero if we went to higher and higher $\beta$, meaning that our extrapolation is simply an artefact for the range of $\beta$ that we have considered. However, our highest value is $\beta \approx 25\beta_c$, a temperature well above the critical point, and it seems unlikely that any dramatically different behaviour will set in at a temperature higher than this. A sceptic could also argue that a lattice with $N_t = 2$ could show behaviour quite distinct from that in the continuum; but quite apart from the agreement with the lattice prediction of $\alpha_0$ for this value of $N_t$, a similar survey carried out for $SU(2)$ gauge theory for a range of $N_t \leq 5$ has found similar agreement with perturbation theory[43].

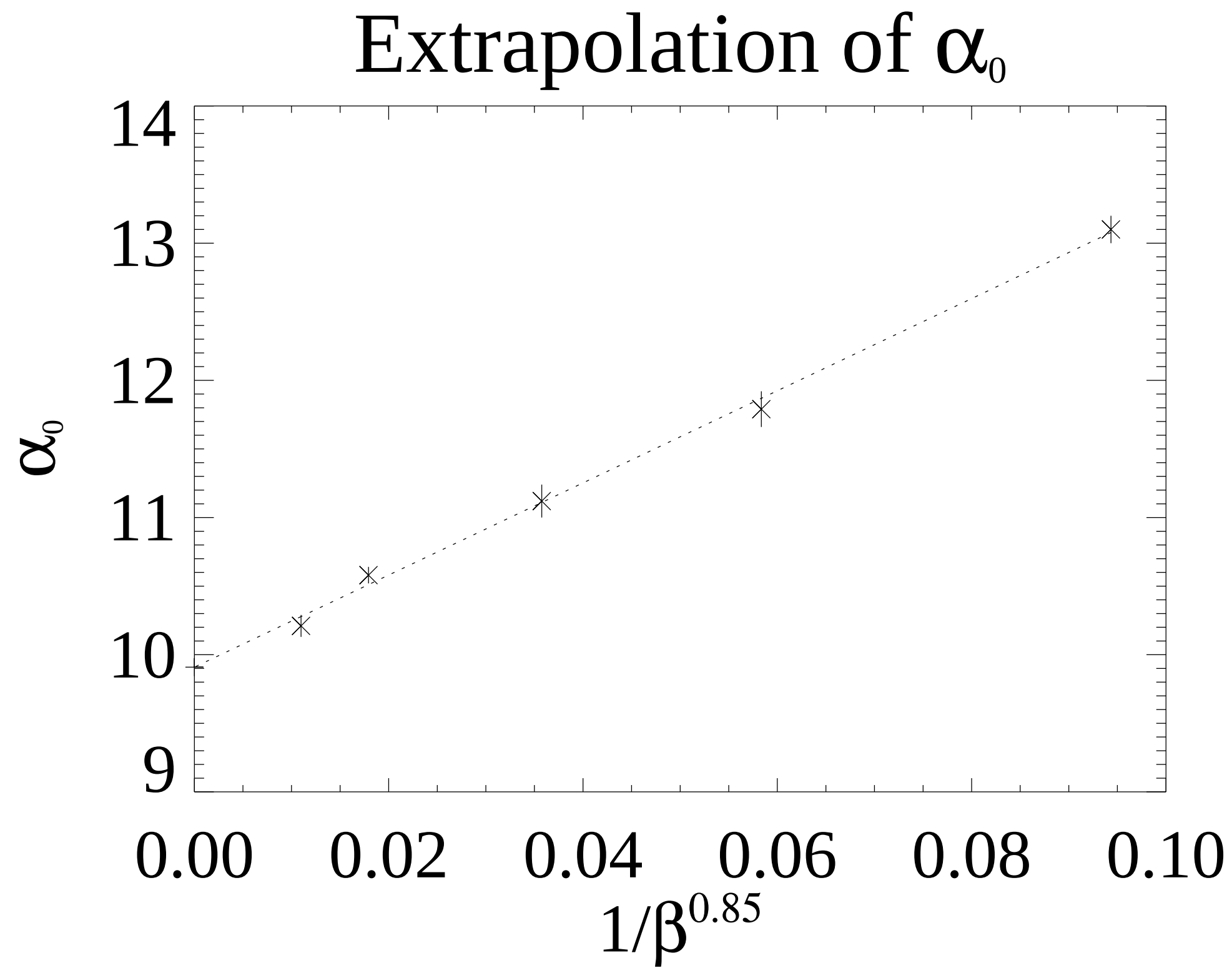

**Fig. 3.8:** Extrapolation of $\alpha_0$ for preferred power of $1/\beta$ as $\beta \to \infty$.

## 3.5. The Debye Mass

To measure the Debye mass on the computer, we need to study the correlations of fluctuations in the gauge fields. The behaviour of these correlations will give us information about the Debye screening of sources, as in (1.18). We need to find a "good" operator, *viz.* one whose correlator rapidly converges to a decaying exponential of only the lowest mass state. This will allow us to obtain, in a straightforward manner, an estimate for this lowest mass. A simple plaquette correlator does not fit our requirements, as it depends on flux loops between the plaquettes being considered, and thus has to be considered over a long range of separation before a single exponential dominates. However, a good operator turns out to be given by a transverse average of Polakov lines, as shown in fig. 3.5.

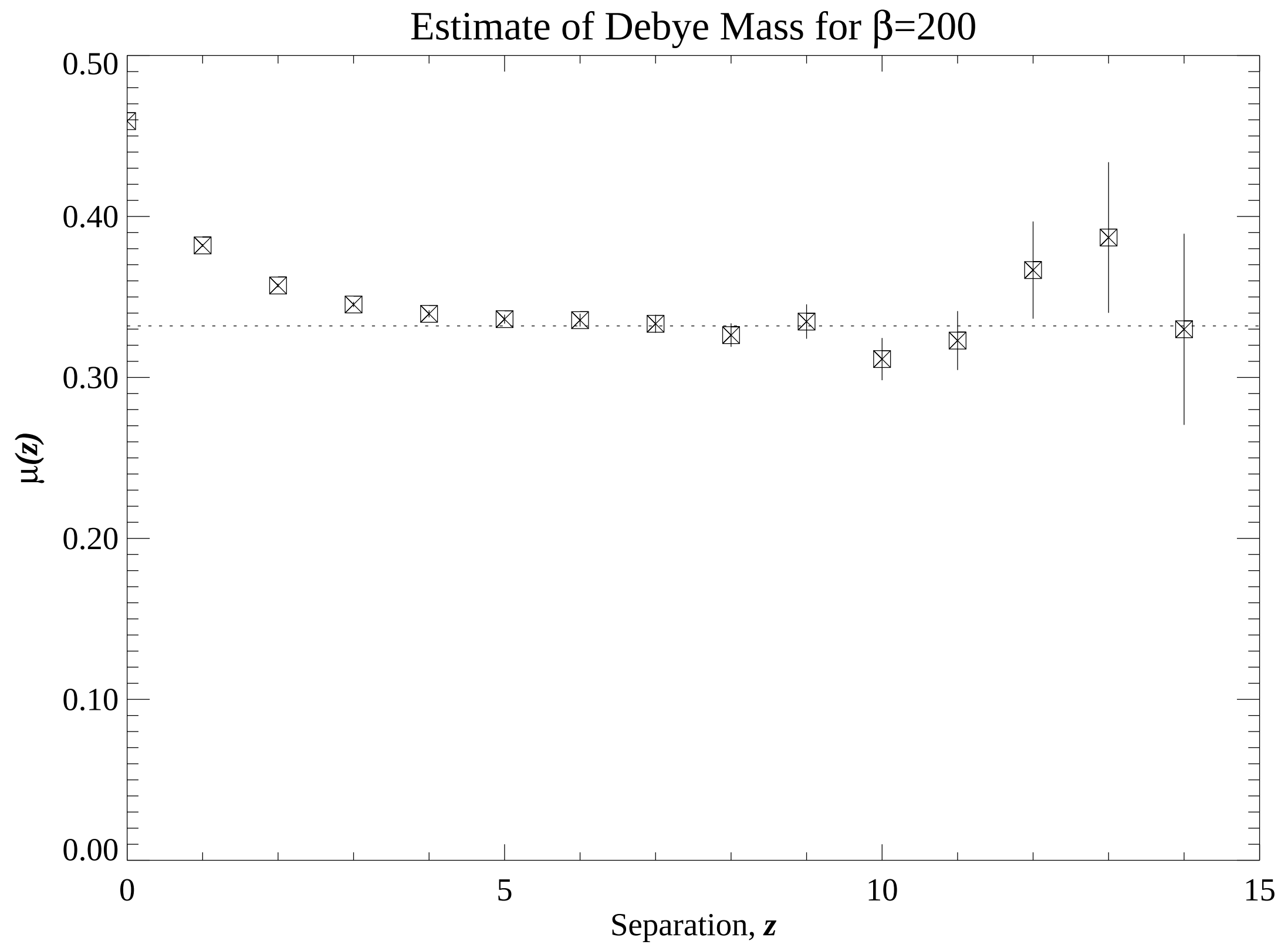

**Fig. 3.9:** Estimation of the Debye mass from $\mu(z)$ for untwisted data at $\beta = 200$.

We measure the correlation function of this operator along the untwisted lattice, as it happens to be easier to obtain accurate results from this system than from the twisted lattice, where the $Z(3)$ interface acts as a dominating source term and the Debye mass can be estimated from the shape of its tail. Owing to the periodic boundary conditions in the space directions, we expect a fluctuation at one value of $z$ to influence another point further along the lattice by correlations going in the $+z$ *and* $-z$ direction. This gives correlations, at long distances, proportional to $e^{-2mz}$ (as in section 2.4) *and* $e^{-2m(L_z-z)}$. Therefore, we estimate $2m$ by a cosh fit, an improvement on the following function which fits simply to a decaying exponential:

$$\mu(z) = \ln\left(\frac{< L^\dagger(z-1)L(0) > - < \overline{L} >^2}{< L^\dagger(z)L(0) > - < \overline{L} >^2}\right),$$

where each $< L(z) >$ represents an average of $L(z)$ over all configurations (sweeps), and $< \overline{L} >$ is a further average over all $z$. From the definition of the Debye mass, we expect

$$\mu(z \to \infty) \to 2m.$$

In fig. 3.9, a typical set of data is shown for $\mu(z)$ at high $\beta$. For each similar set of data, we extracted a value of $\mu(z)$ corresponding to the level at which the function flattened out before errors began to increase wildly. These values are given in the following table for the simulations listed in table 3.1, in units of the inverse lattice spacing.

| Table 3.4: Debye (Electric) Mass (units of $a^{-1}$) | |
|---|---|
| $\beta$ | Twice Debye mass, $\mu = 2m$ |
| 16 | 0.80(1) |
| 28.125 | 0.690(5) |
| 50 | 0.555(5) |
| 112.5 | 0.410(5) |
| 200 | 0.332(4) |

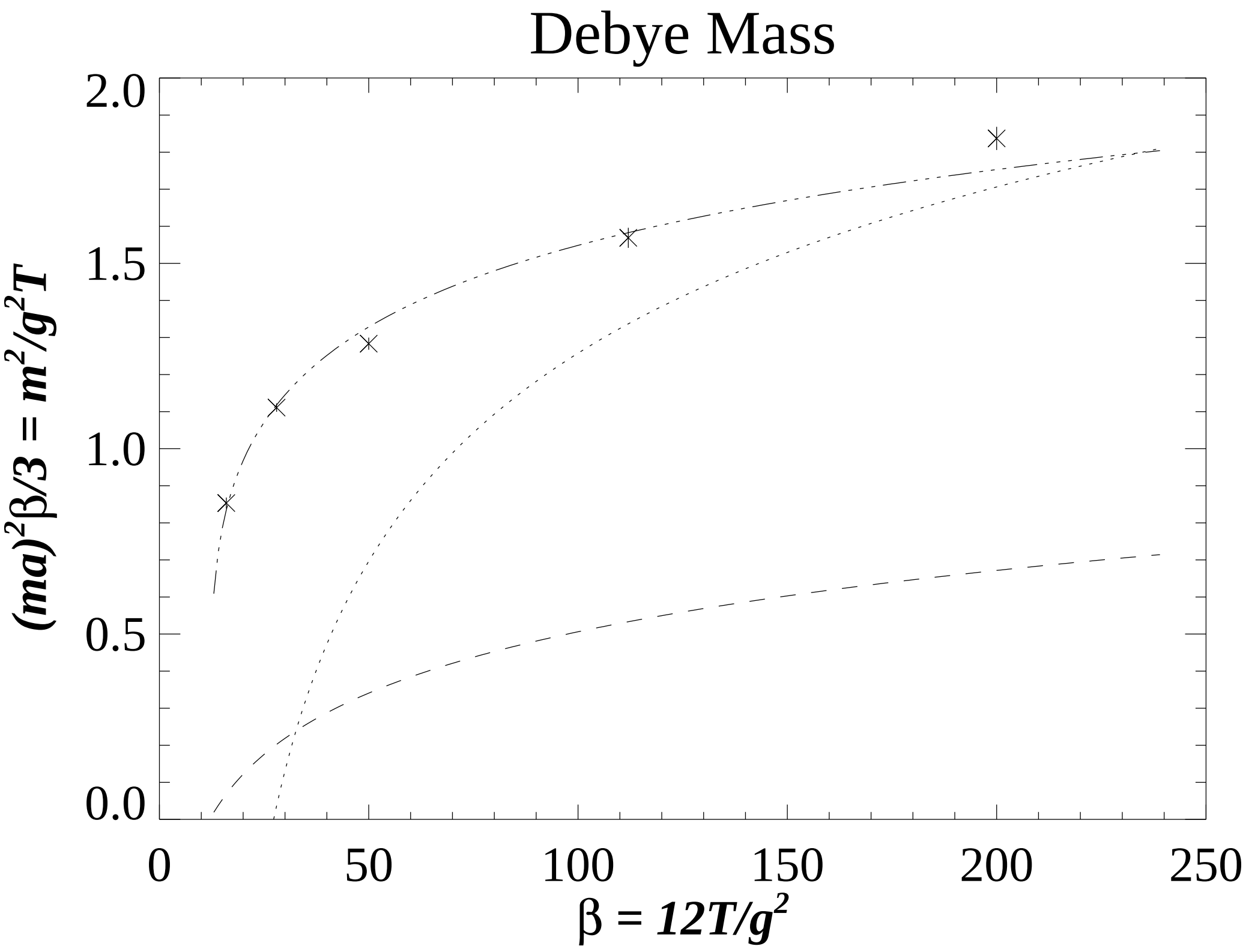

**Fig. 3.10:** Results for the Debye mass. The prediction of (2.21) is plotted to leading-order only (dashes, C.F. fig. 2.4); with an additional correction of $\ln(\ln(\beta/12))$ (dots); and with an alternative additional correction of $0.14\ln(\ln(\beta/12)) + 0.94$ (dot-dashes).

The values are plotted in fig. 3.10, together with the theoretical prediction from fig. 2.4 (dashed line), which is expected to be valid in the high-temperature continuum limit. The results show some qualitative agreement with the leading-order prediction of (2.20) and (2.21), as plotted previously in fig. 2.4. However, although the shape of the rise of the mass with temperature looks correct, the results differ from the prediction by a factor of two or so. The reason for this becomes clearer when we consider the size of the $O\left[\ln(\ln(\beta/12))\right]$ correction compared to the leading-order $\frac{3}{4\pi}\ln(\beta/12)$ that we plot: for our range $16 < \beta < 200$, the leading-order term is between 0.1 and 0.7, but the correction is $O(-1.2)$ to $O(1.0)$! Thus, we shall find it very difficult to draw any firm conclusions from the data to support or deny the prediction.

To show the effect of the correction, the dotted curve in fig. 3.10 is $\frac{3}{4\pi}\ln(\beta/12) + \ln(\ln(\beta/12))$. Our data certainly look more reasonable when this correction is included, but, of course, this estimate is only of the order of the correction, not of an exact expression. A free fit to a correction of the form "$c_1 \ln(\ln(\beta/12)) + c_2$" gives the dot-dashed line in the fig. 3.10, with $c_1 \approx 0.14$ and $c_2 \approx 0.94$, which shows extremely good agreement with our data. Although we have two free parameters and only five data points, the fit is strikingly good, and this gives us some reason to hope that our data is consistent with the theoretical prediction. Of course, no firm conclusions can be drawn without a more explicit expression for the $O\left[\ln(\ln(\beta/12))\right]$ correction. However, the ratio of coefficients of the free fit does at least confirm that most of the temperature dependence is contained within the leading $\frac{3}{4\pi}\ln(\beta/12)$ term.

# Chapter 4.

# Interface Properties Near the Critical Temperature

## 4.1. Interface Behaviour Approaching Criticality

In the previous chapter, we examined some properties of the $Z(3)$ interface at high temperatures, $\beta \gg \beta_c \approx 8.175$. It was seen that the presence of a twist in our simulations is sufficient to put the left- and right-hand longitudinal ends of the lattice in different $Z(3)$ phases, with one physical $Z(3)$ interface separating the two. At these high temperatures, the phases are very well defined, with only small fluctuations seen in the value of the order parameter (the Polyakov line) across their extent. The interface is similarly well defined, with a fairly rigid shape. Both of these aspects are due to the suppression of fluctuations in phase resulting from the increased energy penalty at high temperatures.

Consider, however, the phase structure at the critical temperature itself, $\beta_c$, where the deconfinement transition occurs. At and below this temperature, the three ordered, $Z(3)$-broken phases cease to exist, being replaced by a single disordered phase with symmetry restored. Obviously, therefore, no $Z(3)$ interface can exist in this regime either. The order-disorder, confinement-deconfinement transition itself has been studied in the past. In [8], for instance, the surface tensions of order-order and order-disorder interfaces were compared in $3+1$ dimensions near the phase transition, and various thermodynamic properties were examined. Also, the prediction of complete wetting, mentioned in the introductory chapter, was tested: *viz.* that an order-order ($Z(3)$) interface consists of two order-disorder ones stuck back to back, so that $\alpha_{oo} \approx 2\alpha_{od}$. A question that has not been addressed sufficiently in the past, though, is that of the behaviour of the interface: how does this change from the rigid structure we see at high temperatures, as we reduce the temperature towards the critical value? This chapter aims to rectify this gap in our knowledge by studying various properties of the $Z(3)$ interface at temperatures just above the critical value.

So, then, what changes can we predict as the temperature drops? To begin with, a decrease in temperature causes a corresponding reduction in the energy penalty suppressing fluctuations in phase. It follows that more and larger fluctuations, bubbles of different phase, will appear within the two main $Z(3)$ domains on the lattice, following from the reduction in free energy of the bubble walls. Since the phase transition at the critical point is second-order, one also expects the expectation values of Polyakov lines in the two regions to converge smoothly on zero as the critical point is approached, this being the value taken in the disordered phase below $\beta_c$. This means that the "height" of the interface (the change in the real part of the Polyakov line as one moves across it) will also decrease. These two consequences present a double bind, with the interface decreasing in height just as the surrounding fluctuations increase. As the interface becomes less and less discernible amidst the phase turbulence, one might expect a detailed study of its structure to become well nigh impossible near to the critical temperature.

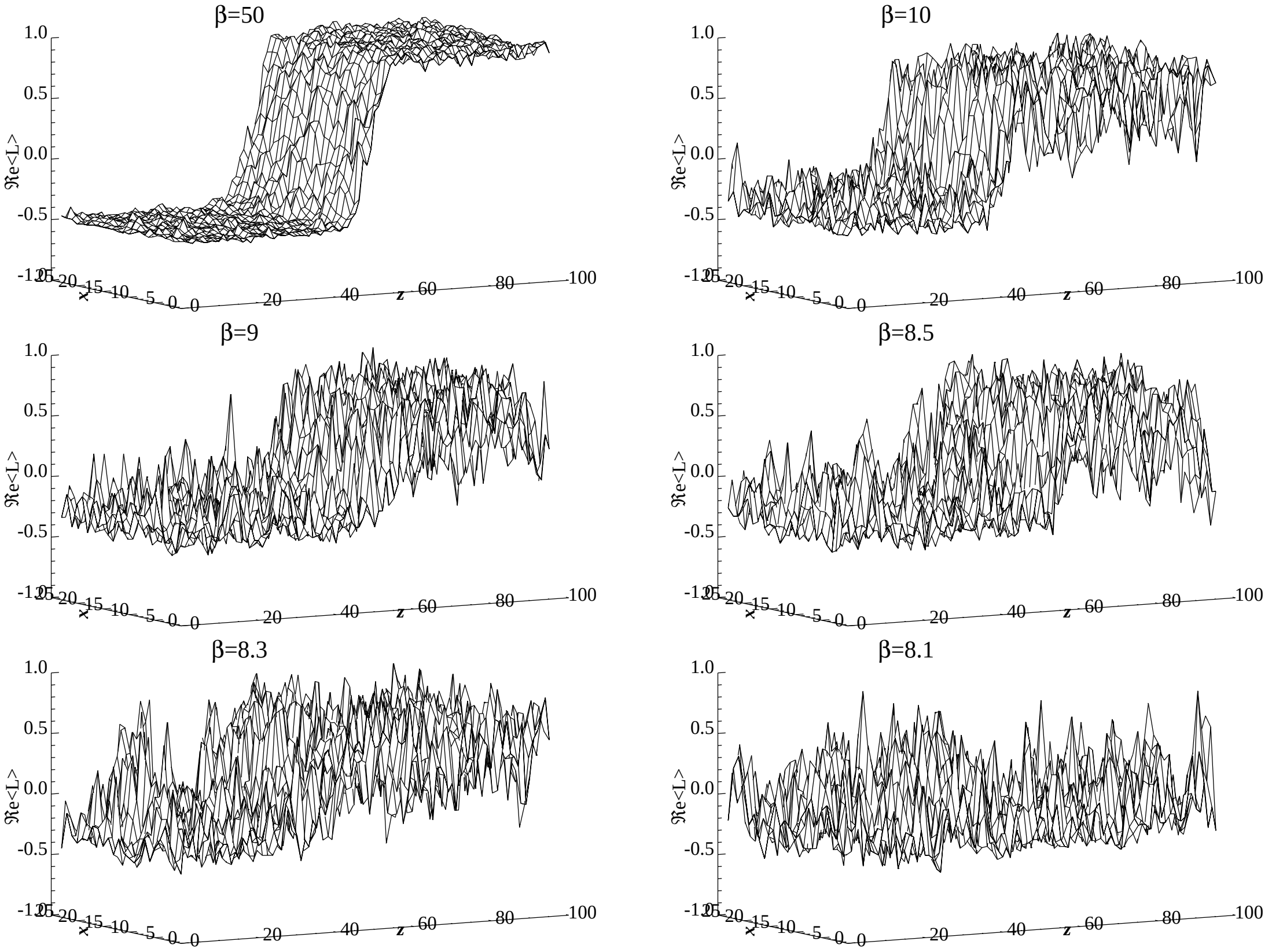

**Fig. 4.1:** Polyakov line (real part) profiles of the same $Z(3)$ interface at different temperatures. The top left picture is identical to that of fig. 3.3, with $\beta = 50$ and a $2 \times 24 \times 96$ lattice. Going from left to right, and then top to bottom, the other pictures illustrate an interface between the same phases ($< L > \approx e^{2\pi i/3}$ on the left and the $< L > \approx 1$ on the right) for $\beta = 10, 9, 8.5, 8.3$ and $8.1$ respectively.

The pictures of Polyakov lines shown in fig. 4.1 appear to confirm these fears. One particular $Z(3)$ interface is shown at a series of different temperatures, which drop towards the critical value. By the time $\beta \sim 10$, it is looking much more ragged than at the high temperatures of the last chapter, illustrated in the first picture. By the time $\beta < 9$, the interface is difficult to spot at all, much less study in detail. In the last picture, just below the critical temperature, all traces of the interface have vanished, as expected, and the Polyakov lines are fluctuating about zero.

The first question to address is that of the qualitative behaviour of the interface as the temperature drops. Given that it must vanish at $\beta_c$, several possibilities seem to present themselves:

1. The interface maintains the rigid, taut structure seen at high temperatures. It keeps its "area", defined as the Polyakov "height" multiplied by the lattice width, to a minimum, by avoiding fluctuations in its shape. Its height shrinks steadily until it reaches zero at the critical temperature, where the interface consequently vanishes. Before this, though, it becomes invisible amidst the growing fluctuations.

2. The interface spreads out in the longitudinal direction, following the behaviour of correlation lengths (such as the Debye screening length) on the lattice. These diverge as the temperature approaches the critical value, meaning that regions of phase separated by a large distance on the lattice have an increasing effect on each other. The interface thus reveals more and more disordered phase within, in line with the prediction of complete wetting, and as its height decreases and its width increases, talk of the interface as a physical object becomes less and less meaningful.

3. In addition to small bubbles of phase, many more entire interfaces form across the lattice as the energy penalty drops. These, together with the original interface, interact with each other and with the fluctuating bubbles of phase, which are also forming in increasing numbers. These interactions, collisions and buddings off, eventually wash out all phase structure, leaving disordered phase throughout the lattice, with the Polyakov line averaging to zero. We would expect this behaviour to be suppressed by increases in the transverse size of the lattice, and use a variety of such sizes later in this chapter.

4. The interface maintains a relatively narrow width, behaving essentially as a one-dimensional object in the spatial dimensions — a string, stretching across the lattice — but it suffers increasingly violent transverse fluctuations, contorting its shape and thereby increasing its area (*i.e.* its length, disregarding the time dimension). The fluctuations diverge towards the critical temperature, and the interface eventually breaks into pieces to leave only disorder.

As mentioned above, it is fairly clear from fig. 4.1 that the interface has vanished by the final picture ($\beta = 8.1$), but the manner of its collapse remains ambiguous. The fluctuations wash out the interface so much that we cannot really rule out any of the possible behaviours at this stage. Clearly, we need a better way to keep track of the interface as the temperature drops, and this is the matter addressed in the next section.

## 4.2. Tracking the Interface

One way to deal with the increase in phase fluctuations is to cut out the highest frequency modes altogether. This can be achieved most simply by a "box-car" average, where the Polyakov lines of each three-by-three spatial array of points are averaged over to give a new value, allocated to the central point. Note that the periodic boundary conditions give the spatial dimensions the topology of a torus, so we have no edge effects to worry about. The expectation value of the Polyakov line in a region is unaffected by this averaging procedure.

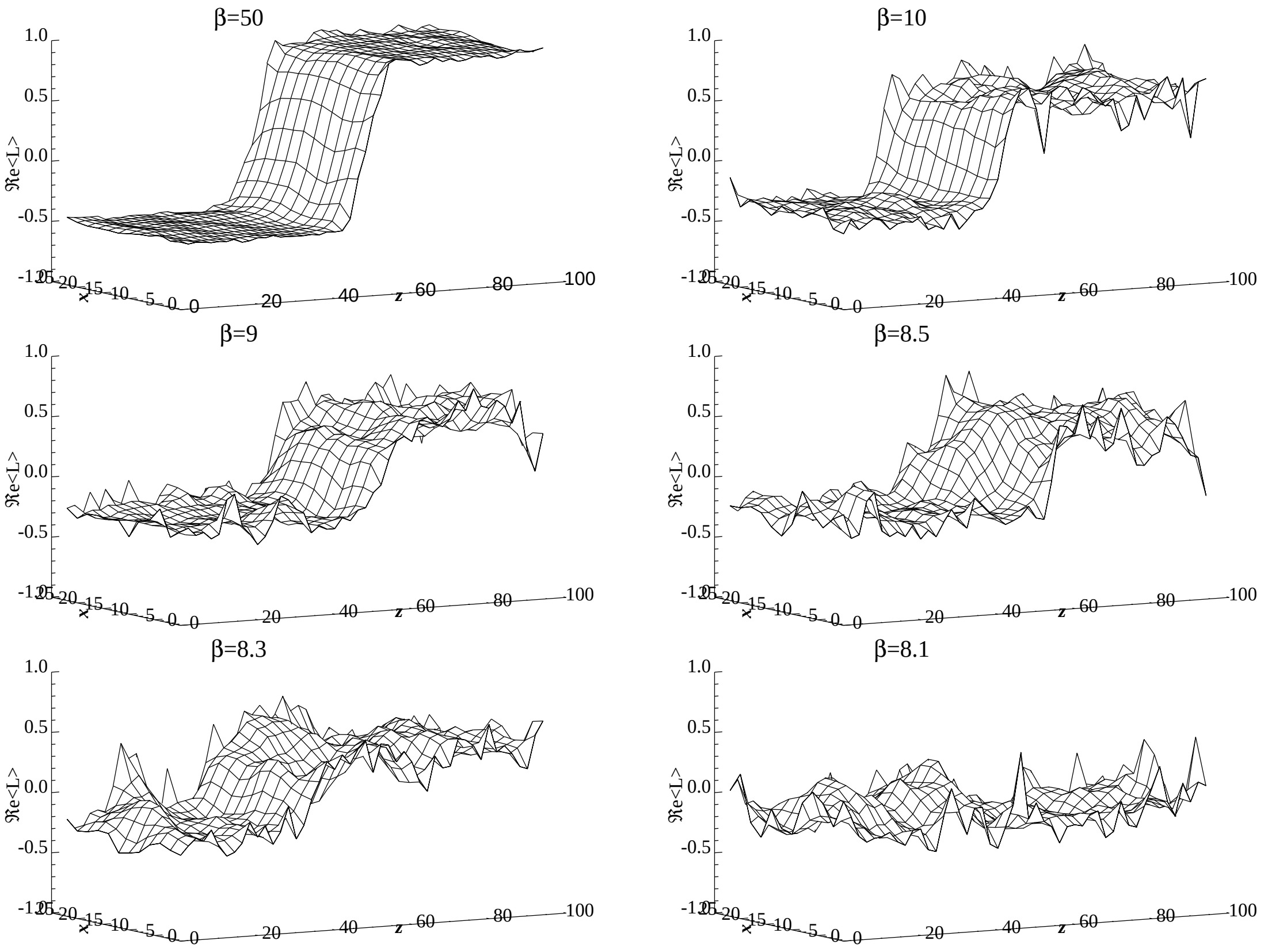

**Fig. 4.2:** Smoothed Polyakov line profiles, produced by processing the raw pictures of fig. 4.1. As in that figure, $\beta = 50, 10, 9, 8.5, 8.3$ and $8.1$ respectively.

The effect of smoothing out the fluctuations is shown dramatically in the pictures of fig. 4.2. These have been produced simply by applying the box-car average to the pictures of fig. 4.1. To make things even clearer, the Polyakov lines have first been re-binned so that the $24 \times 96$ spatial dimensions become $12 \times 48$, with each two-by-two array being averaged to one point. Now the $Z(3)$ interface can clearly be seen for all but the last picture, *i.e.* for all temperatures above $\beta_c$. The fact that the interface can still be seen for temperatures just above the critical value gives encouragement to the belief that a study of its structure should still be possible quite close to the phase transition, and that the shape of the interface can be monitored.

However, there is a drawback with this smoothing technique: it seems less than ideal as a mechanism for study of the interface, as it averages out a whole class of fluctuations which may be important in the interface collapse. Thus, we would like to find a better way to keep track of the interface. We shall, though, use smoothing later on in conjunction with a more reliable method of tracking the interface, and a comparison of the results will allow us to judge how much distortion is introduced by smoothing.

A more satisfactory method is to produce a contour map from the real part of the Polyakov lines on the lattice. At its simplest, consider following a contour whose height is mid-way between that of the average Polyakov line at the left-hand end of the lattice and that at the right-hand end. This will pick out the mid-height of the interface as it goes across the lattice, as well as any bubbles of fluctuating phase which are large enough to cross the contour height. The crucial point to note is that *only* the $Z(3)$ interface will cross the entire lattice, *i.e.* the only contour that will wrap once around the lattice in the transverse direction is that corresponding to the interface. In a sense, it has unit "winding number", whereas any contour representing a bubble of phase will join up with itself without a net crossing of any lattice boundary (winding number zero). This gives us a way to tell the interface apart from any fluctuations, and to identify the position of its mid-height. By following contours at various heights between the two extremes, we can study more of the structure of the interface. Also, the contours allow us to measure any fluctuations in the shape of the interface, as well as its width in lattice units. The advantage of using contours is that no prior processing of the Polyakov line data is required. However, we can study contours of the box-car-averaged data as well if we wish, as this may be useful very close to the critical temperature.

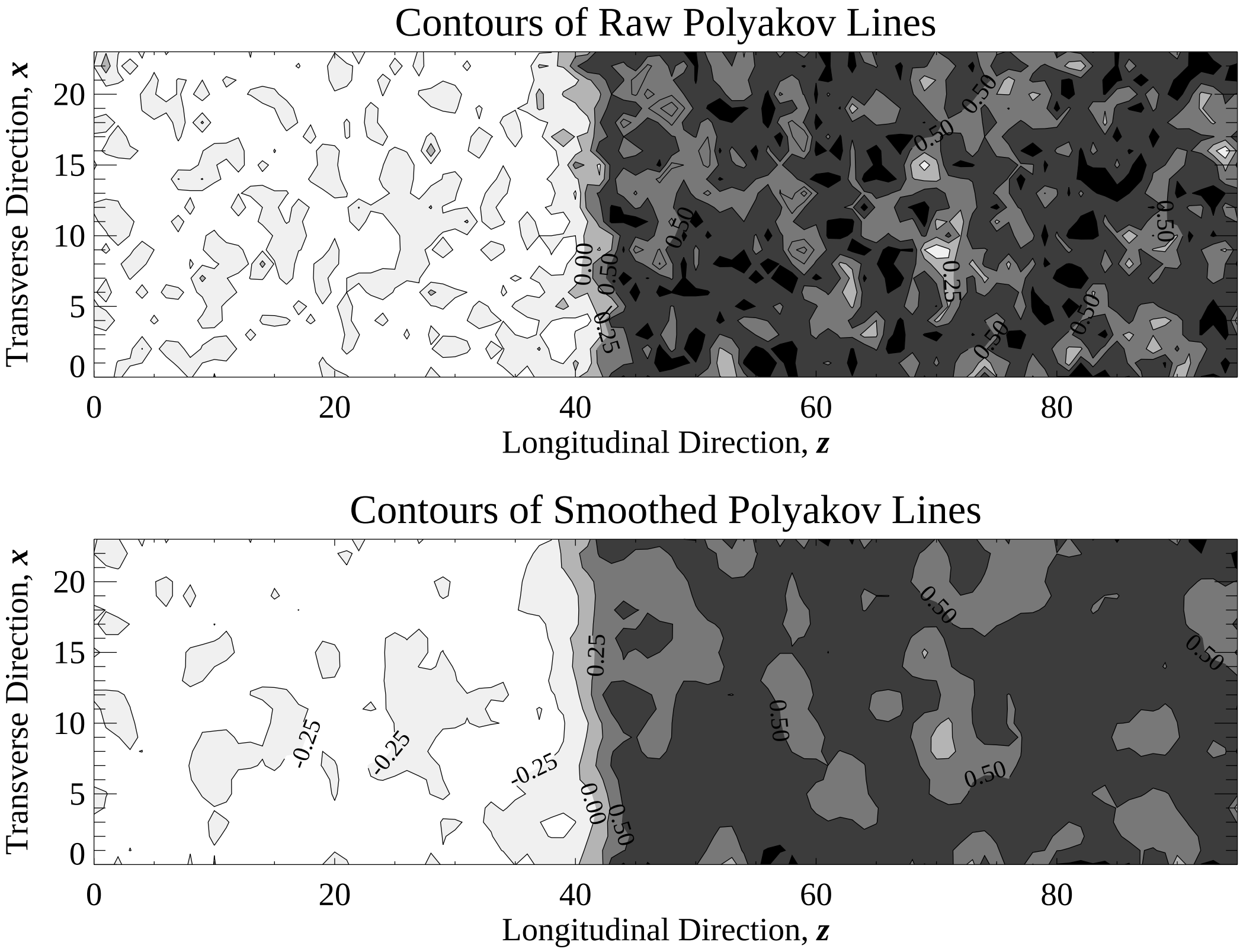

**Fig. 4.3:** Contour maps of raw and processed Polyakov lines for $\beta = 10$ on a $2 \times 24 \times 96$ lattice. These maps correspond to the $\beta = 10$ pictures of fig. 4.1 and fig. 4.2 respectively. The contour levels are at $-0.25, 0, 0.25, 0.5$ and $0.75$, darker regions corresponding to higher values.

An example of a contour map is shown in fig. 4.3, for the same gauge configuration as featured in the prior two diagrams for $\beta = 10$. Notice the interface in the centre of the maps, represented by the only contours going all the way across the lattice. The first picture shows the raw data of fig. 4.1; the second, the smoothed data of fig. 4.2, but without first re-binning into a $12 \times 48$ lattice in this case.

## 4.3. Fixing the Interface Position

The ideas behind contour following also solve another problem for us: how to keep the interface fixed in place. In the previous chapter, we allowed it to execute a random walk along the lattice, as permitted by translation invariance. This allowed the system to tunnel into different $Z(3)$ vacua, producing a different $Z(3)$ interface each time. However, this is unsuitable for our present purposes, as we wish to study the profile of one particular $Z(3)$ interface without interference from other vacua. Of course, the properties that we find will apply equally to all the $Z(3)$ interfaces, but our measurement techniques cope best with one particular interface. We need to stop the interface wandering through the twist, as this is the process by which the system tunnels. Ideally, the interface should be held in place well away from the twist, so that no part of it, even its wildest fluctuations, passes through the twist.

How, then, can we pin the twist down in the longitudinal centre of the lattice? One method might be to use a partial anti-twist, *i.e.* premultiply just one or two plaquettes at $z = L/2$ by the inverse of the factor used in the twist at $z = 0$. If we inserted a full anti-twist, it would just act as another change of variables, taking the Polyakov lines which have been flipped by the twist back to their original value, without physical consequence; no $Z(3)$ interface would need to form. However, by flipping just one or two plaquettes rather than the whole set across the lattice, we make it energetically favourable for the Polyakov lines to flip at that position, whilst still forcing the appearance of a physical interface because of the untwisted state of most of the plaquettes.

The problems with this suggestion are twofold. First, an objection of principle: this method of pinning is distinctly artificial, breaking the translation invariance of the interface. How, then, can we be sure that the behaviour of the interface is not be affected by the procedure? The presence of the anti-twisted plaquettes, together with the periodic boundary conditions in the transverse direction, fixes the part of the interface at the boundary, suppressing the even (cosinusoidal) fluctuations in the interface in favour of the odd (sinusoidal) ones. Second, an objection of practice: the anti-twist makes it energetically favourable for the interface to form at its location, but we have already seen that energetic considerations become less and less important as the temperature drops towards the critical value, *i.e.* the penalty for the interface to wander away from the anti-twist becomes less and less. Thus, we need to introduce more and more anti-twisted plaquettes as the temperature drops, in order to keep the interface pinned in place. This causes more and more distortion in the fluctuations of the interface. In computational trials, we found that this technique was simply not effective at the sort of temperatures in which we were interested.

A far more appealing approach is to locate the position of the interface after each sweep of the lattice, and then slide the whole gauge configuration along the lattice until the interface is re-centred. Any link variables which are shifted through the twist are multiplied by the appropriate factor, so that there is no physical effect from the procedure. This is clearly superior to using the anti-twist: there is no physical effect and no violation of translation invariance, and it can be used close to the critical temperature, for as long as the interface can be located. This, therefore, is the method used in all following simulations, unless stated otherwise.

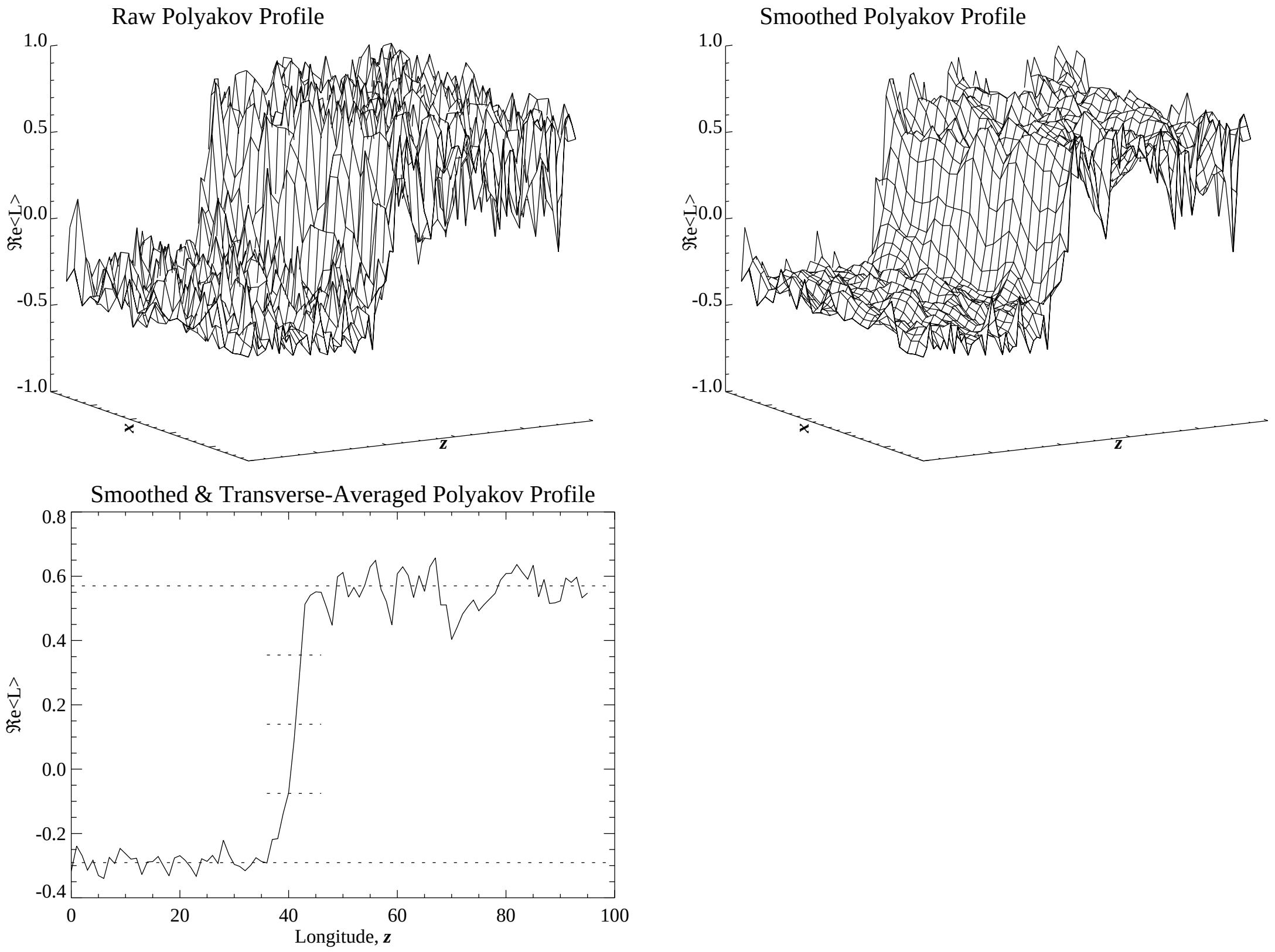

**Fig. 4.4:** The procedure for locating the interface, illustrated for the $\beta = 10$ configuration of previous figures, on a $2 \times 24 \times 96$ lattice. The raw Polyakov line profile is smoothed, and then averaged in the transverse direction.

## 4.4. Locating & Profiling the Interface

To find the position of the interface after each sweep of the lattice, we construct a smoothed Polyakov profile using box-car averaging, as above, and then average this profile in the transverse direction. This process is illustrated in fig. 4.4. The Polyakov line averages of this last profile are then averaged separately at opposite ends, to give the long dotted lines marking the top and bottom levels of the interface. Three evenly spaced contours are then chosen between these two, as illustrated by the short dotted lines crossing the interface. Starting from the (longitudinal) centre of the lattice, the algorithm searches for the nearest point to pass through the middle contour level in the correct direction (increasing from left to right). It is assumed that the Polyakov lines interpolate linearly between neighbouring lattice sites. The point found is defined to be the centre of the interface, and the configuration is then shifted to reposition this at the centre of the lattice. Given that this procedure is repeated after each sweep, it is extremely unlikely that the algorithm will be misled by a large fluctuation passing through the middle contour level; it will almost certainly find the true interface first, as this will be closest to the centre of the lattice.

To find a contour at a particular level, we return to the raw Polyakov line data. For $x = 0$, the algorithm follows a procedure similar to that above: starting from the centre, $z = L_z/2$, it looks for the nearest point passing through the contour level. Then, it considers the square of points formed by the two neighbouring lattice sites at $x = 0$ and the corresponding sites at $x = 1$.

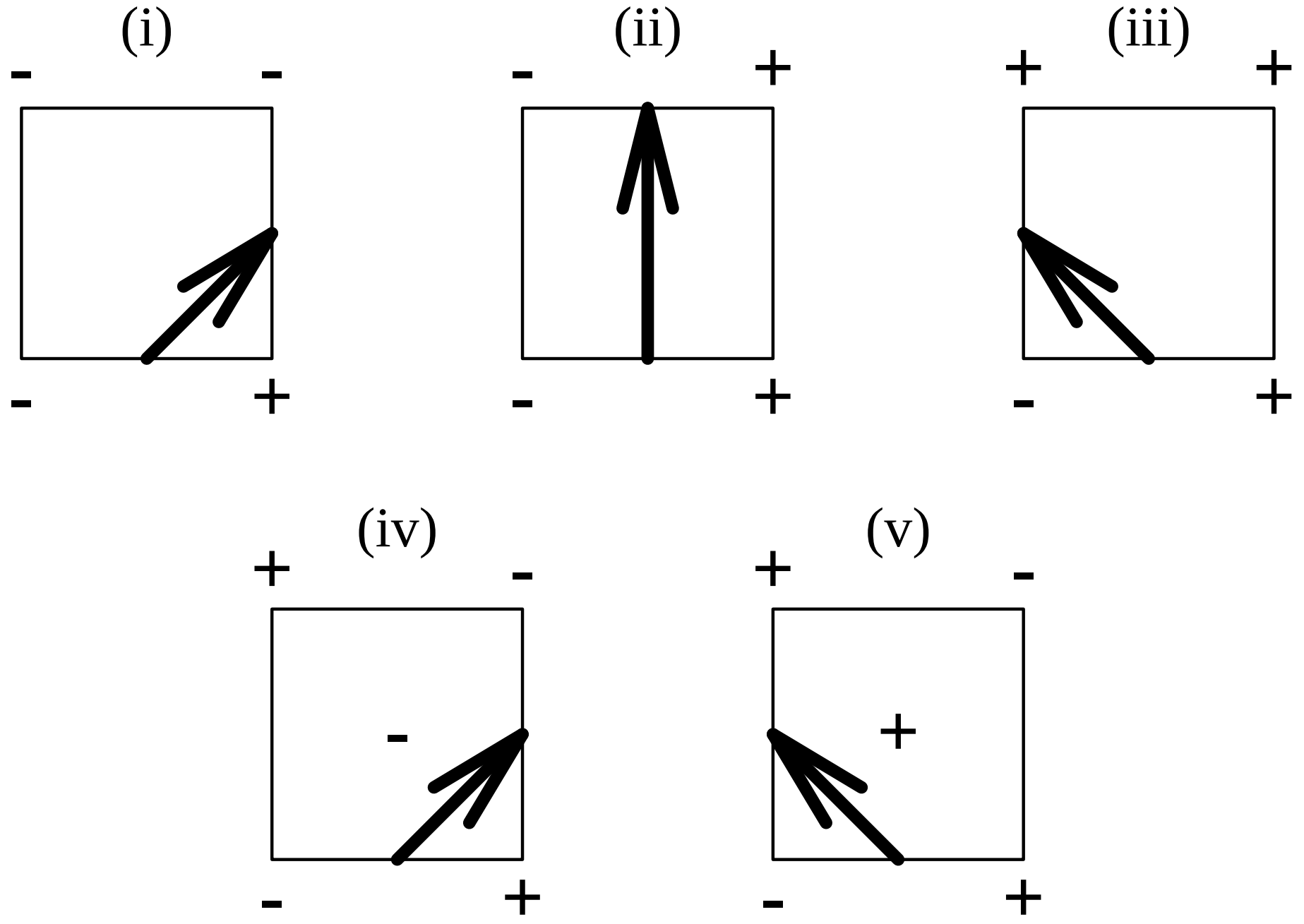

**Fig. 4.5:** Options for the contour-following algorithm. Given a point at the desired contour level on the lower side, the Polyakov lines at the upper two corners determine the progress of the contour in cases (i)-(iii). A plus (minus) sign indicates a Polyakov line value greater (less) than the contour level. The same corner configurations of (iv) and (v) give two possible contour directions; we choose between them by considering the central average of the corner values. These pictures exhaust all possibilities for the values of the lower corners given, but any general configuration can be rotated and reflected into one of these.

The algorithm decides to which side of the square the contour should go, as in fig. 4.5. Denoting Polyakov line values which are greater or less than the contour level by plus and minus signs respectively, the five possible configurations illustrated determine the progress of the contour. The last two are ambiguous, with a central average of the four corner values used to decide which direction of the two possibilities should be chosen, as shown. Once the destination side is determined, the point at which the contour crosses that side is found by extrapolating between the two relevant corners, as usual. Then, starting with this side, the whole procedure continues. Eventually, the contour must join up with itself. If it has wrapped across the lattice exactly once in this process, it must be the contour representing the interface; if not, it has merely identified a bubble of phase, and can be discarded. Repeated application of this technique for different contour levels determines the structure of the interface to the desired precision. The same procedure can be applied to the smoothed Polyakov line configurations, to obtain smoothed contour maps.

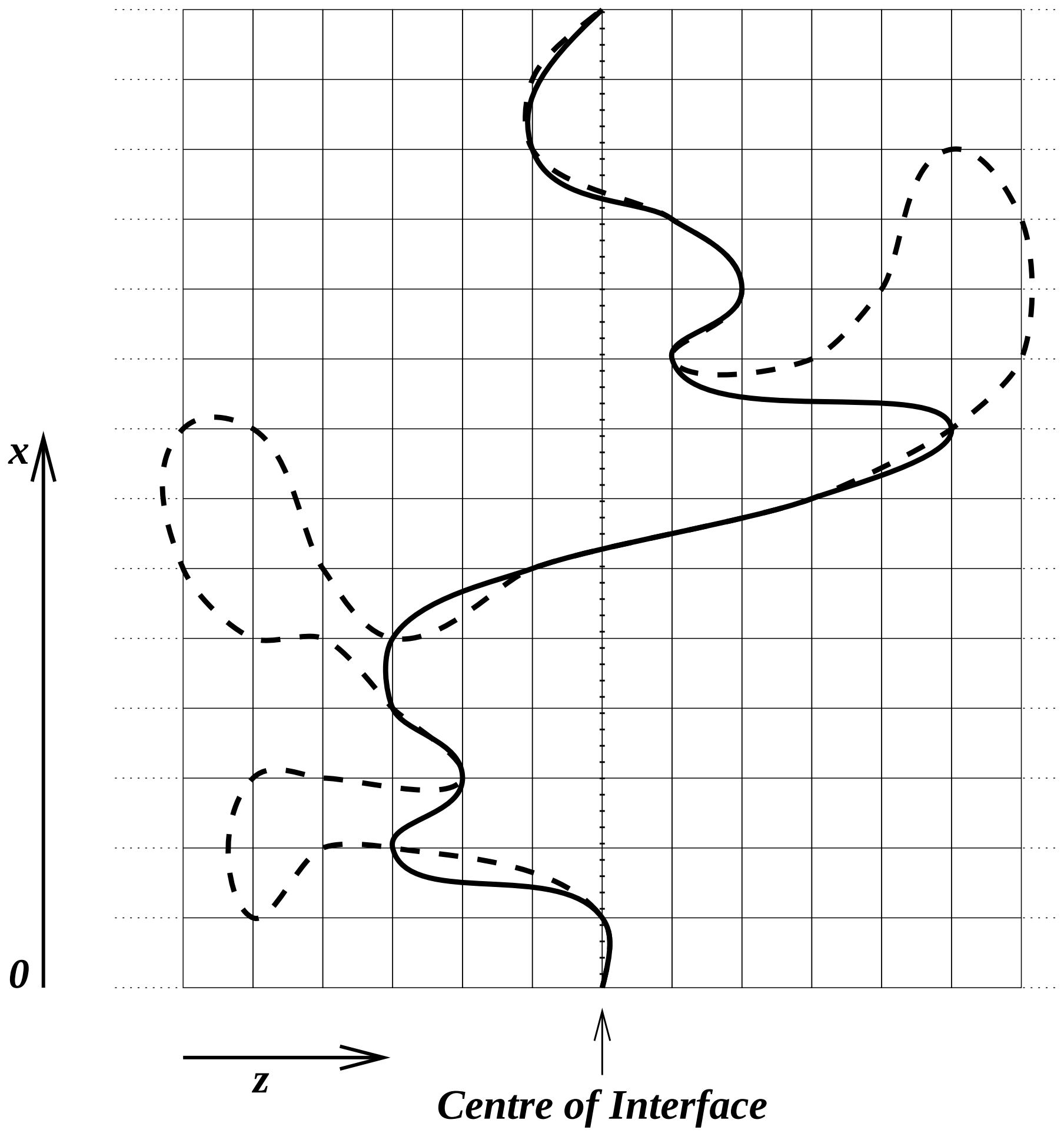

**Fig. 4.6:** An interface contour is illustrated crossing the lattice: the dashed line. The dotted line marks the average position of the contour, at the longitudinal centre of the lattice. As the contour twists and turns to cross some transverse lattice positions more than once, we shall approximate it, where necessary, by the solid contour, which crosses each transverse lattice position only once, at the closest approach of the dashed contour to the central (dotted) line.

In a tiny fraction of cases, the algorithm above may fail. For instance, cases (iv) and (v) of fig. 4.5 cannot be distinguished with certainty as Polyakov lines are only defined *on* the lattice sites. Also, extremely close to the phase transition, the interface may break up into several pieces, so that no contour goes all the way across the lattice. In the event of such failure, a reserve method is used to find the rough positions of the contours; in fact, this is the same method as originally used to locate the centre of the interface, but using raw rather than smoothed Polyakov lines: for each transverse position, $x$, the nearest point to the centre of the lattice is found at which the Polyakov line passes through the desired contour level in the correct direction (*i.e.* increasing or decreasing with $z$, as appropriate for the interface).

Later, we shall need to define the interface just once for each transverse lattice site, but at very low temperatures, the contour sometimes twists and turns so much that it crosses some transverse values several times. In these cases, we take the interface to be at that crossing point which is closest to the average longitudinal position of the interface across the lattice. This procedure is illustrated in fig. 4.6, with the solid contour being the approximation taken to the dashed contour.

## 4.5. Qualitative Results for Interface Behaviour

To address the question posed in section 4.1 of how the interface behaves near the critical temperature, we use the contour techniques described above to follow the changes in its structure as the temperature drops. We start by considering some results from our simulations on a $2 \times 36 \times 72$ lattice. The length of the lattice, seventy-two sites, was chosen after studies with various lengths, to ensure that not even the wildest fluctuations of the interface would reach as far as the twist during our full-length runs. The simulations consisted, in each case, of 2k heat-bath sweeps and 100k main sweeps. Since we were interested in the evolution of the interface, contour measurements were taken after every sweep of the lattice. Approximately 100—200 hours of CPU time on a DEC 2100 A500MP machine were needed for each value of $\beta$, as shown in the following table (occasionally, parts of simulations were performed on DEC 3000 machines; these times have been adjusted to compensate):

Table 4.1: Simulation Times for $2 \times 36 \times 72$ lattice

| $\beta$ | CPU time for 2k+100k Sweeps (hrs) |
|---|---|
| 8.25 | 96 |
| 8.35 | 134 |
| 8.50 | 152 |
| 8.75 | 184 |
| 9.00 | 197 |

In fig. 4.7, a contour snapshot of the interface is shown for each temperature studied, at the same point in each simulation: just over halfway through. Three contours have been followed in each case. We can see that the contours stay close together across most of the interface, indicating that the interface remains relatively narrow as the temperature drops.

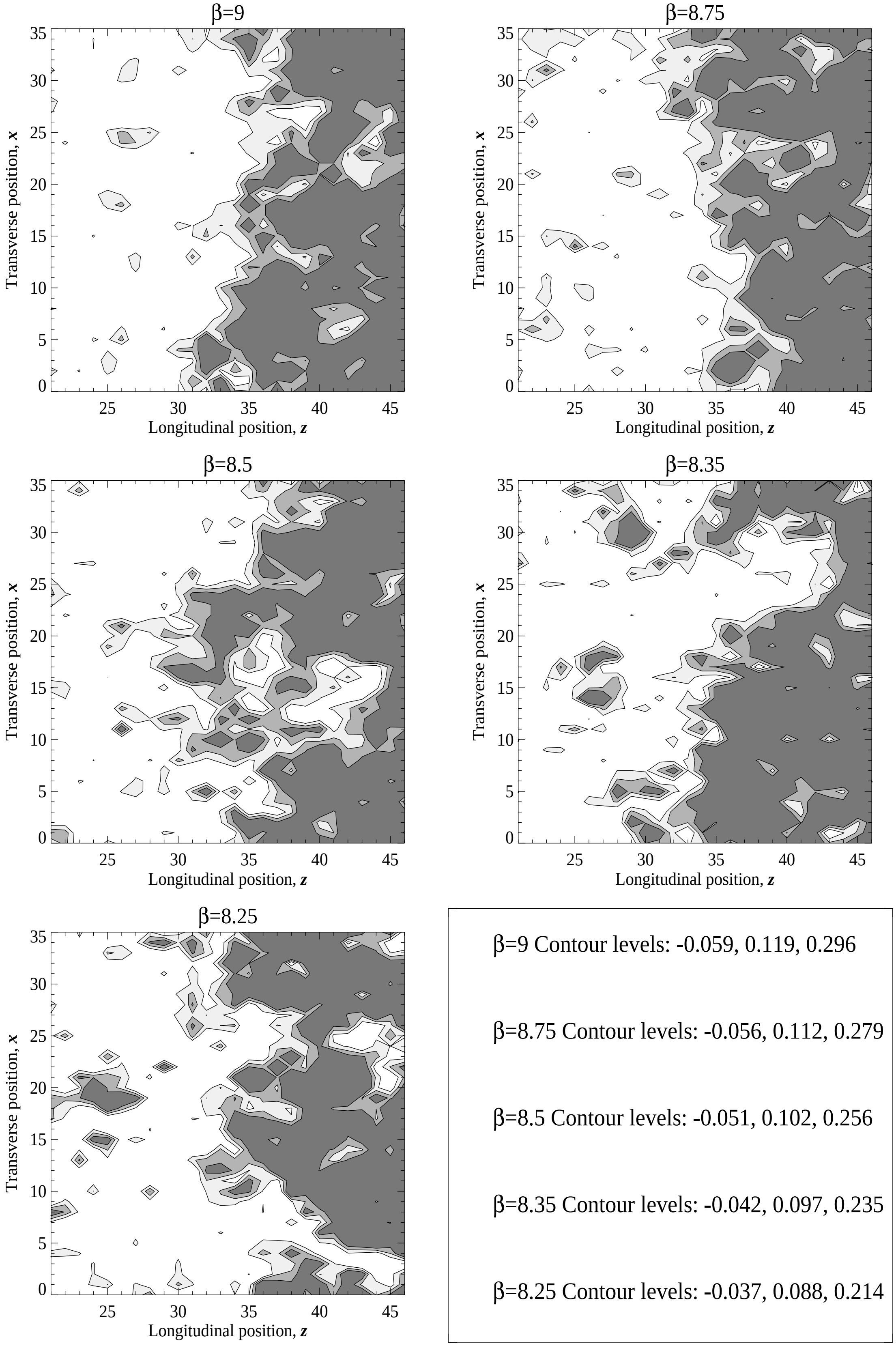

**Fig. 4.7:** Contour maps of the interface on a $2 \times 36 \times 72$ lattice, for decreasing temperature. Moving left to right, and then top to bottom, the picture illustrates the interface structure around the longitudinal centre of the lattice after 2k heat-bath and 60k main sweeps for $\beta = 9, 8.75, 8.5, 8.35$ and $8.25$.

The quantitative behaviour of the interface width will be studied more closely later in this chapter. However, the shape of the interface appears to fluctuate rather more wildly with the falling temperature. Again, a detailed study of the fluctuations will be undertaken later in the chapter.

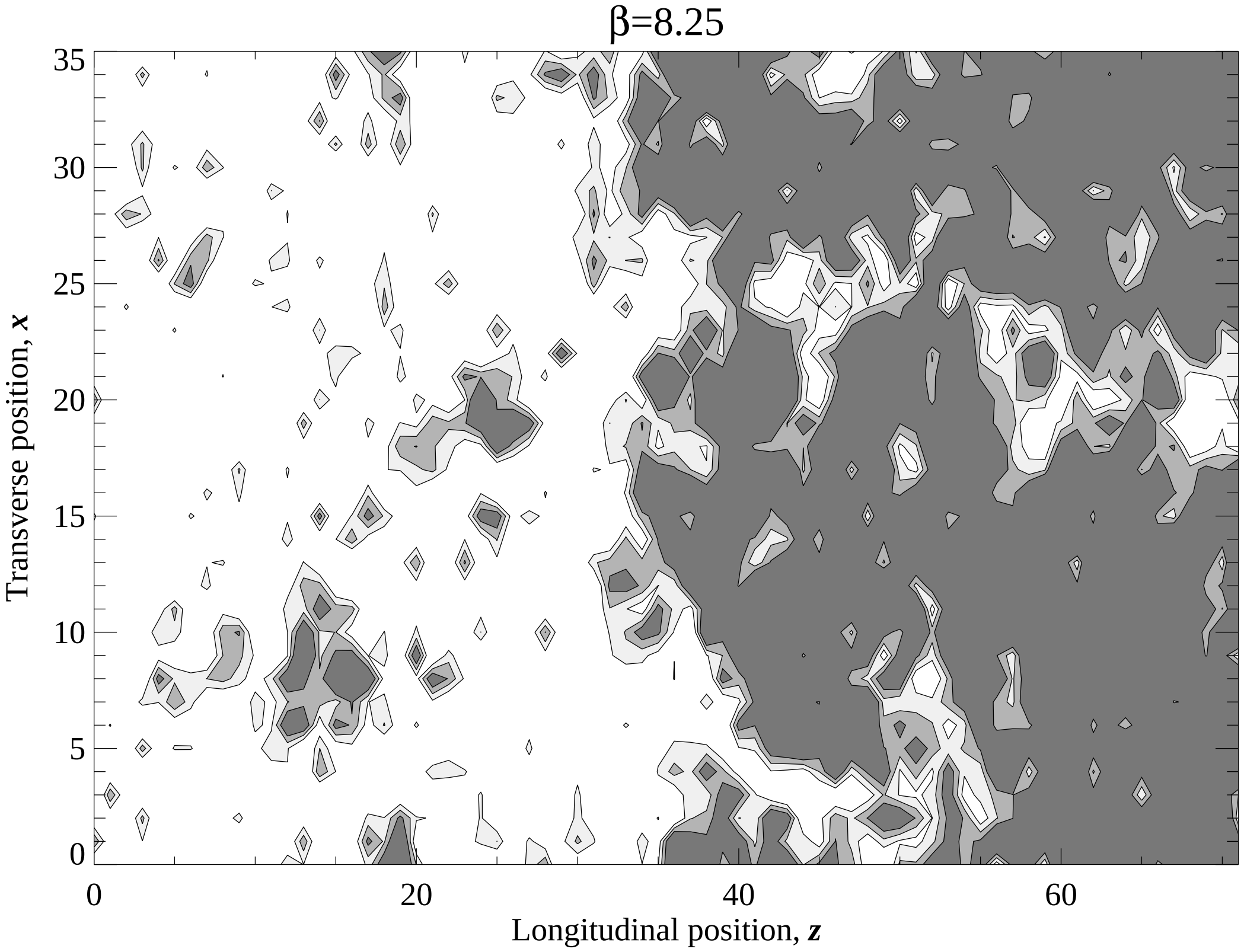

**Fig. 4.8:** A contour map of the whole $2\times36\times72$ lattice for $\beta = 8.25$, corresponding to the last picture of fig. 4.7, showing that only one interface exists.

These observations suggest a behaviour of the fourth type in section 4.1, with the interface retaining a relatively narrow width as it approaches the critical temperature but suffering increasingly violent fluctuations in shape. In fig. 4.7, only the central longitudinal portion of the lattice is shown, to show the interface contours more clearly. However, lest one wonder whether more interfaces are appearing across the lattice further away from the one shown, we show in fig. 4.8 the full structure corresponding to the last picture of fig. 4.7. This demonstrates that even at $\beta = 8.25$, no other interfaces appear. Only bubbles of phase form, increasing in size and occurrence as the temperature drops and occasionally budding from the main interface or recombining with it. As already mentioned, the interface does not spread out noticeably in width, which we can define to be the average difference in position between the lowest and highest contour across the interface, and so it appears to be relatively unaffected by the divergence of physical correlation lengths such as the Debye screening length. Neither does it remain in the sort of rigid shape of minimal "area" that we saw at higher temperatures, following from the reduction in free energy per unit area of the interface.

The snapshots of fig. 4.9 for $\beta = 8.5$ give an idea of how the interface continually changes shape as it evolves in computer time. The initial configuration is shown in the first picture: the lattice is set up to be in two definite $Z(3)$ phases, with a sharp boundary between them midway along the lattice, as discussed in section 3.1. The configuration after the equilibration sweeps have occurred is shown in the second picture, with the physical $Z(3)$ interface broadening into an equilibrated shape. Pictures after this are spaced at intervals of 20k main sweeps, showing some snapshots of the interface as it writhes and twists. It is clear that the it maintains its narrow width, but fluctuates a great deal as bubbles of phase appear and disappear around it. Note that the interface remains centred, and that the fluctuations are not large enough to go beyond our field of view in the centre of the lattice. This reassures us that the interface does not bump into the twist at any time.

We also follow the interface evolution by contouring the smoothed (box-car-averaged) configurations after each sweep, as illustrated in fig. 4.10; the pictures shown are those from fig. 4.9, but from the smoothed Polyakov lines rather than the raw ones. Notice that we can identify the interface a great deal more clearly, at the price of losing a great deal of phase structure. We continue to monitor the smoothed configurations, as the greater clarity of the interface may allow us to follow its behaviour closer to the critical temperature than the raw data alone. Later in this chapter, we shall have the opportunity to check how results from the smoothed data compare with those from the raw data, and thus gauge the usefulness of this procedure. The increase in computer time to keep track of the smoothed configurations is extremely small.

To provide further evidence for the behaviour that we are suggesting above, we take a transverse-averaged Polyakov profile of the lattice after each sweep. Averaging this profile over all 100k main sweeps, we obtain an average profile such as that of fig. 4.11. Notice that this average one-dimensional profile of the interface fits the expected shape for an instanton interpolating between two vacua, namely a tanh function, remarkably well. This is true even at $\beta = 8.25$, though the divergence from the tanh profile is more noticeable at the lower temperatures; this is discussed briefly in the next section but one.

We can determine the Debye mass of the interface from the coefficient of the fitted tanh function, since we expect the interface width to be of order the Debye screening length, *viz.* the inverse of the Debye mass. We can also estimate the intrinsic width of the interface, characterising the width at any time and point of the interface, rather than the Debye screening length, where all fluctuations are washed out to give the width of a "fuzzy" average interface. The intrinsic width is estimated by taking the average distance between the lowest contour, at 25% of the height of the interface, and the highest, at 75%, all the way across the lattice, after each sweep. The result for this case is shown in fig. 4.11, and full results for each type of width are given later in this chapter.

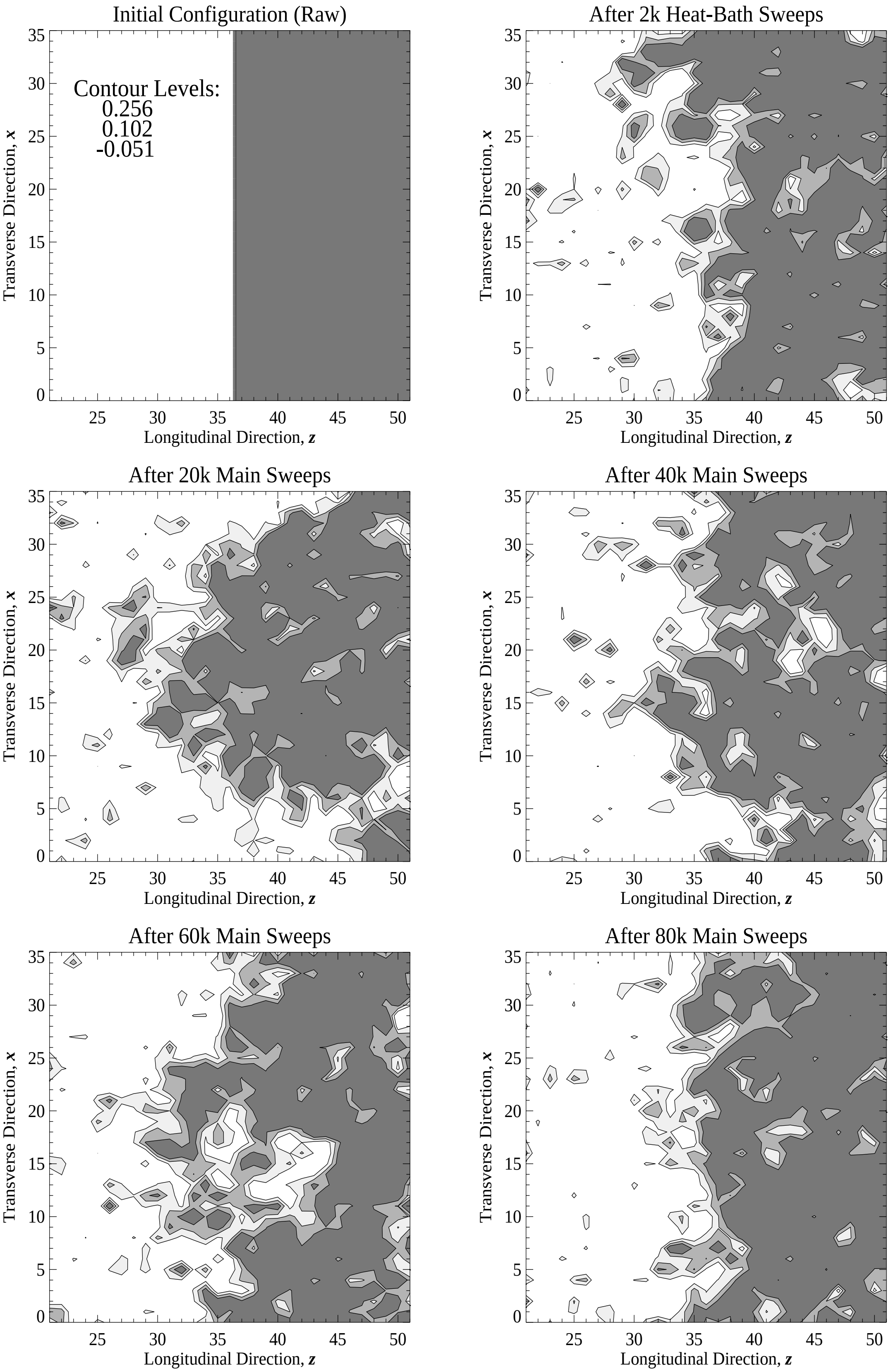

**Fig. 4.9:** Contour snapshots of the interface at various stages of a $\beta = 8.5$ run on a $2 \times 36 \times 72$ lattice, showing just the central longitudinal portion of the lattice. Three contour levels were chosen by the method of fig. 4.4, with values as given in the first picture.

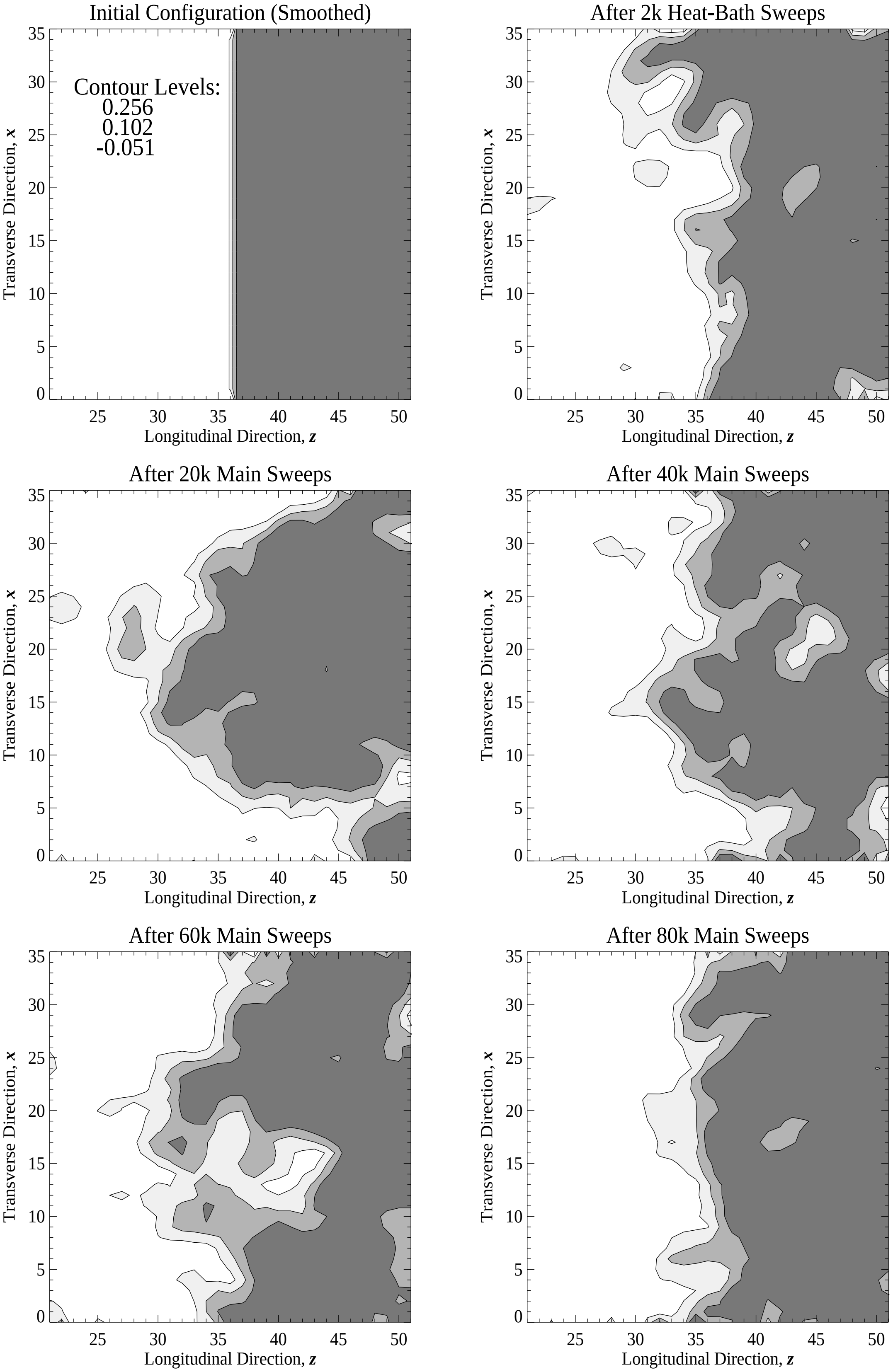

**Fig. 4.10:** The same contour snapshots as in fig. 4.9, but this time using the smoothed Polyakov lines.

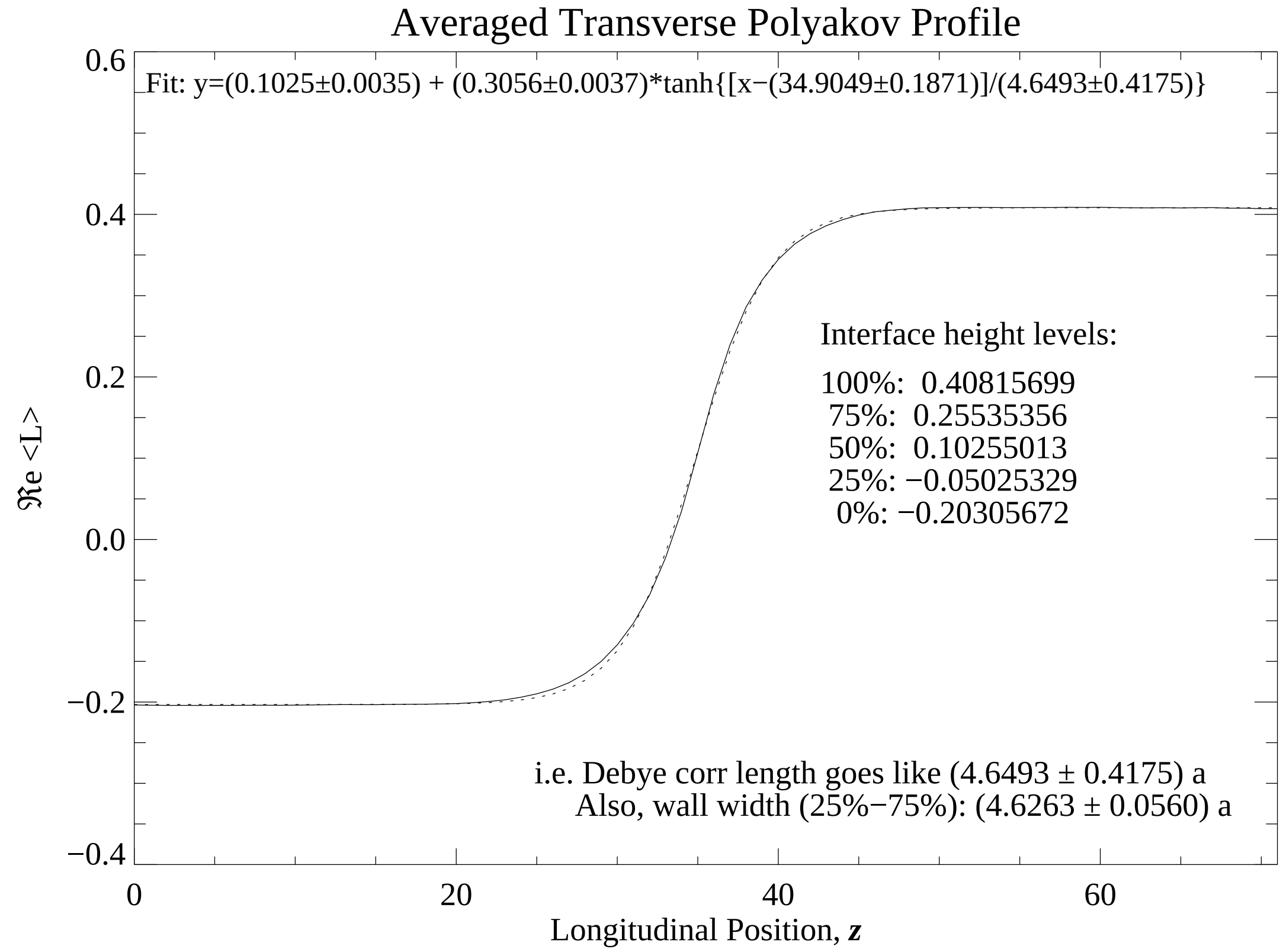

**Fig. 4.11:** Polyakov line profile for $\beta = 8.5$ on a $2 \times 36 \times 72$ lattice, produced by taking an average over all sweeps of the transverse-averaged profile (see text). The dotted line represents the best fit to a tanh function, giving the Debye screening length (inverse mass); the value shown is actually $(2m)^{-1}$, as discussed in section 2.4. The average interface width, between the lowest (25%) and highest (75%) contours, is also shown.

## 4.6. Simulation Parameters

Having established the qualitative features of the behaviour of the interface near the critical temperature, we now proceed with a detailed quantitative analysis of our results. Since we wish to examine aspects of the interface behaviour as the temperature drops to the critical value, we perform simulations for the various values of $\beta$ mentioned previously; these are tabulated below for easy reference. It will be seen later that it is difficult for us to obtain reliable results closer to the critical temperature than $\beta = 8.25$.

| Table 4.2: Simulation Temperatures, $\beta$ |
|---|
| 8.25 |
| 8.35 |
| 8.50 |
| 8.75 |
| 9.00 |

For each value of $\beta$ in table 4.2, we carry out simulations with 2k heat-bath and 100k main sweeps on $2 \times N_x \times 72$ lattices with a range of transverse lattice sizes, $N_x$. This will allow us to examine the dependence of the interface behaviour on its size, *i.e.* the lattice width.

| Table 4.3: Simulation Lattice Widths, $L_x$ |
|---|
| 18 |
| 24 |
| 30 |
| 36 |
| 42 |
| 48 |
| 54 |

## 4.7. Interface Wetting

Before moving on, we pause briefly to consider the question of interface wetting. As mentioned earlier, a $Z(3)$ interface is believed to consist of two order-disorder interfaces stuck back to back, enclosing a narrow slice of disordered (low-temperature) phase. Evidence to support this theory has been provided by simulations in 3+1 dimensions[8], measuring the surface tension of both types of interface at the critical temperature, and it seems reasonable that the same behaviour should be seen in $2+1$ dimensions. As we do not measure the properties of order-disorder interfaces in this survey, it is worth considering whether we can see any indication of the wetting behaviour in our study of the $Z(3)$ interface alone. At the critical temperature itself, complete wetting would suggest that the interface profile should split in two to include an area of disordered ($< L >= 0$) phase. Although average profiles akin to fig. 4.11 show slightly more divergence from the tanh form at lower $\beta$, it is not possible to see such a splitting clearly in our system. However, we would expect any area of disordered phase within the interface to remain extremely small until $\beta \approx \beta_c = 8.175$[44], and so we cannot take this as evidence against wetting in $2+1$ dimensions.

We cannot obtain reliable average profiles much below $\beta = 8.25$ because of the difficulty of identifying the interface so close to collapse, in order to keep it centred for the average. Another approach to the problem which can be followed at lower $\beta$, though, is artificially to split the interface into two by changing the initial conditions to include a large area of disordered phase between the two ordered domains. The persistence, or otherwise, of this interface structure can then be examined at various $\beta \approx \beta_c$. In fig. 4.12, instantaneous profiles are shown at fifty-sweep intervals on a long lattice at $\beta = 8.15$ (actually just above $\beta_c$ for our finite-sized system). Disordered phase is present only where both left- *and* right-hand plots (*i.e.* real *and* imaginary Polyakov profiles) coincide with the dotted line ($< L >= 0$: disordered phase). The disordered region given by the initial conditions is seen not to persist, but rather quickly dissolves into many bubbles of $Z(3)$ phase. Again, this cannot be taken as strong evidence against complete wetting, because of the imprecise nature of the observation; but neither does there seem anything in our observations which could be taken as positive evidence of the wetting.

## 4.8. Estimates of the Critical Coupling and the Magnetisation Exponent

We now proceed to more concrete measurements of interface properties. We know that the expectation values of the Polyakov lines in the main phases to either side of the interface must tend to zero as the temperature descends to its critical value, since the disordered phase which exists below that temperature has the $Z(3)$ vacuum symmetry restored: $< L >= 0$. This means that by monitoring the average values of $< L >$ at either end of the lattice, which can be read off from graphs such as fig. 4.11, one can obtain an estimate for the critical temperature itself, and also the magnetisation exponent, $\beta_M$, determining the fall in the Polyakov expectation:

$$< L >\sim (\beta - \beta_c)^{\beta_M}, \qquad \beta \geq \beta_c.$$

Our estimates of the Polyakov line expectations at opposite ends of the lattice are given in the following tables for each simulation performed, along with an average value thence obtained for each value of $\beta$:

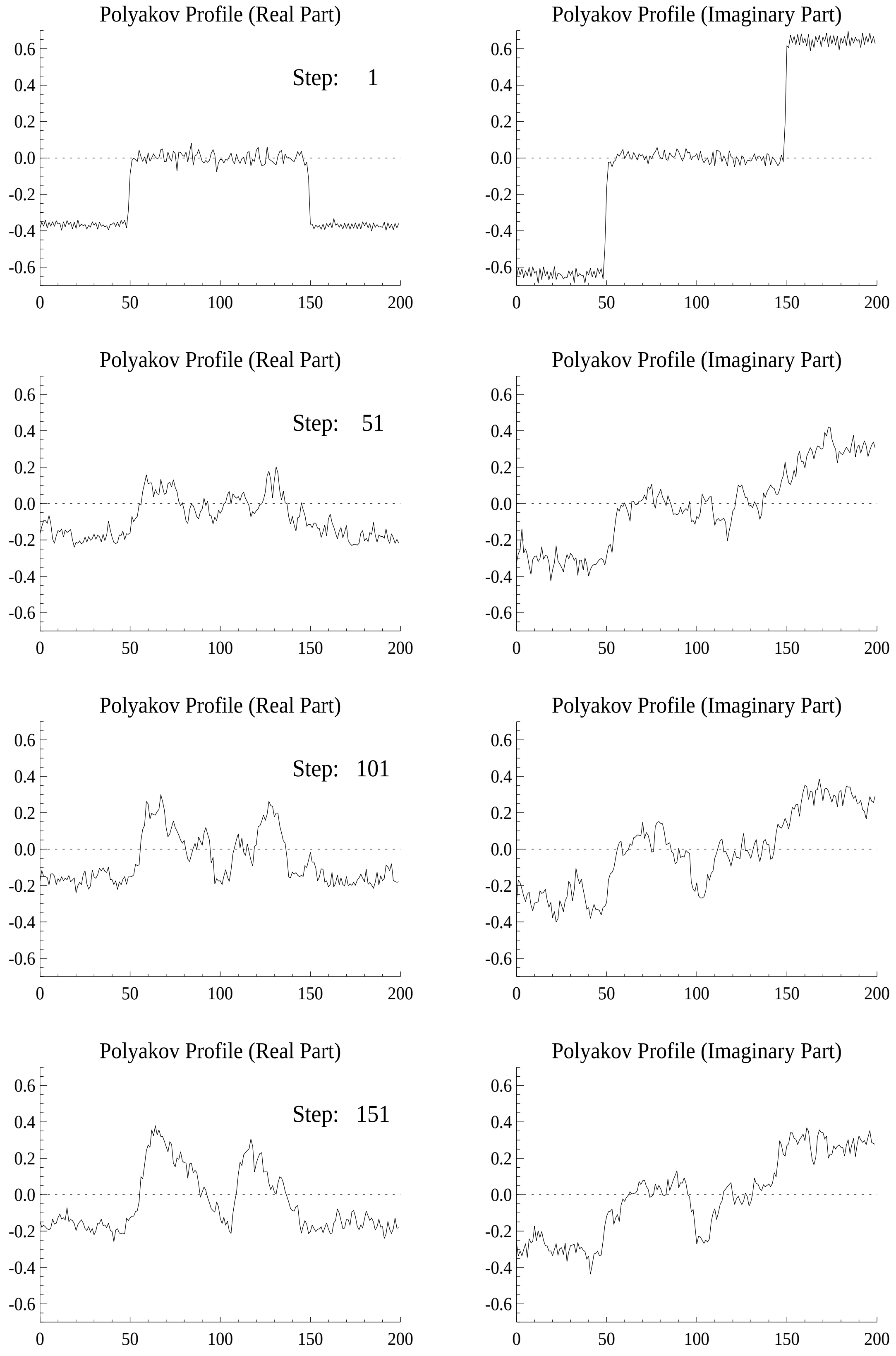

**Fig. 4.12:** Polyakov line profiles for $\beta = 8.15$ on a $2 \times 48 \times 200$ lattice, produced by taking instantaneous transverse averages at fifty-sweep intervals.

Table 4.4: Estimates of $< L >_L$

| $\beta \ddots L_x$ | *18* | *24* | *30* | *36* | *42* | *48* | *54* | *Average* |
|---|---|---|---|---|---|---|---|---|
| *8.25* | -0.15414 | -0.15975 | -0.15421 | -0.16121 | -0.16205 | -0.16053 | -0.16360 | -0.1594(37) |
| *8.35* | -0.18135 | -0.18355 | -0.18407 | -0.18655 | -0.18585 | -0.18511 | -0.18647 | -0.1847(19) |
| *8.50* | -0.20172 | -0.20304 | -0.20348 | -0.20306 | -0.20409 | -0.20379 | -0.20457 | -0.2034(9) |
| *8.75* | -0.22241 | -0.22277 | -0.22247 | -0.22296 | -0.22306 | -0.22286 | -0.22340 | -0.2228(3) |
| *9.00* | -0.23644 | -0.23666 | -0.23687 | -0.23685 | -0.23724 | -0.23715 | -0.23711 | -0.2369(3) |

Table 4.5: Estimates of $< L >_R$

| $\beta \ddots L_x$ | *18* | *24* | *30* | *36* | *42* | *48* | *54* | *Average* |
|---|---|---|---|---|---|---|---|---|
| *8.25* | 0.33667 | 0.33570 | 0.32767 | 0.34012 | 0.33955 | 0.33600 | 0.33651 | 0.3360(41) |
| *8.35* | 0.37655 | 0.37486 | 0.37494 | 0.37491 | 0.37511 | 0.37453 | 0.37612 | 0.3753(7) |
| *8.50* | 0.40925 | 0.40898 | 0.40797 | 0.40816 | 0.40895 | 0.40908 | 0.40855 | 0.4087(5) |
| *8.75* | 0.44595 | 0.44575 | 0.44549 | 0.44665 | 0.44629 | 0.44620 | 0.44635 | 0.4461(4) |
| *9.00* | 0.47399 | 0.47416 | 0.47412 | 0.47340 | 0.47391 | 0.47452 | 0.47407 | 0.4740(3) |

The figures from each simulation come from the transverse-averaged Polyakov profiles, averaging over all 100k main sweeps, and then averaging over the six longitudinal points at each end of the lattice ($0 \leq z \leq 5$ and $66 \leq z \leq 71$). The errors from a given simulation are $O(10^{-6})$, decreasing to $O(10^{-7})$ for the larger $\beta$ values, and are not shown, being negligible compared to the error between different simulations, which is included with the average.

An accurate fit to the form

$$< L >= a(\beta - \beta_c)^{\beta_M}$$

is difficult, since we only have five points and must fit three parameters. Using the gradient expansion fitting of *IDL*'s function "CURVEFIT", we plot the average values from the tables in fig. 4.13. The independent estimates from each of the two $Z(3)$ phases give the following results for the critical temperature and magnetisation exponent:

$$\beta_c = 8.170(5),$$

$$\beta_M = 0.159(3).$$

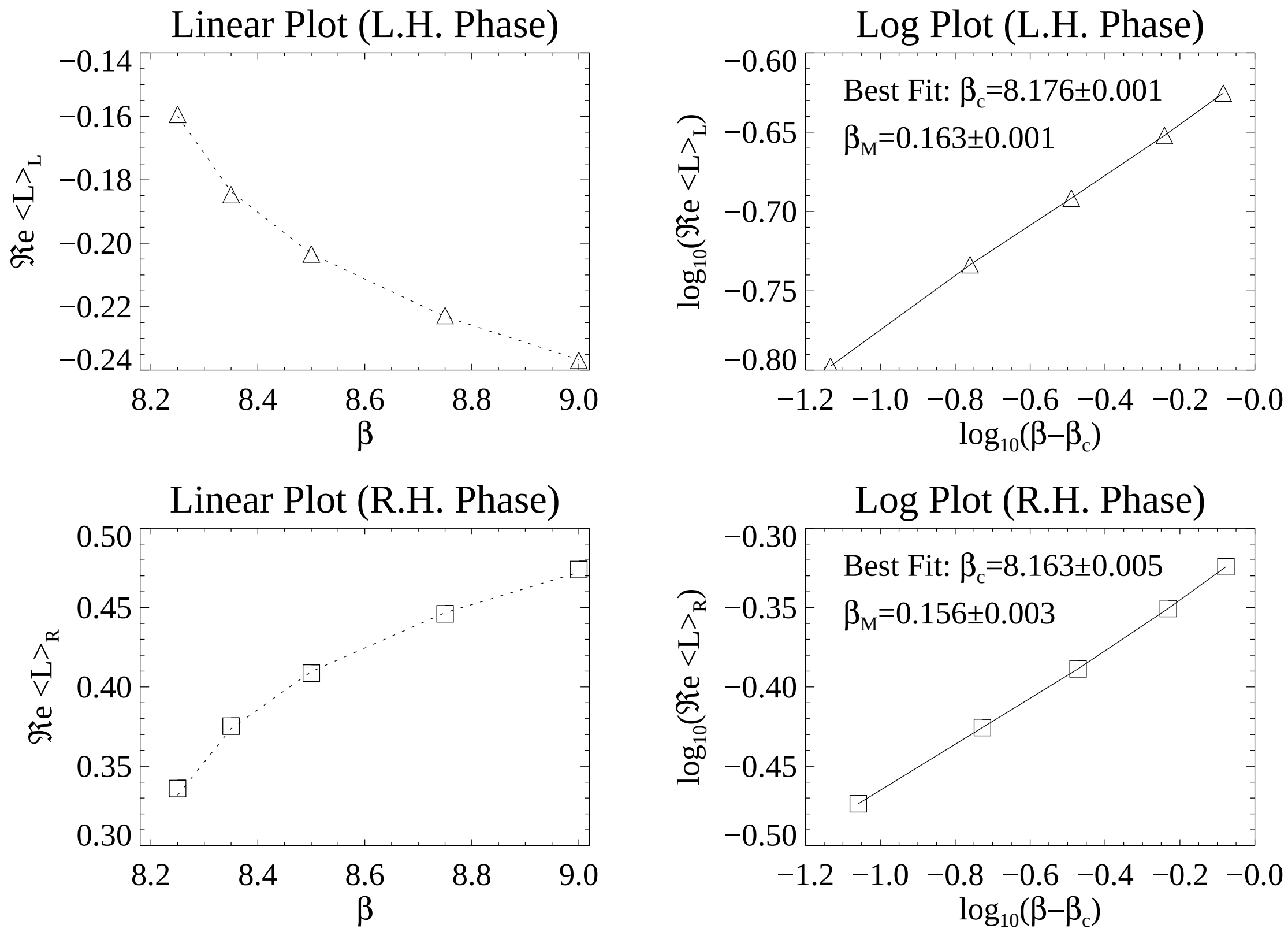

**Fig. 4.13:** A three-parameter fit for the critical temperature, $\beta_c$, and the magnetisation exponent, $\beta_M$, from the expectation value of the Polyakov line deep in the phases at the right- and left-hand ends of the lattice (denoted by "R.H." and "L.H." respectively), as found from averaged profiles such as fig. 4.11.

These are some distance from the values of $\beta_c \approx 8.14$ and $\beta_M \approx 0.13$ obtained in [45], the estimate for $\beta_c$ being closer to the continuum estimate of 8.175 than to the value of 8.14 which broadly corresponds to our lattice sizes. As a check, fig. 4.14 shows our estimates for $\beta_M$ if we assume that $\beta_c = 8.14$; the estimates are little changed.

## 4.9. Quantitative Results

Having established that the interface remains as a thin, localised object even very close to the critical temperature, we now wish to examine its behaviour in more quantitative detail. To do this, we need a way to quantify the fluctuations in the interface. The phase transition is second-order, with the Polyakov lines (and the interface) vanishing smoothly at $\beta_c$; thus, we expect the fluctuations to diverge as $\beta \to \beta_c$, and we wish to quantify this divergence.

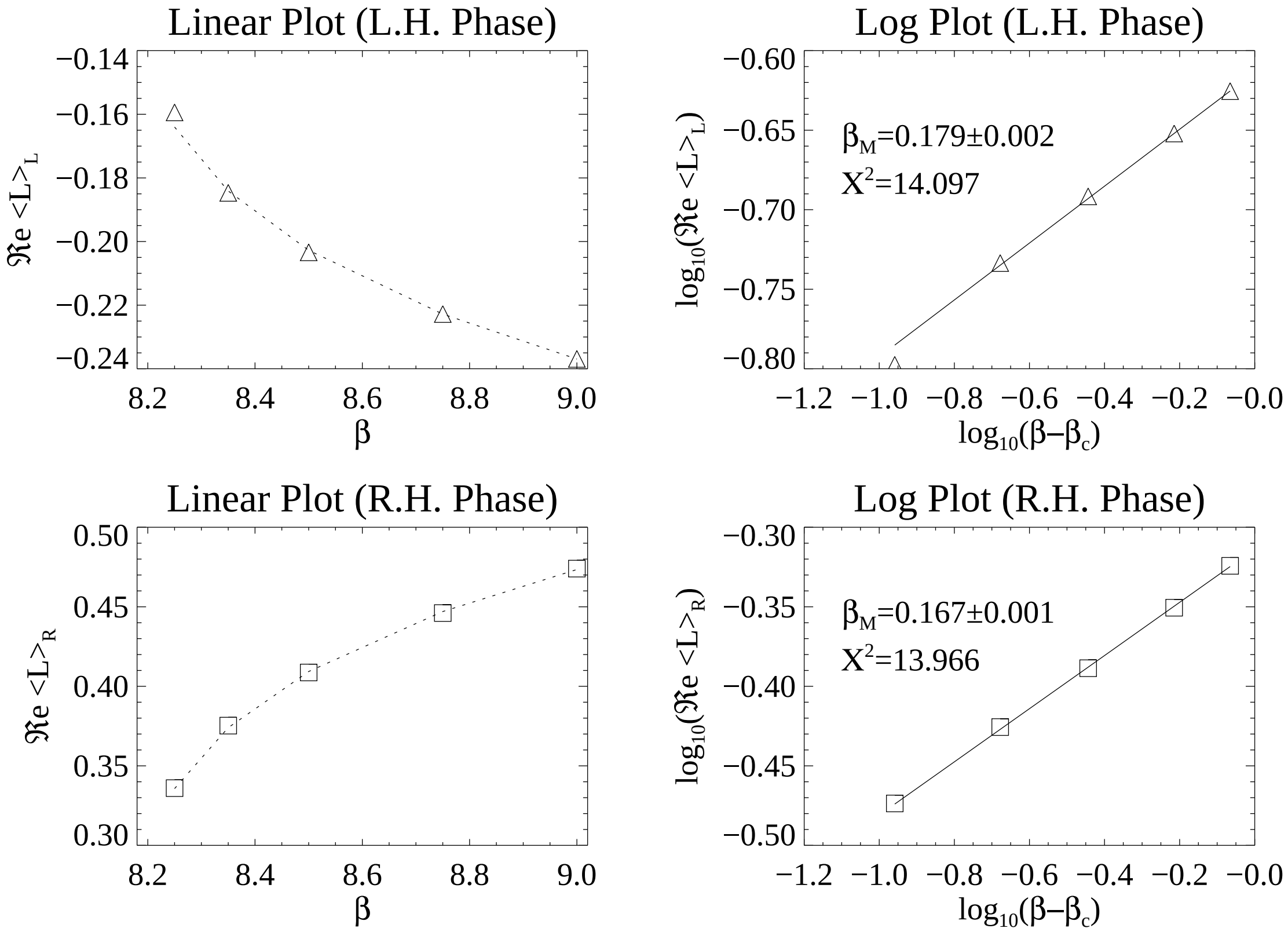

**Fig. 4.14:** A two-parameter fit for the magnetisation exponent, $\beta_M$, using $\beta_c = 8.14$.

Illustrated in fig. 4.15 is a contour going across the lattice, with displacements from its average position at transverse location $x$ given by $\phi(x)$. To characterise the fluctuations, we monitor the fluctuation "moments", where the $n$th moment is defined to be $< \phi(x)^n >$. We track the first six moments ($1 \leq n \leq 6$) in our simulations; higher moments are subject to unacceptably large errors. We take a separate average for each transverse position across the lattice, and, because of the translation invariance of the interface, we expect the moments to be roughly constant across the lattice after a large number of sweeps.

First of all, we need to examine the qualitative behaviour of the fluctuation moments. As mentioned above, we expect the moments to be essentially constant across the lattice, owing to the periodic boundary conditions in the transverse direction and the translation invariance of the interface. Of course, we would expect the moments to approach constancy more closely with increasing simulation length.

The following graphs show some moments for $\beta = 8.5$, averaged over the full 100k sweeps. Separate sets of data are shown for each of the three contours followed, at 25%, 50% and 75% of the height of the interface respectively. The qualitative behaviour is roughly the same for each contour. The upper and lower contours are subject to rather more fluctuation than the middle one, as we expect since they are closer to the outside of the interface and therefore more susceptible to bubbles of phase forming nearby and distorting the edge of the interface. The even moments are fairly constant across the lattice, whilst the odd moments are very much smaller and show more variation. Note that the first moment must average to zero across the lattice, by definition, and we likewise expect all odd moments to be comparatively close to zero, becoming closer with even longer runs.

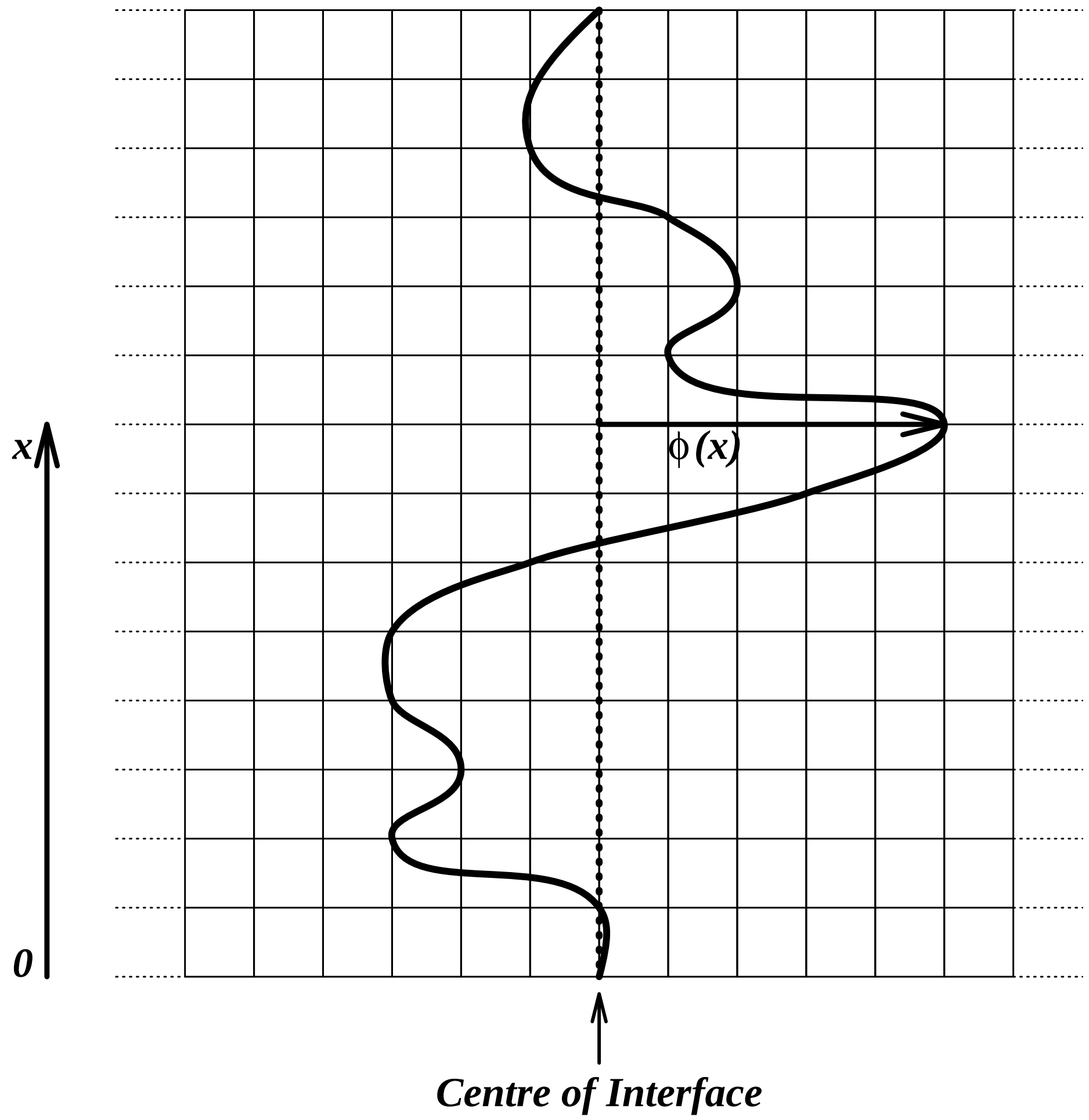

**Fig. 4.15:** An interface contour is illustrated crossing the lattice: the bold line. The dotted line marks the average position of the contour, at the longitudinal centre of the lattice. We quantify the fluctuations in the interface shape by the displacement, $\phi(x)$, from the average position at each point $x$ across the lattice.

In fig. 4.17, we show the moment data from the smoothed contours measured in the same simulation as fig. 4.2. One can see that the behaviour is very similar to the raw data, both qualitatively and quantitatively. Since this may indicate that the smoothed data has not lost a significant amount of information about the fluctuations, we shall present our data later on for both raw and smoothed moments, in order to compare the two.

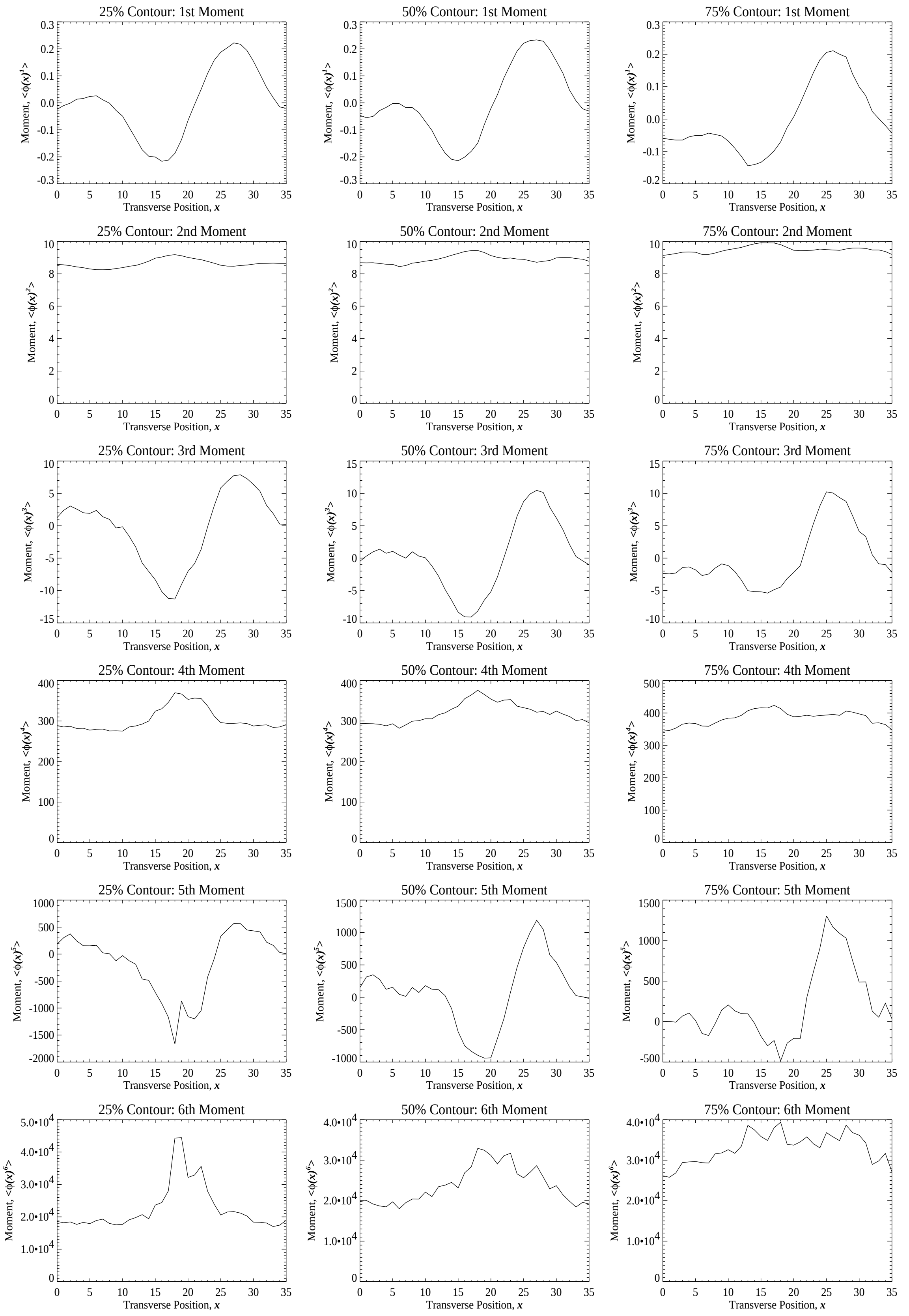

**Fig. 4.16:** The first six moments for $\beta = 8.5$ on a $2 \times 36 \times 72$ lattice, shown separately for each of the three contour levels followed. These are averages over 100k sweeps for the raw Polyakov data.

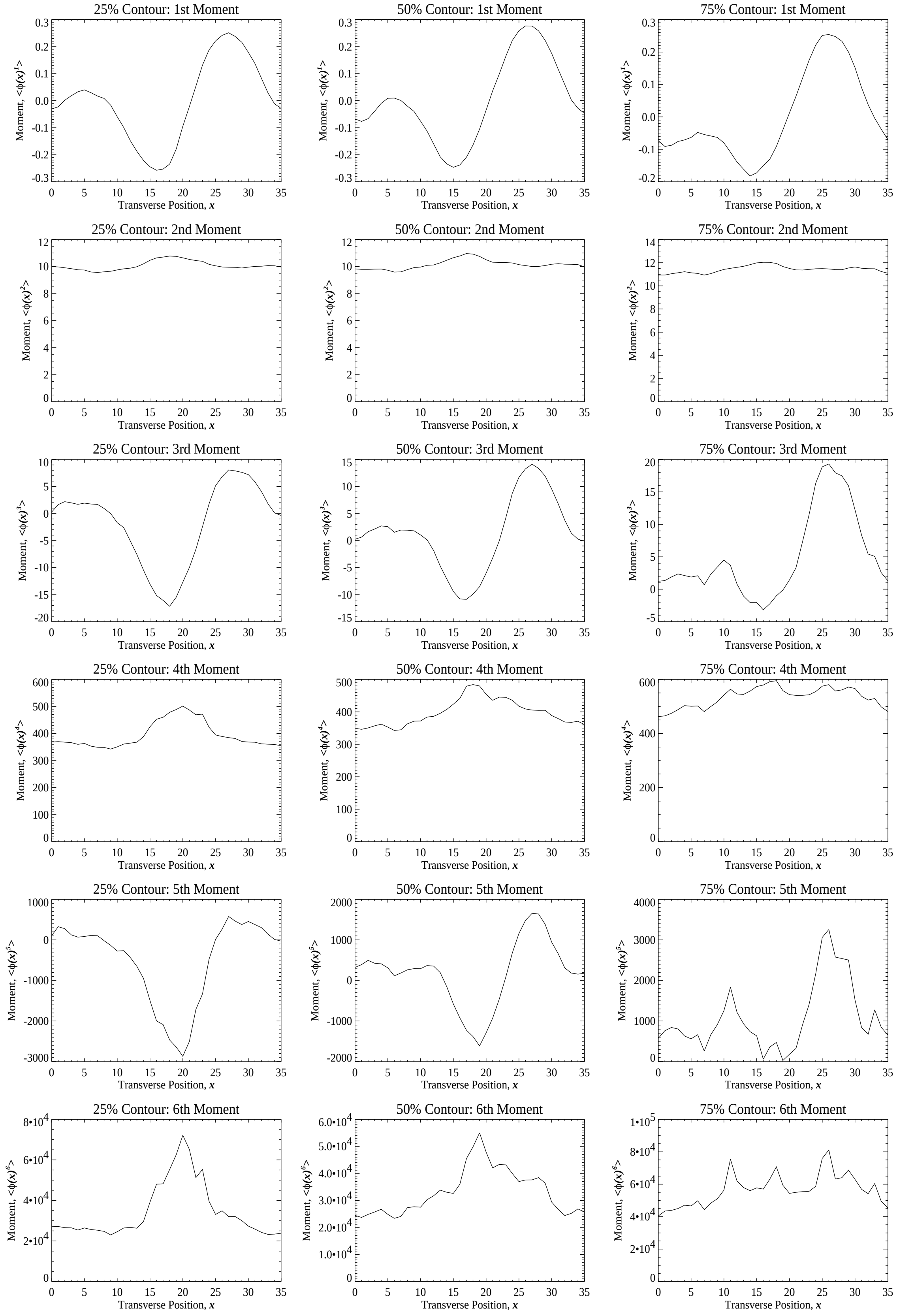

**Fig. 4.17:** The first six moments for $\beta = 8.5$ on a $2 \times 36 \times 72$ lattice, shown separately for each of the three contour levels followed. These are averages over 100k sweeps for the smoothed Polyakov data corresponding to fig. 4.16.

## 4.10. A Toy Model: A Scalar Field Theory with Associated Feynman Rules

Since the interface has been shown to behave as a fluctuating, narrow object, winding across the lattice, it may be helpful to think about it as a one-dimensional object, *viz.* a string. At high temperatures, the string stretches tightly across the lattice, its length being equal to the transverse size of the lattice. At low temperatures, though, if coils and twists ever more violently, and loops form which break off and recombine with it, representing bubbles of $Z(3)$ phase. If we take this picture further, we can consider the possibility of describing the string in terms of a one-dimensional field theory. The scalar field in this case is simply the displacement of the string from equilibrium at a position $x$ along its length: $\phi(x)$. At high temperatures, we essentially have a free field theory; at low temperatures, we need an interaction term to account for the breaking off and recombination of loops of string.

To preserve the translation invariance of the interface, the Lagrangian cannot depend directly on $\phi(x)$, but instead must depend only on the derivative $\partial_x\phi(x)$. We keep the normal kinetic term for a scalar field theory, and add a quartic interaction term; a cubic interaction is ruled out by "parity", as we expect the interface to be invariant under $\phi \to -\phi$. Thus, we have a Lagrangian of the form

$$\mathcal{L} = \tfrac{1}{2}\gamma(\partial_x\phi)^2 + \tfrac{\lambda}{4!}(\partial_x\phi)^4, \tag{4.1}$$

where $\gamma$ and $\lambda$ are both unknown functions of $\beta$. By examining the dependence of the fluctuations on the temperature, we shall be able to determine these functions. Since we expect the fluctuations to diverge as $\beta \to \beta_c$, this implies that $\gamma$ and $\lambda$ will both be functions of $(\beta - \beta_c)$.

To test how well this toy model describes the actual behaviour of the $Z(3)$ interface, we need to calculate the correlation functions of $\phi$, which we can then compare with the behaviour seen. The $n$-point connected vacuum correlation function at point $x$, $<\phi(x)^n>$, corresponds to the $n$th Wick-subtracted fluctuation moment at the same point. The Wick subtracted moments, denoted by the subscript "Ws", are given, as is well known, by the following relations:

$$\begin{aligned}
&\text{2nd Moment:} \quad && <\phi^2>_{Ws} = <\phi^2>,\\
&\text{4th Moment:} \quad && <\phi^4>_{Ws} = <\phi^4> -3<\phi^2>^2,\\
&\text{6th Moment:} \quad && <\phi^6>_{Ws} = <\phi^6> -15<\phi^2>^3 -15<\phi^2><\phi^4>_{Ws}\,.
\end{aligned}$$

For the Lagrangian above, the odd correlation functions are zero, so we expect all the odd moments to be zero. To calculate the even moments, we need to calculate the appropriate vacuum diagrams.

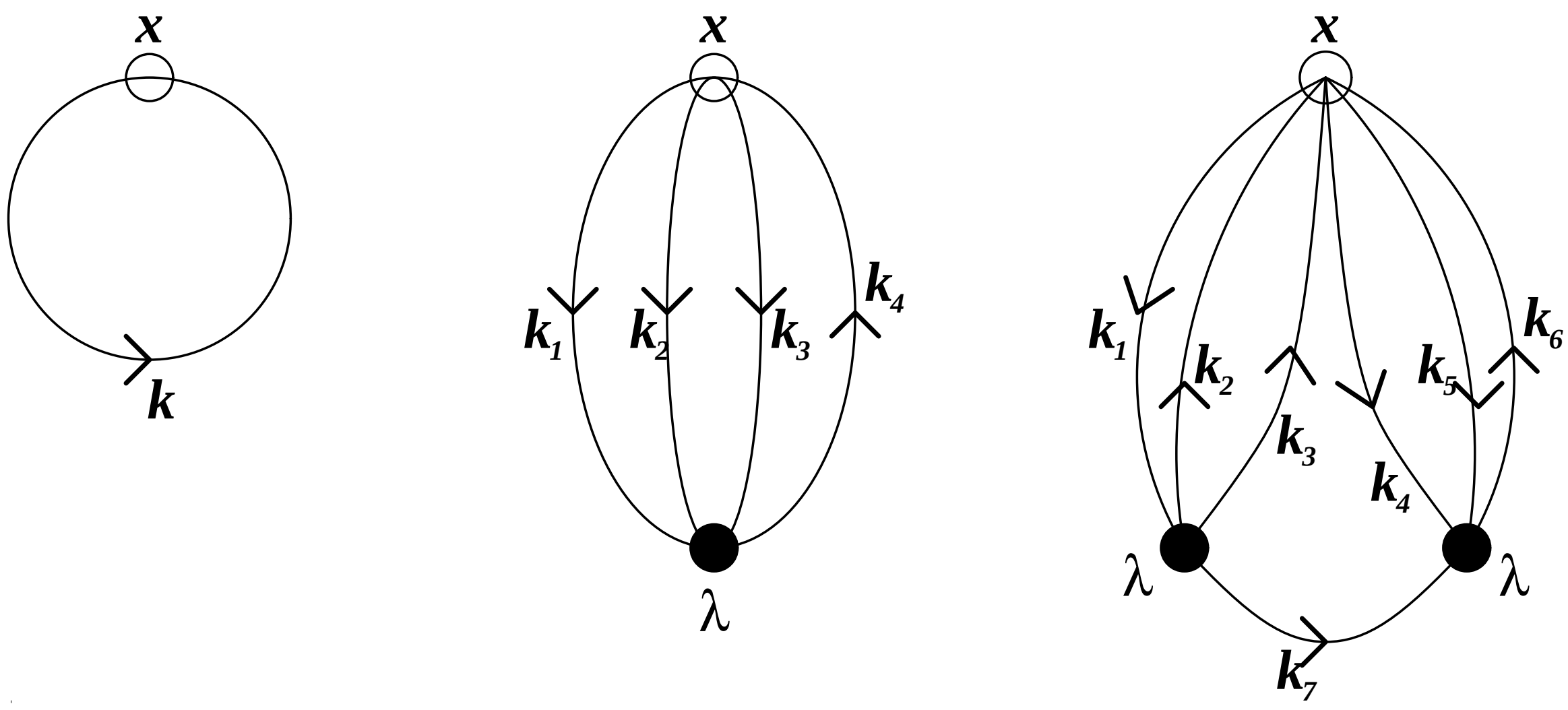

**Fig. 4.18:** The vacuum diagrams corresponding to the 2nd, 4th and 6th connected vacuum correlation functions respectively. The open circle simply denotes point $x$, the endpoint of each external leg; the solid circle denotes an interaction vertex, $\lambda$; and the momentum travelling along a leg is denoted by $k$.

These diagrams are shown in fig. 4.18, with the momenta denoted by the $k$'s, and the four-way vertices marked by $\lambda$, corresponding to the interaction term in the Lagrangian. The diagrams shown are the lowest-order Wick-subtracted diagrams with two, four and six legs respectively leaving the point $x$. To perform the calculations, we need to derive Feynman rules for a $(\partial_x\phi)^4$ theory, rather than the usual $\phi^4$, and we do this now.

The path integral, for our one spatial dimension only, is

$$W[J] = N\int \mathcal{D}\phi e^{\int dx(\mathcal{L}(\phi,\partial_x\phi)+J\phi)},$$

where $N$ is the normalisation and $\mathcal{L}$ is given by (4.1), written in the form

$$\mathcal{L} = \mathcal{L}_0 + \mathcal{L}_{int}(\partial_x\phi), \qquad \mathcal{L}_0 = \tfrac{1}{2}\gamma(\partial_x\phi)^2, \qquad \mathcal{L}_{int} = \tfrac{\lambda}{4!}(\partial_x\phi)^4.$$

In the absence of the interaction,

$$W[J] = W_0[J] = Ne^{\frac{1}{2}\int dxdyJ(x)\Delta_F(x-y)J(y)},$$

with the Feynman propagator for the scalar field,

$$\Delta_F(x-y) = \int \frac{dp}{2\pi\gamma}\frac{e^{ip(x-y)}}{p^2}.$$

To include the interaction, we expand $W[J]$ as a power series in $\lambda$:

$$\begin{aligned}
W[J] &= e^{dx\mathcal{L}_{int}(\frac{\delta}{\delta J(x)})}W_0[J] \\
&= \left[1 + \int dx\mathcal{L}_{int}(\frac{\delta}{\delta J(x)}) + \dots\right]W_0[J] \\
&= W_0[J] + \frac{\lambda}{4!}\int dx\left(\frac{\partial}{\partial x}\frac{\delta}{\delta J(x)}\right)^4 W_0[J] + \dots.
\end{aligned}$$

Now,

$$\frac{\partial}{\partial x}\frac{\delta W_0}{\delta J(x)} = \int dy \partial_x[\Delta_F(x-y)]J(y)W_0[J],$$

and further differentiation of this yields

$$\begin{aligned} W[J] = \Bigg\{1 + \frac{\lambda}{4!}\int dx\Bigg[ & \, 3(\partial_x^2\Delta_F(0))^2 \\ & +6\int dy_1 dy_2 \partial_x^2[\Delta_F(0)\partial_x[\Delta_F(x-y_1)\partial_x[\Delta_F(x-y_2)]]]J(y_1)J(y_2) \\ & +\int dy_1 dy_2 dy_3 dy_4 \partial_x[\Delta_F(x-y_1)\partial_x[\Delta_F(x-y_2) \\ & \quad \times \partial_x[\Delta_F(x-y_3)\partial_x[\Delta_F(x-y_4)]]]]J(y_1)J(y_2)J(y_3)J(y_4) \\ & +\ldots\Bigg]\Bigg\}W_0[J]. \end{aligned}$$

The free-field Green's functions are defined by

$$\mathcal{G}^{(n)}(x_1,\ldots,x_n) = \frac{\delta^n W[J]}{\delta J(x_1)\ldots\delta J(x_n)}\Big|_{J=0},$$

giving the propagator

$$\mathcal{G}^{(2)}(x_1,x_2) = \Delta_F(x_1-x_2).$$

The connected Green's functions are defined by

$$G^{(n)}(x_1,\ldots,x_n) = \frac{\delta^n X[J]}{\delta J(x_1)\ldots\delta J(x_n)}\Big|_{J=0}, \qquad X[J] = \ln W[J],$$

giving interaction term

$$\begin{aligned} G^{(4)}(x_1,x_2,x_3,x_4) &= \lambda\int dx \partial_x[\Delta_F(x-x_1)\partial_x[\Delta_F(x-x_2)\partial_x[\Delta_F(x-x_3)\partial_x[\Delta_F(x-x_4)]]]] \\ &= \lambda\int dx(-\partial_{x_1}[\Delta_F(x-x_1)])(-\partial_{x_2}[\Delta_F(x-x_2)]) \\ &\qquad \times(-\partial_{x_3}[\Delta_F(x-x_3)])(-\partial_{x_4}[\Delta_F(x-x_4)]). \end{aligned}$$

Converting to momentum space,

$$\begin{aligned} &\widetilde{G}^{(4)}(p_1,p_2,p_3,p_4)(2\pi)^4\delta(p_1+p_2+p_3+p_4) \\ &\qquad = \int dx_1 dx_2 dx_3 dx_4 G^{(4)}(x_1,x_2,x_3,x_4)e^{-i(p_1x_1+p_2x_2+p_3x_3+p_4x_4)}, \end{aligned}$$

and, integrating by parts,

$$\int dx_1 \partial_{x_1}\Delta_F(x_1-x)e^{-ip_1x_1} = ip_1\int \Delta_F(x_1-x)e^{-ip_1x_1}.$$

Thus, each vertex picks up a factor of

$$\widetilde{G}^{(4)}(p_1,p_2,p_3,p_4) = \lambda p_1p_2p_3p_4. \qquad \text{where} \qquad p_1+p_2+p_3+p_4 = 0.$$

## 4.11. Field Theoretic Predictions for Fluctuation Moment Behaviour

In the continuum, therefore, the Feynman diagrams of fig. 4.18 are given by the following formulae:

$$
\begin{aligned}
< \phi(x)^2 > &= \frac{1}{2\pi\gamma}\int \frac{dk}{k^2}, \\
< \phi(x)^4 >_{Ws} &= \frac{\lambda}{4!}\frac{1}{(2\pi\gamma)^4}\int \frac{dk_1}{k_1^2}\cdots\frac{dk_4}{k_4^2}e^{-i(k_1+k_2+k_3-k_4)x} \\
&\qquad \times 2\pi k_1k_2k_3k_4\delta(k_1+k_2+k_3-k_4), \\
&= \frac{\lambda}{4!\gamma^4}\frac{1}{(2\pi)^3}\int \frac{dk_1}{k_1}\frac{dk_2}{k_2}\frac{dk_3}{k_3}\frac{1}{(k_1+k_2+k_3)}, \\
< \phi(x)^6 >_{Ws} &= \frac{\lambda^2}{2!}\frac{1}{(2\pi\gamma)^7}\int \frac{dk_1}{k_1^2}\cdots\frac{dk_7}{k_7^2}e^{-i(k_1-k_2-k_3+k_4+k_5-k_6)x} \\
&\qquad \times (2\pi)^2k_1k_2k_3k_4k_5k_6k_7^2\delta(k_1-k_2-k_3-k_7)\delta(k_4+k_5-k_6+k_7) \\
&= \frac{\lambda^2}{2!\gamma^7}\frac{1}{(2\pi)^5}\int \frac{dk_2}{k_2}\frac{dk_3}{k_3}\frac{dk_4}{k_4}\frac{dk_5}{k_5}dk_7\frac{1}{(k_2+k_3+k_7)}\frac{1}{(k_4+k_5+k_7)}.
\end{aligned}
$$

Note that there is no exponential in the first formula, as the propagator starts and finishes at the same point, $x$. Also, the symmetry factor in the last formula is from $\frac{1}{2!}\times(\frac{1}{4!})^2\times(4\times3\times2)^2=\frac{1}{2!}$, from the two vertices, the four legs per vertex, the number of legs connected to $x$, and the number connected within the loop respectively.

Since we are working on a finite space of extent $L_x$, we have to replace the integrals over $k$ with sums over integers $n$, putting $k=2\pi n/L_x$:

$$\frac{1}{2\pi}\int dk \rightarrow \frac{1}{L_x}\sum_{n=1}^{\infty}.$$

Using the numerically-calculated results

$$\sum_n \frac{1}{n^2}=\frac{\pi^2}{6}, \qquad \sum_{n_a\ldots n_c}\frac{1}{n_a}\frac{1}{n_b}\frac{1}{n_c}\frac{1}{(n_a+n_b+n_c)}\approx 6.49(1)$$

$$\text{and} \qquad \sum_{n_a\ldots n_e}\frac{1}{n_a}\frac{1}{n_b}\frac{1}{n_c}\frac{1}{n_d}\frac{1}{(n_a+n_b+n_c)}\frac{1}{(n_a+n_d+n_e)}\approx 60(1),$$

we obtain the following expressions:

$$
\begin{aligned}
< \phi(x)^2 > &= \frac{L_x}{(2\pi)^2\gamma}\sum_n\frac{1}{n^2}=\frac{L_x}{24\gamma}, \\
< \phi(x)^4 >_{Ws} &= (6.49(1))\frac{\lambda}{4!}\frac{L_x}{(2\pi)^4\gamma^4}, \\
< \phi(x)^6 >_{Ws} &= (60(1))\frac{\lambda^2}{2!}\frac{L_x}{(2\pi)^6\gamma^7}.
\end{aligned}
$$

Note that the above expressions for the correlation functions do not depend on the position $x$, so we now write $\phi(x)$ simply as $\phi$. This confirms our original assertion that the fluctuation moments should be constant across the lattice. In future, we shall work with the average of a fluctuation moment across the lattice. Now, since the action, $\mathcal{S} = \int \mathcal{L} dx$, must be a dimensionless quantity, and assuming $\gamma$ to be dimensionless, we find that $\phi$ must have space dimension $\frac{1}{2}$, and coupling $\lambda$ dimension 1. Thus, as the only fundamental length in our space is $L_x$, this implies that $\lambda \sim L_x$, or $\lambda = \lambda_0 L_x$. Substituting this dependence into the previous expressions, we obtain the following predictions:

$$< \phi^2 > \sim L_x, \qquad < \phi^4 >_{Ws} \sim L_x^2 \qquad \text{and} \qquad < \phi^6 >_{Ws} \sim L_x^3.$$

Checking the $L_x$-dependence of our results will thus help to verify whether the scalar field theory really is a good model for the interface fluctuations. However, the above calculations have all been for the continuum, so we first need to calculate some corrections.

We have neglected the fact that we are working on a lattice with site spacing $a$ and periodic boundary conditions. To take this into account, we need to replace $k^2$ with $4\sin^2(ak/2)$. This is equivalent to the following change in our sums over $n$:

$$\frac{1}{k^2} = \left(\frac{L_x}{2\pi n}\right)^2 \to \frac{1}{4\sin^2(\pi n a/L_x)} = \frac{1}{4\sin^2(\pi n/N_x)}.$$

We now calculate the corrections to our previous expressions which result from this change.

First, we tackle the two-point function. To do this, we calculate the sum from an integral, working with an expansion in the small parameter $\xi$:

$$\int_{\frac{\pi}{N_x}}^{\frac{(N_x-1)\pi}{N_x}} \frac{1}{\sin^2 x} dx = 2\int_{\frac{\pi}{N_x}}^{\frac{\pi}{2}} \frac{1}{\sin^2 x} dx = 2\sum_{n=1}^{\frac{N_x}{2}-1} \int_0^{\frac{\pi}{N_x}} \frac{1}{\sin^2(\frac{\pi n}{N_x}+\xi)} d\xi.$$

Consider the $n = 1$ term:

$$\int_0^{\frac{\pi}{N_x}} \frac{1}{\sin^2(\frac{\pi}{N_x}+\xi)} d\xi = \frac{\pi/N_x}{\sin^2(\pi/N_x)} - \frac{N_x}{2\pi} + \mathcal{O}[(\pi/N_x)^3].$$

Similarly, for the $n > 1$ terms,

$$\begin{aligned}
\sum_{n=2}^{\frac{N_x}{2}-1} \int_0^{\frac{\pi}{N_x}} \frac{1}{\sin^2(\frac{\pi n}{N_x}+\xi)} d\xi &= \sum_{n=2}^{\frac{N_x}{2}-1} \int_0^{\frac{\pi}{N_x}} \bigg[\csc^2(\frac{\pi n}{N_x}) - 2\xi\cot(\frac{\pi n}{N_x})\csc^2(\frac{\pi n}{N_x}) \\
&\quad + 3\xi^2\left(\cot^2(\frac{\pi n}{N_x})\csc^2(\frac{\pi n}{N_x}) + \frac{1}{3}\csc^2(\frac{\pi n}{N_x})\right) + \dots\bigg] d\xi \\
&= \frac{\pi}{N_x}\sum_{n=2}^{\frac{N_x}{2}-1}\csc^2(\frac{\pi n}{N_x}) \quad - \quad \left(\frac{\pi}{N_x}\right)^2 \sum_{n=2}^{\frac{N_x}{2}-1}\cot(\frac{\pi n}{N_x})\csc^2(\frac{\pi n}{N_x}) \\
&\quad + \left(\frac{\pi}{N_x}\right)^3 \sum_{n=2}^{\frac{N_x}{2}-1}\left(\cot^2(\frac{\pi n}{N_x})\csc^2(\frac{\pi n}{N_x}) + \frac{1}{3}\csc^2(\frac{\pi n}{N_x})\right) + \dots \\
&= \frac{\pi}{N_x}\sum_{n=2}^{\frac{N_x}{2}-1}\left(\csc^2(\frac{\pi n}{N_x}) + \frac{N_x}{\pi}(-\tfrac{1}{n^3} + \tfrac{1}{n^4} - \dots)\right) + \mathcal{O}[(\pi/N_x)^3].
\end{aligned}$$

Now,

$$\sum_{n=2}^{\frac{N_x}{2}-1}(-\tfrac{1}{n^3}+\tfrac{1}{n^4}-\ldots)=\sum_{n=2}^{\frac{N_x}{2}-1}\tfrac{-1}{n^2(n+1)}=\tfrac{1}{2}-(\tfrac{\pi^2}{6}-1)+\mathcal{O}[(2/N_x)^3].$$

Thus, our result for the original integral, dropping $\mathcal{O}[(\pi/N_x)^3]$ terms and above, is

$$\frac{2\pi}{N_x}\sum_{n=1}^{\frac{N_x}{2}-1}\csc^2(\frac{\pi n}{N_x})-\frac{2N_x}{\pi}(\frac{\pi^2}{6}-1)=\frac{\pi}{N_x}\sum_{n=1}^{N_x-2}\csc^2(\frac{\pi n}{N_x})+\frac{N_x}{\pi}(3-\frac{\pi^2}{3}).$$

This series expansion in $N_x^{-1}$ must be equal to the value of the original integral,

$$[-\cot x]_{\frac{\pi}{N_x}}^{\frac{(N_x-1)\pi}{N_x}}=2\cot\frac{\pi}{N_x}=2(\frac{N_x}{\pi}-\frac{\pi}{3N_x}+\ldots),$$

so our final result is that

$$\frac{1}{N_x}\sum_{n=1}^{N_x-2}\csc^2(\frac{\pi n}{N_x})\approx\frac{N_x}{\pi^2}\left(\frac{\pi^2}{3}-1\right)-\frac{2}{3N_x}.$$

The correction to our original prediction, that $<\phi^2>\sim N_x$, is negligible for our purposes (about 0.3% for $N_x=18$). We assume that the corrections to the higher moments are similarly negligible, as the calculation of these lattice corrections is considerably more difficult.

## 4.12. Observed Behaviour of Fluctuation Moments

We now compare these predictions with the results that we obtain from our computer simulations. Tables 4.6-8 list a sample set of results for a contour halfway up the interface, using raw (*i.e.* unprocessed) data. An estimate of each even moment is presented for each value of $\beta$ and $L_x$, together with the calculated gradients of log–log plots of the moments against $(\beta-\beta_c)$ or $L_x$ as appropriate, which give the power-law dependence of the moments on these quantities. To examine the behaviour of the interface more fully, we tracked fluctuations in a total of three contours, defined to be equally spaced between the top and bottom of the interface. For each of these three contours, we also measured the fluctuations in the interface after smoothing, to see whether the procedure would affect results. Appendix B lists complete results for all three contours used, and both raw and smoothed Polyakov data. In this chapter, tables 4.9-11 summarise the power-law dependence of the moments from each dataset on $L_x$, while tables 4.12-14 give the equivalent dependence on $\beta-\beta_c$.

Table 4.6: Estimates of $<\phi^2>$ from Raw Binned Data, Middle (50%) Contour

| $\beta \ddots L_x$ | *18* | *24* | *30* | *36* | *42* | *48* | *54* | *Slope* |
|---|---|---|---|---|---|---|---|---|
| *8.25* | $6.87^{+0.22}_{-0.14}$ | $8.27^{+0.12}_{-0.03}$ | $12.3^{+0.2}_{-0.2}$ | $14.4^{+0.1}_{-0.1}$ | $17.8^{+0.1}_{-0.1}$ | $20.8^{+0.1}_{-0.2}$ | $23.1^{+0.2}_{-0.1}$ | $1.20 \pm 0.03$ |
| *8.35* | $5.02^{+0.20}_{-0.01}$ | $6.71^{+0.05}_{-0.05}$ | $9.14^{+0.09}_{-0.05}$ | $11.4^{+0.1}_{-0.1}$ | $13.4^{+0.1}_{-0.1}$ | $16.5^{+0.1}_{-0.1}$ | $18.2^{+0.1}_{-0.1}$ | $1.21 \pm 0.01$ |
| *8.50* | $3.93^{+0.03}_{-0.03}$ | $5.19^{+0.03}_{-0.04}$ | $7.17^{+0.05}_{-0.05}$ | $8.89^{+0.07}_{-0.05}$ | $10.0^{+0.1}_{-0.1}$ | $11.9^{+0.1}_{-0.1}$ | $13.8^{+0.1}_{-0.1}$ | $1.15 \pm 0.01$ |
| *8.75* | $3.07^{+0.02}_{-0.02}$ | $4.21^{+0.03}_{-0.03}$ | $5.32^{+0.04}_{-0.03}$ | $6.34^{+0.05}_{-0.04}$ | $7.63^{+0.06}_{-0.05}$ | $8.85^{+0.07}_{-0.05}$ | $9.57^{+0.06}_{-0.06}$ | $1.05 \pm 0.01$ |
| *9.00* | $2.62^{+0.01}_{-0.02}$ | $3.44^{+0.02}_{-0.02}$ | $4.38^{+0.03}_{-0.02}$ | $5.27^{+0.03}_{-0.03}$ | $6.29^{+0.05}_{-0.03}$ | $7.37^{+0.05}_{-0.05}$ | $7.89^{+0.04}_{-0.06}$ | $1.03 \pm 0.01$ |
| *Slope* | $-0.41 \pm 0.01$ | $-0.37 \pm 0.02$ | $-0.44 \pm 0.02$ | $-0.43 \pm 0.03$ | $-0.44 \pm 0.02$ | $-0.45 \pm 0.03$ | $-0.46 \pm 0.04$ | |

Table 4.7: Estimates of $<\phi^4>_{Ws}$ from Raw Binned Data, Middle (50%) Contour

| $\beta \ddots L_x$ | *18* | *24* | *30* | *36* | *42* | *48* | *54* | *Slope* |
|---|---|---|---|---|---|---|---|---|
| *8.25* | $2270^{+480}_{-180}$ | $334^{+125}_{-1.}$ | $967^{+343}_{-86}$ | $463^{+141}_{-21}$ | $492^{+58}_{-24}$ | $585^{+16}_{-13}$ | $690^{+31}_{-10}$ | $-0.54 \pm 0.27$ |
| *8.35* | $775^{+349}_{-12}$ | $63.1^{+2.4}_{-1.9}$ | $111^{+5}_{-2}$ | $166^{+6}_{-5}$ | $199^{+6}_{-6}$ | $344^{+24}_{-14}$ | $299^{+11}_{-8}$ | $1.31 \pm 0.49$ |
| *8.50* | $19.5^{+0.7}_{-0.6}$ | $28.3^{+0.9}_{-0.8}$ | $59.7^{+2.1}_{-1.7}$ | $82.5^{+3.3}_{-1.6}$ | $88.0^{+3.2}_{-2.6}$ | $126^{+8}_{-1}$ | $163^{+7}_{-4}$ | $1.95 \pm 0.05$ |
| *8.75* | $9.92^{+0.30}_{-0.27}$ | $16.7^{+0.7}_{-0.5}$ | $22.3^{+0.8}_{-0.6}$ | $30.3^{+1.2}_{-0.7}$ | $43.2^{+1.9}_{-1.5}$ | $48.6^{+2.4}_{-1.3}$ | $49.5^{+2.1}_{-1.5}$ | $1.54 \pm 0.04$ |
| *9.00* | $5.85^{+0.14}_{-0.17}$ | $8.80^{+0.27}_{-0.23}$ | $13.1^{+0.5}_{-0.3}$ | $15.6^{+0.6}_{-0.4}$ | $23.3^{+0.9}_{-0.6}$ | $28.6^{+1.2}_{-0.9}$ | $28.6^{+1.4}_{-1.0}$ | $1.54 \pm 0.03$ |
| *Slope* | $-2.40 \pm 0.44$ | $-1.33 \pm 0.12$ | $-1.53 \pm 0.17$ | $-1.47 \pm 0.08$ | $-1.29 \pm 0.05$ | $-1.26 \pm 0.12$ | $-1.32 \pm 0.11$ | |

Table 4.8: Estimates of $<\phi^6>_{Ws}$ from Raw Binned Data, Middle (50%) Contour

| $\beta \ddots L_x$ | *18* | *24* | *30* | *36* | *42* | *48* | *54* | *Slope* |
|---|---|---|---|---|---|---|---|---|
| *8.25* | 5720000 $+1370000$ $-510000$ | 391000 $+441000$ $+45000$ | 1280000 $+720000$ $-180000$ | 297000 $+320000$ $-16000$ | 199000 $+295000$ $-23000$ | 24900 $+13600$ $-3800$ | 71700 $+112000$ $-7300$ | $-4.40 \pm 0.39$ |
| *8.35* | 1880000 $+970000$ $-20000$ | 1830 $+260$ $-170$ | 3710 $+640$ $-330$ | 5150 $+700$ $-560$ | 6180 $+1300$ $-980$ | 42800 $+41100$ $-2700$ | 2640 $+2570$ $-1400$ | $-2.87 \pm 1.05$ |
| *8.50* | $289^{+44}_{-29}$ | $380^{+47}_{-47}$ | $1260^{+210}_{-140}$ | $2160^{+510}_{-160}$ | $939^{+312}_{-210}$ | $1660^{+880}_{+50}$ | $968^{+686}_{-399}$ | $1.82 \pm 0.26$ |
| *8.75* | $109^{+15}_{-12}$ | $167^{+40}_{-24}$ | $110^{+29}_{-27}$ | $98.0^{+79.2}_{-43.7}$ | $285^{+127}_{-81}$ | $77.5^{+161.1}_{-61.2}$ | $-625^{+111}_{-94}$ | $0.44 \pm 0.25$ |
| *9.00* | $27.1^{+3.7}_{-3.2}$ | $48.2^{+7.9}_{-7.9}$ | $39.4^{+15.0}_{-11.8}$ | $-19.7^{+26.3}_{-18.2}$ | $-65.2^{+34.7}_{-27.2}$ | $-213^{+44}_{-36}$ | $-125^{+67}_{-45}$ | $1.06 \pm 0.14$ |
| *Slope* | $-5.57 \pm 0.98$ | $-3.16 \pm 0.55$ | $-3.95 \pm 0.47$ | $-3.33 \pm 0.37$ | $-3.15 \pm 0.23$ | $-2.27 \pm 0.82$ | $-2.98 \pm 0.29$ | |

Table 4.9: Power-Law Dependence of $<\phi^2>$ on $L_x$

| Contour | Data | $\beta=8.25$ | $\beta=8.35$ | $\beta=8.50$ | $\beta=8.75$ | $\beta=9.00$ |
|---|---|---|---|---|---|---|
| *Lower* | *Raw* | 1.30±0.02 | 1.22±0.01 | 1.13±0.01 | 1.04±0.01 | 1.02±0.01 |
| *Lower* | *Smooth* | 1.21±0.02 | 1.16±0.02 | 1.14±0.02 | 1.09±0.01 | 1.10±0.01 |
| *Middle* | *Raw* | 1.20±0.03 | 1.21±0.01 | 1.15±0.01 | 1.05±0.01 | 1.03±0.01 |
| *Middle* | *Smooth* | 1.26±0.01 | 1.23±0.01 | 1.17±0.01 | 1.11±0.01 | 1.13±0.01 |
| *Higher* | *Raw* | 0.91±0.03 | 1.08±0.02 | 1.04±0.01 | 0.92±0.01 | 0.89±0.01 |
| *Higher* | *Smooth* | 1.23±0.01 | 1.12±0.01 | 1.03±0.01 | 0.94±0.01 | 0.95±0.01 |

Table 4.10: Power-Law Dependence of $<\phi^4>_{Ws}$ on $L_x$

| Contour | Data | $\beta=8.25$ | $\beta=8.35$ | $\beta=8.50$ | $\beta=8.75$ | $\beta=9.00$ |
|---|---|---|---|---|---|---|
| *Lower* | *Raw* | 0.27±0.25 | 1.84±0.47 | 1.99±0.05 | 1.58±0.03 | 1.51±0.03 |
| *Lower* | *Smooth* | −0.18±0.15 | −0.11±0.38 | 1.65±0.07 | 1.21±0.05 | 1.19±0.05 |
| *Middle* | *Raw* | −0.54±0.27 | 1.31±0.49 | 1.95±0.05 | 1.54±0.04 | 1.54±0.03 |
| *Middle* | *Smooth* | −0.56±0.17 | −0.02±0.31 | 1.68±0.07 | 1.17±0.06 | 1.27±0.05 |
| *Higher* | *Raw* | −1.01±0.24 | 0.88±0.54 | 1.62±0.04 | 1.32±0.03 | 1.28±0.02 |
| *Higher* | *Smooth* | 0.08±0.16 | 0.24±0.33 | 1.36±0.05 | 1.06±0.04 | 1.02±0.03 |

Table 4.11: Power-Law Dependence of $<\phi^6>_{Ws}$ on $L_x$

| Contour | Data | $\beta=8.25$ | $\beta=8.35$ | $\beta=8.50$ | $\beta=8.75$ | $\beta=9.00$ |
|---|---|---|---|---|---|---|
| *Lower* | *Raw* | −4.38±0.24 | 0.04±1.18 | 1.60±0.29 | 0.73±0.17 | −0.52±0.11 |
| *Lower* | *Smooth* | −1.62±0.20 | −5.38±1.17 | 1.19±0.32 | 1.02±0.09 | 0.90±0.08 |
| *Middle* | *Raw* | −4.40±0.39 | −2.87±1.05 | 1.82±0.26 | 0.44±0.25 | 1.06±0.14 |
| *Middle* | *Smooth* | −2.55±0.24 | −5.29±1.10 | 1.66±0.22 | 0.92±0.08 | ?±? |
| *Higher* | *Raw* | −4.40±0.39 | −3.69±0.99 | 1.54±0.27 | 0.94±0.26 | 0.60±0.19 |
| *Higher* | *Smooth* | −2.60±0.19 | −4.14±0.86 | 1.56±0.18 | 1.21±0.10 | 1.03±0.07 |

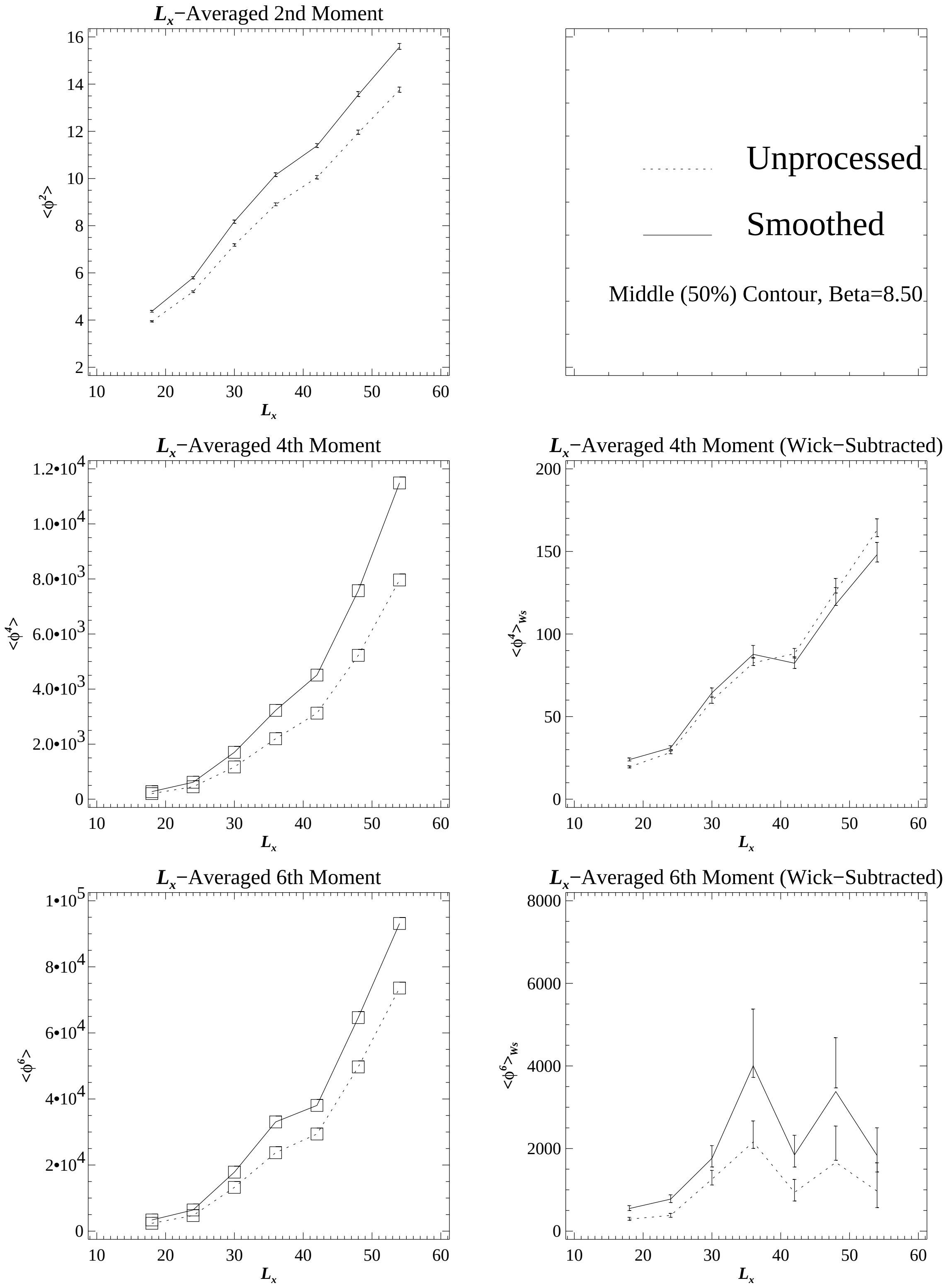

**Fig. 4.19:** Graphs of even moments against transverse size, $L_x$, for $\beta = 8.5$. Data are shown for both raw (unprocessed) and smoothed Polyakov configurations. For the 4th and 6th moments, the left-hand graphs are as measured initially; the right-hand ones are Wick-subtracted.

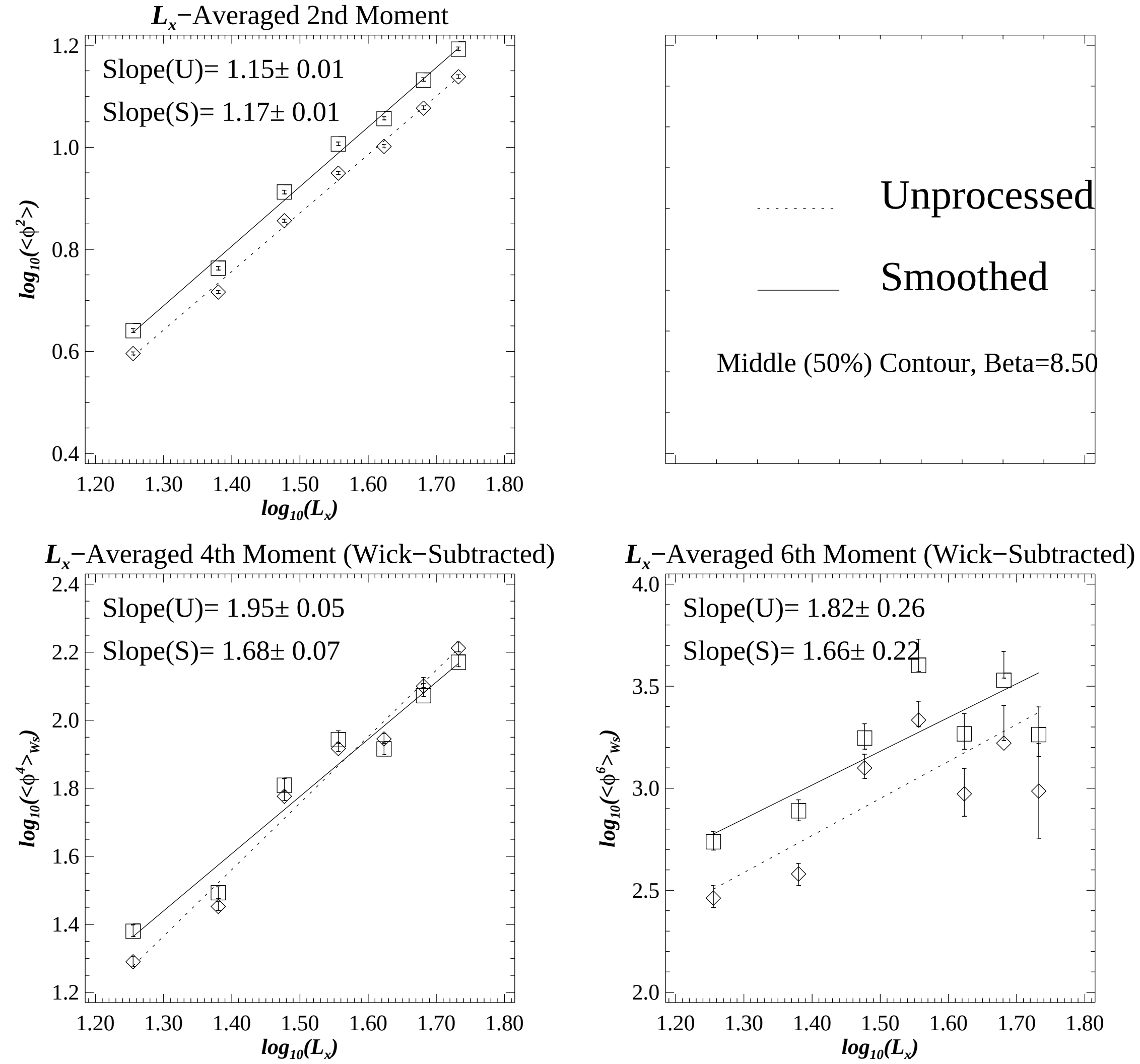

**Fig. 4.20:** Log plots corresponding to the Wick-subtracted data of fig. 4.19, with best fits to a power-law dependence on the lattice width.

A typical set of results is shown in fig. 4.19, showing the dependence of $< \overline{\phi(x)^n} >$ on the lattice width (interface length), $L_x$, for a particular value of $\beta$. Both raw and smoothed data are shown, and these are in good qualitative agreement, though there is a small quantitative difference between them. In fig. 4.20, we obtain best estimates for the power-law dependences of the even Wick-subtracted moments on the lattice width. These are displayed in the graphs, matching the values given in the earlier tables, and show reasonable agreement with our theoretical predictions for the 2nd and 4th moments. The agreement is distinctly less good for the 6th, but the nature of this Wick-subtracted moment, the small difference between two large numbers derived from powers of a fluctuation, may go a long way towards explaining this discrepancy. Our results are similar for the other values of $\beta$, as shown in tables 4.12-14. Though not in perfect agreement with our predictions for the higher moments, these measurements encourage us to believe that our toy Lagrangian is a reasonable model for the interface behaviour.

It should further be pointed out that some of the results at low $\beta$ and small $L_x$ seem to suffer from anomalous large "spikes" in the data, which can skew the estimates of the higher moments severely, and with them the slopes. This problem is addressed at the end of the next chapter, and oddities such as the negative slopes of the higher moments against $L_x$ for small $\beta$ then disappear from the corrected results given there.

Having obtained reasonable results for our model, we now turn to the question of the unknown functions, $\gamma$ and $\lambda$. To tell anything about these functions, we need to produce similar plots to fig. 4.19 and fig. 4.20, but showing the dependence on $\beta$ instead of $L_x$. In fig. 4.21, we show the divergence of the non-Wick-subtracted even moments as $\beta \to \beta_c$. The data shown are for lattices with $L_x = 36$. To find the degree of divergence, we need to fit this behaviour to a power law dependence on $(\beta - \beta_c)$.

In fig. 4.22, we fit our results to just such a power-law divergence, taking $\beta_c = 8.175$. We obtain good fits even for the sixth moment, giving another indication that our model is reasonable. For the data shown in the graphs of this chapter ($L_x = 36$, middle contour, raw data), the fitted slopes for the second, fourth and sixth moments are:

$$< \phi^2 >\sim (\beta - \beta_c)^{-0.43(3)}, \quad < \phi^4 >_{Ws}\sim (\beta - \beta_c)^{-1.47(8)}, \quad < \phi^6 >_{Ws}\sim (\beta - \beta_c)^{-3.33(37)}.$$

Referring back to our formulae for the moments, we see that the first two imply the following for our unknown functions of $\beta$:

$$\gamma \sim (\beta - \beta_c)^{0.43(3)}, \qquad \lambda \sim (\beta - \beta_c)^{0.25(14)}.$$

These two together predict, from our previous formulae, that

$$< \phi^6 >_{Ws}\sim (\beta - \beta_c)^{-2.51(21)},$$

within roughly one standard deviation of the observed behaviour. Our other sets of data give similar results for $\gamma$ and $\lambda$.

It can be seen from the tables that the different contours give similar results, as hoped. For this reason, we use the middle contour, as it should be less susceptible to the influence of nearby bubbles of phase. Also, the results after smoothing are close to those from the raw data, but the discrepancies increase for the higher moments. Smoothing appears to be a useful technique for keeping track of the interface at low $\beta$, indeed lower $\beta$ than is possible in its absence. It appears not to alter the results significantly for the lower moments, though the discrepancies increase fast enough for the results to disagree somewhat with the unprocessed results over the higher moments.

Table 4.12: Power-Law Dependence of $<\phi^2>$ on $(\beta - \beta_c)$

| Contour | Data | $L_x = 18$ | $L_x = 24$ | $L_x = 30$ | $L_x = 36$ | $L_x = 42$ | $L_x = 48$ | $L_x = 54$ |
|---|---|---|---|---|---|---|---|---|
| *Lower* | *Raw* | $-0.33 \pm 0.02$ | $-0.32 \pm 0.02$ | $-0.39 \pm 0.02$ | $-0.40 \pm 0.03$ | $-0.41 \pm 0.01$ | $-0.42 \pm 0.02$ | $-0.44 \pm 0.03$ |
| *Lower* | *Smooth* | $-0.51 \pm 0.03$ | $-0.46 \pm 0.03$ | $-0.51 \pm 0.03$ | $-0.50 \pm 0.03$ | $-0.50 \pm 0.02$ | $-0.51 \pm 0.02$ | $-0.53 \pm 0.03$ |
| *Middle* | *Raw* | $-0.41 \pm 0.01$ | $-0.37 \pm 0.02$ | $-0.44 \pm 0.02$ | $-0.43 \pm 0.03$ | $-0.44 \pm 0.02$ | $-0.45 \pm 0.03$ | $-0.46 \pm 0.04$ |
| *Middle* | *Smooth* | $-0.50 \pm 0.03$ | $-0.48 \pm 0.03$ | $-0.54 \pm 0.03$ | $-0.52 \pm 0.04$ | $-0.52 \pm 0.02$ | $-0.53 \pm 0.03$ | $-0.56 \pm 0.03$ |
| *Higher* | *Raw* | $-0.41 \pm 0.03$ | $-0.37 \pm 0.02$ | $-0.42 \pm 0.02$ | $-0.39 \pm 0.03$ | $-0.40 \pm 0.02$ | $-0.41 \pm 0.03$ | $-0.42 \pm 0.04$ |
| *Higher* | *Smooth* | $-0.40 \pm 0.03$ | $-0.42 \pm 0.03$ | $-0.47 \pm 0.03$ | $-0.48 \pm 0.03$ | $-0.49 \pm 0.02$ | $-0.51 \pm 0.02$ | $-0.54 \pm 0.03$ |

Table 4.13: Power-Law Dependence of $<\phi^4>_{Ws}$ on $(\beta - \beta_c)$

| Contour | Data | $L_x = 18$ | $L_x = 24$ | $L_x = 30$ | $L_x = 36$ | $L_x = 42$ | $L_x = 48$ | $L_x = 54$ |
|---|---|---|---|---|---|---|---|---|
| *Lower* | *Raw* | $-2.00 \pm 0.43$ | $-1.25 \pm 0.10$ | $-1.40 \pm 0.06$ | $-1.23 \pm 0.08$ | $-1.19 \pm 0.04$ | $-1.24 \pm 0.10$ | $-1.29 \pm 0.11$ |
| *Lower* | *Smooth* | $-2.69 \pm 0.34$ | $-1.63 \pm 0.15$ | $-1.84 \pm 0.14$ | $-1.67 \pm 0.10$ | $-1.54 \pm 0.06$ | $-1.65 \pm 0.12$ | $-1.76 \pm 0.11$ |
| *Middle* | *Raw* | $-2.40 \pm 0.44$ | $-1.33 \pm 0.12$ | $-1.53 \pm 0.17$ | $-1.47 \pm 0.08$ | $-1.29 \pm 0.05$ | $-1.26 \pm 0.12$ | $-1.32 \pm 0.11$ |
| *Middle* | *Smooth* | $-2.76 \pm 0.25$ | $-1.80 \pm 0.17$ | $-1.88 \pm 0.17$ | $-1.74 \pm 0.10$ | $-1.61 \pm 0.04$ | $-1.62 \pm 0.13$ | $-1.82 \pm 0.11$ |
| *Higher* | *Raw* | $-2.33 \pm 0.52$ | $-1.20 \pm 0.25$ | $-1.33 \pm 0.24$ | $-1.21 \pm 0.07$ | $-1.11 \pm 0.05$ | $-1.18 \pm 0.07$ | $-1.26 \pm 0.06$ |
| *Higher* | *Smooth* | $-2.28 \pm 0.34$ | $-1.44 \pm 0.13$ | $-1.38 \pm 0.15$ | $-1.43 \pm 0.05$ | $-1.34 \pm 0.04$ | $-1.45 \pm 0.06$ | $-1.50 \pm 0.02$ |

Table 4.14: Power-Law Dependence of $<\phi^6>_{Ws}$ on $(\beta - \beta_c)$

| Contour | Data | $L_x = 18$ | $L_x = 24$ | $L_x = 30$ | $L_x = 36$ | $L_x = 42$ | $L_x = 48$ | $L_x = 54$ |
|---|---|---|---|---|---|---|---|---|
| *Lower* | *Raw* | $-4.99 \pm 1.05$ | $-3.49 \pm 0.57$ | $-4.08 \pm 0.37$ | $-3.11 \pm 0.38$ | $-2.75 \pm 0.12$ | $-1.98 \pm 0.47$ | $-3.00 \pm 0.32$ |
| *Lower* | *Smooth* | $-5.40 \pm 0.94$ | $-3.30 \pm 0.47$ | $-3.96 \pm 0.37$ | $-3.25 \pm 0.48$ | $-3.45 \pm 0.33$ | $-3.47 \pm 0.51$ | $-3.55 \pm 0.22$ |
| *Middle* | *Raw* | $-5.57 \pm 0.98$ | $-3.16 \pm 0.55$ | $-3.95 \pm 0.47$ | $-3.33 \pm 0.37$ | $-3.15 \pm 0.23$ | $-2.27 \pm 0.82$ | $-2.98 \pm 0.29$ |
| *Middle* | *Smooth* | $-5.78 \pm 0.92$ | $-3.35 \pm 0.59$ | $-3.79 \pm 0.52$ | $-3.24 \pm 0.32$ | $-3.39 \pm 0.25$ | $? \pm ?$ | $-3.71 \pm 0.17$ |
| *Higher* | *Raw* | $-5.36 \pm 1.01$ | $-2.22 \pm 0.67$ | $-2.58 \pm 0.70$ | $-2.39 \pm 0.24$ | $-1.99 \pm 0.07$ | $-2.36 \pm 0.23$ | $-2.85 \pm 0.11$ |
| *Higher* | *Smooth* | $-5.25 \pm 0.99$ | $-2.74 \pm 0.43$ | $-2.43 \pm 0.57$ | $-2.44 \pm 0.22$ | $-2.29 \pm 0.10$ | $-2.46 \pm 0.17$ | $-2.36 \pm 0.19$ |

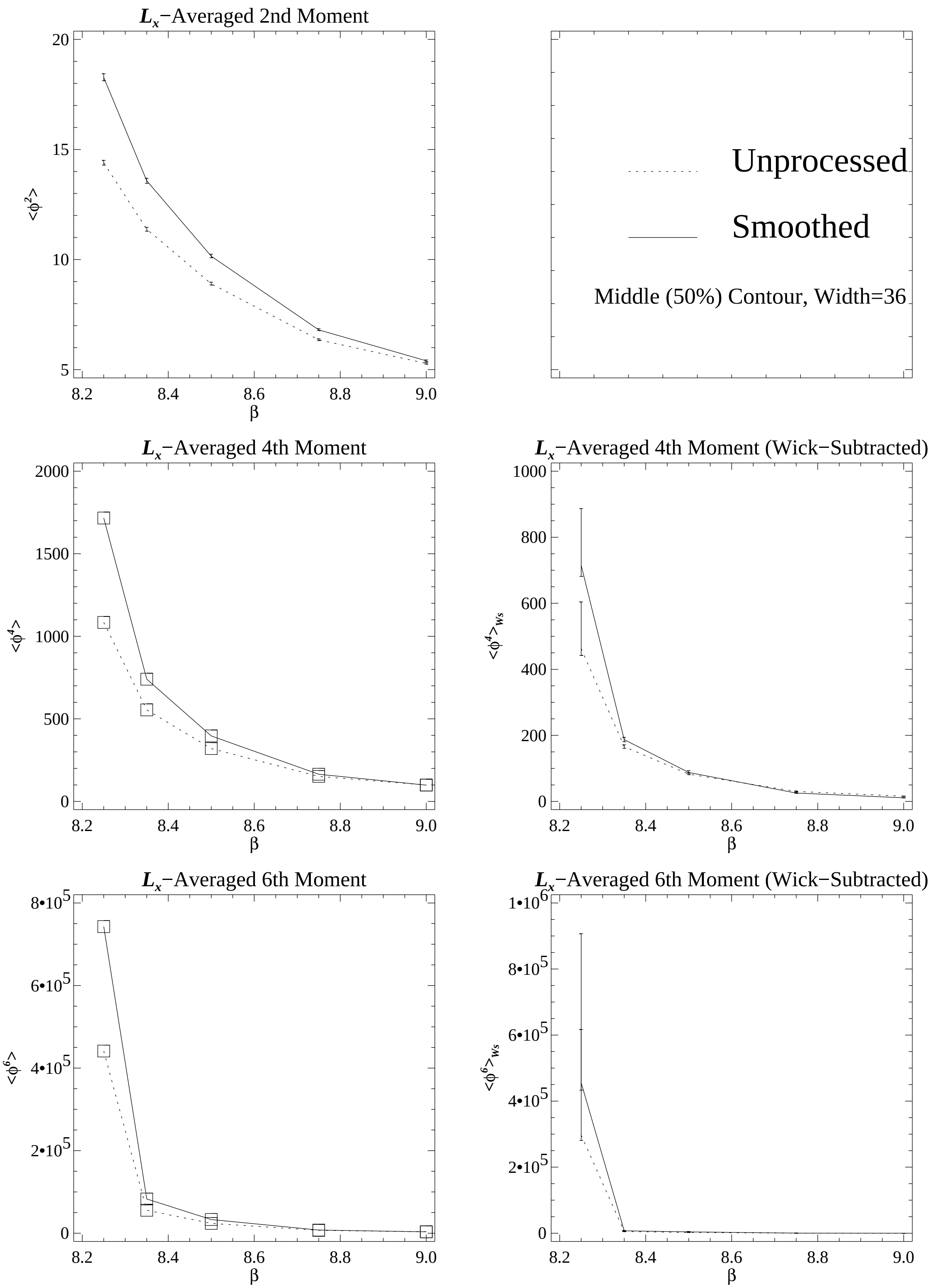

**Fig. 4.21:** Graphs of even moments against $\beta$, for $L_x = 36$. Data are shown for both raw (unprocessed) and smoothed Polyakov configurations. For the 4th and 6th moments, the first graphs are as measured by the computer, and the second graphs are Wick-subtracted.

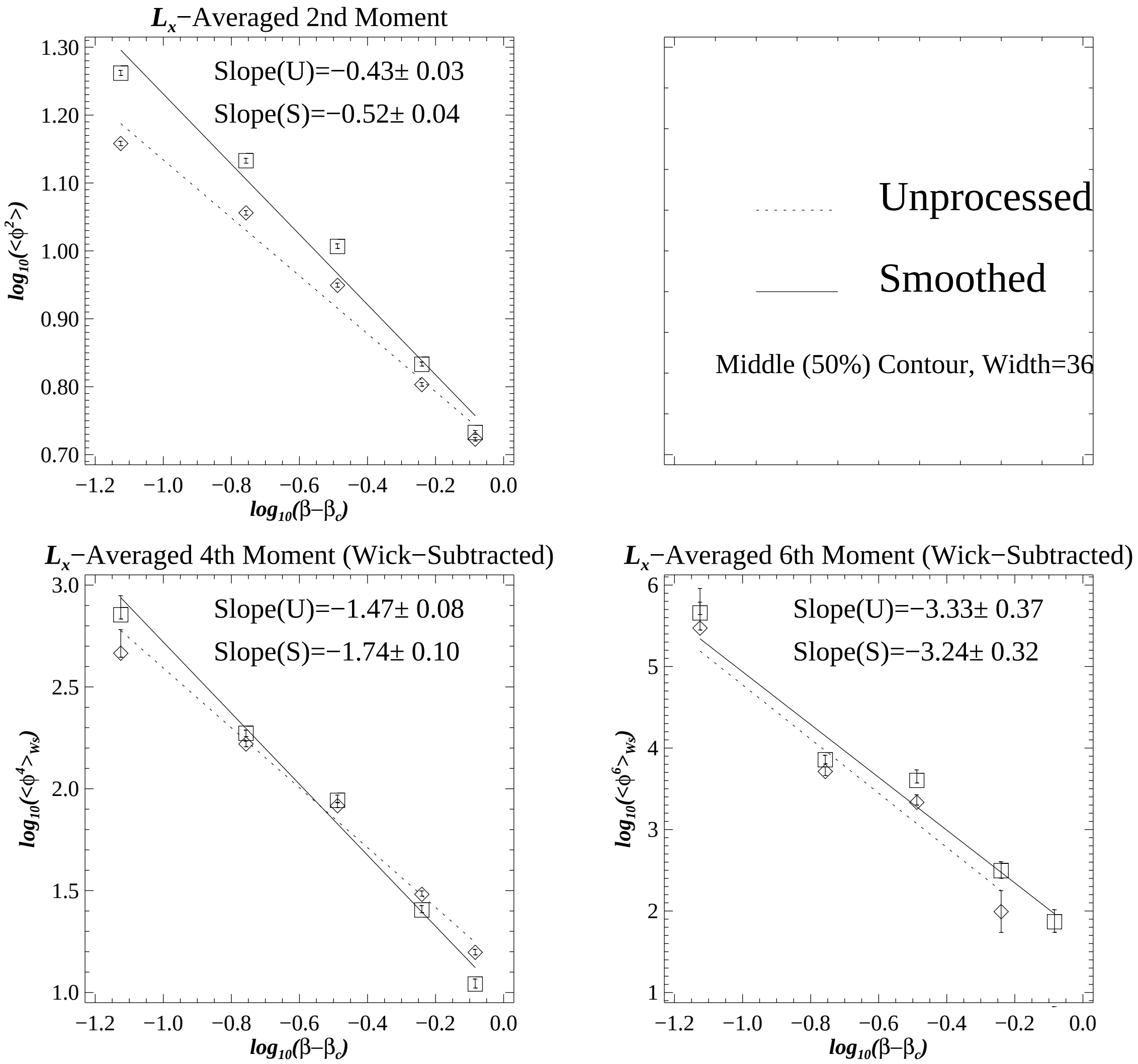

**Fig. 4.22:** Log plots corresponding to the Wick-subtracted data of fig. 4.21, with best fits to a power-law dependence on $(\beta - \beta_c)$.

## 4.13. Interface Width: Intrinsic & Screening

In addition to our survey of fluctuation moments, we also measure the Debye electric screening mass and interface width for each value of $\beta$ and $L_x$, as illustrated in fig. 4.11. As touched on previously, we can measure the "intrinsic" width of the interface by taking the distance between the lower and upper contours at every point across the lattice, and then averaging this quantity across the lattice, and then over all sweeps. By contrast, we obtain the screening mass by averaging the transverse-averaged Polyakov profile of the interface, obtained after each sweep, over all sweeps, and fitting the resulting function to a tanh with coefficient proportional to the Debye mass. The inverse of this mass is the Debye screening length, giving the screening width of the interface, in contrast to the intrinsic width mentioned before.

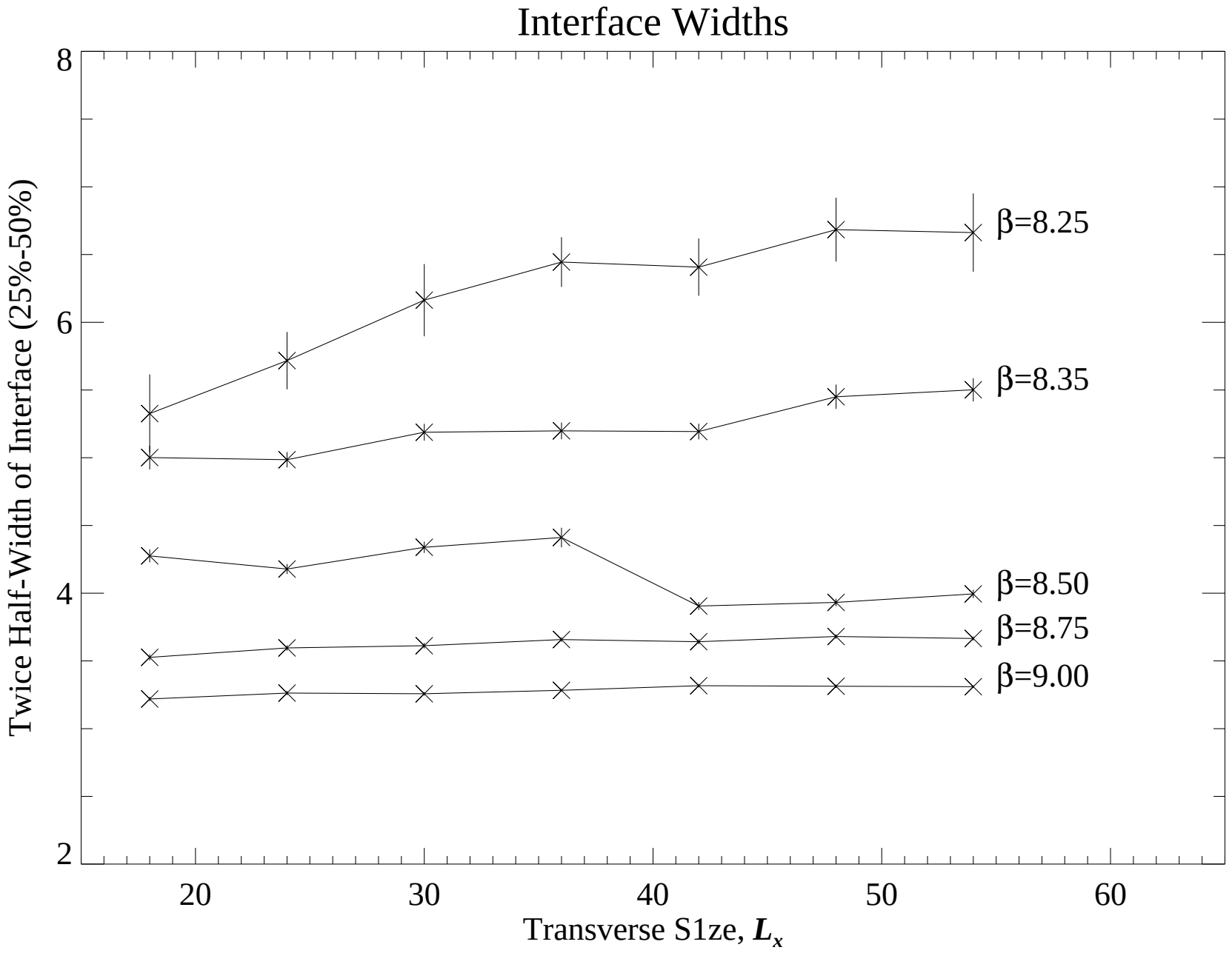

**Fig. 4.23:** Measurements of the average interface width in units of the lattice spacing. For reasons of accuracy, twice the half-width (the average separation of the 25% and 50% contours) is plotted rather than the width (25%—75%). This remains roughly constant as the width of the lattice increases, and rises somewhat as the temperature drops.

We can return to our picture of the interface as a string. The intrinsic width is the width of the string itself, which should be independent of the length of the string ($L_x$), and which we would not expect to change much with temperature. To visualise the screening width, however, we need to imagine letting the string fluctuate very fast while we look at it, so that the different shapes blur into one another; at very high temperatures, where the fluctuations are suppressed, this blurred string should look only a little wider than the stationary string, but as the temperature drops and the fluctuations diverge, we expect the blurred string to appear wider and wider, also diverging in width towards the critical temperature. Since the mean square fluctuations (the second moments) grow like $L_x$, we might expect the screening width of the interface to grow like $\sqrt{L_x}$.

The results for the intrinsic width, shown in fig. 4.23, show that the width does indeed remain constant as $L_x$ increases, and that it increases only moderately as $\beta \to \beta_c$. Presumably, this increase is due to the less rigid shape of the interface as the energy penalty drops.

Our results for the Debye screening length are shown in fig. 4.24 and fig. 4.25. Rather surprisingly, the best fit is for a linear dependence on $L_x$, with a constant term to which it tends at very high temperatures when the interface is rigid. The length also appears to diverge somewhat as $\beta \to \beta_c$, as expected. The best fit of this divergence is to a curve of the form $a + b \ln(\beta - \beta_c)$ for each $L_x$, as predicted for completely wet $Z(3)$ interfaces in $3 + 1$ dimensions[44].

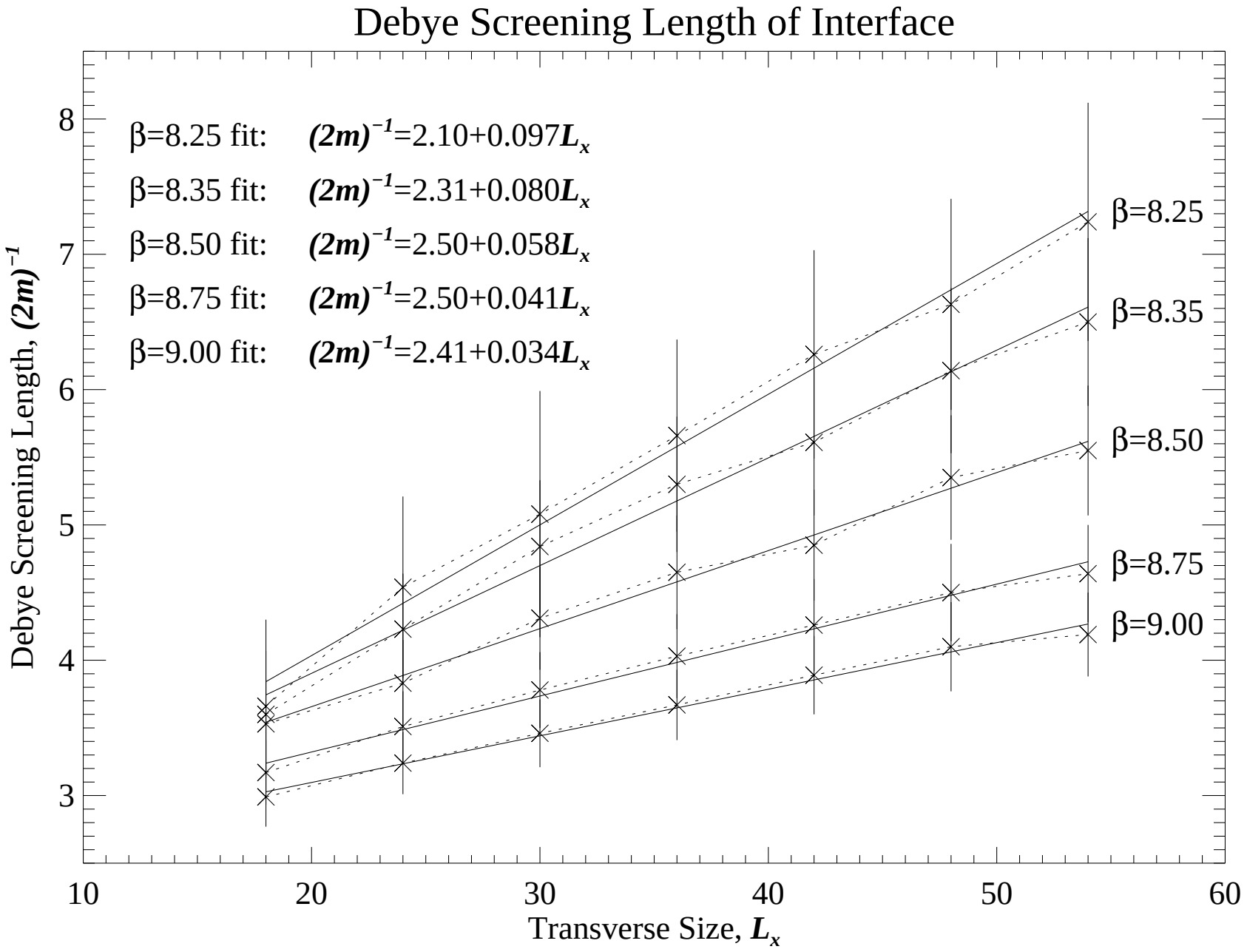

**Fig. 4.24:** Measurements of the Debye screening length for each value of $(\beta, L_x)$, showing a roughly linear dependence on $L_x$.

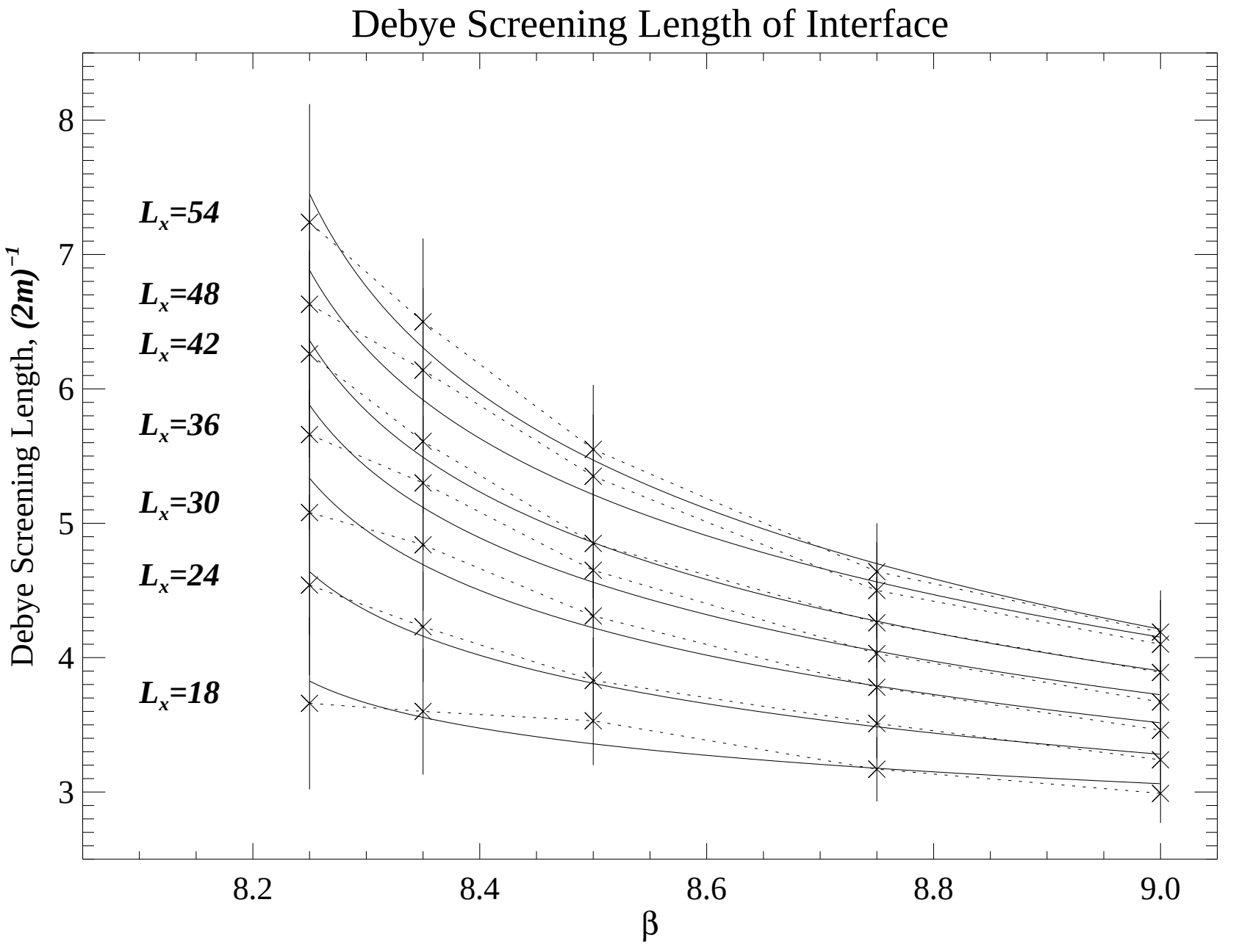

**Fig. 4.25:** Measurements of the Debye screening length for each value of $(\beta, L_x)$, showing a mild divergence as $\beta \to \beta_c$, which we fit to a curve of the form $a + b \ln(\beta - \beta_c)$ for each $L_x$.

## 4.14. Critical Acceleration of the Random Walk

As a curiosity, we can also measure the speed of the random walk of the interface near the critical temperature if we turn off the mechanism that constantly re-centres the interface. As the temperature drops and more and larger bubbles of phase froth within the main phase domains, the interface will combine with, and bud off, more and more of these bubbles. Therefore, we would expect it to wander more and more rapidly along the lattice, moving in bursts as it merges with bubbles in "front" of it or buds off bubbles "behind" it.

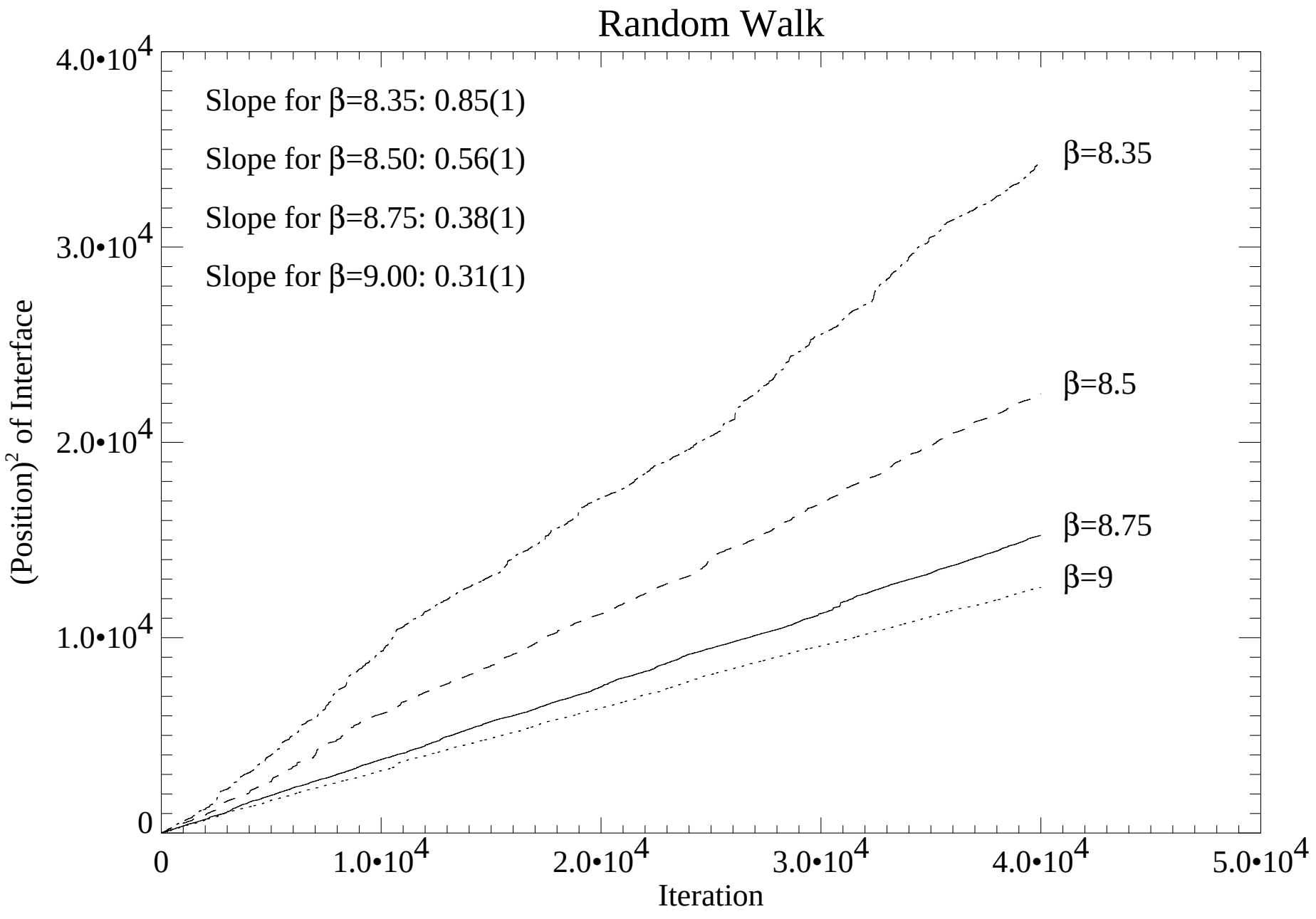

**Fig. 4.26:** This shows the random wandering of the interface, for temperatures near the critical temperature, over 40k sweeps after an initial 400 heat-bath sweeps.

In fig. 4.26, we do indeed see an increase in speed as the temperature drops. In fact, in common with many other parameters, the speed appears to diverge as $\beta \to \beta_c$. Of course, it could be argued that this is simply an artefact of the Monte-Carlo heatbath algorithm, with no relevance to a "real" interface at all. In the absence of a physical "time" to measure and use as a yardstick in the Monte-Carlo procedure, the best way to test this hypothesis would seem to be to to examine the behaviour of the interface under another Monte-Carlo scheme, with different autocorrelation behaviour as $\beta$ drops; we do not do this here.

In fig. 4.27, we fit the speed to a function of the form

$$\frac{\partial z^2}{\partial t} \sim (\beta - \beta_c)^{\beta_{RW}}.$$

Our results suggest values of

$$\beta_c = 8.14(3), \qquad \beta_{RW} = -0.71(6).$$

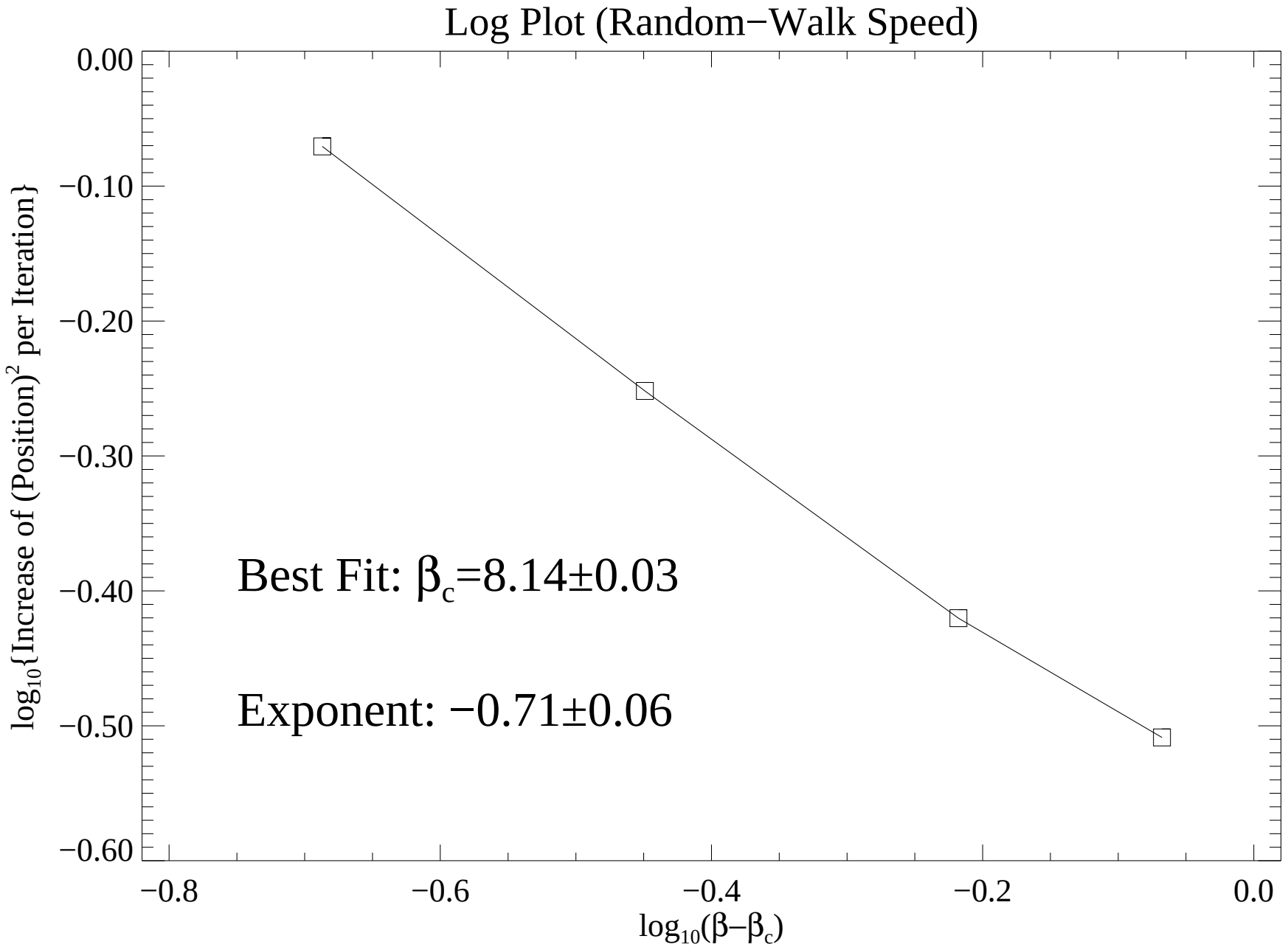

**Fig. 4.27:** This uses the divergence of the speed of the random walk in fig. 4.26 to estimate the critical temperature and exponent of divergence.

This value for $\beta_c$ is certainly consistent with our previous estimates, and the dynamical exponent governing the critical acceleration of the random walk is given by $\beta_{RW}$.

## 4.15. Interface Equilibration

It is interesting to note that studies of discrete models[46] and continuum growth equations[47] in condensed matter make predictions for the behaviour of the width of an interface. The "width" here is usually defined to be the square-root of our second moment,

$$W(L_x,t) = \sqrt{\overline{\phi^2} - \overline{\phi}^2},$$

where $t$ is the time since formation of the interface (initial width zero).

Models generally predict that for times much smaller than some critical time, $t_X(L_x)$, determined by the interface size, an exponent $\beta_G$ will govern the growth of the width:

$$W(t) \sim t^{\beta_G}, \qquad t \ll t_X(L_x).$$

For times much larger than this, as might be expected, the width is predicted to saturate at a value whose relationship to the interface size is governed by $\alpha_G$, the roughness exponent:

$$W_{sat}(L_x) \sim L^{\alpha_G}.$$

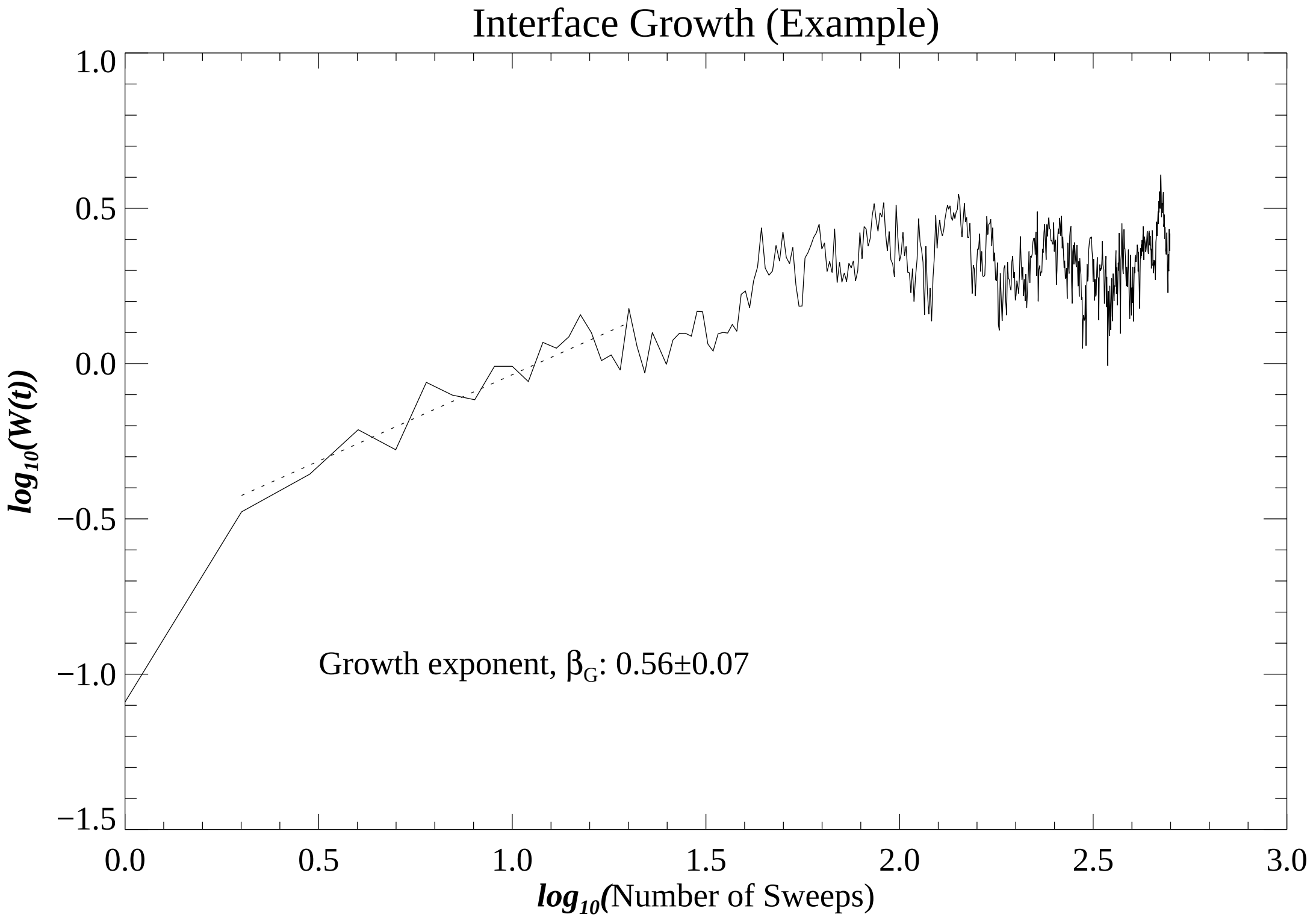

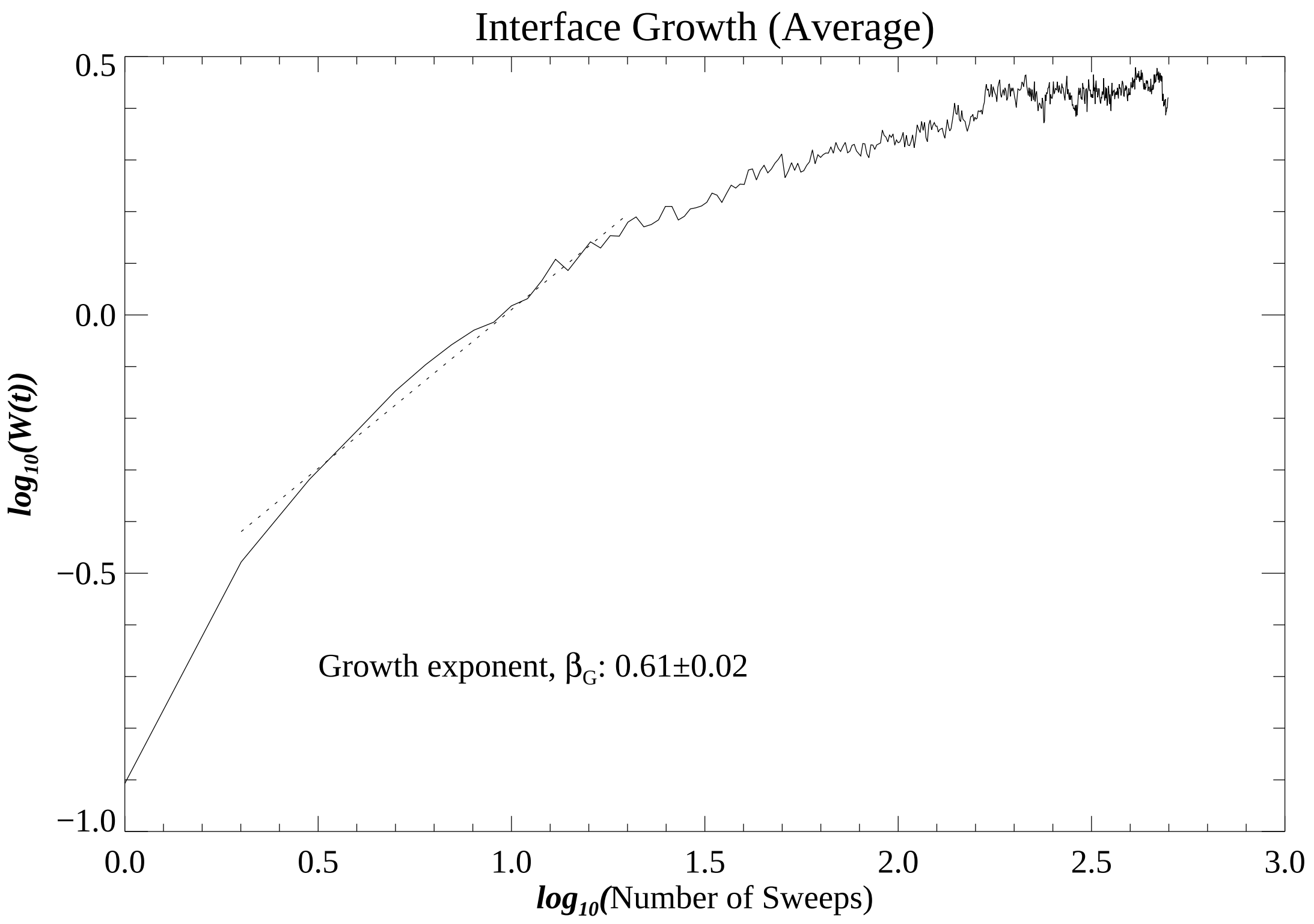

**Fig. 4.28:** An example of the growth in size of an interface for $\beta = 8.5, L_x = 36$, and a growth profile averaged over all $L_x$ and $\beta$. In each case, $\beta_G$ is estimated from the slope of the dotted line.

For the $Z(3)$ interface, we have already seen behaviour corresponding to the latter of these two, since

$$<\phi^2>\sim L_x^{1.13(1)} \Rightarrow \alpha_G = 0.56(1).$$

It can be seen again in fig. 4.28, an example of the growth of a $Z(3)$ interface. At high $t$, the size saturates as usual, but at low $t$, we see growth governed by an exponent, as in the condensed matter models. For our interface, the average profile yields

$$\beta_G = 0.61(2).$$

This, again, could be dismissed as an artefact of the simulation technique, but the manner of formation simulated by the Monte-Carlo technique seems eminently plausible for a "real" interface, and it is suggestive that the interface shares many characteristics with more familiar interfaces in condensed matter physics.

# Chapter 5.

# Statistical Discussion: Bootstrap Techniques

## 5.1. The Bootstrap

SINCE IT is not clear that the Wick-subtracted moments will be normally distributed, we need a reliable method to construct confidence intervals for them which does not rely on our knowing their distribution at the start. A suitable method is given by the "bootstrap" principle[48], and we now discuss some aspects of it as applied to our analysis.

Let $\mathbf{x} = (x_1, x_2, \ldots, x_n)$ be a vector of samples from a normal distribution with unknown mean and variance. Further, sample the random variable $T$ from the t-distribution with $n-1$ degrees of freedom, and define the quartile $t_{n-1,\alpha}$, shortened in future to $t_\alpha$, by $\text{Prob}(T \leq t_\alpha) = \alpha$. If $\overline{x}_n$ and $s_n^2$ are the sample mean and variance, then

$$(\overline{x}_n + s_n t_{0.975} \quad , \quad \overline{x}_n - s_n t_{0.025})$$

is the shortest unbiased exact 95% confidence interval for the mean. To obtain the standard deviation (a 68% confidence interval for the normal distribution), we simply change the subscripts on $t$ to 0.84 and 0.18.

How, then, can we construct confidence intervals for an unknown distribution? This is where the bootstrap principle comes to our aid, as it enables us, as its name suggests, to estimate confidence intervals just from our observed data, with no prior knowledge of the distribution. Let us denote the population parameter, in our case corresponding to one of the moments, by $\theta$. Given a random sample from the population, $\mathbf{x}$, we should like to be able to construct an estimate, $\hat{\Theta} \equiv \hat{\theta}(\mathbf{x})$, for $\theta$. The idea of the bootstrap is to learn about the relationship between $\theta$ and $\hat{\theta}(\mathbf{x})$ by examining that between $\hat{\theta} \equiv \hat{\theta}(\mathbf{x}_{obs})$ and $\hat{\theta}(\mathbf{x}^*)$, where $\mathbf{x}^*$ is a resample with replacement from the observed data, $\mathbf{x}_{obs}$ (our results for the particular moment). This is known as the "nonparametric bootstrap".

In the simplest case, we choose to consider the statistic $W = \hat{\Theta} - \theta$, invoking the bootstrap principle to estimate its quantiles by those of $W^* = \hat{\Theta}^* - \hat{\theta}$. We follow the procedure given here:

- Fix $\theta$ at the value given by $\hat{\theta} \equiv \hat{\theta}(\mathbf{x}_{obs})$;
- Sample with replacement from $\mathbf{x}_{obs}$ to obtain $\mathbf{x}^*$;
- From this, make an estimate of $\hat{\theta}(\mathbf{x}_{obs})$, denoted $\hat{\theta}^* \equiv \hat{\theta}(\mathbf{x}^*)$;
- Repeat the previous steps $(m-1)$ times to obtain $m$ estimates for $\hat{\theta}(\mathbf{x}_{obs})$;
- Sort the $\hat{\theta}^*$ into order, denoting the $p$th ordered estimate by $\hat{\theta}^*_{(p)}$;
- The $100(1-\alpha)$th percentile of the distribution of $W^*$ is now consistently estimated by $\hat{\theta}^*_{1-\alpha} - \hat{\theta}$, where we define $\hat{\theta}^*_{1-\alpha} = \hat{\theta}^*_{(p)}$, choosing $m$ so that $p = m(1-\alpha)$ is an integer.

It then follows that the approximate 95% confidence interval so obtained is

$$(2\hat{\theta} - \hat{\theta}^*_{0.975} \quad , \quad 2\hat{\theta} - \hat{\theta}^*_{0.025}).$$

There are two approximations behind this. The first is that $\hat{\theta}^*_{\alpha} - \hat{\theta}$ is a consistent estimate of $w^*_{\alpha}$, where $\text{Prob}(W^* \leq w^*_{\alpha}) = \alpha$, and can be reduced as desired by increasing the number of resamplings, $m$. The second relies on $W$ having the same distribution as $W^*$, and will be true if the distribution of $W$ is pivotal, *i.e.* depends on no unknown parameters. Unfortunately, a sizeable error may be introduced if this is not the case.

## 5.2. The Bootstrap t

To overcome this problem, we need to make the statistic $W$ more accurately pivotal.

Such modification is suggested by considering a sample $\mathbf{x} = (x_1, \ldots, x_n)$ from normal distribution with mean $\theta$ and variance $\sigma^2$. The distribution of $\overline{x} - \theta$, our previous choice of $W$, depends on $\sigma^2$, and is hence non-pivotal. However, introducing the sample variance, $s^2 = \frac{1}{n-1}\sum_{i=1}^{n}(x_i - \overline{x})^2$, we can see that the distribution of $\sqrt{n}(\overline{x} - \theta)/s$ *is* pivotal, following a $t$ distribution with $(n-1)$ degrees of freedom. Thus, we "studentise" our previous $W$: we divide by an estimate of the standard error of $\hat{\Theta}$, giving a new statistic:

$$T = \frac{\hat{\Theta} - \theta}{s(\hat{\Theta})}.$$

Using $T$, we proceed exactly as for $W$, invoking the bootstrap principle to estimate the quantiles by those of

$$T^* = \frac{\hat{\Theta}^* - \theta}{s(\hat{\Theta}^*)},$$

and finding the 95% confidence interval to be

$$(\hat{\theta} - s(\hat{\theta})\hat{t}^*_{0.975} \quad , \quad \hat{\theta} - s(\hat{\theta})\hat{t}^*_{0.025}).$$

The approximations are as before, but the error associated with the pivotal quality of the statistic should be much reduced. It has been shown, in fact[49], that the coverage error of the interval reduces from $O(n^{-0.5})$ to $O(n^{-1})$ for many cases. If $T$ is truly pivotal, the error will reduce to zero.

## 5.3. The Non-Parametric Delta Method

FOR our purposes, there is an added complication: whilst we can calculate the sample variance for the second moment, $\overline{\phi^2}$, in the usual manner, the form of the Wick subtraction in $\overline{\phi^4_{Ws}}$ and $\overline{\phi^6_{Ws}}$ makes it impossible to use a similar formula for them. Fortunately, a variance approximation exists for many general statistics $T$ which can take the representation $\theta = t(F)$. In this case, $F$ is the cumulative distribution function $F$ (the integral of the probability density function) for a distribution $\mathbf{X}$. If the sample data is denoted by the usual vector $\mathbf{x}$, then a simple example would be the mean of $\mathbf{X}$, given by $t(F) = \int x dF(x)$.

The method works by applying aa Taylor expansion to $t(F)$:

$$t(G) = t(F) + \int L_t(x; F) dG(x).$$

$L_t$ gives the first derivative of $t$ at $F$ in direction $G$, and is sometimes called the "influence function":

$$L_t(x; F) = \lim_{\epsilon \to 0} \frac{t((1-\epsilon)F + \epsilon H_x) - t(F)}{\epsilon} = \left. \frac{\partial t((1-\epsilon)F + \epsilon H_x)}{\partial \epsilon} \right|_{\epsilon=0}.$$

Here, $H_x(y)$ is the Heaviside step function, stepping from 0 to 1 at $y = x$. The empirical approximations given by the data values, $l_i = L_t(x_i, \hat{F})$, are the "empirical influence values". The "nonparametric delta method" relies on applying this approximation with $G = \hat{F}$, so that

$$t(\hat{F}) = t(F) + \frac{1}{n} \sum_{i=1}^{n} L_t(x_i). \tag{5.1}$$

Now, the difference, $(T - \theta)$, between $t(F)$ and $t(\hat{F})$ will follow a normal distribution, by the central limit theorem, with mean zero and variance,

$$\nu_L(F) = \frac{1}{n} \int L_t^2(x) dF(x).$$

We approximate $\nu_L(F)$ by the sample version,

$$\nu_L(\hat{F}) = \frac{1}{n^2} \sum_{i=1}^{n} l_i^2, \tag{5.2}$$

the "nonparametric delta method variance estimate". Note that subsitution of $\hat{F}$ for $F$ in (5.1) implies that $\sum_i l_i = 0$. For our purposes, we also note that we can use the chain rule for statistics which are functions of others, $t(F) = \mathcal{T}\{t_1(f), \ldots, t_m(F)\}$, yielding

$$L_t(x) = \sum_{i=1} m \frac{\partial \mathcal{T}}{\partial t_i} L_{t_i}(x).$$

Let us consider the three even moments, denoting the population means of the *unsubtracted* moments by $\mu_2$, $\mu_4$ and $\mu_6$ respectively. First, the second moment:

$$\begin{aligned} t = \overline{x^2} \Rightarrow L_t(x) &= \partial_\epsilon[(1-\epsilon)\mu_2 + \epsilon x^2]|_{\epsilon=0} \\ &= x^2 - \mu_2 \\ \Rightarrow l(x_i) &= (x_i^2 - \overline{x^2}). \end{aligned}$$

Similarly, the unsubtracted fourth and sixth moments are:

$$\begin{aligned} t = \overline{x^4} &\Rightarrow l(x_i) = (x_i^4 - \overline{x^4}) \\ t = \overline{x^6} &\Rightarrow l(x_i) = (x_i^6 - \overline{x^6}). \end{aligned}$$

Using the chain rule then allows us to calculate the empirical influence values for the subtracted moments:

$$\begin{aligned} t &= \overline{x^4} - 3\overline{x^2}^2 \\ &\Rightarrow l(x_i) = (x_i^4 - \overline{x^4}) - 6\overline{x^2}(x_i^2 - \overline{x^2}) \\ t &= \overline{x^6} - 15\overline{x^2}^3 - 15\overline{x^2}(\overline{x^4} - 3\overline{x^2}^2) \\ &\Rightarrow l(x_i) = (x_i^6 - \overline{x^6}) + 90\overline{x^2}^2(x_i^2 - \overline{x^2}) - 15\overline{x^2}(x_i^4 - \overline{x^4}) - 15\overline{x^4}(x_i^2 - \overline{x^2}). \end{aligned}$$

In each case, the variance is estimated by (5.2); also, as expected, $\sum_i l(x_i) = 0$.

Finally, then, we are able to construct confidence intervals for the subtracted moments, and this is how we obtain the standard errors for the fluctuation moments given in tables and graphs. For a full data vector, $\mathbf{x}$, of 100,000 values, approximately 2,000 resamplings, $\mathbf{x}^*$, are necessary.

## 5.4. Variance Stabilisation

A check on the standard errors obtained in this way is provided by using a numerical approach rather than (5.2). To do this, we perform a second-level bootstrap analysis on each resampled vector, $\mathbf{x}^*$, resampling each vector itself ($\ll 2,000$ times) to obtain $\mathbf{x}^{**}$. The sample variance of the second-level estimates of the mean of the subtracted moment gives the standard deviation for the "parent" first-level estimate. To check for any correlation between the mean and deviation of the first-level estimates, which might skew

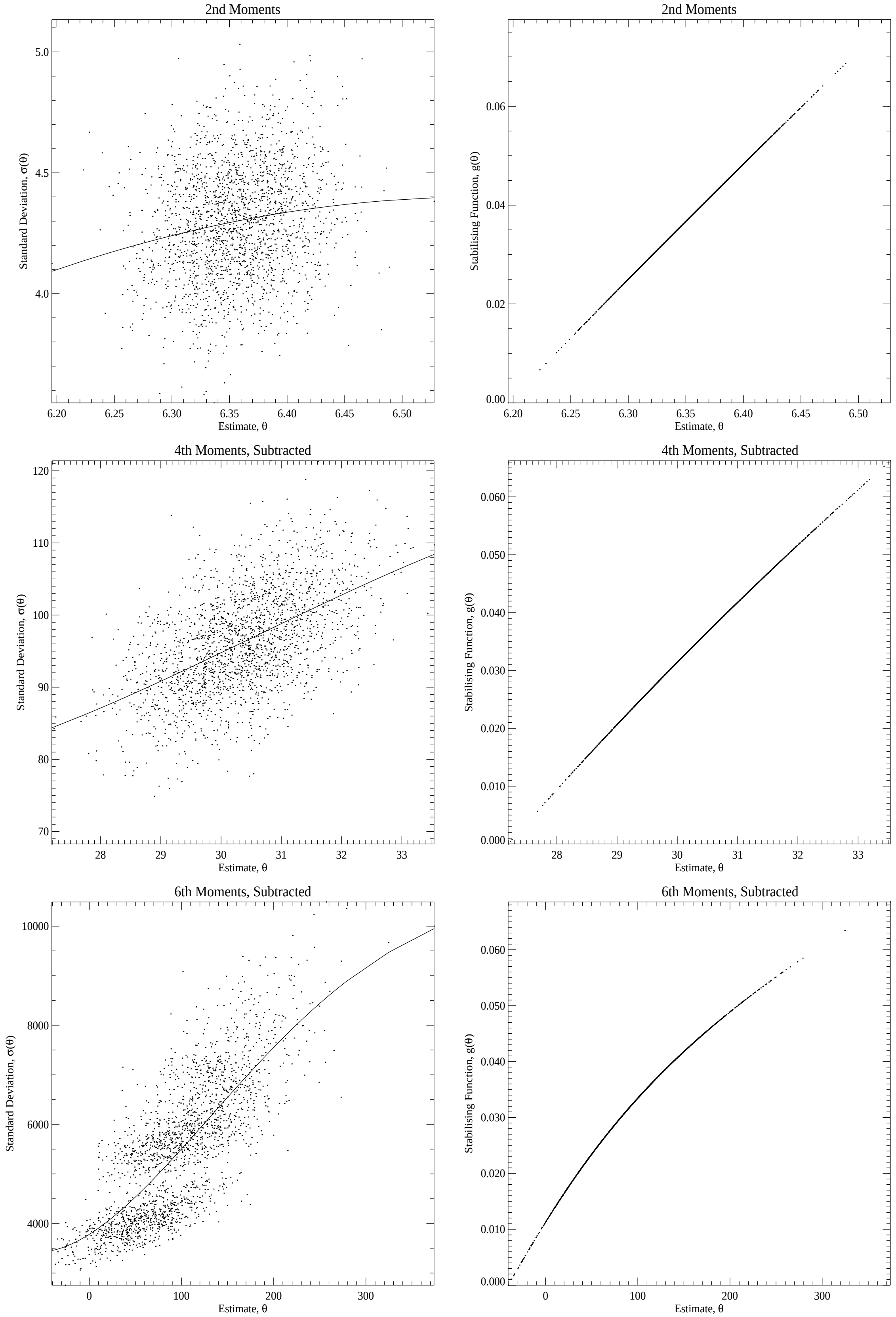

**Fig. 5.1:** Variance stabilisation in numerical second-level bootstrap procedures, illustrated for ($\beta = 8.75, L_x = 36$). The left-hand plots show the second-level standard deviations $\sigma(\hat{\theta})$, against the first-level estimates for the average moment, $\hat{\theta}$, together with a fit to a smooth function. The right-hand plots show $g(\hat{\theta})$, the integral of the function $\sigma(\hat{\theta})^{-1}$. In this case, 200 second-level bootstraps are used,

the confidence intervals, we then plot one against the other. If a trend is seen, we introduce "variance stabilisation" [50], as follows. A smooth curve is fitted to the data, denoted by $\sigma(\hat{\theta})$. We then construct a compensating function by numerically integrating the inverse:

$$g(\hat{\theta}) = \int^{\hat{\theta}} \frac{1}{\sigma(y)} dy,$$

as illustrated in fig. 5.1 for data from a $(\beta = 8.75, L_x = 36)$ run. We rescale $g(\hat{\theta})$ so that its mean and variance are the same as that of the original data vector $\mathbf{x}$, order it, and plot it against the estimates $\hat{\theta}$. Then, we simply read off the values for $\hat{\theta}$ which correspond to the desired percentiles in $g(\hat{\theta})$, giving the desired confidence interval (quoted for 95%):

$$(2\hat{\theta} - \hat{\theta}^*_{0.975} \quad , \quad 2\hat{\theta} - \hat{\theta}^*_{0.025}).$$

The best way to compare the validity of the different methods is to construct "Q-Q plots", where the ordered bootstrap vectors are plotted against quantiles of the standard normal distribution, as illustrated in fig. 5.2. The dotted lines on these graphs show what would be obtained by a normal distribution with mean and standard deviation given by the relevant bootstrap vector. Thus, the most reliable data is that which most closely matches the dotted line over the desired confidence interval. The upper three and middle three sets of plots show respectively that the bootstrap and bootstrap $t$ distributions are close to, but not quite, normal. The variance stabilisation technique is designed to transform to a scale where the variance is constant relative to the mean, and will therefore compensate for any divergence from normality in the bootstrap $t$ distribution; the lower three plots are included merely to show that it does indeed produce curves which compensate for the deviations from normality of the bootstrap $t$ curves above.

As just mentioned, the intervals from variance stabilisation are the most reliable, being produced from accurately normal bootstrap $t$ distributions, but they differ little from the bootstrap $t$ estimates for our data. As an example of this, the following table lists the 95% and 1-$\sigma$ confidence intervals for $\beta = 8.75$ and $L_x = 36$, matching the data of fig. 5.2. Since the variance stabilisation technique requires computationally intensive second-level bootstrap analysis, we therefore use straight bootstrap $t$ intervals for our results.

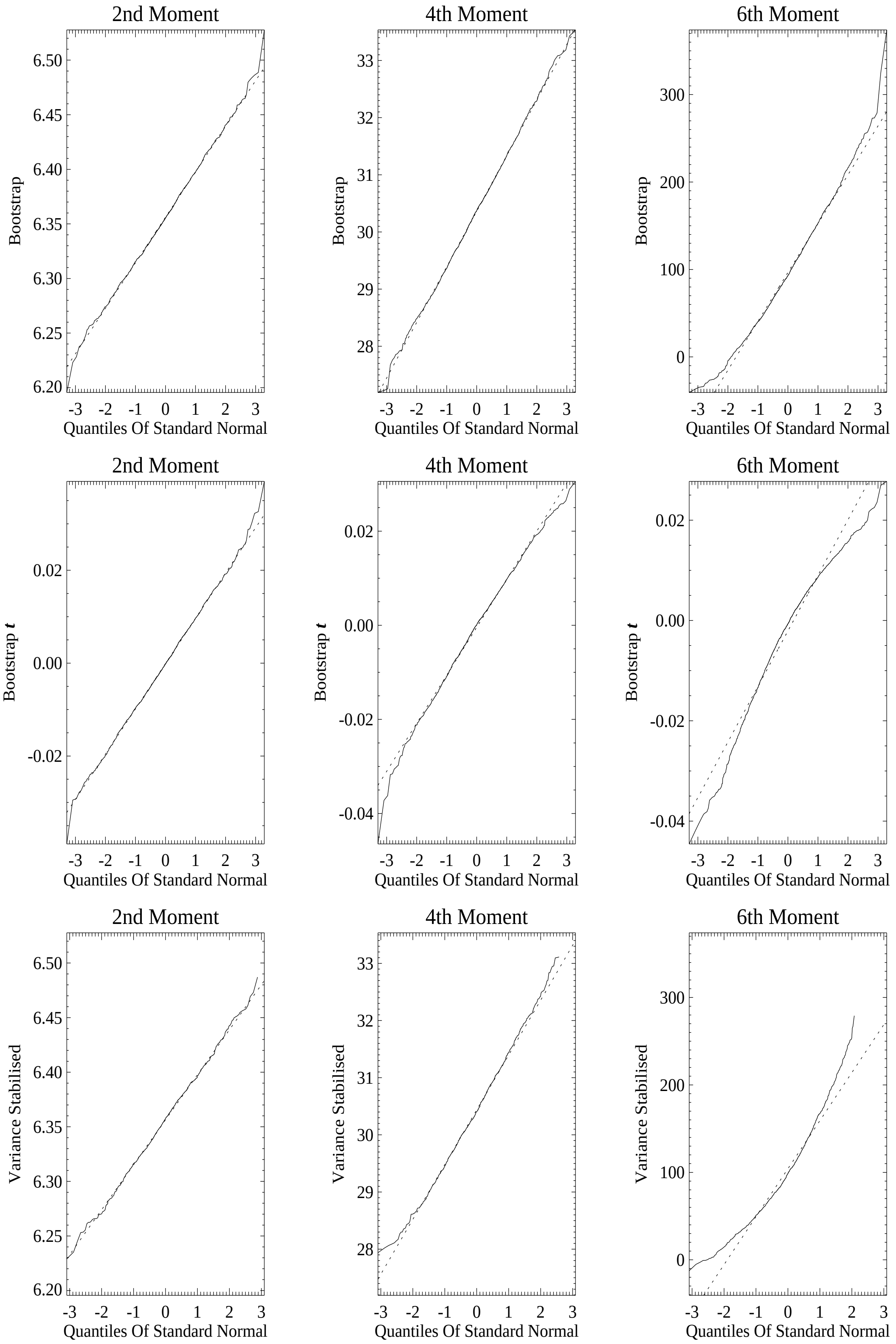

**Fig. 5.2:** Q-Q plots comparing the different methods for obtaining confidence intervals, using data for ($\beta = 8.75, L_x = 36$). Horizontally, the plots refer to the second, fourth and sixth moments respectively. Vertically, the upper plots are for the basic bootstrap, the central ones for the bootstrap $t$, and the lower ones for variance stabilisation. The dotted lines are those which would be obtained for

Table 5.1: Estimates of Confidence Intervals

| Moment | Technique | 95% (Lower) | $\sigma$ (Lower) | $\sigma$ (Upper) | 95% (Upper) |
|---|---|---|---|---|---|
| 2nd | Bootstrap | 6.269 | 6.313 | 6.396 | 6.436 |
| 2nd | Bootstrap $t$ | 6.276 | 6.314 | 6.397 | 6.442 |
| 2nd | V. Stabilisation | 6.270 | 6.316 | 6.395 | 6.437 |
| 4th | Bootstrap | 28.43 | 29.38 | 31.30 | 32.32 |
| 4th | Bootstrap $t$ | 28.47 | 29.43 | 31.34 | 32.44 |
| 4th | V. Stabilisation | 28.63 | 29.46 | 31.44 | 32.38 |
| 6th | Bootstrap | -5.3 | 42.1 | 155.2 | 214.5 |
| 6th | Bootstrap $t$ | 12.4 | 48.8 | 171.6 | 257.8 |
| 6th | V. Stabilisation | 15.7 | 51.1 | 167.1 | 251.3 |

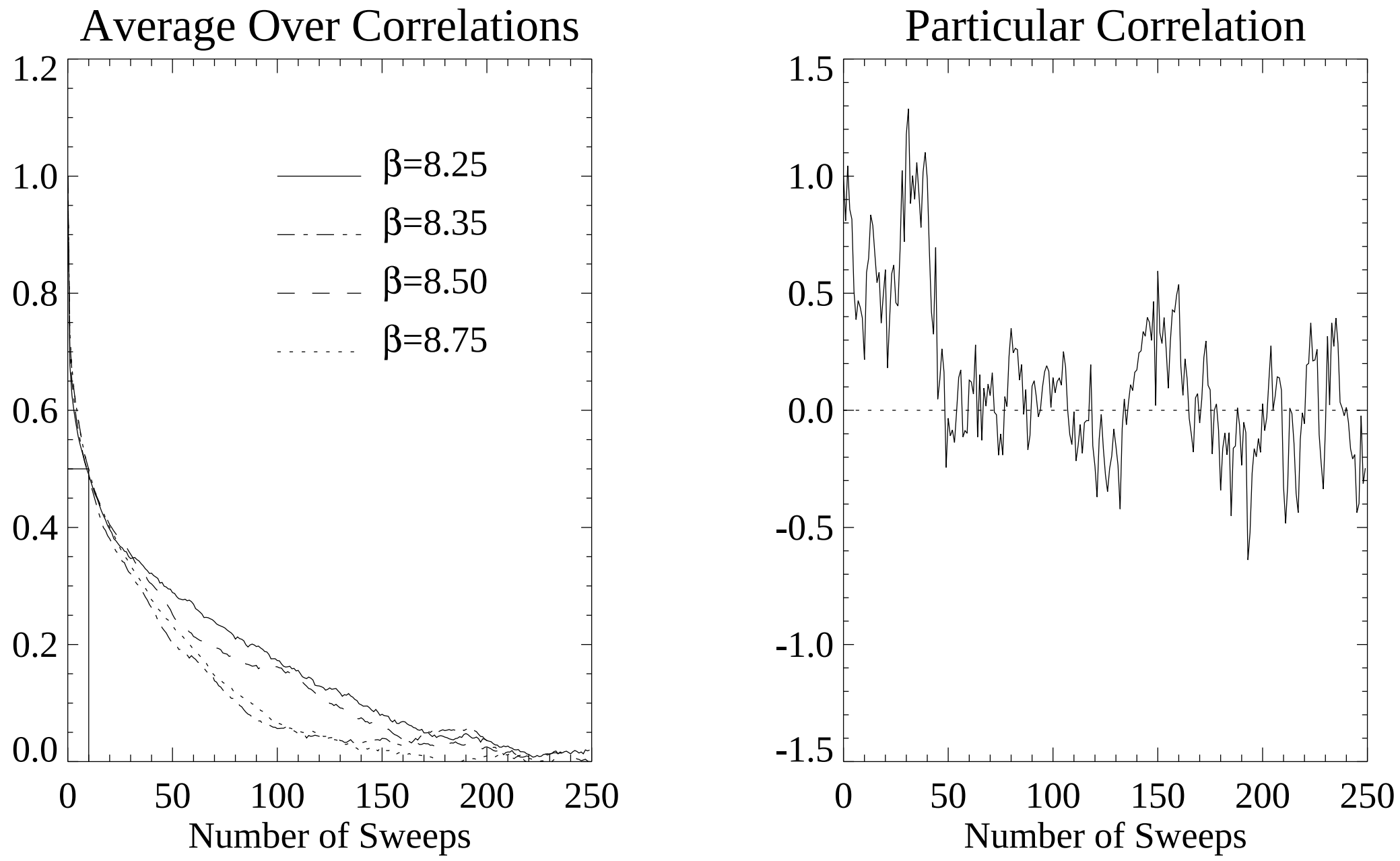

**Fig. 5.3:** Correlations are given between interface contour shapes separated by a certain number of sweeps, for $L_x = 36$. A particular correlation is shown on the right, whilst approximately 4,000 such graphs are averaged on the left, for four values of $\beta$. It is clear that the correlation function drops to one half within approximately ten sweeps, as illustrated on the left.

## 5.5. Correlations and Data Bins

In our simulations, we record the full 100k measurements for each raw, even moment of the central (50%) contour, but bin all data from other contours, and all smoothed data, in groups of one thousand. Thus, we must consider how to ensure that the standard errors are not distorted by the binning used. In fig. 5.3, the correlation between two interface contour shapes is given as a function of the number of sweeps separating them, and it is seen to drop to less than one half after about ten sweeps. Therefore, bins containing far fewer than ten samples may give results tainted by unacceptably large correlations, leading to artificially small errors, whereas those containing far more than ten will lose some information, giving errors which are too large. Thus, we bin our raw data for the central contour into bins of size ten before calculating bootstrap errors.

Illustrated in fig. 5.4 are errors from the non-studentised bootstrap analysis, and from the bootstrap $t$, with more differences apparent between them for the higher moments. The graphs show the effect of rebinning the raw moment data into bins of different size. We take the full 100k measurements for each moment, and bin the results into groups of between one

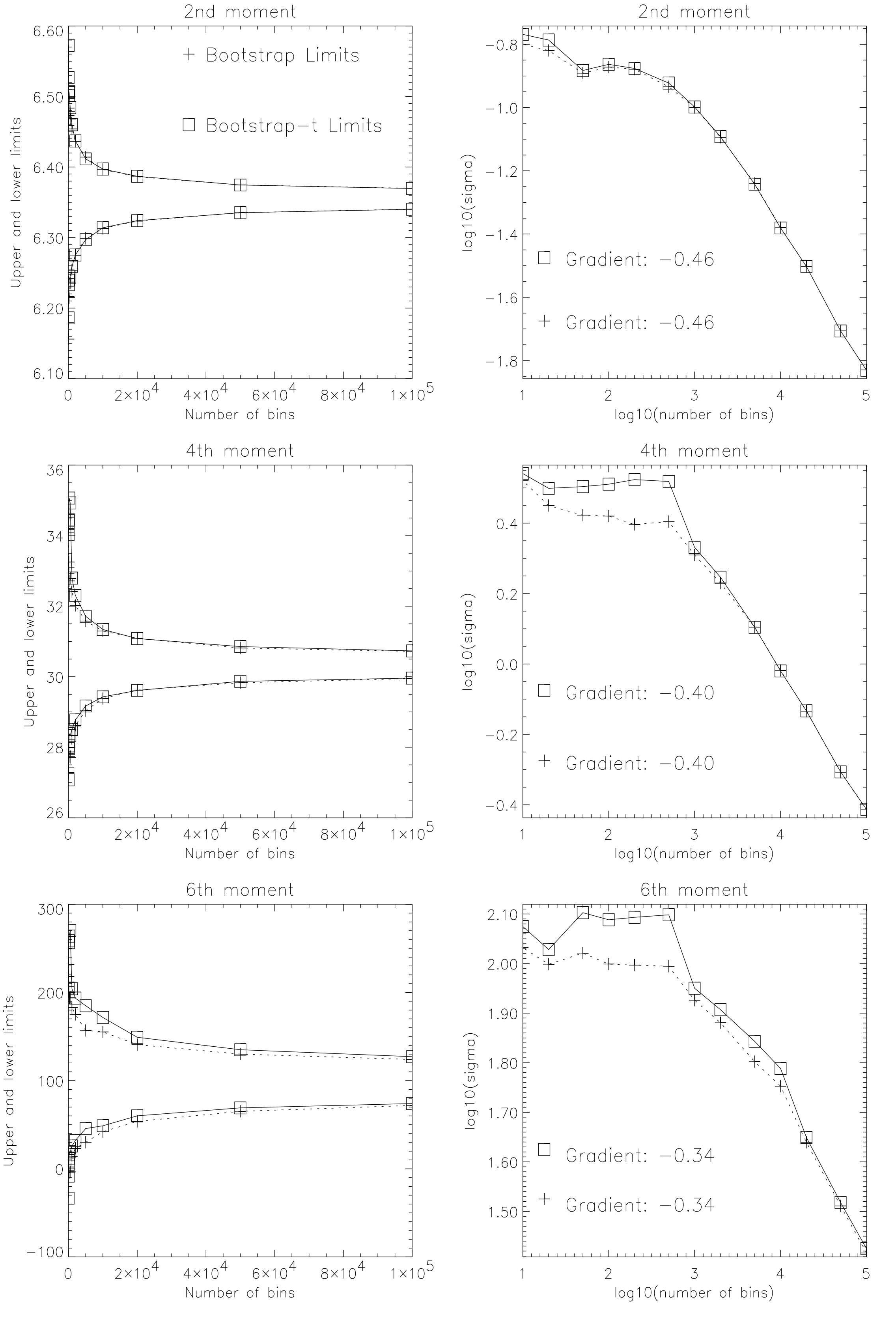

**Fig. 5.4:** The bootstrap errors, both studentised ("bootstrap $t$") and non-studentised, are shown here for $(\beta = 8.75, L_x = 36)$. On the left, the $\pm\sigma$ bounds (68% confidence intervals) are shown for different binnings of the data; on the right, a logarithmic plot of interval sizes against number of bins shows a linear dependence for large numbers of bins, whose gradient gives the corresponding

and ten thousand. Whereas a random variable with normal distribution would have obey a relationship of the form $\sigma \sim (\text{number of bins})^{-0.5}$, the moment errors flatten off for bins containing more than a few hundred measurements. For bins with fewer items than this, an estimated power law dependence of $\sigma$ on the number of bins is shown, and this enables us to trivially adjust the errors obtained from our bins of size one thousand to those that would be obtained from bins of size ten.

## 5.6. Statistical Problems Near Criticality

Our results for $\beta \geq 8.35$ show reasonable statistical distributions for the moments over the 100k sweeps. However, those for $\beta = 8.25$ and small $L_x$ occasionally contain "spikes", observations for the moments which are far larger than we would expect, sometimes by several magnitudes. This can be seen by comparing the distributions in the following two sets of figures: the first, where $L_x = 54$ and no spikes are apparent; and the second, where $L_x = 18$ and spikes have given each distributions an artificially long, flat upper tail. The spikes are unlikely to be caused by wild fluctuations passing through the twist, since similar spikes are not seen for wider lattices where fluctuations are larger. It is most likely that they are an artefact of the greater energetic instability of the smaller interfaces at very low temperatures: if parts of the interface collapse during interaction with nearby bubbles of phase, the interface contour may briefly undergo a drastic distortion or partial collapse, even appearing to pass through the twist because of the change in Polyakov values there.

It is difficult to prevent the spikes occurring, but they have a massively disproportionate effect on the overall averages, especially for the higher moments. Thus, we would like to be able to examine the data with the spikes removed. To do this, we impose an upper cut-off on the statistical distributions, as illustrated by the dashed lines in fig. 5.5 and fig. 5.6. To ensure that only the spikes are removed, we need to define the cut-off in a statistically sensible manner. We take it, for each distribution, to be the median plus 2.5 times the interquartile range of the distribution of *logarithmic* values of data, where the interquartile range is defined to run between the 25th and 75th percentiles of the distribution. This ensures that the cut-off corresponds only to the upper tip (to be precise, the top 0.04%) of a normal distribution overlaid on the right-hand plots of fig. 5.5 and fig. 5.6, ensuring that only data sets containing spikes are affected by the procedure. The exact multiplier "2.5" can, of course, be varied, but we find that a lower number (*e.g.* 2.0, cutting off 0.3% of the distribution), has too great an effect on the normal contribution to the distributions, whereas a higher number (*e.g.* 3.0, cutting off only 0.002% of the normal distribution) permits too much influence from the spikes.

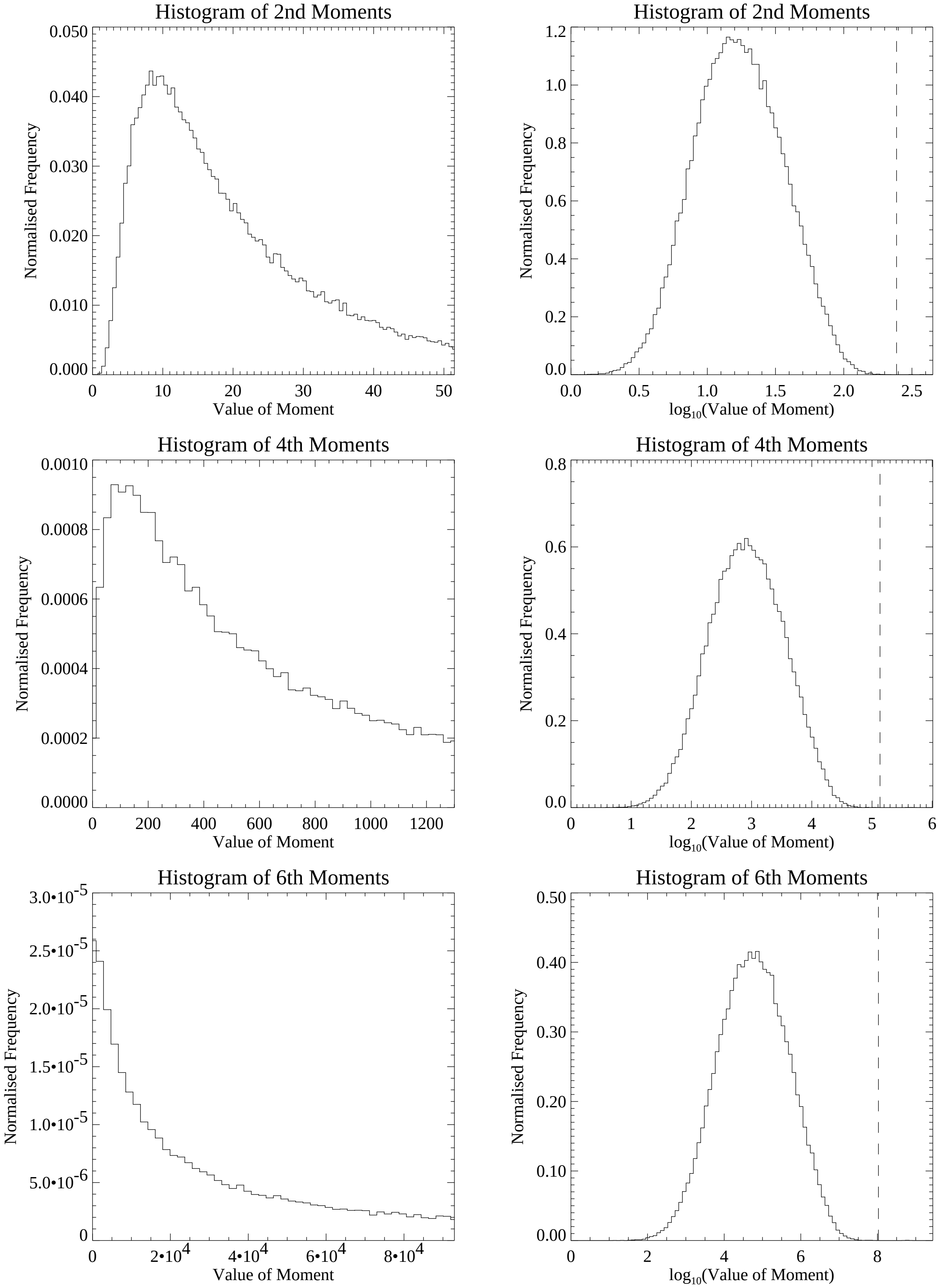

**Fig. 5.5:** Illustrated on the left are the statistical distributions of unsubtracted moments for ($\beta = 8.25, L_x = 54$). The distributions have long upper tails, as expected for moments of a roughly normal distribution. To the right are the corresponding distributions of the *logarithm* of the moments, each resembling a normal distribution. The dashed vertical lines represent a suggested upper cut-off, as discussed in the text.

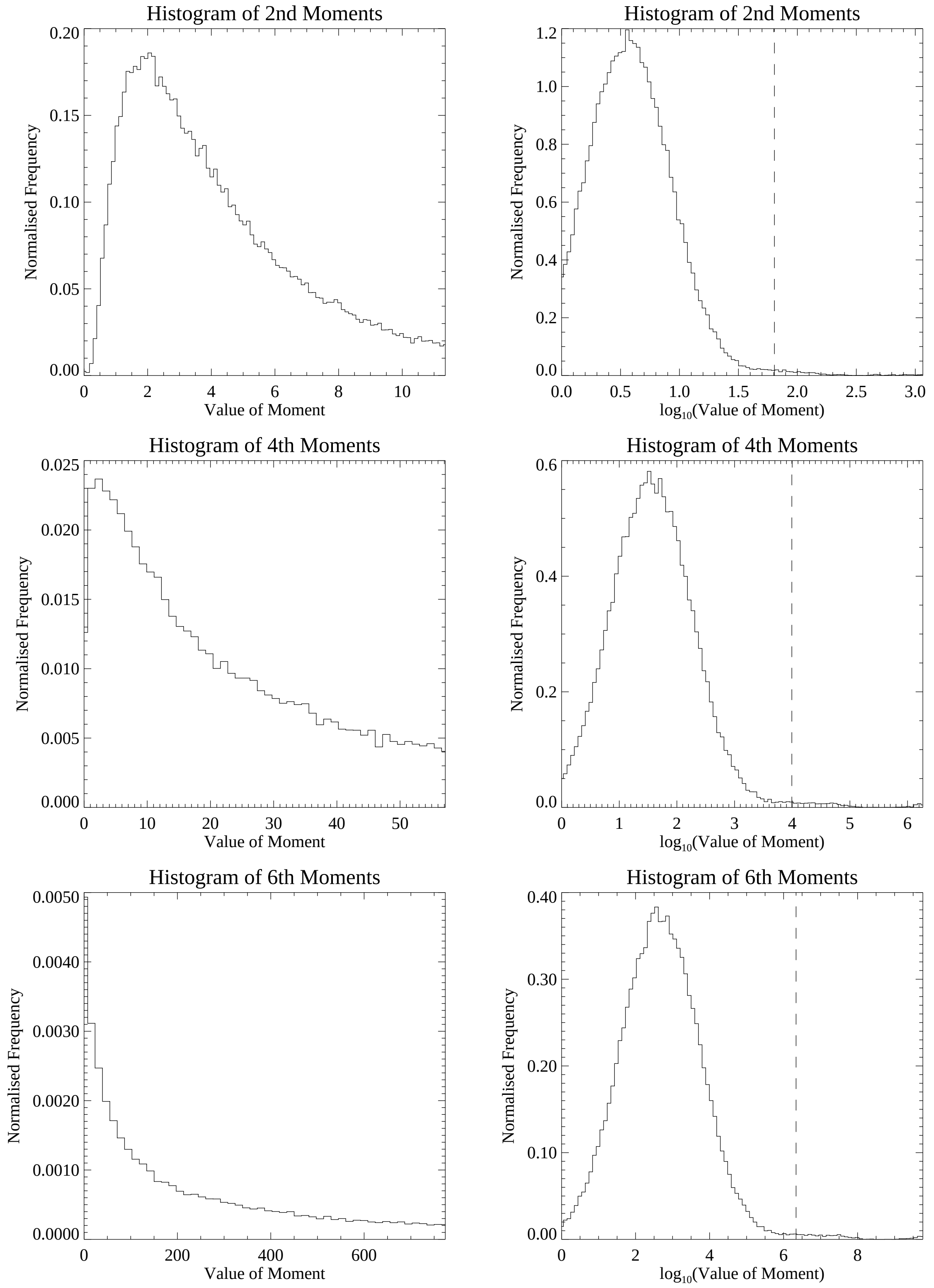

**Fig. 5.6:** The partners to fig. 5.5, with ($\beta = 8.25, L_x = 54$), show unexpectedly long, flat upper tails in the right-hand plots. The effects of these tails can be eliminated by the use of a cut-off defined in the same way as that in fig. 5.5.

## 5.7. Revised Moment Averages (Without Spikes)

IN this section, we give revised results for moment averages after the removal of spikes as discussed, and then some plots for comparison with those of the previous chapter. We quote revised results only for the middle (50%) contour and raw Polyakov data:

Table 5.2: Estimates of $< \phi^2 >$ from Raw Data (Corrected), Middle (50%) Contour

| $\beta \ddots L_x$ | *18* | *24* | *30* | *36* | *42* | *48* | *54* | *Slope* |
|---|---|---|---|---|---|---|---|---|
| *8.25* | $5.44^{+0.05}_{-0.04}$ | $8.09^{+0.06}_{-0.06}$ | $11.9^{+0.1}_{-0.1}$ | $14.3^{+0.1}_{-0.1}$ | $17.8^{+0.1}_{-0.1}$ | $20.8^{+0.2}_{-0.2}$ | $23.1^{+0.2}_{-0.2}$ | $1.33\pm0.02$ |
| *8.35* | $4.52^{+0.03}_{-0.03}$ | $6.69^{+0.05}_{-0.05}$ | $9.13^{+0.07}_{-0.06}$ | $11.4^{+0.1}_{-0.1}$ | $13.4^{+0.1}_{-0.1}$ | $16.5^{+0.1}_{-0.1}$ | $18.2^{+0.1}_{-0.1}$ | $1.27\pm0.01$ |
| *8.50* | $3.93^{+0.03}_{-0.03}$ | $5.19^{+0.03}_{-0.03}$ | $7.17^{+0.05}_{-0.05}$ | $8.89^{+0.07}_{-0.06}$ | $10.0^{+0.1}_{-0.1}$ | $11.9^{+0.1}_{-0.1}$ | $13.8^{+0.1}_{-0.1}$ | $1.15\pm0.01$ |
| *8.75* | $3.06^{+0.02}_{-0.02}$ | $4.21^{+0.03}_{-0.03}$ | $5.32^{+0.04}_{-0.03}$ | $6.34^{+0.04}_{-0.04}$ | $7.62^{+0.05}_{-0.05}$ | $8.85^{+0.06}_{-0.06}$ | $9.57^{+0.06}_{-0.06}$ | $1.05\pm0.01$ |
| *9.00* | $2.62^{+0.02}_{-0.02}$ | $3.44^{+0.02}_{-0.02}$ | $4.38^{+0.03}_{-0.03}$ | $5.27^{+0.03}_{-0.03}$ | $6.29^{+0.04}_{-0.04}$ | $7.37^{+0.05}_{-0.05}$ | $7.89^{+0.05}_{-0.05}$ | $1.03\pm0.01$ |
| *Slope* | $-0.31\pm0.03$ | $-0.36\pm0.03$ | $-0.42\pm0.03$ | $-0.42\pm0.03$ | $-0.44\pm0.02$ | $-0.45\pm0.03$ | $-0.46\pm0.04$ | |

Table 5.3: Estimates of $< \phi^4 >_{Ws}$ from Raw Data (Corrected), Middle (50%) Contour

| $\beta \ddots L_x$ | *18* | *24* | *30* | *36* | *42* | *48* | *54* | *Slope* |
|---|---|---|---|---|---|---|---|---|
| *8.25* | $63.2^{+3.7}_{-3.0}$ | $115^{+7}_{-5}$ | $273^{+14}_{-11}$ | $326^{+9}_{-9}$ | $425^{+15}_{-14}$ | $580^{+14}_{-13}$ | $658^{+15}_{-15}$ | $2.12\pm0.07$ |
| *8.35* | $30.6^{+1.0}_{-1.1}$ | $61.7^{+2.1}_{-2.0}$ | $108^{+3}_{-3}$ | $166^{+6}_{-5}$ | $197^{+6}_{-6}$ | $322^{+1}_{-1}$ | $299^{+10}_{-9}$ | $2.17\pm0.06$ |
| *8.50* | $19.2^{+0.6}_{-0.5}$ | $27.8^{+0.8}_{-0.8}$ | $59.0^{+2.0}_{-1.8}$ | $82.0^{+2.8}_{-2.5}$ | $87.8^{+2.8}_{-2.8}$ | $126^{+5}_{-4}$ | $163^{+6}_{-5}$ | $1.97\pm0.05$ |
| *8.75* | $9.68^{+0.34}_{-0.31}$ | $16.3^{+0.6}_{-0.6}$ | $22.3^{+0.7}_{-0.6}$ | $30.3^{+1.1}_{-1.0}$ | $43.0^{+2.2}_{-2.0}$ | $48.2^{+1.9}_{-1.7}$ | $49.7^{+1.9}_{-1.8}$ | $1.55\pm0.04$ |
| *9.00* | $5.84^{+0.17}_{-0.16}$ | $8.76^{+0.25}_{-0.25}$ | $13.0^{+0.4}_{-0.4}$ | $15.6^{+0.6}_{-0.5}$ | $23.3^{+0.8}_{-0.8}$ | $28.6^{+1.1}_{-1.0}$ | $28.5^{+1.3}_{-1.0}$ | $1.54\pm0.03$ |
| *Slope* | $-0.99\pm0.06$ | $-1.10\pm0.07$ | $-1.28\pm0.07$ | $-1.25\pm0.11$ | $-1.21\pm0.06$ | $-1.27\pm0.11$ | $-1.28\pm0.12$ | |

Table 5.4: Estimates of $< \phi^6 >_{Ws}$ from Raw Data (Corrected), Middle (50%) Contour

| $\beta \ddots L_x$ | *18* | *24* | *30* | *36* | *42* | *48* | *54* | *Slope* |
|---|---|---|---|---|---|---|---|---|
| *8.25* | $3530^{+490}_{-400}$ | $6330^{+1080}_{-990}$ | $23500^{+3500}_{-2800}$ | $18500^{+1600}_{-1500}$ | $18500^{+3400}_{-2600}$ | $14500^{+2800}_{-2300}$ | $15200^{+3400}_{-2900}$ | $1.46\pm0.22$ |
| *8.35* | $712^{+75}_{-55}$ | $1700^{+210}_{-190}$ | $2860^{+440}_{-390}$ | $5180^{+670}_{-580}$ | $4790^{+1050}_{-840}$ | $10200^{+2100}_{-1700}$ | $2640^{+2300}_{-1680}$ | $2.38\pm0.29$ |
| *8.50* | $259^{+29}_{-23}$ | $350^{+85}_{-73}$ | $1090^{+160}_{-150}$ | $1710^{+350}_{-280}$ | $912^{+316}_{-250}$ | $1580^{+430}_{-360}$ | $1020^{+670}_{-510}$ | $1.84\pm0.22$ |
| *8.75* | $82.5^{+9.9}_{-8.6}$ | $142^{+40}_{-35}$ | $102^{+30}_{-24}$ | $61.5^{+46.0}_{-43.4}$ | $244^{+116}_{-102}$ | $17.3^{+101.8}_{-78.0}$ | $-634^{+111}_{-91}$ | ?±? |
| *9.00* | $24.7^{+3.6}_{-3.1}$ | $34.4^{+9.1}_{-9.5}$ | $27.3^{+15.8}_{-13.9}$ | $-35.7^{+19.5}_{-16.2}$ | $-65.2^{+34.2}_{-28.0}$ | $-214^{+47}_{-39}$ | $-139^{+61}_{-47}$ | $0.49\pm0.11$ |
| *Slope* | $-1.98\pm0.11$ | $-2.10\pm0.15$ | $-2.58\pm0.25$ | $-1.94\pm0.60$ | $-2.09\pm0.12$ | $-2.79\pm0.80$ | $-1.87\pm0.07$ | |

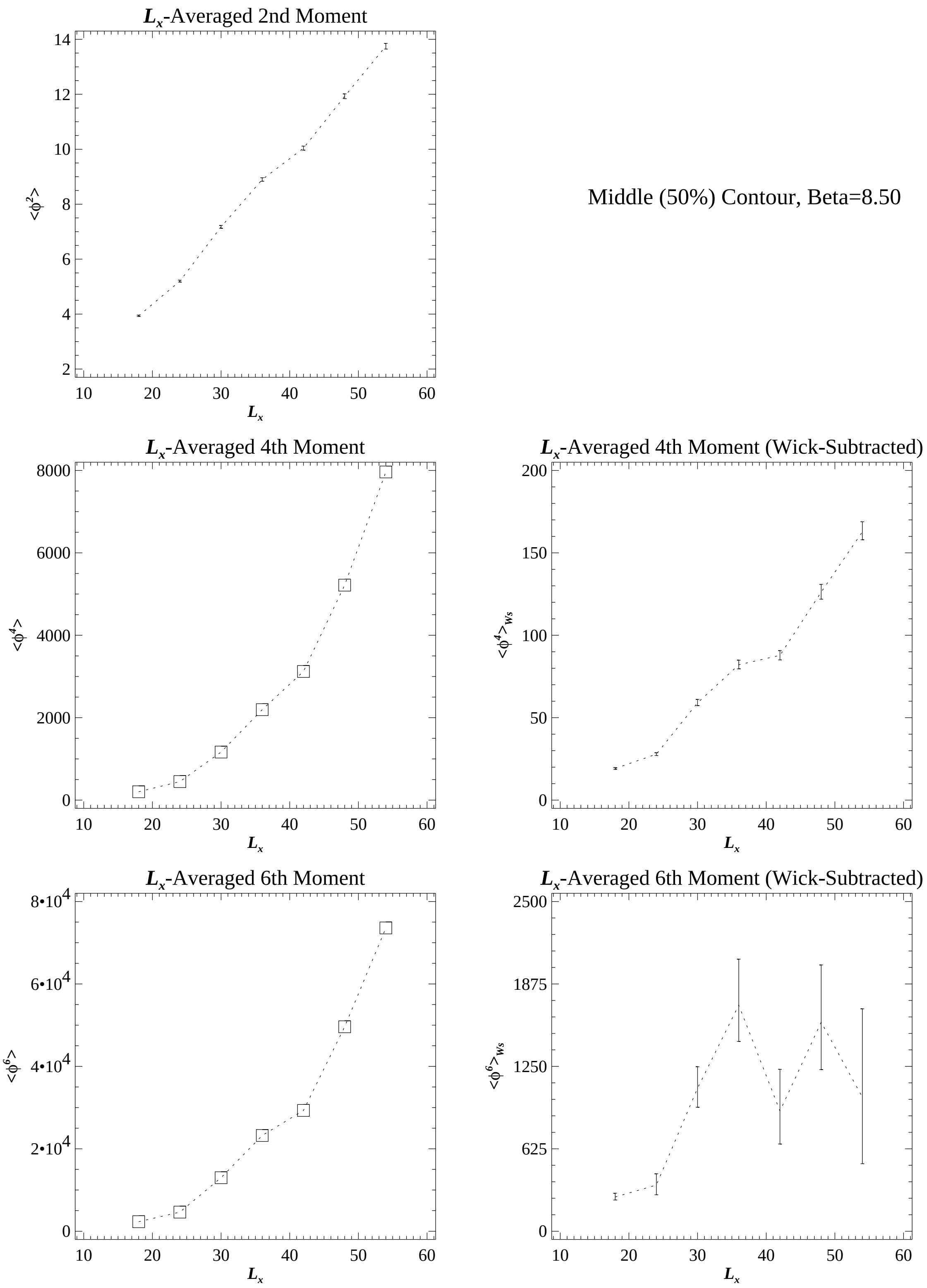

**Fig. 5.7:** Graphs of even moments against transverse size, $L_x$, for $\beta = 8.5$. Data are shown for corrected raw Polyakov configurations.

It can be seen by comparison with tables 4.6-8 that revision of estimates beyond the original errors generally occurs only for low $\beta$ and $L_x$. However, this can have an appreciable effect on the calculated power-law dependence of moments on $L_x$ and $(\beta - \beta_c)$. For comparison, figures corresponding to fig. 4.19, fig. 4.20, fig. 4.21 and fig. 4.22 conclude this chapter, showing the effects of removing data spikes.

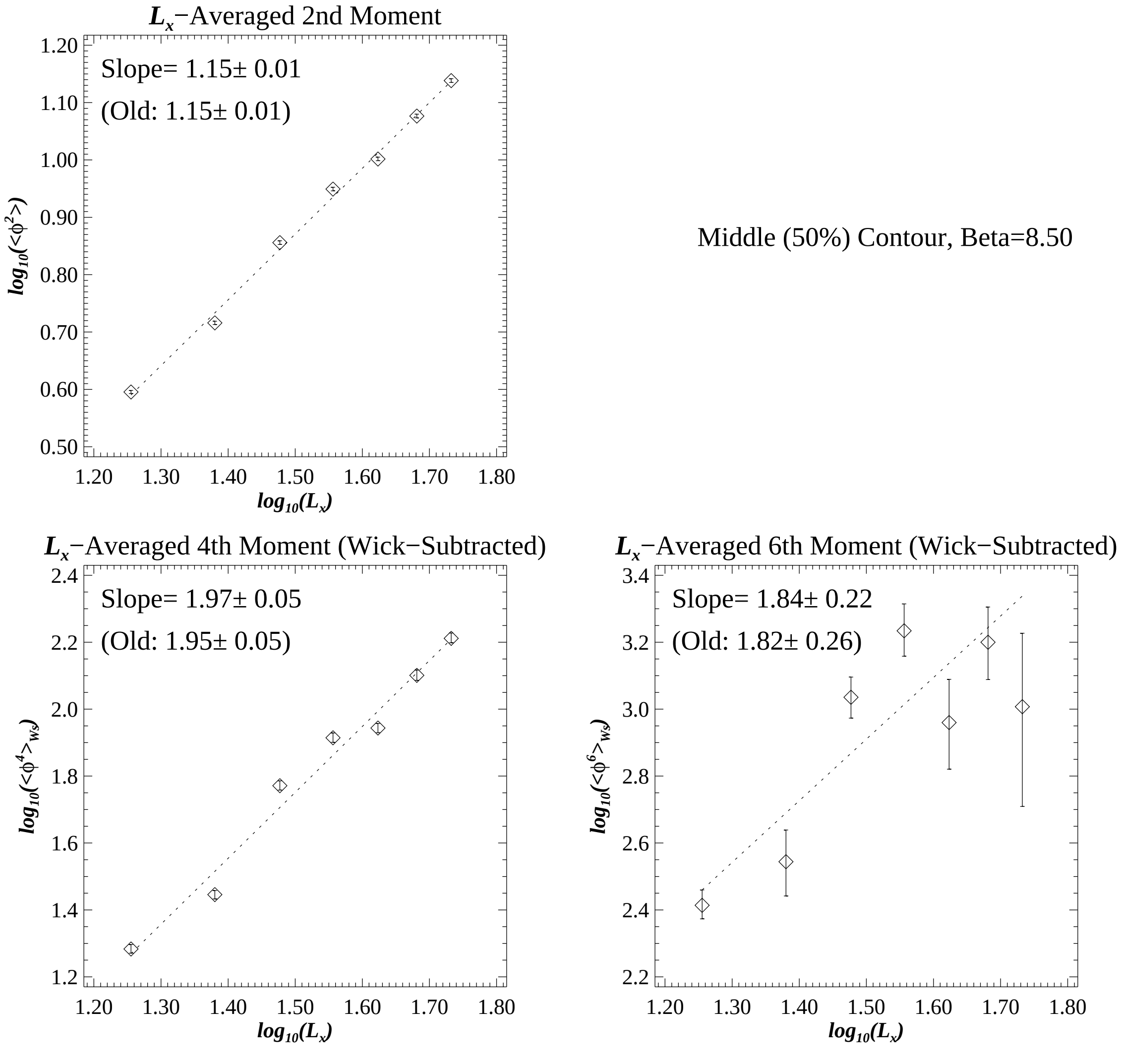

**Fig. 5.8:** Log plots corresponding to the Wick-subtracted data of fig. 5.7, with best fits to a power-law dependence on the lattice width. The previous fits from fig. 5.7 are given for ease of comparison.

For the data shown in the graphs of this chapter ($L_x = 36$, middle contour, raw data), the fitted slopes for the revised second, fourth and sixth moments are:

$$< \phi^2 >\sim (\beta - \beta_c)^{-0.42(3)}, \quad < \phi^4 >_{Ws}\sim (\beta - \beta_c)^{-1.25(11)}, \quad < \phi^6 >_{Ws}\sim (\beta - \beta_c)^{-1.94(60)}.$$

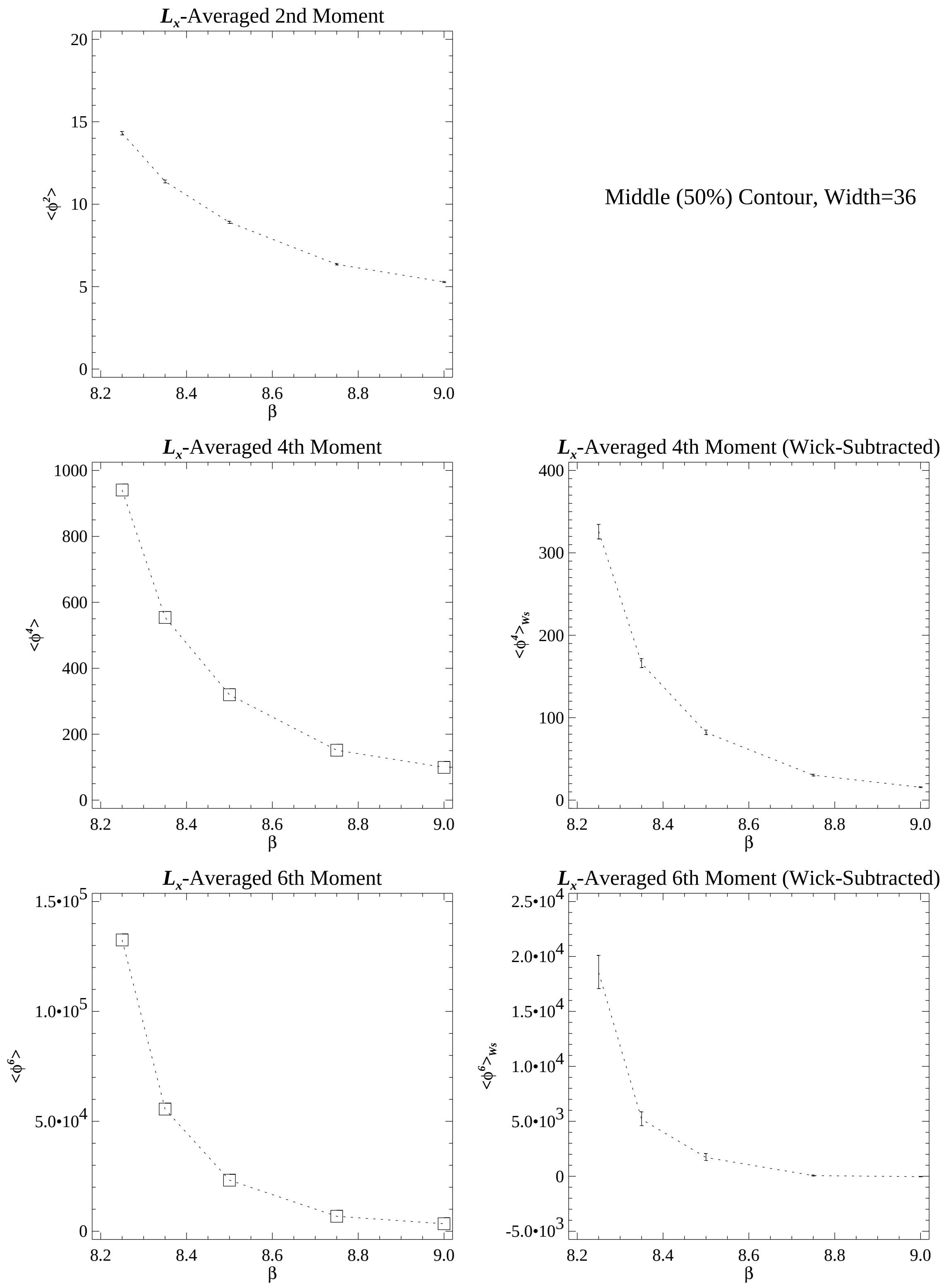

**Fig. 5.9:** Graphs of even moments against $\beta$, for $L_x = 36$. Data are shown for corrected raw Polyakov configurations.

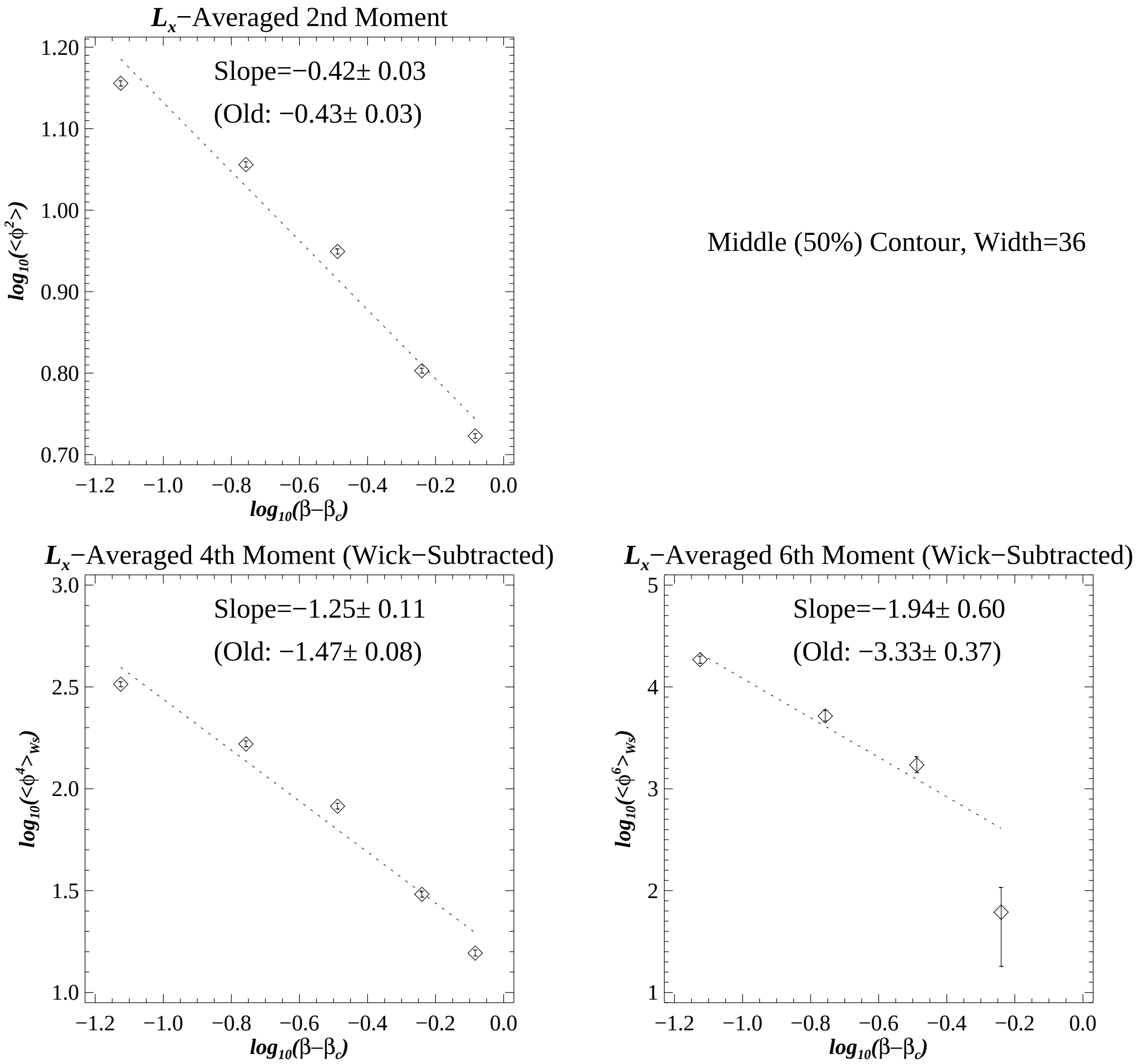

**Fig. 5.10:** Log plots corresponding to the Wick-subtracted data of fig. 5.9, with best fits to a power-law dependence on the lattice width. The previous fits from fig. 5.9 are given for ease of comparison.

Referring back to our formulae for the moments, we see that the first two imply the following for our unknown functions of $\beta$:

$$\gamma \sim (\beta - \beta_c)^{0.42(3)}, \qquad \lambda \sim (\beta - \beta_c)^{0.43(16)}.$$

These two together predict, from our previous formulae, that

$$< \phi^6 >_{Ws} \sim (\beta - \beta_c)^{-2.08(24)},$$

extremely close to the observed behaviour. The other sets of revised data now give very consistent results for $\gamma$ and $\lambda$.

# Chapter 6.

# Conclusions

## 6.1. Conclusions

In this thesis, we have examined the properties of the $Z(3)$ interface in two very different regimes:

At high temperatures, we have measured the interface tension on the lattice, and extrapolated our results to the infinite-temperature limit, where we have found good agreement with the prediction of an instanton calculation. This reinforces the argument for the existence of $Z(3)$-breaking phases in the Euclidean functional integral, although their physical existence in the early universe remains in question, and suggests that the predictions of perturbation theory are indeed reliable in this regime. We have also examined the Debye profile of the interface, but have not been able to compare our results accurately with predictions at the temperatures concerned.

Just above the critical temperature, we have seen that it is possible to follow the fluctuations in profile of the interface even though it appears, at first sight, to disappear in the phase turbulence. We have observed that the interface maintains a narrow width right down to the critical temperature, but undergoes more and more severe fluctuations in shape. We have characterised these fluctuations using moments, and have seen that these diverge at the critical temperature; and we have constructed a toy scalar field theory model which reproduces the observed behaviour to a reasonable approximation. We have also seen that whilst the intrinsic width of the interface increases rather slowly near the critical point, the Debye screening width appears to diverge logarithmically, in line with the predictions for a completely wet interface; but we have not seen direct evidence for this wetting. Finally, we have noted that the formation and movement of the interface on the lattice shares many of the characteristics of interface behaviour in condensed matter systems.

IUCUNDI ACTI LABORES

Cicero

*pltlh!*

(Klingon saying)

# Appendix A: Table of Values for $F(2r, 3)$

Table A.1: Values of $F(2r, 3)$ for $r \leq 30$

| $r$ | $F(2r, 3)$ |
|---|---|
| 1 | 2.9257441147538591e-02 |
| 2 | 2.8588133947184931e-03 |
| 3 | 7.7625872330950078e-04 |
| 4 | 3.1782999811761687e-04 |
| 5 | 1.5992165454570213e-04 |
| 6 | 9.1150725927608572e-05 |
| 7 | 5.6469174319874464e-05 |
| 8 | 3.7115784724567322e-05 |
| 9 | 2.5484489828066094e-05 |
| 10 | 1.8087564731975696e-05 |
| 11 | 1.3170447587156097e-05 |
| 12 | 9.7840479099121027e-06 |
| 13 | 7.3840421441212328e-06 |
| 14 | 5.6428031018154392e-06 |
| 15 | 4.3549604790025167e-06 |
| 16 | 3.3872387617351618e-06 |
| 17 | 2.6505179822474595e-06 |
| 18 | 2.0836256901432198e-06 |
| 19 | 1.6435963519300661e-06 |
| 20 | 1.2996333087139130e-06 |
| 21 | 1.0292617591260539e-06 |
| 22 | 8.1581683728676361e-07 |
| 23 | 6.4676674563294586e-07 |
| 24 | 5.1257056950486260e-07 |
| 25 | 4.0588577520678718e-07 |
| 26 | 3.2100885546335727e-07 |
| 27 | 2.5347419875900234e-07 |
| 28 | 1.9976210969186481e-07 |
| 29 | 1.5708328861159867e-07 |
| 30 | 1.2321764912912764e-07 |

# Appendix B: Fluctuation Moment Results

Table B.1: Estimates of $<\phi^2>$ from Raw Binned Data, Lower (25%) Contour

| $\beta \cdot\cdot L_x$ | *18* | *24* | *30* | *36* | *42* | *48* | *54* | *Slope* |
|---|---|---|---|---|---|---|---|---|
| *8.25* | $5.63^{+0.08}_{-0.05}$ | $7.50^{+0.04}_{-0.01}$ | $10.9^{+0.1}_{-0.1}$ | $13.5^{+0.1}_{-0.1}$ | $16.6^{+0.1}_{-0.1}$ | $19.7^{+0.1}_{-0.2}$ | $21.7^{+0.2}_{-0.1}$ | 1.30±0.02 |
| *8.35* | $4.61^{+0.11}_{-0.01}$ | $6.32^{+0.05}_{-0.04}$ | $8.67^{+0.08}_{-0.04}$ | $10.6^{+0.1}_{-0.1}$ | $12.4^{+0.1}_{-0.1}$ | $15.5^{+0.1}_{-0.1}$ | $17.2^{+0.1}_{-0.1}$ | 1.22±0.01 |
| *8.50* | $3.82^{+0.02}_{-0.03}$ | $5.00^{+0.03}_{-0.03}$ | $6.85^{+0.05}_{-0.05}$ | $8.63^{+0.08}_{-0.05}$ | $9.57^{+0.08}_{-0.06}$ | $11.5^{+0.1}_{-0.0}$ | $13.1^{+0.1}_{-0.1}$ | 1.13±0.01 |
| *8.75* | $3.04^{+0.02}_{-0.02}$ | $4.11^{+0.03}_{-0.03}$ | $5.21^{+0.04}_{-0.03}$ | $6.21^{+0.04}_{-0.03}$ | $7.46^{+0.05}_{-0.04}$ | $8.63^{+0.07}_{-0.04}$ | $9.34^{+0.07}_{-0.06}$ | 1.04±0.01 |
| *9.00* | $2.61^{+0.01}_{-0.02}$ | $3.43^{+0.02}_{-0.02}$ | $4.34^{+0.03}_{-0.02}$ | $5.21^{+0.03}_{-0.03}$ | $6.19^{+0.05}_{-0.04}$ | $7.27^{+0.05}_{-0.05}$ | $7.77^{+0.04}_{-0.05}$ | 1.02±0.01 |
| *Slope* | −0.33±0.02 | −0.32±0.02 | −0.39±0.02 | −0.40±0.03 | −0.41±0.01 | −0.42±0.02 | −0.44±0.03 | |

Table B.2: Estimates of $<\phi^2>$ from Raw Binned Data, Middle (50%) Contour

| $\beta \cdot\cdot L_x$ | *18* | *24* | *30* | *36* | *42* | *48* | *54* | *Slope* |
|---|---|---|---|---|---|---|---|---|
| *8.25* | $6.87^{+0.22}_{-0.14}$ | $8.27^{+0.12}_{-0.03}$ | $12.3^{+0.2}_{-0.2}$ | $14.4^{+0.1}_{-0.1}$ | $17.8^{+0.1}_{-0.1}$ | $20.8^{+0.1}_{-0.2}$ | $23.1^{+0.2}_{-0.1}$ | 1.20±0.03 |
| *8.35* | $5.02^{+0.20}_{-0.01}$ | $6.71^{+0.05}_{-0.05}$ | $9.14^{+0.09}_{-0.05}$ | $11.4^{+0.1}_{-0.1}$ | $13.4^{+0.1}_{-0.1}$ | $16.5^{+0.1}_{-0.1}$ | $18.2^{+0.1}_{-0.1}$ | 1.21±0.01 |
| *8.50* | $3.93^{+0.03}_{-0.03}$ | $5.19^{+0.03}_{-0.04}$ | $7.17^{+0.05}_{-0.05}$ | $8.89^{+0.07}_{-0.05}$ | $10.0^{+0.1}_{-0.1}$ | $11.9^{+0.1}_{-0.1}$ | $13.8^{+0.1}_{-0.1}$ | 1.15±0.01 |
| *8.75* | $3.07^{+0.02}_{-0.02}$ | $4.21^{+0.03}_{-0.03}$ | $5.32^{+0.04}_{-0.03}$ | $6.34^{+0.05}_{-0.04}$ | $7.63^{+0.06}_{-0.05}$ | $8.85^{+0.07}_{-0.05}$ | $9.57^{+0.06}_{-0.06}$ | 1.05±0.01 |
| *9.00* | $2.62^{+0.01}_{-0.02}$ | $3.44^{+0.02}_{-0.02}$ | $4.38^{+0.03}_{-0.02}$ | $5.27^{+0.03}_{-0.03}$ | $6.29^{+0.05}_{-0.03}$ | $7.37^{+0.05}_{-0.05}$ | $7.89^{+0.04}_{-0.06}$ | 1.03±0.01 |
| *Slope* | −0.41±0.01 | −0.37±0.02 | −0.44±0.02 | −0.43±0.03 | −0.44±0.02 | −0.45±0.03 | −0.46±0.04 | |

Table B.3: Estimates of $<\phi^2>$ from Raw Binned Data, Higher (75%) Contour

| $\beta \cdot\cdot L_x$ | *18* | *24* | *30* | *36* | *42* | *48* | *54* | *Slope* |
|---|---|---|---|---|---|---|---|---|
| *8.25* | $9.21^{+0.50}_{-0.30}$ | $9.63^{+0.48}_{-0.08}$ | $13.4^{+0.4}_{-0.3}$ | $14.6^{+0.1}_{-0.1}$ | $17.6^{+0.1}_{-0.1}$ | $20.1^{+0.1}_{-0.1}$ | $22.0^{+0.2}_{-0.1}$ | 0.91±0.03 |
| *8.35* | $6.19^{+0.36}_{-0.04}$ | $7.31^{+0.04}_{-0.05}$ | $9.71^{+0.06}_{-0.04}$ | $11.9^{+0.1}_{-0.1}$ | $13.9^{+0.1}_{-0.1}$ | $16.5^{+0.1}_{-0.1}$ | $18.0^{+0.1}_{-0.1}$ | 1.08±0.02 |
| *8.50* | $4.44^{+0.02}_{-0.02}$ | $5.96^{+0.03}_{-0.04}$ | $7.86^{+0.04}_{-0.05}$ | $9.48^{+0.06}_{-0.05}$ | $10.6^{+0.1}_{-0.1}$ | $12.3^{+0.1}_{-0.1}$ | $14.1^{+0.1}_{-0.1}$ | 1.04±0.01 |
| *8.75* | $3.65^{+0.02}_{-0.02}$ | $4.88^{+0.03}_{-0.03}$ | $5.96^{+0.04}_{-0.02}$ | $6.94^{+0.04}_{-0.03}$ | $8.17^{+0.05}_{-0.05}$ | $9.29^{+0.06}_{-0.04}$ | $9.98^{+0.06}_{-0.05}$ | 0.92±0.01 |
| *9.00* | $3.17^{+0.01}_{-0.02}$ | $4.04^{+0.02}_{-0.02}$ | $4.94^{+0.03}_{-0.02}$ | $5.79^{+0.03}_{-0.03}$ | $6.77^{+0.04}_{-0.03}$ | $7.75^{+0.05}_{-0.05}$ | $8.30^{+0.04}_{-0.05}$ | 0.89±0.01 |
| *Slope* | −0.41±0.03 | −0.37±0.02 | −0.42±0.02 | −0.39±0.03 | −0.40±0.02 | −0.41±0.03 | −0.42±0.04 | |

Table B.4: Estimates of $< \phi^4 >_{Ws}$ from Raw Binned Data, Lower (25%) Contour

| $\beta \ddots L_x$ | *18* | *24* | *30* | *36* | *42* | *48* | *54* | *Slope* |
|---|---|---|---|---|---|---|---|---|
| *8.25* | $1150^{+240}_{-10}$ | $251^{+81}_{+12}$ | $441^{+54}_{-33}$ | $331^{+15}_{-7}$ | $414^{+23}_{-16}$ | $568^{+14}_{-14}$ | $642^{+26}_{-10}$ | 0.27±0.25 |
| *8.35* | $438^{+891}_{-11}$ | $55.5^{+1.9}_{-1.4}$ | $102^{+5}_{-2}$ | $144^{+5}_{-4}$ | $166^{+5}_{-5}$ | $313^{+17}_{-12}$ | $286^{+9}_{-7}$ | 1.84±0.47 |
| *8.50* | $18.3^{+0.6}_{-0.5}$ | $26.7^{+0.8}_{-0.8}$ | $54.6^{+2.0}_{-1.6}$ | $79.5^{+6.3}_{-1.6}$ | $85.0^{+3.0}_{-2.3}$ | $121^{+1}_{+0.}$ | $160^{+8}_{-4}$ | 1.99±0.05 |
| *8.75* | $9.48^{+0.29}_{-0.26}$ | $16.1^{+0.6}_{-0.4}$ | $22.0^{+0.8}_{-0.4}$ | $30.3^{+1.2}_{-0.6}$ | $41.2^{+1.6}_{-1.1}$ | $49.2^{+2.3}_{-1.2}$ | $49.2^{+1.9}_{-1.5}$ | 1.58±0.03 |
| *9.00* | $6.29^{+0.17}_{-0.18}$ | $8.88^{+0.25}_{-0.19}$ | $13.4^{+0.5}_{-0.2}$ | $17.2^{+0.6}_{-0.5}$ | $23.8^{+0.9}_{-0.6}$ | $29.2^{+1.4}_{-1.0}$ | $28.7^{+1.3}_{-0.9}$ | 1.51±0.03 |
| *Slope* | −2.00±0.43 | −1.25±0.10 | −1.40±0.06 | −1.23±0.08 | −1.19±0.04 | −1.24±0.10 | −1.29±0.11 | |

Table B.5: Estimates of $< \phi^4 >_{Ws}$ from Raw Binned Data, Middle (50%) Contour

| $\beta \ddots L_x$ | *18* | *24* | *30* | *36* | *42* | *48* | *54* | *Slope* |
|---|---|---|---|---|---|---|---|---|
| *8.25* | $2270^{+480}_{-180}$ | $334^{+125}_{-1.}$ | $967^{+343}_{-86}$ | $463^{+141}_{-21}$ | $492^{+58}_{-24}$ | $585^{+16}_{-13}$ | $690^{+31}_{-10}$ | −0.54±0.27 |
| *8.35* | $775^{+349}_{-12}$ | $63.1^{+2.4}_{-1.9}$ | $111^{+5}_{-2}$ | $166^{+6}_{-5}$ | $199^{+6}_{-6}$ | $344^{+24}_{-14}$ | $299^{+11}_{-8}$ | 1.31±0.49 |
| *8.50* | $19.5^{+0.7}_{-0.6}$ | $28.3^{+0.9}_{-0.8}$ | $59.7^{+2.1}_{-1.7}$ | $82.5^{+3.3}_{-1.6}$ | $88.0^{+3.2}_{-2.6}$ | $126^{+8}_{-1}$ | $163^{+7}_{-4}$ | 1.95±0.05 |
| *8.75* | $9.92^{+0.30}_{-0.27}$ | $16.7^{+0.7}_{-0.5}$ | $22.3^{+0.8}_{-0.6}$ | $30.3^{+1.2}_{-0.7}$ | $43.2^{+1.9}_{-1.5}$ | $48.6^{+2.4}_{-1.3}$ | $49.5^{+2.1}_{-1.5}$ | 1.54±0.04 |
| *9.00* | $5.85^{+0.14}_{-0.17}$ | $8.80^{+0.27}_{-0.23}$ | $13.1^{+0.5}_{-0.3}$ | $15.6^{+0.6}_{-0.4}$ | $23.3^{+0.9}_{-0.6}$ | $28.6^{+1.2}_{-0.9}$ | $28.6^{+1.4}_{-1.0}$ | 1.54±0.03 |
| *Slope* | −2.40±0.44 | −1.33±0.12 | −1.53±0.17 | −1.47±0.08 | −1.29±0.05 | −1.26±0.12 | −1.32±0.11 | |

Table B.6: Estimates of $< \phi^4 >_{Ws}$ from Raw Binned Data, Higher (75%) Contour

| $\beta \ddots L_x$ | *18* | *24* | *30* | *36* | *42* | *48* | *54* | *Slope* |
|---|---|---|---|---|---|---|---|---|
| *8.25* | $3820^{+540}_{-300}$ | $838^{+246}_{-22}$ | $1370^{+200}_{-100}$ | $483^{+29}_{-11}$ | $516^{+25}_{-16}$ | $705^{+19}_{-17}$ | $908^{+55}_{-20}$ | −1.01±0.24 |
| *8.35* | $1750^{+340}_{-30}$ | $89.2^{+2.1}_{-1.9}$ | $139^{+3}_{-2}$ | $210^{+6}_{-5}$ | $268^{+6}_{-7}$ | $375^{+13}_{-1}$ | $371^{+11}_{-7}$ | 0.88±0.54 |
| *8.50* | $30.8^{+0.6}_{-0.5}$ | $53.4^{+2.7}_{-1.6}$ | $80.7^{+1.6}_{-1.6}$ | $114^{+2}_{-2}$ | $117^{+4}_{-3}$ | $153^{+6}_{-1}$ | $191^{+6}_{-4}$ | 1.62±0.04 |
| *8.75* | $18.5^{+0.4}_{-0.4}$ | $30.1^{+0.7}_{-0.5}$ | $38.7^{+0.8}_{-0.7}$ | $48.9^{+1.0}_{-0.7}$ | $64.4^{+2.0}_{-1.5}$ | $71.8^{+1.8}_{-1.1}$ | $76.8^{+2.2}_{-1.7}$ | 1.32±0.03 |
| *9.00* | $12.3^{+0.2}_{-0.3}$ | $18.0^{+0.3}_{-0.3}$ | $23.8^{+0.7}_{-0.3}$ | $27.8^{+0.6}_{-0.6}$ | $39.0^{+1.0}_{-0.6}$ | $45.2^{+1.2}_{-0.9}$ | $47.1^{+1.3}_{-1.1}$ | 1.28±0.02 |
| *Slope* | −2.33±0.52 | −1.20±0.25 | −1.33±0.24 | −1.21±0.07 | −1.11±0.05 | −1.18±0.07 | −1.26±0.06 | |

Table B.7: Estimates of $<\phi^6>_{Ws}$ from Raw Binned Data, Lower (25%) Contour

| $\beta \ddots L_x$ | *18* | *24* | *30* | *36* | *42* | *48* | *54* | *Slope* |
|---|---|---|---|---|---|---|---|---|
| *8.25* | 3450000 +730000 −310000 | 540000 +699000 +105000 | 654000 +166000 −91000 | 84100 +94600 −3900 | 49200 +5100 −2900 | 34500 +7200 −3500 | 43000 +7900 −2900 | −4.38±0.24 |
| *8.35* | 1370000 +7100000 −20000 | 1580 +420 −160 | 3400 +1050 −300 | 3760 +600 −470 | 3210 +660 −640 | 19200 +6700 −900 | 1700 +1580 −880 | 0.04±1.18 |
| *8.50* | $278^{+52}_{-29}$ | $335^{+48}_{-39}$ | $1090^{+230}_{-140}$ | $2550^{+4340}_{-330}$ | $971^{+407}_{-204}$ | $1310^{+1590}_{+470}$ | $715^{+539}_{-377}$ | 1.60±0.29 |
| *8.75* | $75.2^{+11.1}_{-7.7}$ | $135^{+48}_{-18}$ | $70.9^{+30.3}_{-13.6}$ | $101^{+151}_{-47}$ | $191^{+98}_{-56}$ | $-37.5^{+156.6}_{-60.7}$ | $-637^{+111}_{-95}$ | 0.73±0.17 |
| *9.00* | $39.6^{+5.4}_{-4.5}$ | $34.2^{+6.8}_{-5.9}$ | $33.4^{+12.6}_{-9.1}$ | $19.6^{+27.8}_{-18.1}$ | $-64.2^{+25.5}_{-23.8}$ | $-157^{+56}_{-41}$ | $-149^{+48}_{-39}$ | −0.52±0.11 |
| *Slope* | −4.99±1.05 | −3.49±0.57 | −4.08±0.37 | −3.11±0.38 | −2.75±0.12 | −1.98±0.47 | −3.00±0.32 | |

Table B.8: Estimates of $<\phi^6>_{Ws}$ from Raw Binned Data, Middle (50%) Contour

| $\beta \ddots L_x$ | *18* | *24* | *30* | *36* | *42* | *48* | *54* | *Slope* |
|---|---|---|---|---|---|---|---|---|
| *8.25* | 5720000 +1370000 −510000 | 391000 +441000 +45000 | 1280000 +720000 −180000 | 297000 +320000 −16000 | 199000 +295000 −23000 | 24900 +13600 −3800 | 71700 +112000 −7300 | −4.40±0.39 |
| *8.35* | 1880000 +970000 −20000 | 1830 +260 −170 | 3710 +640 −330 | 5150 +700 −560 | 6180 +1300 −980 | 42800 +41100 −2700 | 2640 +2570 −1400 | −2.87±1.05 |
| *8.50* | $289^{+44}_{-29}$ | $380^{+47}_{-47}$ | $1260^{+210}_{-140}$ | $2160^{+510}_{-160}$ | $939^{+312}_{-210}$ | $1660^{+880}_{+50}$ | $968^{+686}_{-399}$ | 1.82±0.26 |
| *8.75* | $109^{+15}_{-12}$ | $167^{+40}_{-24}$ | $110^{+29}_{-27}$ | $98.0^{+79.2}_{-43.7}$ | $285^{+127}_{-81}$ | $77.5^{+161.1}_{-61.2}$ | $-625^{+111}_{-94}$ | 0.44±0.25 |
| *9.00* | $27.1^{+3.7}_{-3.2}$ | $48.2^{+7.9}_{-7.9}$ | $39.4^{+15.0}_{-11.8}$ | $-19.7^{+26.3}_{-18.2}$ | $-65.2^{+34.7}_{-27.2}$ | $-213^{+44}_{-36}$ | $-125^{+67}_{-45}$ | 1.06±0.14 |
| *Slope* | −5.57±0.98 | −3.16±0.55 | −3.95±0.47 | −3.33±0.37 | −3.15±0.23 | −2.27±0.82 | −2.98±0.29 | |

Table B.9: Estimates of $<\phi^6>_{Ws}$ from Raw Binned Data, Higher (75%) Contour

| $\beta \ddots L_x$ | *18* | *24* | *30* | *36* | *42* | *48* | *54* | *Slope* |
|---|---|---|---|---|---|---|---|---|
| *8.25* | 7170000 +1170000 −620000 | 653000 +821000 −43000 | 1580000 +690000 −180000 | 257000 +323000 −12000 | 59300 +5600 −2800 | 73900 +9500 −6300 | 157000 +46000 −9000 | −4.40±0.39 |
| *8.35* | 3880000 +500000 −40000 | 3750 +390 −260 | 5410 +390 −270 | 9560 +1130 −800 | 13700 +1700 −1500 | 31000 +3400 −100 | 16000 +7100 −2300 | −3.69±0.99 |
| *8.50* | $749^{+55}_{-38}$ | $3930^{+26650}_{-1300}$ | $2180^{+200}_{-140}$ | $3900^{+230}_{-180}$ | $2650^{+390}_{-280}$ | $3650^{+960}_{-30}$ | $2190^{+950}_{-450}$ | 1.54±0.27 |
| *8.75* | $334^{+25}_{-22}$ | $660^{+55}_{-42}$ | $792^{+54}_{-58}$ | $935^{+65}_{-61}$ | $1070^{+170}_{-100}$ | $1020^{+110}_{-80}$ | $338^{+153}_{-134}$ | 0.94±0.26 |
| *9.00* | $173^{+13}_{-12}$ | $315^{+37}_{-27}$ | $378^{+43}_{-29}$ | $341^{+44}_{-32}$ | $536^{+77}_{-64}$ | $274^{+63}_{-46}$ | $207^{+68}_{-54}$ | 0.60±0.19 |
| *Slope* | −5.36±1.01 | −2.22±0.67 | −2.58±0.70 | −2.39±0.24 | −1.99±0.07 | −2.36±0.23 | −2.85±0.11 | |

Table B.10: Estimates of $< \phi^2 >$ from Smooth Binned Data, Lower (25%) Contour

| $\beta \cdot\cdot L_x$ | *18* | *24* | *30* | *36* | *42* | *48* | *54* | *Slope* |
|---|---|---|---|---|---|---|---|---|
| *8.25* | $8.14^{+0.23}_{-0.13}$ | $10.1^{+0.2}_{-0.1}$ | $14.9^{+0.3}_{-0.2}$ | $17.8^{+0.2}_{-0.2}$ | $21.5^{+0.2}_{-0.2}$ | $25.6^{+0.2}_{-0.2}$ | $28.6^{+0.4}_{-0.3}$ | $1.21\pm0.02$ |
| *8.35* | $6.02^{+0.14}_{-0.03}$ | $7.71^{+0.06}_{-0.06}$ | $10.6^{+0.1}_{-0.1}$ | $13.0^{+0.1}_{-0.1}$ | $15.0^{+0.1}_{-0.1}$ | $18.8^{+0.2}_{-0.1}$ | $20.6^{+0.1}_{-0.1}$ | $1.16\pm0.02$ |
| *8.50* | $4.38^{+0.03}_{-0.04}$ | $5.75^{+0.04}_{-0.05}$ | $8.00^{+0.06}_{-0.06}$ | $10.1^{+0.1}_{-0.1}$ | $11.0^{+0.1}_{-0.1}$ | $13.1^{+0.1}_{-0.1}$ | $15.1^{+0.1}_{-0.1}$ | $1.14\pm0.02$ |
| *8.75* | $3.14^{+0.03}_{-0.03}$ | $4.40^{+0.04}_{-0.04}$ | $5.67^{+0.05}_{-0.04}$ | $6.82^{+0.05}_{-0.04}$ | $8.19^{+0.06}_{-0.05}$ | $9.51^{+0.08}_{-0.05}$ | $10.3^{+0.1}_{-0.1}$ | $1.09\pm0.01$ |
| *9.00* | $2.54^{+0.02}_{-0.02}$ | $3.46^{+0.02}_{-0.02}$ | $4.50^{+0.04}_{-0.03}$ | $5.46^{+0.04}_{-0.04}$ | $6.54^{+0.05}_{-0.04}$ | $7.71^{+0.06}_{-0.06}$ | $8.26^{+0.05}_{-0.06}$ | $1.10\pm0.01$ |
| *Slope* | $-0.51\pm0.03$ | $-0.46\pm0.03$ | $-0.51\pm0.03$ | $-0.50\pm0.03$ | $-0.50\pm0.02$ | $-0.51\pm0.02$ | $-0.53\pm0.03$ | |

Table B.11: Estimates of $< \phi^2 >$ from Smooth Binned Data, Middle (50%) Contour

| $\beta \cdot\cdot L_x$ | *18* | *24* | *30* | *36* | *42* | *48* | *54* | *Slope* |
|---|---|---|---|---|---|---|---|---|
| *8.25* | $7.90^{+0.19}_{-0.12}$ | $10.6^{+0.2}_{-0.0}$ | $15.4^{+0.3}_{-0.2}$ | $18.3^{+0.2}_{-0.1}$ | $22.5^{+0.2}_{-0.2}$ | $26.6^{+0.2}_{-0.2}$ | $30.3^{+0.4}_{-0.3}$ | $1.26\pm0.01$ |
| *8.35* | $5.67^{+0.05}_{-0.02}$ | $8.00^{+0.07}_{-0.07}$ | $11.0^{+0.1}_{-0.1}$ | $13.6^{+0.1}_{-0.1}$ | $15.8^{+0.1}_{-0.1}$ | $19.6^{+0.2}_{-0.1}$ | $21.5^{+0.2}_{-0.1}$ | $1.23\pm0.01$ |
| *8.50* | $4.36^{+0.04}_{-0.04}$ | $5.79^{+0.04}_{-0.05}$ | $8.17^{+0.07}_{-0.07}$ | $10.1^{+0.1}_{-0.1}$ | $11.4^{+0.1}_{-0.1}$ | $13.5^{+0.1}_{-0.1}$ | $15.6^{+0.1}_{-0.1}$ | $1.17\pm0.01$ |
| *8.75* | $3.07^{+0.03}_{-0.03}$ | $4.40^{+0.03}_{-0.04}$ | $5.65^{+0.05}_{-0.04}$ | $6.80^{+0.05}_{-0.04}$ | $8.22^{+0.06}_{-0.06}$ | $9.59^{+0.08}_{-0.05}$ | $10.3^{+0.1}_{-0.1}$ | $1.11\pm0.01$ |
| *9.00* | $2.47^{+0.01}_{-0.02}$ | $3.37^{+0.02}_{-0.02}$ | $4.40^{+0.04}_{-0.03}$ | $5.40^{+0.04}_{-0.04}$ | $6.52^{+0.05}_{-0.04}$ | $7.69^{+0.06}_{-0.06}$ | $8.25^{+0.05}_{-0.06}$ | $1.13\pm0.01$ |
| *Slope* | $-0.50\pm0.03$ | $-0.48\pm0.03$ | $-0.54\pm0.03$ | $-0.52\pm0.04$ | $-0.52\pm0.02$ | $-0.53\pm0.03$ | $-0.56\pm0.03$ | |

Table B.12: Estimates of $< \phi^2 >$ from Smooth Binned Data, Higher (75%) Contour

| $\beta \cdot\cdot L_x$ | *18* | *24* | *30* | *36* | *42* | *48* | *54* | *Slope* |
|---|---|---|---|---|---|---|---|---|
| *8.25* | $8.35^{+0.15}_{-0.09}$ | $11.9^{+0.2}_{-0.0}$ | $16.1^{+0.2}_{-0.2}$ | $19.8^{+0.2}_{-0.2}$ | $24.0^{+0.2}_{-0.2}$ | $28.3^{+0.2}_{-0.2}$ | $32.1^{+0.4}_{-0.2}$ | $1.23\pm0.01$ |
| *8.35* | $6.88^{+0.12}_{-0.03}$ | $9.29^{+0.07}_{-0.07}$ | $12.1^{+0.1}_{-0.1}$ | $15.0^{+0.1}_{-0.1}$ | $17.5^{+0.1}_{-0.1}$ | $21.0^{+0.2}_{-0.1}$ | $22.8^{+0.1}_{-0.1}$ | $1.12\pm0.01$ |
| *8.50* | $5.32^{+0.03}_{-0.03}$ | $7.12^{+0.04}_{-0.05}$ | $9.50^{+0.06}_{-0.06}$ | $11.4^{+0.1}_{-0.1}$ | $12.6^{+0.1}_{-0.1}$ | $14.5^{+0.1}_{-0.1}$ | $16.6^{+0.1}_{-0.1}$ | $1.03\pm0.01$ |
| *8.75* | $4.07^{+0.03}_{-0.03}$ | $5.54^{+0.04}_{-0.04}$ | $6.79^{+0.05}_{-0.03}$ | $7.92^{+0.05}_{-0.04}$ | $9.35^{+0.06}_{-0.05}$ | $10.6^{+0.1}_{-0.0}$ | $11.4^{+0.1}_{-0.1}$ | $0.94\pm0.01$ |
| *9.00* | $3.34^{+0.01}_{-0.03}$ | $4.33^{+0.03}_{-0.02}$ | $5.40^{+0.04}_{-0.03}$ | $6.37^{+0.03}_{-0.04}$ | $7.50^{+0.05}_{-0.04}$ | $8.63^{+0.05}_{-0.06}$ | $9.23^{+0.05}_{-0.06}$ | $0.95\pm0.01$ |
| *Slope* | $-0.40\pm0.03$ | $-0.42\pm0.03$ | $-0.47\pm0.03$ | $-0.48\pm0.03$ | $-0.49\pm0.02$ | $-0.51\pm0.02$ | $-0.54\pm0.03$ | |

Table B.13: Estimates of $< \phi^4 >_{Ws}$ from Smooth Binned Data, Lower (25%) Contour

| $\beta \ddots L_x$ | *18* | *24* | *30* | *36* | *42* | *48* | *54* | *Slope* |
|---|---|---|---|---|---|---|---|---|
| *8.25* | $1730^{+220}_{-120}$ | $628^{+10}_{-20}$ | $1340^{+170}_{-100}$ | $981^{+478}_{-65}$ | $936^{+135}_{-80}$ | $1050^{+80}_{-50}$ | $1130^{+360}_{-50}$ | $-0.18\pm0.15$ |
| *8.35* | $729^{+57}_{-9}$ | $71.8^{+2.9}_{-2.2}$ | $132^{+9}_{-3}$ | $166^{+6}_{-5}$ | $182^{+6}_{-6}$ | $456^{+261}_{-46}$ | $291^{+11}_{-8}$ | $-0.11\pm0.38$ |
| *8.50* | $24.1^{+0.9}_{-0.8}$ | $31.3^{+1.3}_{-1.2}$ | $62.1^{+2.7}_{-2.1}$ | $89.2^{+15.8}_{-3.4}$ | $82.5^{+3.7}_{-3.0}$ | $112^{+11}_{+1.}$ | $148^{+7}_{-4}$ | $1.65\pm0.07$ |
| *8.75* | $10.4^{+0.5}_{-0.4}$ | $16.1^{+0.8}_{-0.5}$ | $20.6^{+1.0}_{-0.5}$ | $26.3^{+1.6}_{-0.8}$ | $34.9^{+1.8}_{-1.2}$ | $38.2^{+2.4}_{-1.3}$ | $33.7^{+2.2}_{-1.4}$ | $1.21\pm0.05$ |
| *9.00* | $6.04^{+0.26}_{-0.22}$ | $7.91^{+0.32}_{-0.24}$ | $11.5^{+0.5}_{-0.3}$ | $13.5^{+0.7}_{-0.5}$ | $18.7^{+0.9}_{-0.6}$ | $20.5^{+1.4}_{-1.0}$ | $18.0^{+1.3}_{-1.1}$ | $1.19\pm0.05$ |
| *Slope* | $-2.69\pm0.34$ | $-1.63\pm0.15$ | $-1.84\pm0.14$ | $-1.67\pm0.10$ | $-1.54\pm0.06$ | $-1.65\pm0.12$ | $-1.76\pm0.11$ | |

Table B.14: Estimates of $< \phi^4 >_{Ws}$ from Smooth Binned Data, Middle (50%) Contour

| $\beta \ddots L_x$ | *18* | *24* | *30* | *36* | *42* | *48* | *54* | *Slope* |
|---|---|---|---|---|---|---|---|---|
| *8.25* | $2230^{+270}_{-160}$ | $861^{+163}_{-12}$ | $1620^{+330}_{-140}$ | $715^{+172}_{-33}$ | $816^{+132}_{-60}$ | $912^{+55}_{-32}$ | $1300^{+340}_{-60}$ | $-0.56\pm0.17$ |
| *8.35* | $497^{+28}_{-9}$ | $79.8^{+4.0}_{-3.2}$ | $132^{+7}_{-3}$ | $187^{+7}_{-7}$ | $223^{+8}_{-9}$ | $422^{+87}_{-25}$ | $300^{+16}_{-10}$ | $-0.02\pm0.31$ |
| *8.50* | $24.0^{+1.1}_{-0.8}$ | $31.1^{+1.3}_{-1.2}$ | $64.4^{+2.9}_{-2.5}$ | $87.7^{+5.4}_{-2.4}$ | $82.3^{+3.9}_{-3.2}$ | $118^{+10}_{-1.}$ | $148^{+7}_{-4}$ | $1.68\pm0.07$ |
| *8.75* | $10.3^{+0.5}_{-0.4}$ | $16.1^{+0.9}_{-0.6}$ | $20.0^{+1.0}_{-0.7}$ | $25.3^{+1.3}_{-0.8}$ | $35.9^{+2.2}_{-1.6}$ | $36.5^{+2.7}_{-1.3}$ | $32.2^{+2.0}_{-1.6}$ | $1.17\pm0.06$ |
| *9.00* | $5.04^{+0.20}_{-0.20}$ | $7.11^{+0.34}_{-0.25}$ | $10.1^{+0.6}_{-0.3}$ | $11.0^{+0.6}_{-0.5}$ | $16.8^{+0.9}_{-0.7}$ | $19.1^{+1.3}_{-1.0}$ | $17.3^{+1.5}_{-1.0}$ | $1.27\pm0.05$ |
| *Slope* | $-2.76\pm0.25$ | $-1.80\pm0.17$ | $-1.88\pm0.17$ | $-1.74\pm0.10$ | $-1.61\pm0.04$ | $-1.62\pm0.13$ | $-1.82\pm0.11$ | |

Table B.15: Estimates of $< \phi^4 >_{Ws}$ from Smooth Binned Data, Higher (75%) Contour

| $\beta \ddots L_x$ | *18* | *24* | *30* | *36* | *42* | *48* | *54* | *Slope* |
|---|---|---|---|---|---|---|---|---|
| *8.25* | $2180^{+310}_{-170}$ | $909^{+92}_{-3}$ | $1280^{+230}_{-10}$ | $967^{+57}_{-29}$ | $989^{+76}_{-44}$ | $1300^{+60}_{-40}$ | $1620^{+10}_{-40}$ | $0.08\pm0.16$ |
| *8.35* | $1060^{+90}_{-20}$ | $138^{+5}_{-4}$ | $192^{+5}_{-3}$ | $284^{+9}_{-8}$ | $366^{+11}_{-10}$ | $522^{+24}_{-19}$ | $444^{+15}_{-11}$ | $0.24\pm0.33$ |
| *8.50* | $43.0^{+1.2}_{-0.9}$ | $68.8^{+2.7}_{-2.0}$ | $102^{+2}_{-2}$ | $143^{+3}_{-3}$ | $131^{+5}_{-3}$ | $169^{+8}_{-1}$ | $196^{+7}_{-4}$ | $1.36\pm0.05$ |
| *8.75* | $24.3^{+0.5}_{-0.5}$ | $37.7^{+1.0}_{-0.8}$ | $45.9^{+1.1}_{-1.0}$ | $56.0^{+1.2}_{-0.9}$ | $71.5^{+2.7}_{-1.8}$ | $74.5^{+2.0}_{-1.2}$ | $74.0^{+2.1}_{-2.0}$ | $1.06\pm0.04$ |
| *9.00* | $15.4^{+0.4}_{-0.4}$ | $21.6^{+0.5}_{-0.5}$ | $27.1^{+0.9}_{-0.5}$ | $30.0^{+0.8}_{-0.7}$ | $41.4^{+1.1}_{-0.8}$ | $43.6^{+1.3}_{-0.9}$ | $45.0^{+1.7}_{-1.2}$ | $1.02\pm0.03$ |
| *Slope* | $-2.28\pm0.34$ | $-1.44\pm0.13$ | $-1.38\pm0.15$ | $-1.43\pm0.05$ | $-1.34\pm0.04$ | $-1.45\pm0.06$ | $-1.50\pm0.02$ | |

Table B.16: Estimates of $< \phi^6 >_{Ws}$ from Smooth Binned Data, Lower (25%) Contour

| $\beta \ddots L_x$ | *18* | *24* | *30* | *36* | *42* | *48* | *54* | *Slope* |
|---|---|---|---|---|---|---|---|---|
| *8.25* | $3450000^{+330000}_{-180000}$ | $540000^{+170000}_{-54000}$ | $654000^{+260000}_{-150000}$ | $84100^{+507000}_{-38000}$ | $49200^{+67000}_{-39000}$ | $34500^{+166000}_{-57000}$ | $43000^{+359000}_{-30000}$ | $-1.62\pm0.20$ |
| *8.35* | $1370000^{+51000}_{-6000}$ | $1580^{+630}_{-270}$ | $3400^{+9720}_{-1340}$ | $3760^{+820}_{-640}$ | $3210^{+1130}_{-1020}$ | $19200^{+606000}_{-13000}$ | $1700^{+2940}_{-1650}$ | $-5.38\pm1.17$ |
| *8.50* | $562^{+83}_{-49}$ | $804^{+110}_{-90}$ | $1840^{+290}_{-190}$ | $5580^{+13520}_{-860}$ | $1760^{+500}_{-290}$ | $2020^{+1510}_{+370}$ | $1010^{+530}_{-370}$ | $1.19\pm0.32$ |
| *8.75* | $156^{+21}_{-15}$ | $247^{+64}_{-28}$ | $222^{+54}_{-14}$ | $303^{+159}_{-61}$ | $485^{+105}_{-53}$ | $338^{+169}_{-59}$ | $-67.7^{+116.7}_{-86.0}$ | $1.02\pm0.09$ |
| *9.00* | $67.0^{+11.0}_{-7.4}$ | $97.5^{+15.4}_{-10.8}$ | $119^{+19}_{-13}$ | $164^{+46}_{-31}$ | $121^{+34}_{-34}$ | $137^{+69}_{-48}$ | $208^{+53}_{-47}$ | $0.90\pm0.08$ |
| *Slope* | $-5.40\pm0.94$ | $-3.30\pm0.47$ | $-3.96\pm0.37$ | $-3.25\pm0.48$ | $-3.45\pm0.33$ | $-3.47\pm0.51$ | $-3.55\pm0.22$ | |

Table B.17: Estimates of $< \phi^6 >_{Ws}$ from Smooth Binned Data, Middle (50%) Contour

| $\beta \ddots L_x$ | *18* | *24* | *30* | *36* | *42* | *48* | *54* | *Slope* |
|---|---|---|---|---|---|---|---|---|
| *8.25* | $5720000^{+660000}_{-370000}$ | $391000^{+910000}_{-80000}$ | $1280000^{+850000}_{-290000}$ | $297000^{+452000}_{-22000}$ | $199000^{+406000}_{-51000}$ | $24900^{+250000}_{-40000}$ | $71700^{+610000}_{-39000}$ | $-2.55\pm0.24$ |
| *8.35* | $1880000^{+28000}_{-8000}$ | $1830^{+560}_{-360}$ | $3710^{+940}_{-500}$ | $5150^{+960}_{-790}$ | $6180^{+2600}_{-1900}$ | $42800^{+109400}_{-5700}$ | $2640^{+5970}_{-2450}$ | $-5.29\pm1.10$ |
| *8.50* | $546^{+70}_{-49}$ | $775^{+103}_{-83}$ | $1760^{+310}_{-210}$ | $4000^{+1370}_{-280}$ | $1850^{+470}_{-290}$ | $3380^{+1300}_{+90}$ | $1830^{+670}_{-400}$ | $1.66\pm0.22$ |
| *8.75* | $207^{+38}_{-24}$ | $286^{+51}_{-37}$ | $295^{+42}_{-36}$ | $313^{+88}_{-60}$ | $574^{+147}_{-98}$ | $484^{+147}_{-53}$ | $-157^{+106}_{-97}$ | $0.92\pm0.08$ |
| *9.00* | $49.7^{+9.0}_{-6.5}$ | $76.4^{+11.9}_{-9.5}$ | $90.2^{+21.3}_{-13.3}$ | $73.8^{+29.8}_{-19.0}$ | $53.7^{+47.2}_{-31.3}$ | $19.3^{+40.2}_{-35.5}$ | $178^{+65}_{-50}$ | n/a |
| *Slope* | $-5.78\pm0.92$ | $-3.35\pm0.59$ | $-3.79\pm0.52$ | $-3.24\pm0.32$ | $-3.39\pm0.25$ | n/a | $-3.71\pm0.17$ | |

Table B.18: Estimates of $< \phi^6 >_{Ws}$ from Smooth Binned Data, Higher (75%) Contour

| $\beta \ddots L_x$ | *18* | *24* | *30* | *36* | *42* | *48* | *54* | *Slope* |
|---|---|---|---|---|---|---|---|---|
| *8.25* | $7170000^{+750000}_{-420000}$ | $653000^{+710000}_{-40000}$ | $1580000^{+670000}_{-240000}$ | $257000^{+322000}_{-26000}$ | $59300^{+21000}_{-13000}$ | $73900^{+57000}_{-32000}$ | $157000^{+65000}_{-20000}$ | $-2.60\pm0.19$ |
| *8.35* | $3880000^{+110000}_{-30000}$ | $3750^{+2700}_{-1500}$ | $5410^{+800}_{-600}$ | $9560^{+3400}_{-2200}$ | $13700^{+4800}_{-3500}$ | $31000^{+10500}_{-2800}$ | $16000^{+5700}_{-3200}$ | $-4.14\pm0.86$ |
| *8.50* | $1460^{+130}_{-80}$ | $3610^{+1460}_{-450}$ | $5230^{+1300}_{-440}$ | $9660^{+800}_{-510}$ | $7010^{+770}_{-560}$ | $7880^{+1360}_{-60}$ | $5700^{+1070}_{-690}$ | $1.56\pm0.18$ |
| *8.75* | $707^{+46}_{-43}$ | $1380^{+120}_{-90}$ | $1860^{+140}_{-130}$ | $2290^{+320}_{-160}$ | $2640^{+300}_{-180}$ | $3080^{+240}_{-170}$ | $2140^{+250}_{-200}$ | $1.21\pm0.10$ |
| *9.00* | $411^{+50}_{-36}$ | $713^{+93}_{-61}$ | $812^{+68}_{-46}$ | $1070^{+80}_{-70}$ | $1390^{+160}_{-110}$ | $1060^{+110}_{-70}$ | $1420^{+120}_{-10}$ | $1.03\pm0.07$ |
| *Slope* | $-5.25\pm0.99$ | $-2.74\pm0.43$ | $-2.43\pm0.57$ | $-2.44\pm0.22$ | $-2.29\pm0.10$ | $-2.46\pm0.17$ | $-2.36\pm0.19$ | |